\documentclass[12pt]{article}

\newlength{\abstractwidth}
\usepackage{epsf}
\usepackage{color}
\usepackage{graphicx}
\usepackage{hyperref}
\usepackage{dsfont}
\usepackage{setspace}

\usepackage{amsmath, nccmath}
\usepackage{amsmath, amsthm}
\usepackage{amssymb}
\usepackage{amsfonts}
\usepackage{latexsym}
\usepackage{mathtools}
\usepackage{shuffle}

\usepackage{tikz, pgf}
\usepackage{tkz-fct}
\usepackage{pgfplots}
\usetikzlibrary{shapes.misc}
\usetikzlibrary{shapes,snakes}
\usetikzlibrary{decorations.pathmorphing}	
\usetikzlibrary{decorations.markings}
\usetikzlibrary{arrows.meta}

\usetikzlibrary{arrows,shapes,positioning}
\usetikzlibrary{decorations.markings}
\usepackage[rightcaption]{sidecap}
\tikzstyle arrowstyle=[scale=1]
\tikzstyle directed=[postaction={decorate,decoration={markings,
    mark=at position .65 with {\arrow[arrowstyle]{stealth}}}}]
\tikzstyle reverse directed=[postaction={decorate,decoration={markings,
    mark=at position .65 with {\arrowreversed[arrowstyle]{stealth};}}}]
\usetikzlibrary{positioning}

\renewcommand{\thefootnote}{\fnsymbol{footnote}}
\renewcommand{\thanks}[1]{\footnote{#1}}
\newcommand{\starttext}{
\setcounter{footnote}{0}
\renewcommand{\thefootnote}{\arabic{footnote}}}

\numberwithin{equation}{section}
\newcommand{\bea}{\begin{eqnarray}}
\newcommand{\eea}{\end{eqnarray}}
\newcommand{\be}{\begin{eqnarray}}
\newcommand{\ee}{\end{eqnarray}}
\def\beq{\begin{equation}}
\def\eeq{\end{equation}}

\newcommand{\bma}{\begin{matrix}}
\newcommand{\ema}{\end{matrix}}

\def\cG{{\cal G}}

\def\cI{{\cal I}}

\def\cK{{\cal K}}

\def\cO{{\cal O}}

\def\cR{{\cal R}}

\def\cV{{\cal V}}
\def\cW{{\cal W}}

\def\bD{{\bf D}}

\def\bz{{\bf z}}

\def\mA{\mathfrak{A}}
\def\mB{\mathfrak{B}}

\def\mJ{\mathfrak{J}}

\def\mL{\mathfrak{L}}

\def\mN{\mathfrak{N}}

\def\mg{\mathfrak{g}}

\def\CC{{\mathbb C}}

\def\RR{{\mathbb R}}
\def\ZZ{{\mathbb Z}}

\def\Im{{\rm Im \,}}

\def\half{{1\over 2}}
\def\thalf{\tfrac{1}{2}}
\def\p{\partial}

\def\b{\beta}

\def\tet{\vartheta}
\def\ep{\varepsilon}
\def\om{\omega}

\def\pbx{\p _{\bar x}}
\def\pby{\p _{\bar y}}

\def\Sp{{\text{Sp}}}

\def\MM{\cM}
\def\LL{\cL}

\def\DD{D_\delta}
\def\CD{C_\delta}
\def\DDD{\bD_\delta}

\def\qq{q}
\def\Ber{\text{Ber}}
\def\KK{{\bf K}}
\def\OO{{\bf W}}
\def\XX{{\bf X}}

\def\QQ{\beta}
\def\MM{\mathfrak{M}}

\def\LL{{L _\delta}}
\def\newH{\mathfrak{D}}

\def\no{\nonumber}
\def\sm{\smallskip}
\def\hp{\omega}
\def\ad{{\rm ad}}

\definecolor{Cyan}{cmyk}{1.,0,0,0}
\definecolor{Magenta}{cmyk}{0,1.,0,0}
\definecolor{Yellow}{cmyk}{0,0,1.,0}
\definecolor{White}{cmyk}{0,0,0,0}
\definecolor{Orange}{cmyk}{0,0.61,0.87,0}
\definecolor{RedOrange}{cmyk}{0,0.77,0.87,0}
\definecolor{Red}{cmyk}{0,1.,1.,0}
\definecolor{Purple}{cmyk}{0.45,0.86,0,0}
\definecolor{Violet}{cmyk}{0.79,0.88,0,0}
\definecolor{Blue}{cmyk}{1,0.5,0,0}
\definecolor{ProcessBlue}{cmyk}{0.96,0,0,0}
\definecolor{GreenYellow}{cmyk}{0.6,0,1.,0}
\definecolor{Black}{cmyk}{0,0,0,1}
\newtheorem{thm}{Theorem}[section]
\newtheorem{lem}[thm]{Lemma}
\newtheorem{prop}[thm]{Proposition}
\newtheorem{cor}[thm]{Corollary}

\begin{document}
\starttext
\setcounter{footnote}{0}

\begin{flushright}
2025 August 20  \\
UUITP--15/25
\end{flushright}

\vskip 0.1in

\begin{center}

{\Large \bf Worldsheet fermion correlators, modular tensors }

\vskip 0.1in

{\Large \bf and higher genus integration kernels}

\vskip 0.2in

{\large Eric D'Hoker${}^{a}$, Oliver Schlotterer${}^{b}$} 

\vskip 0.15in

{ \sl ${}^{a}$Mani L. Bhaumik Institute for Theoretical Physics}\\
{\sl  Department of Physics and Astronomy}\\
{\sl University of California, Los Angeles, CA 90095, USA}

\vskip 0.1in

{\sl ${}^b$Department of Physics and Astronomy,} \\
  { \sl Department of Mathematics,} \\
  { \sl Centre for Geometry and Physics,} \\ 
  {\sl Uppsala University, 75120 Uppsala, Sweden}

\vskip 0.15in 

{\tt \small dhoker@physics.ucla.edu, oliver.schlotterer@physics.uu.se}

\vskip 0.2in

\begin{abstract}
\vskip 0.1in

The cyclic product of an arbitrary number of Szeg\"o kernels for even spin structure $\delta$ on a compact higher-genus Riemann surface $\Sigma$ may be decomposed via a descent procedure which systematically separates the dependence on the points  $z_i \in \Sigma$ from the dependence on the spin structure $\delta$. In this paper, we prove two different, but complementary, descent procedures to achieve this decomposition. In the first procedure, the dependence on the points $z_i \in \Sigma$ is expressed via the meromorphic multiple-valued Enriquez kernels of e-print 1112.0864 while the  dependence on $\delta$ resides in multiplets of  functions that are independent of $z_i$, locally holomorphic in the moduli of $\Sigma$ and generally do not have simple modular transformation properties. The $\delta$-dependent constants are expressed as multiple convolution integrals over homology cycles of~$\Sigma$, thereby generalizing a similar representation of the individual Enriquez kernels. In the second procedure, which was proposed without proof in e-print 2308.05044, the dependence on $z_i$ is expressed in terms  the single-valued, modular invariant, but non-meromorphic DHS kernels introduced in e-print 2306.08644 while the dependence on $\delta$ resides in modular tensors that are independent of $z_i$  and are generally non-holomorphic in the moduli of $\Sigma$. Although the individual building blocks of these decompositions have markedly different properties, we show that the combinatorial structure of the two decompositions is virtually identical, thereby extending the striking correspondence observed earlier between the roles played by Enriquez and DHS kernels. Both decompositions are further generalized to the case of linear chain products of Szeg\"o kernels.

\end{abstract}
\end{center}

\newpage

\setcounter{tocdepth}{2} 
\begin{spacing}{0.98}
\tableofcontents
\end{spacing}

\baselineskip=15pt
\setcounter{equation}{0}
\setcounter{footnote}{0}

\newpage

\section{Introduction}
\setcounter{equation}{0}
\label{sec:1}

Recent years have witnessed an increasing symbiosis between progress in string perturbation theory
and mathematical developments at the interface of number theory and algebraic geometry (see for example \cite{Berkovits:2022ivl,DHoker:2024cup}). Some of these results are  intimately  related to advances in the modern  methods of quantum field theory amplitudes \cite{Bourjaily:2022bwx,Abreu:2022mfk,Blumlein:2022qci, Dorigoni:2022iem}.

\sm

The perturbative expansion of string amplitudes involves conformal field theory correlators on compact Riemann surfaces $\Sigma$ of arbitrary genus, suitably integrated over multiple copies of $\Sigma$ and over the complex structure moduli of $\Sigma$. The study of genus zero superstring amplitudes (see \cite{Mafra:2022wml} for a review) has advanced in fruitful exchange with number-theoretic progress on multiple zeta values \cite{Brown:2011ik, Brown:2019wna} and twisted de Rham theory \cite{Mizera:2017cqs, Mizera:2017rqa}. The low energy expansion of closed superstring amplitudes at genus one may be organized in terms of modular graph functions and forms \cite{DHoker:2015wxz,DHoker:2016mwo}, which naturally generalize Eisenstein series  and were reformulated in algebraic geometry terms  in \cite{Brown:2017qwo, Brown:2017qwo2}. Open-string amplitudes at genus one in turn offer applications of and new perspectives on elliptic polylogarithms \cite{BrownLevin}, elliptic multiple zeta values \cite{Enriquez:Emzv} and iterated integrals of holomorphic modular forms \cite{Brown:2014mmv, Broedel:2015hia}.
The low energy expansion of closed superstring amplitudes at genus two were found to involve the number-theoretic invariants of Kawazumi \cite{kawazumi2008johnson} and Zhang \cite{zhang2010gross}  and to produce an infinite family of natural generalizations thereof \cite{DHoker:2013fcx,DHoker:2014oxd}. These results have further motivated the generalization to arbitrary genus of modular graph functions  in \cite{DHoker:2017pvk} and  of modular graph tensors in \cite{DHoker:2020uid}.

\sm

The overarching goal of the present project is to disentangle, organize and formalize the structure of the various spaces of functions out of which string amplitudes are built. In the Ramond-Neveu-Schwarz formulation of superstrings, the worldsheet fields are scalars, spin $\half$ fermions, ghosts and super ghosts. The summation over all possible $2^{2h}$ spin structures of the spin $\half$ fermions and super ghosts at genus $h$ implements the Gliozzi-Scherk-Olive projection which, amongst other roles, ensures the presence of space-time supersymmetry in the different superstring theories.  In practice, the summation over spin structures can be carried out fairly effectively at genus one where well-known identities between Jacobi $\tet$-functions suffice, see for instance \cite{Tsuchiya:1988va, Stieberger:2002wk, Bianchi:2006nf, Tsuchiya:2012nf} for external NS states and  \cite{Atick:1986rs, Lin:1988xb, Lee:2017ujn} for external R states. At higher genus, however, carrying out the spin structure summations requires a major effort (see for example \cite{DHoker:2005vch,DHoker:2021kks} for the calculations of the genus two four- and five-point amplitudes of massless NS-NS states, whose results were matched with calculations in the pure spinor formulation  \cite{Berkovits:2005ng, DHoker:2020prr}).

\sm

In this work, we shall focus on disentangling the correlators of the spin $\half$ worldsheet fermions for even spin structure $\delta$ and generic moduli. They enter string  amplitudes whose external states are all NS  through cyclic products of Szeg\"o kernels $S_\delta(x,y)$,\footnote{For odd spin structures and for even spin structures at genus $h\geq 3$ and non-generic moduli there exist holomorphic $(\half, 0)$ form zero modes which modify the correlators of the worldsheet fermions.}
\bea
C_\delta (z_1,\cdots, z_n) = S_\delta(z_1,z_2) S_\delta(z_2,z_3) \cdots S_\delta (z_{n-1}, z_n) S_\delta(z_n,z_1)
\label{intro.01}
\eea
as well as through linear chain products of Szeg\"o kernels,   
\bea
L_\delta (x;z_1,\cdots, z_n;y) = S_\delta(x,z_1) S_\delta(z_1,z_2) \cdots S_\delta (z_{n-1}, z_n) S_\delta(z_n,y)
\label{intro.02}
\eea
and products thereof. The Szeg\"o kernel $S_\delta(x,y)$ is  a meromorphic $(\half, 0)$ form in $x,y \in \Sigma$ with a single pole at $x=y$,  so that $C_\delta$ and $L_\delta$ are meromorphic  $(1,0)$ forms in the points $z_1 , \cdots, z_n \in \Sigma$ while $L_\delta$ is a meromorphic $(\half,0)$ form in $x,y \in \Sigma$. The cyclic products $C_\delta$ result from the correlators of spin $\half $ fermions that occur in the NS vertex operators, while linear chain products $L_\delta$ are needed for correlators that end on a worldsheet supercurrent or on a worldsheet  stress tensor \cite{Friedan:1985ge, DHoker:1988pdl, DHoker:2005dys}. 

\sm

In this paper we shall establish, for a Riemann surface of arbitrary genus,  a decomposition of $C_\delta$ which completely separates the dependence on the points $z_1, \cdots, z_n$ from that on the spin structures $\delta$. More specifically, $C_\delta$ will be expressed as a sum of binary products in which one factor contains all the dependence on $z_1, \cdots, z_n$ but is independent of $\delta$, while the other factor contains all the dependence on $\delta$ but is independent of $z_1, \cdots, z_n$. An analogous decomposition will hold for $L_\delta$ but in this case both factors will depend on the end points $x,y$ as well. In both cases, the dependence on $z_1,\cdots,z_n$ will be expressed in a well-controlled function space whose mathematical significance for integration on Riemann surfaces will be elaborated on below.

\sm

For the case of genus one, the spin structure independent components admit a natural formulation in terms of the Kronecker-Eisenstein coefficients   \cite{Broedel:2014vla, Tsuchiya:2017joo} that serve as integration kernels for elliptic polylogarithms  \cite{Levin:2007, BrownLevin, Broedel:2017kkb}. More general correlators that enter heterotic string amplitudes may similarly be decomposed in terms of Kronecker-Eisenstein kernels \cite{Dolan:2007eh, Gerken:2018}. Accordingly,  Kronecker-Eisenstein kernels furnish a universal function space out of which the integrands of genus one superstring amplitudes 
\cite{Berg:2016wux, Mafra:2016nwr,  Mafra:2018qqe, Gerken:2019cxz, Balli:2024wje}
and ambitwistor-string theories \cite{Adamo:2013tsa, Geyer:2015jch, He:2017spx} may be built. An attractive property of the Kronecker-Eisenstein kernels and their associated polylogarithms is that they form a space that is closed under addition, multiplication, differentiation, and integration. 

\sm

For the case of genus two, a complete solution to the decomposition problem for $C_\delta$ was obtained in \cite{DHoker:2022xxg} using the fact that every genus two Riemann surface is hyperelliptic, and that every point in moduli space is generic. An analogous decomposition was obtained for the linear chain products $L_\delta$ in unpublished work by the authors. Actually, all the spin structure dependence, for any number $n$ of points, may be reduced to that of the cases $n \leq 4$, thereby permitting a systematic summation over all even spin structures against the genus two superstring measure obtained in \cite{DHoker:2001qqx}. For genus greater than two, however, hyperelliptic surfaces are not generic and the above results do not generalize.

\sm

For arbitrary genus, it is the structure provided by integration kernels and their associated polylogarithms, already discussed above for the case of genus one,  that systematically provides the proper ingredients for the decomposition of both the cyclic and linear chain products of Szeg\"o kernels. Integration kernels and polylogarithms for Riemann surfaces of higher genus are conveniently constructed from flat connections that take values in certain freely generated Lie algebras. On general grounds, any two such flat connections may be related to one another by the composition of a gauge transformation and an automorphism of the freely generated Lie algebra \cite{DHoker:2025szl}. Different flat connections may present themselves, however, in different guises depending on their analyticity, monodromy, and modular properties. On a compact Riemann surface $\Sigma$ of genus $h$, we shall consider,
\begin{itemize}
\itemsep 0in
\item the meromorphic  integration kernels $g^{I_1 \cdots I_r}{}_J(x,y)$  with $I_1, \cdots, I_r, J \in \{ 1, \cdots, h\}$ for $r\geq 1$,  which are multiple-valued in $x,y \in \Sigma$ with prescribed monodromies and are not modular tensors of $\Sp(2h,\ZZ)$. They were  introduced by Enriquez in \cite{Enriquez:2011} through their functional properties, which will  be reviewed in section \ref{sec:2.1}. They may be expressed as multiple $\mA$ periods of combinations of Abelian differentials and the prime form \cite{DHoker:2025dhv} or,  on hyperelliptic surfaces, as Poincar\'e series~\cite{Baune:2024biq}. 
\item the real-analytic  integration kernels $f^{I_1 \cdots I_r}{}_J(x,y)$ with $I_1, \cdots, I_r, J \in \{ 1, \cdots, h\}$ for $r \geq 1$ are single-valued in $x,y \in \Sigma$ and transform as tensors under the modular group $\Sp(2h,\ZZ)$.  They were introduced by D'Hoker-Hidding-Schlotterer (DHS) in \cite{DHS:2023} and will be reviewed in section~\ref{sec:4.1}. They may be expressed as multiple  integrals over $\Sigma$ involving the Arakelov Green function and Abelian differentials.
\end{itemize}
The relation between these families of integration kernels and their associated polylogarithms was exhibited and proven in \cite{DHoker:2025szl}.


\subsection{Summary of results and organization}
\label{sec:intro.1}

A first main result of this work is the construction of the decomposition of the cyclic product of Szeg\"o kernels $\CD(z_1, \cdots, z_n)$ of (\ref{intro.01}) on a Riemann surface of arbitrary genus~$h$ with even spin structure $\delta$ into a sum of a finite number of terms,\footnote{Throughout we shall use the Einstein summation convention for a repeated pair of upper and lower indices whenever no confusion is expected to arise.} 
\bea
C_\delta(z_1,\cdots,z_n) = {\cal W} (z_1,\cdots,z_n) 
+ \sum_{r=2}^{n}  {\cal W}_{I_1  \cdots I_{r}}(z_1,\cdots, z_n) \, \DD^{I_1  \cdots I_{r}}
\label{intro.03}
\eea
where the various components have the following properties.
\begin{itemize}
\itemsep=0in
\item The functions $\cW_{I_1 \cdots I_r} (z_1,\cdots, z_n)$ are independent of the spin structure $\delta$; single-valued and meromorphic $(1,0)$ forms in the points $z_1, \cdots, z_n$; expressed solely in terms of the Enriquez kernels; and cyclically symmetric in the indices $I_1, \cdots, I_r$.
\item The functions $\DD^{I_1  \cdots I_{r}} $ are independent of the points $z_1, \cdots, z_n$; depend non-trivially on the spin structure $\delta$; are locally holomorphic in the complex structure moduli of $\Sigma$; and cyclically symmetric in the indices $I_1, \cdots, I_r$. 
\end{itemize}
In general, neither $\cW_{I_1 \cdots I_r} (z_1,\cdots, z_n)$ nor $\DD^{I_1  \cdots I_{r}} $ transforms as a tensor under the modular group $\Sp(2h,\ZZ)$ or under one of its congruence subgroups. For this reason we shall refer to them simply as \textit{multiplets}, allowing for the possibility that they are not tensors.

\sm

The proof of this result will be given in Theorem \ref{2.thm:cdec} of section~\ref{sec:5} with the help of the descent procedure stated  in Theorem~\ref{2.thm:3} of section~\ref{sec:2}, and subsequently proven in section~\ref{sec:2} through~\ref{sec:4}, using the tools of generating functions developed in section~\ref{sec:3}. Various technical proofs are relegated to appendices~\ref{sec:Ba} and~\ref{sec:B}. Furthermore, it will be shown in section \ref{sec:4} that the coefficients $\DD^{I_1  \cdots I_{r}}$ may be expressed as multiple $\mA$ periods, thereby generalizing the analogous representation for Enriquez kernels obtained in \cite{DHoker:2025dhv}.

\sm

A second main result is to provide a proof of the corresponding decomposition of $\CD(z_1, \cdots, z_n)$ of (\ref{intro.01}) on a Riemann surface of arbitrary genus~$h$ with even spin structure~$\delta$ into DHS kernels $f$, 
\bea
C_\delta(z_1,\cdots,z_n) = {\cal V} (z_1,\cdots,z_n) 
+ \sum_{r=2}^{n}  {\cal V}_{I_1  \cdots I_{r}}(z_1,\cdots, z_n) \, C_\delta^{I_1  \cdots I_{r}}
\label{intro.04}
\eea
which had already been proposed without proof in an earlier paper \cite{DHoker:2023khh}, and
where the various components have the following properties,
\begin{itemize}
\itemsep=0in
\item The multiplets $\cV_{I_1 \cdots I_r} (z_1,\cdots, z_n)$ are independent of the spin structure $\delta$; single-valued and meromorphic $(1,0)$ forms in the points $z_1, \cdots, z_n$; cyclically symmetric in $I_1, \cdots, I_r$; expressed solely in terms of  DHS kernels;  and tensors under $\Sp(2h,\ZZ)$.
\item The multiplets $C_\delta^{I_1  \cdots I_{r}}$ are independent of $z_1, \cdots, z_n$; cyclically symmetric in $I_1 \cdots, I_r$; depend non-trivially on $\delta$;  are tensors under the principal congruence subgroup $\Gamma_h(2) \subset \Sp(2h,\ZZ)$; but are generally not locally holomorphic in the moduli of $\Sigma$.
\end{itemize}
The proof of these results is the subject of section \ref{sec:6}. It is worth highlighting 
that the same modular tensors $C_\delta^{I_1  \cdots I_{r}}$ and meromorphic multiplets
$D_\delta^{I_1  \cdots I_{r}}$ at rank $r$ universally enter the decompositions (\ref{intro.03})
and (\ref{intro.04}) for any number $n\geq r$ of Szeg\"o kernels.

\sm

A third main result is a remarkable correspondence between the decompositions of $\CD(z_1, \cdots, z_n)$ given in (\ref{intro.03}) and (\ref{intro.04}), despite the fact that the analyticity, monodromy, and modular properties of their respective building blocks are virtually opposite to one another. The precise correspondence may be formulated in terms of the following map,  
\bea
D_\delta^{I_1\cdots I_r} & \longleftrightarrow & C_\delta^{I_1\cdots I_r}
\no \\
g^{I_1 \cdots I_r}{}_J(x,y) & \quad \longleftrightarrow \quad & f^{I_1 \cdots I_r}{}_J(x,y) 
\no \\
\cW_{I_1 \cdots I_r} (z_1,\cdots, z_n) & \longleftrightarrow & \cV_{I_1 \cdots I_r} (z_1,\cdots, z_n)
\label{int.cor}
\eea
The correspondence just between $f$ and $g$ was already observed to hold between various higher genus  generalizations of the Fay identities and interchange lemmas in \cite{DHoker:2024ozn}. The correspondence may be further extended, modulo some subtle qualifiers related to regularization that will be addressed in section \ref{sec:4},  to integral representations for various functions, including $C_\delta^{I_1\cdots I_r}$ and $D_\delta^{I_1\cdots I_r}$, by the following formal map,
\bea
\oint _{\mA^I} dt & \quad  \longleftrightarrow  \quad & \int _\Sigma d^2t \, \bar \om^I(t)
\eea 
between line integrals on homology cycles and surface integrals on $\Sigma$.

\sm 

As our fourth main result, both types of decompositions are generalized in section~\ref{sec:ch} to the case of
linear chain products $L_\delta(x;z_1,\cdots,z_n;y)$ of (\ref{intro.02}). Their entire dependence on the 
points $z_1,\cdots,z_n$ will then be carried by functions $\cW_{I_1 \cdots I_r} (x;z_1,\cdots, z_n;y)$ and
$\cV_{I_1 \cdots I_r} (x;z_1,\cdots, z_n;y)$ that are
single-valued and meromorphic $(1,0)$ forms in $z_1,\cdots,z_n$. The counterparts
$M_\delta^{I_1 \cdots I_r}(x,y)$ and
$L_\delta^{I_1 \cdots I_r}(x,y)$ of the constants $\DD^{I_1  \cdots I_{r}}$ and
$C_\delta^{I_1  \cdots I_{r}}$ are spinors in $x,y$ which compensate for the
monodromies of $\cW_{I_1 \cdots I_r} (x;z_1,\cdots, z_n;y)$ in $x,y$
and the non-meromorphic dependence of $\cV_{I_1 \cdots I_r} (x;z_1,\cdots, z_n;y)$
on $x,y$, respectively. The correspondence of (\ref{int.cor}) is also generalized to
the case of linear chain products.

\sm

The tradeoff between meromorphicity and single-valuedness is familiar from the different constructions for genus one polylogarithms \cite{Levin:2007, BrownLevin, Broedel:2014vla, Broedel:2017kkb, Enriquez:2023} 
and will be reviewed in section \ref{sec:h1} to illustrate the case of arbitrary genus in the more familiar elliptic setting. 

\sm

An expected virtue of both decompositions (\ref{intro.03}) and (\ref{intro.04}) of cyclic products 
and their counterparts for linear chain products is a significant simplification of spin structure sums
in string amplitudes. For example, since the quantities $C_\delta^{I_1\cdots I_r}$ and $D_\delta^{I_1\cdots I_r}$ no longer depend on any points on the surface, their spin-structure sums for arbitrary chiral measures can be performed at the level of constants on $\Sigma$ which  depend only on the moduli of $\Sigma$.  Moreover, the differentials $f^{I_1\cdots I_r}{}_J(x,y) $ and $g^{I_1\cdots I_r}{}_J(x,y)$ carrying the entire dependence on the points $z_1, \cdots, z_n$  in (\ref{intro.03}) and (\ref{intro.04}) are amenable to the integration techniques of higher-genus polylogarithms and modular graph tensors and thereby facilitate low-energy expansions of the associated string amplitudes. For chiral amplitudes at genus two, the combinations of integration kernels in (\ref{intro.03}) and (\ref{intro.04}) complete the classification of the admissible $z_i$ dependences
arising from the contributions of even spin structures for arbitrary multiplicity that was initiated in \cite{DHoker:2007csw}.


\subsection*{Acknowledgments}

We are grateful to Benjamin Enriquez, Martijn Hidding and Federico Zerbini for collaboration on related topics. ED gratefully acknowledges the hospitality extended to him by the Center for Geometry and Physics and the Division of Theoretical Physics  at Uppsala University, while this work was being completed. The research of ED is supported in part by NSF grant PHY-22-09700. The research of OS is supported by the strength area ``Universe and mathematical physics'' which is funded by the Faculty of Science and Technology at Uppsala University.
OS is grateful to the Mainz Institute for Theoretical Physics (MITP) of the Cluster of Excellence PRISMA$^{+}$ (Project ID 390831469), for its hospitality and partial support during advanced stages of this project.

\newpage

\section{Descent in terms of Enriquez kernels}
\setcounter{equation}{0}
\label{sec:2}

The connection $d - \cK_\text{E}$ introduced by Enriquez on an arbitrary compact Riemann surface $\Sigma$ of genus $ h \geq 1$  is meromorphic with simple poles on the universal cover $ \tilde \Sigma$ of  $ \Sigma$ and takes values in the Lie algebra $\mg$ that is freely generated by $2h$ elements $a \, \cup \, b$ where $a=\{ a^1 , \cdots , a^h \}$ and  $b = \{ b_1 , \cdots, b_h \}$ \cite{Enriquez:2011}.  The Enriquez connection form $\cK_\text{E}$ depends on two points   $x, y \in \tilde \Sigma$ and may be expressed as,
\bea
\cK_\text{E}(x,y;a,b) = \KK_J(x,y;B) \, a^J
\label{kekja}
\eea 
where $B_I$ is defined by $B_I X = [b_I, X]$ for all $X \in \mg$ and $\KK_J(x,y;B)$  may be expanded in powers of the generators $B_I$ as follows, 
\bea
\label{2.KJ}
\KK_J(x,y;B) = \sum _{r=0}^\infty g^{I_1 \cdots I_r} {}_J (x,y) B_{I_1} \cdots B_{I_r}
\eea
The coefficient functions $g^{I_1 \cdots I_r}{}_J(x,y)$  will be referred to as \textit{meromorphic integration kernels} or simply as \textit{Enriquez kernels}.\footnote{They are related to those introduced in \cite{Enriquez:2011} by $g^{I_1\cdots I_r}{}_J(x,y) = (-2\pi i)^r \omega^{I_1\cdots I_r}{}_J(x,y)$ and we shall set $g^\emptyset {}_J (x,y) = \om_J(x)$ throughout, where $\om_J$ are the holomorphic Abelian differentials normalized in (\ref{A.abel}). \label{foot:2}} A key motivation for the Enriquez connection and associated Enriquez kernels is their role in the construction of meromorphic polylogarithms on Riemann surfaces of arbitrary genus \cite{Baune:2024biq, DHoker:2024ozn, Enriquez:next}. Alternative constructions
of higher-genus polylogarithms can be derived from the meromorphic flat connections of \cite{Enriquez:2021, Enriquez:2022} or the modular connection \cite{DHS:2023} built from the DHS kernels of section \ref{sec:6}. 

\sm

After giving a summary of the properties of the Enriquez kernels in section~\ref{sec:2.1}, we shall devote the remainder of this section to constructing a descent procedure that expresses the cyclic product  of Szeg\"o kernels $\CD(1,2, \cdots, n)$, already previewed in  (\ref{intro.01}) of the Introduction, in terms of Enriquez kernels.

\sm

The descent procedure will be presented for the cases of $n=2,3$ and $4$ in sections~\ref{sec:2.2}, \ref{sec:2.2a} and \ref{sec:2.3}, respectively,  and for arbitrary $n$ in section~\ref{sec:2.4} as Theorem~\ref{2.thm:3}. This theorem is one of the core results of this paper, but its full proof is quite  involved and for that reason will be carried out  in two parts.  The proof of the first part of Theorem~\ref{2.thm:3} will be given in section~\ref{sec:2.5} with the help of Lemmas~\ref{2.lem:4} and \ref{2.lem:5} which, in turn, are proven in appendices~\ref{sec:Ba} and \ref{sec:B}, respectively. The proof of the second part of Theorem~\ref{2.thm:3} will be greatly facilitated through the use of the generating functions that will be introduced in section~\ref{sec:3}. For this reason, the second part of the  proof of Theorem \ref{2.thm:3} is relegated to the subsequent section, namely section~\ref{sec:4}.

\subsection{Properties of Enriquez kernels}
\label{sec:2.1}

The Enriquez kernel $g^{I_1 \cdots I_r}{}_J(x,y)$,  for $r \geq 0$ and $I_1, \cdots, I_r, J \in \{ 1, \cdots, h\}$,  is a meromorphic $(1,0)$-form\footnote{Throughout, we shall use the conventions of \cite{DHS:2023, DHoker:2023khh, DHoker:2024ozn} in which a differential $(1,0)$-form $\phi$ is expressed in a local complex coordinate $x$ on $\Sigma$ or $\tilde \Sigma$ as $\phi = \phi(x) dx$. By a slight abuse of terminology, we shall refer also to the coefficient function $\phi(x)$ as a $(1,0)$-form. Similarly, a $(0,1)$-form will be denoted $\bar \phi = \bar \phi(x) d\bar x$. Thus, in form notation, the holomorphic Abelian differentials are $\om_I=  \om_I(x) dx$; their complex conjugates $\bar \om _I = \bar \om_I(x) d \bar x$; the Enriquez kernels are $g^{I_1 \cdots I_r}{}_J(x,y)dx $; the DHS kernels are $f^{I_1 \cdots I_r}{}_J(x,y) dx$; and the Szeg\"o kernel is $S_\delta(x,y) \sqrt{dx \, dy}$. Note, however, that in the conventions of \cite{DHoker:2025szl, DHoker:2025dhv} the differentials $dx$ or $d \bar x$  were included as part of the forms.} in $x \in \tilde \Sigma$ and a $(0,0)$-form in $y \in \tilde \Sigma$ which is locally holomorphic in the moduli of $\Sigma$, with prescribed monodromies \cite{Enriquez:2011,DHoker:2025szl}. Its monodromies in $x$ and $y$ around $\mA$ cycles are trivial, while those around $\mB$ cycles are given by,
 \bea
\label{2.4.mon}
\Delta _L ^{(x)} g^{I_1 \cdots I_r}{}_J(x,y)  & = & 
 \sum_{k=1}^r { (-2\pi i)^k \over k!} \, \delta ^{I_1 \cdots I_k} _L \, g^{I_{k+1} \cdots I_r}{}_J(x,y)
\no \\ 
\Delta _L ^{(y)}  g^{I_1 \cdots I_r}{}_J(x ,y)  & = &   
\delta ^{I_r}_J \sum_{k=1}^r { (2\pi i)^k \over k !} \, g^{I_1 \cdots I_{r-k}} {}_L(x,y) \, \delta^{I_{r-k+1} \cdots I_{r-1}}_L 
\eea
The generalized Kronecker symbol is defined by,
\bea
\label{2.kron}
\delta^{I_{1} I_2 \cdots I_k}_L=\delta^{I_{1}}_L \delta^{I_{2}}_L \cdots \delta^{ I_k}_L
\eea 
and specializes to $\delta_L^\emptyset =1$ in the $k=1$ term on the second line of (\ref{2.4.mon}).
The monodromy of an arbitrary  function $\phi(x)$ around a cycle $ \mB_L$ is denoted by,
 \bea
 \label{2.mondef}
 \Delta _L^{(x)} \phi(x) = \phi(\mB_L \cdot x) - \phi(x)
 \eea
 where $\mB_L \cdot x$ denotes the action of the element $\mB_L \in \pi _1(\Sigma , q)$ on the point $x \in \Sigma$.

\sm

Since the kernels $g^{I_1 \cdots I_r}{}_J(x,y)$ have prescribed monodromies, we may define them for $x,y$ in a fundamental domain $D$ for $\Sigma$, which is obtained by cutting $\Sigma$ open along $2h$ loops $\mA^I$ and $\mB_I$ with common base point $q$, as shown in figure \ref{fig:1}.

\sm

One may choose a \textit{preferred fundamental domain} $D$ such that $g^{I_1 \cdots I_r}{}_J(x,y)$ for $x,y$ in the interior $D^o$ of $D$ is holomorphic in $x$ and $y$ for $r \geq 2$,  $g^{I}{}_J(x,y)$ has a single simple pole in $x$ at $y$ with residue $\delta ^I_J$ and a single simple pole in $y$ at $x$ with residue $- \delta^I_J$,
\bea
\label{2.0.pole}
g^I{}_J(x,y) = \frac{\delta^I_J }{x-y} + {\rm regular}
\eea
and is given by $g^\emptyset {}_J(x,y) = \om_J(x)$ for $r=0$ as already stated in footnote \ref{foot:2}. For $x'$ and/or $y'$ outside $D^o$, the kernels $g^{I_1 \cdots I_r}{}_J(x',y')$ are obtained from $g^{I_1 \cdots I_r}{}_J(x,y)$ with $x,y \in D^o$ by mapping $(x',y') \to (x,y)$ by an element in $\pi_1(\Sigma, q)$ and then using the monodromy relations of (\ref{2.4.mon}). These cases where either $x$ and/or $y$ are in the boundary $\p D$ may be obtained by considering limits of interior points. As a result, the forms $g^{I_1 \cdots I_r}{}_J(x,y)$ may acquire simple poles in $x$ at $\pi^{-1}(y)$ for all $r \geq 1$, where $\pi$ is the canonical projection $\pi : \tilde \Sigma \to \Sigma$.

\begin{figure}[htb]
\begin{center}
\tikzpicture[scale=0.8]
\scope[xshift=2.2cm,yshift=0cm, scale=0.8, rotate=90]
\draw[ultra thick] (1.5,1.2) .. controls (2.5, 1.2) .. (3.5,2);
\draw[ultra thick] (1.8,1.2) .. controls (2.5, 1.8) .. (3.2,1.75);
\draw[ultra thick] (6.5,2) .. controls (7.5, 1.6) .. (8.5,2);
\draw[ultra thick] (6.8,1.9) .. controls (7.5, 2.2) .. (8.2,1.87);
\draw[very thick, color=blue] plot [smooth] coordinates {(5,2) (4,2.2) (3, 2.3) (2, 2.15) (1.3,1.5) (1.5,1) ( 2, 0.85) (3,1.1) (5,2) };
\draw[very thick, color=blue] plot [smooth] coordinates {(5,2) (6,2.3) (7, 2.6) (8, 2.6) (8.8,2.2) (8.5,1.5)  (7,1.3) (5,2) };
\draw[very thick, color=red] plot [smooth] coordinates { (5,2) (4,1.8) (3,1.45) (2.7,1.4)};
\draw[very thick, color=red, dotted] plot [smooth] coordinates { (2.7,1.4) (2.5,1.2) (2.3,0.6) (2.5, 0.275) (2.9, -0.18)};
\draw[very thick, color=red] plot [smooth] coordinates { (2.9,-0.18) (4,0.7) (5,2)};
\draw[very thick, color=red] plot [smooth] coordinates { (5,2) (6,1.8) (7,1.6) (7.3,1.7)};
\draw[very thick, color=red, dotted] plot [smooth] coordinates { (7.3,1.7) (7.9,1) (8,0.4) (7.7, 0.2)};
\draw[very thick, color=red] plot [smooth] coordinates { (7.7, 0.15) (6,1.05) (5,2)};
\draw[ultra thick] plot [smooth] coordinates 
{(0.7,0.7) (0.45,2) (1,3) (2.5, 3.8)  (5,3.2) (7.5, 4) (9,3.4) (9.7,1.6) 
(9,0.3) (8,0.1) (7, 0.3) (5,0.6) (3, -0.2) (2,-0.2) (1.1,0.1) (0.7,0.7) };
\draw [color=red] (3.05,0.6) node{\small $\mA^1$};
\draw [color=red] (7.25,0.8) node{\small $\mA^2$};
\draw [color=blue] (1.3,2.2) node{\small $\mB_1$};
\draw [color=blue]  (8.9,1.25) node{\small $\mB_2$};
\draw (9.4,-0.2) node{{$\Sigma$}};
\draw (5,2) node{$\bullet$};
\draw (5,2.38) node{ $\qq$};
\draw (2,2.9) node{ $\bullet$};
\draw (2.4,2.9) node{ $x$};
\draw (7,2.95) node{ $\bullet$};
\draw (7.4,2.97) node{ $y$};
\endscope
\scope[xshift=10cm,yshift=4.2cm, scale=4.2]
\draw  [very  thick, color=red, <-, domain=182:198] plot ({cosh(0.6)*cos(0)+sinh(0.6)*cos(\x)}, {cosh(0.6)*sin(0)+sinh(0.6)*sin(\x)});
\draw  [very  thick, color=red, domain=166:182] plot ({cosh(0.6)*cos(0)+sinh(0.6)*cos(\x)}, {cosh(0.6)*sin(0)+sinh(0.6)*sin(\x)});
\draw  [very  thick, color=red, <-, domain=261.5:282] plot ({cosh(0.65)*cos(85)+sinh(0.65)*cos(\x)}, {cosh(0.65)*sin(85)+sinh(0.65)*sin(\x)});
\draw  [very  thick, color=red,  domain=241:261.5] plot ({cosh(0.65)*cos(85)+sinh(0.65)*cos(\x)}, {cosh(0.65)*sin(85)+sinh(0.65)*sin(\x)});
\draw  [very  thick, color=red, <-, domain=363:396] plot ({cosh(0.5)*cos(175)+sinh(0.5)*cos(\x)}, {cosh(0.5)*sin(175)+sinh(0.5)*sin(\x)});
\draw  [very  thick, color=red,  domain=330:363] plot ({cosh(0.5)*cos(175)+sinh(0.5)*cos(\x)}, {cosh(0.5)*sin(175)+sinh(0.5)*sin(\x)});
\draw  [very  thick, color=red, <-, domain=77:100] plot ({cosh(0.4)*cos(265)+sinh(0.4)*cos(\x)}, {cosh(0.4)*sin(265)+sinh(0.4)*sin(\x)});
\draw  [very  thick, color=red,  domain=54:77] plot ({cosh(0.4)*cos(265)+sinh(0.4)*cos(\x)}, {cosh(0.4)*sin(265)+sinh(0.4)*sin(\x)});
\draw  [very  thick, color=blue, <-, domain=220:239] plot ({cosh(0.7)*cos(40)+sinh(0.7)*cos(\x)}, {cosh(0.7)*sin(40)+sinh(0.7)*sin(\x)});
\draw  [very  thick, color=blue, domain=201:220] plot ({cosh(0.7)*cos(40)+sinh(0.7)*cos(\x)}, {cosh(0.7)*sin(40)+sinh(0.7)*sin(\x)});
\draw  [very  thick, color=blue, <-, domain=293.5:323] plot ({cosh(0.5)*cos(125)+sinh(0.5)*cos(\x)}, {cosh(0.5)*sin(125)+sinh(0.5)*sin(\x)});
\draw  [very  thick, color=blue, domain=264:293.5] plot ({cosh(0.5)*cos(125)+sinh(0.5)*cos(\x)}, {cosh(0.5)*sin(125)+sinh(0.5)*sin(\x)});
\draw  [very  thick, color=blue, <-, domain=45:73] plot ({cosh(0.7)*cos(225)+sinh(0.7)*cos(\x)}, {cosh(0.7)*sin(225)+sinh(0.7)*sin(\x)});
\draw  [very  thick, color=blue, domain=17:45] plot ({cosh(0.7)*cos(225)+sinh(0.7)*cos(\x)}, {cosh(0.7)*sin(225)+sinh(0.7)*sin(\x)});
\draw  [very  thick, color=blue, <-, domain=141.5:169] plot ({cosh(0.7)*cos(315)+sinh(0.7)*cos(\x)}, {cosh(0.7)*sin(315)+sinh(0.7)*sin(\x)});
\draw  [very  thick, color=blue, domain=114:141.5] plot ({cosh(0.7)*cos(315)+sinh(0.7)*cos(\x)}, {cosh(0.7)*sin(315)+sinh(0.7)*sin(\x)});
\draw  [thick, color=black, dashed, ->, domain=238:220] plot ({cosh(0.9)*cos(40)+sinh(0.9)*cos(\x)}, {cosh(0.9)*sin(40)+sinh(0.9)*sin(\x)});
\draw  [thick, color=black, dashed,  domain=220:203] plot ({cosh(0.9)*cos(40)+sinh(0.9)*cos(\x)}, {cosh(0.9)*sin(40)+sinh(0.9)*sin(\x)});
\draw (0.148,-0.74) [fill=black] circle(0.02cm) ;
\draw  [thick, color=black, dashed, ->, domain=68:42] plot ({cosh(0.9)*cos(225)+sinh(0.9)*cos(\x)}, {cosh(0.9)*sin(225)+sinh(0.9)*sin(\x)});
\draw  [thick, color=black, dashed,  domain=42:19] plot ({cosh(0.9)*cos(225)+sinh(0.9)*cos(\x)}, {cosh(0.9)*sin(225)+sinh(0.9)*sin(\x)});
\draw  [thick, color=black, dashed, ->, domain=281:261] plot ({cosh(0.85)*cos(85)+sinh(0.85)*cos(\x)}, {cosh(0.85)*sin(85)+sinh(0.85)*sin(\x)});
\draw  [thick, color=black, dashed, domain=261:243] plot ({cosh(0.85)*cos(85)+sinh(0.85)*cos(\x)}, {cosh(0.85)*sin(85)+sinh(0.85)*sin(\x)});
\draw  [thick, color=black, dashed, ->, domain=99:77] plot ({cosh(0.6)*cos(265)+sinh(0.6)*cos(\x)}, {cosh(0.6)*sin(265)+sinh(0.6)*sin(\x)});
\draw  [thick, color=black, dashed, domain=77:63] plot ({cosh(0.6)*cos(265)+sinh(0.6)*cos(\x)}, {cosh(0.6)*sin(265)+sinh(0.6)*sin(\x)});
\draw (0.57,0.16) [fill=black] circle(0.02cm) ;
\draw (0.251,0.53) [fill=black] circle(0.02cm) ;
\draw (-0.7,0.4) [fill=black] circle(0.02cm) ;
\draw (-0.235,0.61) [fill=black] circle(0.02cm) ;
\draw (-0.675,-0.165) [fill=black] circle(0.02cm) ;
\draw (-0.165,-0.67) [fill=black] circle(0.02cm) ;
\draw (0.57,-0.19) [fill=black] circle(0.02cm) ;

\draw (0.66, -0.21) node{\small $\qq_1$};
\draw (0.68, 0.2) node{\small $\qq_2$};
\draw (0.36, 0.6) node{\small $\qq_3$};
\draw (-0.24, 0.71) node{\small $\qq_4$};
\draw (-0.77, 0.48) node{\small $\qq_5$};
\draw (-0.77, -0.22) node{\small $\qq_6$};
\draw (-0.26, -0.73) node{\small $\qq_7$};
\draw (0.24, -0.8) node{\small $\qq_8$};

\draw [color=red] (0.68,0) node{\footnotesize $\mA^1$};
\draw [color=blue] (0.5,0.38) node{\footnotesize $\mB_1$};
\draw [color=red] (0.1,0.65) node{\footnotesize $(\mA^1)^{-1}$};
\draw [color=blue] (-0.5,0.56) node{\footnotesize $\mB_1^{-1}$};
\draw [color=red] (-0.75,0.1) node{\footnotesize $\mA^2$};
\draw [color=blue]  (-0.5,-0.42) node{\footnotesize $\mB_2$};
\draw [color=red] (-0.02,-0.85) node{\footnotesize $(\mA^2)^{-1}$};
\draw [color=blue]  (0.42,-0.46) node{\footnotesize $\mB_2^{-1}$};
\draw (0,0) node{{\Large $D$}};
\draw (0.12,0.24) node{{$\bullet$}};
\draw (0.18,0.2) node{{$x$}};
\draw (-0.4,0.1) node{{$\bullet$}};
\draw (-0.34,0.06) node{{$y$}};
\endscope

\endtikzpicture
\caption{The left panel shows a compact genus two Riemann surface $\Sigma$ and a choice of canonical homology cycles $\mA^1, \mA^2, \mB_1, \mB_2$ with a common base point $q$. A fundamental domain $D$, contained in the universal cover $\tilde \Sigma$ of $\Sigma$, for the action of $ \pi_1(\Sigma, \qq)$ on $\Sigma$ is obtained in the right panel by cutting $\Sigma$ along the cycles in the left panel. The surface $\Sigma$ may be recovered from $D$ by pairwise identifying inverse boundary components with one another under the dashed arrows. The vertices $q_i \in \tilde \Sigma$ project to $ q = \pi(q_i) \in \Sigma$ for $i=1,\cdots, 8$ under the canonical projection $\pi: \tilde \Sigma \to \Sigma$. 
 \label{fig:1}}
\end{center}
\end{figure}

\sm

The periods around $\mA^L$ cycles on the boundary of the fundamental domain $D$ in figure~\ref{fig:1} are given in terms of Bernoulli numbers $\Ber_r$ by,\footnote{Throughout, we shall denote the Bernoulli numbers by $\Ber_r$ instead of the customary $B_r$ in order to avoid confusion with the Lie algebra generators $B_I$ used, for example, in (\ref{2.KJ}). Recall that the $\Ber_r$ are generated by $\frac{x}{e^x-1} = \sum_{r=0}^{\infty} \frac{x^r}{r!} \, {\rm Ber}_r$. Furthermore, we shall systematically denote integration variables by $t$, or $t_1,\cdots,t_r$ in the case of multiple integrations. \label{foot:5}}  
\bea
\label{2.intg}
\oint _{\mA^L}  dt \,  g^{I_1 \cdots I_r} {}_J (t,y) = (-2\pi i)^r { \Ber_r \over r!} \, \delta ^{I_1 \cdots I_r L}_J
\eea

\sm

The $y$-dependence of $g^{I_1\cdots I_r}{}_J(x,y)$  is concentrated in the trace part with respect to the last two indices \cite{Enriquez:2011}, which leads us to introduce the following decomposition,
\bea
\label{2.dec}
g^{I_1\cdots I_r}{}_J(x,y) = \varpi^{I_1\cdots I_r}{}_J(x) - \delta^{I_r}_J \chi^{I_1\cdots I_{r-1}}(x,y)
\eea
where $\varpi$ is traceless, $\varpi^{I_1\cdots I_s J}{}_J(x) = 0$,  independent of $y$, and holomorphic in $x \in \tilde \Sigma$. Hence, the simple pole (\ref{2.0.pole}) of $g^{I}{}_J(x,y) $ is entirely carried by $\chi(x,y) = - 1/(x-y)  + {\rm regular}$.
As a consequence of the unique double pole of $\p_y \chi (x,y)$ in $x$ at $y$, we deduce that the combination $\p_y \chi(x,y) + \p_y \p_x \ln E(x,y)$ is holomorphic in $x$ and $y$ and single-valued for $x,y \in\Sigma$.  Since its $\mA^L$ periods vanish, we conclude that the following equalities hold,
\bea
\label{2.chi0}
\p_y \chi(x,y) & = & - \p_y \p_x \ln E(x,y)
\no \\
\chi(x,y) - \chi(x,z) & = & - \p_x \ln  E(x,y) + \p_x \ln E(x,z)
\eea
A summary of definitions and useful properties of the prime form $E(x,y)$, the Szeg\"o kernel and the Enriquez kernels is provided in appendix \ref{sec:A}.

\subsection{The case $n=2$}
\label{sec:2.2}

By using the Fay trisecant  identity,\footnote{The Fay trisecant identity between four points was introduced  in equation (45) of \cite{Fay:1973}. The versions involving two or three points, used here,  may be obtained therefrom by taking the limit of coincident points and has appeared, for example, in equations (A.26) and (A.27) of \cite{DHoker:2005vch}. \label{foot:6}} the case $n=2$ may be recast in the following form,\footnote{Throughout, it will often be convenient to abbreviate the arguments $z_1, \cdots , z_n$ of various functions simply by their subscripts so that we set, for example, $\CD(1,\cdots, n) = \CD(z_1, \cdots, z_n)$.}  
\bea
\label{2.1.a}
C_\delta(1,2) =  \om_I(1) \DD^{I} (2) + \p_2 \chi (1,2) 
\eea
where $\DD^I(2)$ is given in terms of a multiplet $\DD^{IJ}$ that is constant on $\Sigma$, 
\bea
\label{2.1.b}
\DD^I(2) = \om_J(2) \DD^{IJ} 
\hskip 1in
\DD ^{IJ} = - { \p^I \p^J \tet[\delta](0) \over \tet [\delta](0)}
\eea
and we have used the relation $\p_2 \chi(1,2) = - \p_1 \p_2 \ln E(1,2)$ of (\ref{2.chi0}). As promised in the Introduction, all dependence on the spin structure $\delta$ is concentrated in the constant multiplet $\DD^{IJ}$, while the remaining $\delta$-independent part is expressed in terms of (the trace part of) an Enriquez kernel. We note that, while $\CD(1,2) $ is modular invariant when the spin structure $\delta = [\delta' , \delta '' ]$ is  transformed to $\tilde \delta = [\tilde \delta' , \tilde \delta '' ]$ as follows \cite{Fay:1973},
\bea
 \left ( \begin{matrix}  \tilde \delta ''   \\  \tilde \delta '  \end{matrix} \right )  = 
 \left ( \begin{matrix} A &-B  \\  -C &D \end{matrix} \right ) 
  \left ( \begin{matrix}    \delta ''   \\ \delta '  \end{matrix} \right )  
  + \frac{1}{2} \, {\rm diag}  \left ( \begin{matrix}   AB^t  \\ CD^t \end{matrix} \right )  
  \hskip 0.65in
  \left ( \bma  A &B \\  C &D \ema \right) \in \Sp(2h,\ZZ)
  \quad
  \label{trfdelta}
  \eea
neither $\p_2 \chi(1,2)$ nor $ \om_J(2) \DD^{J}(2)$ is modular invariant, and $\DD^{IJ}$ is \textit{not} a modular tensor. As we shall see shortly, the symmetry $\DD^{JI}=\DD^{IJ}$, which is manifest in (\ref{2.1.b}), will be of crucial importance for the descent procedure to succeed at multiplicity $n = 3$ and beyond.

\subsection{The case $n=3$}
\label{sec:2.2a}

The case $n=3$ may be handled by inspecting the poles of $\CD(1,2,3)$, collected in the following cyclic orbit of differential relations, 
\bea
\label{2.1.c}
\bar \p_1 C_\delta(1,2,3) & = & \pi \big ( \delta (1,2) - \delta (1,3) \big ) C_\delta(2,3)
\no \\
\bar \p_2 C_\delta(1,2,3) & = & \pi \big ( \delta (2,3) - \delta (2,1) \big ) C_\delta(3,1)
\no \\
\bar \p_3 C_\delta(1,2,3) & = & \pi \big ( \delta (3,1) - \delta (3,2) \big ) C_\delta(1,2)
\eea
Using the relation $\bar \p_1 \chi (1,2) = - \pi \delta (1,2)$ and the fact that $\chi(1,2) - \chi(1,3)$ is single-valued in $z_1$, we see that the combination $\CD(1,2,3) +\big ( \chi(1,2) - \chi(1,3) \big )  \, C_\delta(2,3)$ is holomorphic and single-valued in $z_1$. Therefore, it must be a linear combination of the holomorphic Abelian differentials $\om_I(1)$ with coefficients $\DD^I(2,3)$ that depend on the points $z_2$ and $z_3$ but not on $z_1$. As a result, we obtain the following decomposition, 
\bea
\label{2.1.d}
 \om_I (1) \DD^I(2,3) = C_\delta(1,2,3) + \big ( \chi(1,2) - \chi(1,3) \big ) \, C_\delta(2,3)
\eea
Since $\CD(2,3)$ and $\CD(1,2,3)$ are single-valued in their arguments and $\chi$ has trivial $\mA$ monodromy the coefficients $\DD^I(2,3)$ have trivial $\mA$ monodromy, while their $\mB$ monodromy follows from the second line in (\ref{A.chi}) and is given by,\footnote{We note that the monodromies of $\DD^I(2,3)$ have a double pole in $z_2-z_3$. This is due to the fact that, even though the prefactor of the second term on the right of (\ref{2.1.d}) vanishes as $z_2 \to z_3$ when $z_2$ and $z_3$ are in the same fundamental domain, this factor does not vanish when $z_2$ and $z_3$ are in different fundamental domains, as brought about by performing a $\mB_L$ transformation on one of the points.} 
\bea
\label{2.1.e}
\Delta _L^{(2)} \DD^I(2, 3) & = &  - 2 \pi i \, \delta ^I_L C_\delta(2,3)
\no \\
\Delta _L^{(3)}  \DD^I(2, 3 ) & = &  2 \pi i \, \delta ^I_L C_\delta(2,3)
\eea
The $\bar \p_2$ and $\bar \p_3$ derivates of $\DD^I(2,3)$ are readily evaluated using (\ref{2.1.c}), (\ref{2.1.d})  and the derivatives of $\chi$. For $z_2$ and $z_3$ in the same fundamental domain  we find,
\bea
\label{2.1.f}
\bar \p_2 \DD^I(2,3) & = & + \pi \delta (2,3) \DD^I(3)
\no \\
\bar \p_3 \DD^I(2,3) & = & - \pi \delta (2,3) \DD ^I (2)
\eea
with $D_\delta^I(3)$ given by (\ref{2.1.b}). Using the first equation in (\ref{2.1.f}) and $\bar \p_2 \chi(2,3)= - \pi \delta(2,3)$, we see that the combination,
\bea
\label{2.1.ga}
V_\delta ^I(2,3) = \DD^I(2,3) + \chi(2,3) \DD^I(3)
\eea 
is holomorphic in $z_2$. However, combining the first equation of (\ref{2.1.e}) with the monodromy of $\chi$ given in (\ref{A.chi}), we find that this combination has non-trivial $\mB$ monodromy, 
\bea
\label{2.1.g}
\Delta _L^{(2)} V_\delta ^I(2,3) = - 2 \pi i \, \delta ^I_L \, \p_3 \chi(2,3) - 2 \pi i \, \delta ^I_L \, \om_J(2) \DD^J (3) 
+ { 2 \pi i \over h} \om_L(2) \DD^I (3)
\eea
so it remains to find a counterterm holomorphic in $z_2$ with the opposite monodromies. Indeed,
the monodromy of $\p_3 \chi^I(2,3)$ reproduces the first term of (\ref{2.1.g}) thanks to the third line in (\ref{A.chi}),
and the remaining terms match the monodromy of $\varpi^I{}_J(2) \DD^J (3)$  by virtue of (\ref{A.varpi}).
Hence, the combination $\DD^I(2,3) + \chi(2,3) \DD^I(3) - \p_3 \chi^I(2,3) - \varpi ^I{}_J (2) \DD^J (3)$ is holomorphic and single-valued in $z_2$ and therefore must be a linear combination of $\om_J(2)$ with coefficients $\DD^{IJ}(3)$ that are independent of $z_2$, 
\bea
\om_J(2) \DD ^{IJ}(3) = \DD^I(2,3) - \p_3 \chi^I(2,3) - g ^I{}_J (2,3) \DD^J(3) 
\label{2.1.dnext}
\eea
where we have combined the contributions from $\chi(2,3) \DD^I(3)$ in (\ref{2.1.ga}) with $ \varpi ^I{}_J (2) \DD^J (3)$ into $g ^I{}_J (2,3) \DD^J(3) $. Using the second line in (\ref{2.1.f}), one verifies that $\DD^{IJ}(3)$ is holomorphic in $z_3$ while its $\mA$ monodromy is trivial and its $\mB$ monodromy is given by,
\bea
\Delta _L ^{(3)} \DD^{IJ} (3) = 2 \pi i   \delta ^I_L  \DD^{J}(3)     - 2 \pi i   \delta ^J_L \DD^{I} (3)
\eea
The next step in the descent procedure consists in constructing a linear combination involving the holomorphic form $\varpi^A{}_B  (3)$ and the constant multiplet $\DD^{MN}$ that matches the $\mB$ monodromy of $\DD^{IJ} (3)$ and allows us to express it in terms of a constant multiplet $\DD^{IJK}$.  Consulting the first line in (\ref{A.varpi}), we observe that the combination $- \varpi ^J{}_B (3) \DD^{IB} + \varpi ^I{}_B (3) \DD^{JB}$ partially compensates the monodromy of $\DD^{IJ}(3)$, 
\bea
 \Delta _L ^{(3)} \Big ( \DD^{IJ}(3)- \varpi ^J{}_B (3) \DD^{IB} + \varpi ^I{}_B (3) \DD^{JB} \Big )
 =  {2 \pi i \over h}   \om _L(3) \big ( \DD^{JI} - \DD^{IJ} \big )
 \eea
It is at this point that we use the symmetry $ \DD^{JI} = \DD^{IJ}$, which is manifest from its expression in (\ref{2.1.b}), to ensure that the $\mB$ monodromy of the above combination indeed cancels, so that its dependence on $z_3 $ is holomorphic and single-valued and may be expressed as a linear combination of the holomorphic Abelian differentials $\om_K(3)$. The results for the case $n=3$ are summarized by the proposition below. 

{\prop
\label{2.thm:1}
The descent of the $n=3$ case is given by the following relations, 
\bea
\label{2.thm.1}
 \om_I (1) \DD^I(2,3) & = & C_\delta(1,2,3) + \big ( \chi(1,2) - \chi(1,3) \big ) \, C_\delta(2,3)
\no \\
\om_J(2) \DD ^{IJ}(3) & = &  \DD^I(2,3)  - \p_3 \chi^I(2,3) - g ^I{}_K (2,3) \DD^K(3) 
\no \\
\om_K(3) \DD^{IJK} & = & \DD^{IJ} (3)  + \varpi^I{}_K (3) \DD^{JK} - \varpi ^J{}_K(3) \DD^{IK}
\eea
where the multiplet $\DD^{IJK}$ is cyclically symmetric in $I,J,K$, constant on $\Sigma$, and locally holomorphic  in the moduli of $\Sigma$.}

\sm

The validity of the three relations in (\ref{2.thm.1}) was already established in the paragraphs that precede the proposition, so it only remains to prove the cyclic symmetry relation, 
\bea
\label{2.cycl.3}
\DD^{IJK} = \DD^{JKI}
\eea
 One way to proceed is to take  the difference between the first relation of (\ref{2.thm.1}) and its version for cyclically permuted points, then eliminate $\DD^I(2,3)$, $\DD^{IJ}(3)$ and their cyclic permutations in the points $z_1,z_2,z_3$ using the second and third relations. The difference may be simplified using the cyclic invariance of $\CD(1,2,3)$ in the points $z_1, z_2, z_3$; the $z$-derivative of the Fay identity in (9.24) of  \cite{DHoker:2024ozn} for $x=z_1, y=z_2, z=z_3$; the interchange lemma in (9.11) of  \cite{DHoker:2024ozn};  and Theorem 9.4 for $r=1$ of  \cite{DHoker:2024ozn}. The resulting relation reduces to the vanishing of $(\DD^{IJK} - \DD^{JKI} ) \om_I(1) \om_J(2) \om_K(3)$, which implies the cyclic property of $\DD^{IJK}$ and completes the proof of the proposition. In section \ref{sec:4} a more streamlined proof of (\ref{2.cycl.3}) will be presented that applies to the case of arbitrary $n$ with equal ease.  Similar computations lead to the reflection relation $\DD^{IJK}= -\DD^{KJI}$.

\subsection{The case $n=4$}
\label{sec:2.3}

The result is given by  the following proposition.

{\prop
\label{2.thm:2}
The descent equations for the case $n=4$ are given by,
\bea
\label{5.h}
\om_I (1) \DD^I(2,3,4) & = & C_\delta(1,2,3,4)  + \big ( \chi(1,2) - \chi(1,4) \big )  C_\delta(2,3,4)
\no \\
\om_J(2) \DD^{IJ}  (3,4)   & = & \DD^I (2,3,4)
+ \big ( \chi^I(2,3) - \chi^I(2,4) \big ) C_\delta (3,4)  - g^I{}_J(2,3) \DD^J  (3,4) 
\no \\ 
\om_K(3) \DD^{IJK}  (4) & = &  \DD^{IJ}(3,4)  - \p_4 \chi^{JI}(3,4) 
- g^J{}_K(3,4) \DD^{IK} (4) - g^{JI}{}_K(3,4) \DD^K  (4)
\no \\
 \om_L(4) \DD^{IJKL}  & = & \DD^{IJK} (4) 
- \varpi ^K{}_L (4) \DD^{IJL} + \varpi ^I{}_L(4) \DD^{JKL} + \varpi^{KI}{}_L(4) \DD^{JL}
\no \\ && \hskip 0.65in
 - \varpi^{KJ}{}_L(4) \DD^{IL} + \varpi^{IK}{}_L(4) \DD^{JL} - \varpi^{IJ}{}_L(4) \DD^{KL}
\eea
The multiplet $\DD^{IJKL}$ is cyclically symmetric in $I,J,K,L$,  constant on $\Sigma$ and locally holomorphic  
in the moduli of $\Sigma$.}

\sm

The proof of this proposition proceeds as for the case $n=3$ and starts with, 
\bea
\label{2.thm.4}
\bar \p_1 C_\delta(1,2,3,4) = \pi \big ( \delta (1,2) - \delta (1,4) \big ) C_\delta(2,3,4)
\eea
and its three cyclic permutations. The term $\om_I(1) \DD^I(2,3,4)$ is obtained by verifying that the poles and monodromies in $z_1$ of the right side on the first line of (\ref{5.h}) cancel, so it must be a linear combination of $\om_I(1)$ with $z_1$-independent coefficients $\DD^I(2,3,4)$. The differential equations for $\DD^I(2,3,4)$ are obtained by differentiating the first line of (\ref{5.h}) using (\ref{2.thm.4}) and its cyclic permutations, as well as the relations of (\ref{2.thm.1}) and we find, 
\bea
\label{2.thm.5}
\bar \p_2 \DD^I(2,3,4) & = &  \pi  \delta (2,3)  \DD ^I(3,4)
\no \\
\bar \p_3 \DD^I(2,3,4) & = &  \pi \big (  \delta (3,4)  - \delta (3,2) \big ) \DD ^I(2,4)
\no \\
\bar \p_4 \DD^I(2,3,4) & = & - \pi  \delta (4,3)  \DD ^I(2,3)
\eea
with $\DD^I(3,4)$ defined by the $n=3$ case in (\ref{2.1.d}). The $\mA$ monodromies vanish while the $\mB$ monodromies may be computed by evaluating the $\mB$ monodromies of the first equation in (\ref{5.h}) and are found to be given by, 
\bea
\label{2.thm.6}
\Delta ^{(2)}_L \DD^I(2, 3,4) & = & - 2 \pi i \, \delta ^I_L C_\delta(2,3,4)
\no \\
\Delta ^{(3)}_L \DD^I(2, 3,4) & = & 0 
\no \\
\Delta ^{(4)}_L \DD^I(2, 3, 4) & = &  2 \pi i \, \delta ^I_L C_\delta(2,3,4)
\eea
Next, we show that the poles and monodromies in  $z_2$ of the right side of the second equation in (\ref{5.h}) cancel so that it must be a linear combination of $\om_J(2)$  whose $z_2$-independent coefficients $\DD^{IJ}(3,4)$ in turn satisfy the differential equations,
\bea
\label{2.thm.8}
\bar \p_3 \DD^{IJ} (3,4) & = & + \pi \delta (3,4) \DD^{IJ} (4)
\no \\
\bar \p_4 \DD^{IJ}  (3,4) & = & - \pi \delta (4,3) \DD^{IJ} (3)
\eea
(see (\ref{2.1.dnext}) for $\DD^{IJ} (4)$), have vanishing $\mA$ monodromies and the following $\mB$ monodromies (see (\ref{2.kron}) for the generalized Kronecker delta $\delta ^{IJ}_L$), 
\bea
\label{2.thm.9}
\Delta ^{(3)}_L  \DD^{IJ}  (3,4) & = & 
- 2 \pi i \, \delta ^J_L \, \DD^I  (3,4) - 2 \pi^2 \, \delta ^{IJ}_L \, C_\delta (3,4)
\no \\
\Delta ^{(4)}_L  \DD^{IJ} (3,4) & = &  2 \pi i \, \delta ^I_L \, \DD^J  (3,4) - 2 \pi^2 \, \delta ^{IJ}_L \, C_\delta (3,4)
\eea
Using (\ref{2.thm.8}) and (\ref{2.thm.9}) we show that the right side of the third equation of (\ref{5.h}) is holomorphic and has vanishing monodromies in $z_3$. The coefficients $\DD^{IJK}(4)$ of $\omega_K(3)$ are holomorphic in $z_4$, have vanishing $\mA$ monodromies and their $\mB$ monodromies are given~by,
\bea
\Delta ^{(4)}_L \DD^{IJK}  (4)
& = &  
- 2 \pi i \, \delta ^K_L \Big ( \om_M(4) \DD ^{IJM} - \varpi^I{}_M(4) \DD^{JM} + \varpi ^J {}_M(4) \DD^{IM} \Big ) 
\no \\ && 
+ 2 \pi i \, \delta ^I_L \Big ( \om_M(4) \DD ^{JKM} - \varpi^J{}_M(4) \DD^{KM} + \varpi ^K {}_M(4) \DD^{JM} \Big ) 
\no \\ && 
- 2 \pi^2 \om_M(4) \Big ( \delta ^{JK}_L \, \DD^{IM}  + \delta ^{IJ}_L \, \DD^{KM}  - 2 \delta ^{IK} _L \, \DD^{JM}  \Big ) 
\eea
The next step in the descent procedure consists in constructing a linear combination involving the holomorphic forms $\varpi^A{}_B  (4)$, $\varpi ^{AB}{}_C(4)$ and the constant multiplets $\DD^{MN}$ and $\DD^{MNP}$ that matches the $\mB$ monodromy of $\DD^{IJK} (4)$ and allows us to express it in terms of a constant multiplet $\DD^{IJKL}$.  Consulting (\ref{A.varpi}), we observe that the combination on the right side of the last equation of (\ref{5.h}) has the following monodromy, 
\bea
&& \Delta _L ^{(3)} \Big ( 
\DD^{IJK} (4)  - \varpi ^K{}_L (4) \DD^{IJL}  + \varpi^{KI}{}_L(4) \DD^{JL} - \varpi^{IJ}{}_L(4) \DD^{KL}
 \\ && \hskip 1.02in 
+ \varpi ^I{}_L(4) \DD^{JKL}  + \varpi^{IK}{}_L(4) \DD^{JL}  - \varpi^{KJ}{}_L(4) \DD^{IL}   \Big )
 \no \\ && \hskip 0.2in
 =  {2 \pi i \over h}   \om _L(4) \big ( \DD^{JKI} - \DD^{IJK} \big )
 +{ 2 \pi^2 \over h} \Big ( \delta ^K_L \big ( \DD^{JI} - \DD^{IJ} \big ) - \delta ^I_L \big ( \DD^{KJ} - \DD^{JK} \big ) 
 \Big )
\no
 \eea
It is at this point that we use the cyclic symmetries $ \DD^{JI} = \DD^{IJ}$ and $\DD^{JKI} = \DD^{IJK}$, established earlier, to ensure that the $\mB$ monodromy of the above combination indeed cancels, so that its dependence on $z_4 $ is holomorphic and single-valued and may be expressed as a linear combination of the holomorphic Abelian differentials $\om_L(4)$ in the last equation of (\ref{5.h}). Finally, cyclic invariance of $\DD^{IJKL}$ itself may be  established using the interchange lemmas and Fay identities of \cite{DHoker:2024ozn} in analogy with the $n=3$ case.  

\sm

As was already noted for the case $n=3$, a more streamlined proof of the cyclic invariance of $\DD^{IJKL}$ will be presented in section \ref{sec:4} which applies to the case of arbitrary $n$ with equal ease.  Similar computations lead to the reflection relation $\DD^{IJKL}= \DD^{L KJI}$, see appendix \ref{sec:refDs} for a proof by direct computation and appendix \ref{sec:moreprf} for a proof based on general arguments that applies to arbitrary $n$.

\subsection{Formulation of the case of arbitrary $n$}
\label{sec:2.4}

The case of arbitrary $n$ may be constructed by extending the pattern observed for the cases $n=2,3,4$. The lowest rank cases may be calculated \textit{by hand} and are given by,
\bea
\label{2.desc.0}
\om_J(1) \DD^{J} (2, \cdots,n) 
& = & C_\delta  (1, \cdots, n) 
+ \big ( \chi(1,2) - \chi(1,n) \big ) C_\delta (2,\cdots, n) 
 \\ 
\om _J (2) \DD ^{I_1J} (3, \cdots, n) & = & 
 \DD^{I_1} (2, \cdots,n) +  \big ( \chi^{I_1} (2,3) {-}   \chi^{I_1} (2,n) \big ) C_\delta (3,\cdots, n)
 \no \\ && 
- g^{I_1}{}_J(2,3) \DD ^{J} (3, \cdots , n)  
\no \\
\om _J (3) \DD ^{I_1 I_2 J} (4, \cdots, n)  
& = &
\DD ^{I_1  I_2 } (3, \cdots , n) 
 + \big (   \chi^{I_2   I_1} (3,4) {-}   \chi^{I_2   I_1} (3,n) \big ) C_\delta (4,\cdots, n) 
 \no \\ &&
 - g^{I_2}{}_J (3,4) \DD^{I_1 J} (4, \cdots, n)  
  - g^{I_2I_1}{}_J (3,4) \DD^{J} (4, \cdots, n) 
\no
\eea
The first line is valid for $n \geq 3$, the second for $n \geq 4$ and the third for  $n \geq 5$. To extend the pattern to arbitrary rank, it will be convenient to rearrange the relations so that the number of points $z_i \in\Sigma$ involved in each relation equals $n$. Throughout, we set,
\bea
\label{2.desc.1}
\DD^\emptyset (1,\cdots, n) = C_\delta(1,\cdots, n)
\eea
The resulting \textit{meromorphic descent equations} are given by the theorem below.

{\thm
\label{2.thm:3} 
The  multiplet $\DD^{I_1 \cdots I_{s+1}}{}(2, \cdots, n)$ is determined uniquely in terms of the multiplets $\DD^{I_1 \cdots I_{r}} (1, \cdots,m)$ with $0 \leq r \leq s$ and $m \leq n$ as well as Enriquez kernels by the following descent equations. 
\vskip 0.1in
\noindent (a) For $n \geq 3$,
\bea
\label{2.desc.2}
\om_J(1) \DD ^{I_1 \cdots I_r J} (2, \cdots , n) 
& = &
\DD ^{I_1 \cdots I_r } (1, \cdots , n) 
+ \Big (   \chi^{I_r \cdots  I_1} (1,2) {-}   \chi^{I_r \cdots  I_1} (1,n) \Big ) C_\delta (2,\cdots, n)   
\no \\ &&
- \sum_{i=0}^{r-1} g^{I_r  \cdots I_{i+1}}{}_J(1,2) \DD ^{I_1 \cdots I_iJ} (2, \cdots , n)  
\eea
(b) For $n =2$,
\bea
\label{2.desc.3}
\om_J(1) \DD^{ I_{1} \cdots I_r J}(2)
=
\DD ^{I_1 \cdots I_r }(1,2) 
- \partial_2 \chi^{I_r \cdots I_1}(1,2)
- \sum_{i=0}^{r-1}   g^{I_r \cdots I_{i+1} }{}_J(1,2) \DD^{ I_{1} \cdots I_i J}(2)
\quad
\eea
(c) The constant multiplets, defined for $r \geq 2$ by,
\bea
\DD^{I_1 \cdots I_r} & = & \sum_{k=0}^{r-1} \sum_{\ell=0}^{r-1-k} { (-)^k ( 2 \pi i )^{k+\ell} \over (k+\ell+1) k ! \ell !}
\sum_J  \delta ^{I_1 \cdots I_\ell}_J\,  \delta ^{I_{r-k} \cdots I_r}_J\, \oint _{\mA^J} dt \, \DD^{  I_{\ell+1} \cdots I_{r-1-k} } (t) \, 
\label{defcsts}
\eea
and for $r=0,1$ to vanish, are invariant under cyclic permutations of the indices, 
\bea
\label{2.desc.5cyc}
\DD^{I_1 I_2 \cdots I_r} = \DD^{I_2 \cdots I_r I_1}
\eea
(d) The constant multiplets $\DD^{I_1 \cdots I_r}$ defined by (\ref{defcsts})
satisfy the following descent equation
\bea
\label{2.desc.4}
\om_J(1) \DD ^{I_1 \cdots I_r J} 
& = &
\DD^{I_1 \cdots I_r }(1) 
- \sum_{{0 \leq i < j \leq r \atop (i,j) \not= ( 0,r)} } (-1)^i 
\varpi ^{ I_1 \cdots I_i \, \shuffle \, I_r  \cdots  I_{j+1}  }{}_J(1) \DD^{I_{i+1} \cdots I_j J} 
\quad
\eea
involving the shuffle product $\shuffle$\footnote{The 
shuffle product $\shuffle$ is a binary operation on a pair of sequences (or words) $I_1\cdots I_r$ and $J_1\cdots J_s$ which is commutative and associative with neutral element the empty set.  For arbitrary non-empty sequences with $r,s\geq 1$, the shuffle product may be defined through the recursion relation, 
\[
I_1\cdots I_r \shuffle J_1\cdots J_s = I_1(I_2 \cdots I_r \shuffle J_1\cdots J_s) + J_1(I_1\cdots I_r \shuffle J_2\cdots J_s)
\]
along with the operation of the neutral element for $s=0$, namely $I_1\cdots I_r \shuffle \emptyset = \emptyset \shuffle I_1\cdots I_r = I_1\cdots I_r$.} which extends (\ref{2.desc.2}) and (\ref{2.desc.3}) to the case of $n=1$.}

\sm
 
The proof of items $(a)$ and $(b)$ of the theorem will be carried out in section~\ref{sec:2.5} below, while the proof of items $(c)$ and $(d)$ are relegated to section~\ref{sec:4}. As will be detailed in section \ref{sec:5}, iterative use of items $(a)$, $(b)$ and $(d)$ of the theorem leads to the advertised decomposition
(\ref{intro.03}) of cyclic products of Szeg\"o kernels that separates their spin structure dependence from
their dependence on the points.

\subsection{Proof of items $(a)$ and $(b)$ of Theorem \ref{2.thm:3}}
\label{sec:2.5}

The proof of items $(a)$ and $(b)$ of Theorem \ref{2.thm:3} proceeds by induction in the rank $r$ for $n \geq 2$. We shall show that the descent equations hold for $r=0$ as the initial step in the induction procedure in $r$. The second step in the induction procedure is to show that the descent equations for $r=s$ follow from the assumption that the descent equations (\ref{2.desc.2}) and (\ref{2.desc.3}) hold for all $r \leq s-1$.

\sm

For $r=0$ and $n \geq 3$, the recursion relation of (\ref{2.desc.2}) coincides with the first relation in (\ref{2.desc.0}). To establish its validity, we use the $r=0$ relation, 
\bea
\label{2.desc.5pol}
\bar \p_k \DD^\emptyset (1,\cdots, n) = \pi \, \big ( \delta (k,k+1) - \delta (k,k-1) \big ) \, \DD^\emptyset (1, \cdots, \hat k, \cdots, n)
\eea
for the cyclic product (\ref{2.desc.1}) to verify that the right side is holomorphic in $z_1$ (since the residues of the poles in $z_1$ at $z_2$ and $z_n$ cancel) and single-valued in $z_1$ since the difference $\chi(1,2)-\chi(1,n)$ is single-valued in $z_1$ in view of the first equation of (\ref{A.chi}). Therefore the left side must be a linear combination of the holomorphic Abelian differentials $\om_J(1)$ whose coefficients are independent of $z_1$ and are denoted by $\DD^{I_1 \cdots I_r J}(2,\cdots,n)$. For $r=0$ and $n=2$, the recursion relation (\ref{2.desc.3})  coincides with (\ref{2.1.a}) whose validity was already proven in section~\ref{sec:2.2}. Thus, items $(a)$ and $(b)$ of Lemma~\ref{2.thm:3} hold true for $r=0$. 

\sm

The induction step from ranks $r \leq s-1$ to rank $r=s$ will be proven in 
sections \ref{restprfa} and \ref{restprfb} using two lemmas for the properties of $\DD^{I_1 \cdots I_r} (1, \cdots, n)$:  the first giving differential equations and the second giving  monodromies.

{\lem 
\label{2.lem:4}
Assuming the relations (\ref{2.desc.2}) and (\ref{2.desc.3})  for all $r \leq s-1$,  the multiplets $\DD^{I_1 \cdots I_r} (1,\cdots, n)$  satisfy the following equations  for all $1 \leq r \leq s$ and $2 \leq k \leq n-1$,
\bea
\label{2.desc.6}
\bar \p_1 \DD^{I_1 \cdots I_r} (1, \cdots, n) & = & \pi \, \delta (1,2) \DD^{I_1 \cdots I_r} (2, \cdots , n)
\no \\
\bar \p _k \DD^{I_1 \cdots I_r} (1, \cdots, n) 
& = & \pi \, \Big ( \delta (k,k+1)-\delta(k,k-1) \Big ) \DD^{I_1 \cdots I_r} (1, \cdots, \hat k, \cdots, n) 
\no \\
\bar \p _n \DD^{I_1 \cdots I_r} (1, \cdots, n) 
& = & - \pi \, \delta(n,n-1)  \DD^{I_1 \cdots I_r} (1, \cdots, n-1) 
\eea
where the middle equation above is  absent when $n=2$. For $1 \leq r \leq s$ and $n=1$ we have, 
\bea
\label{2.desc.7}
\bar \p_1 \DD^{I_1 \cdots I_r}(1) & = & 0
\eea}

The proof of this lemma is relegated to appendix \ref{sec:Ba}. 

\sm

{\lem 
\label{2.lem:5}
The cyclic product $\CD(1,\cdots, n)$ is single-valued in $z_1, \cdots, z_n$ for all $n$.  Assuming that the relations (\ref{2.desc.2}) and (\ref{2.desc.3}) hold for all $r \leq s-1$, the  multiplets $\DD^{I_1 \cdots I_r}(1,\cdots, n)$ have vanishing $\mA$-monodromies in $z_1, \cdots ,z_n$, while their $\mB$-monodromies are given as follow for all values $1 \leq r \leq s$ and $n \geq 2$ with $2 \leq k \leq n-1$, 
\bea
\label{2.desc.8}
\Delta _L^{(1)} \DD^{I_1 \cdots I_r} (1, \cdots, n) & = &
\sum _{\ell=1}^{r} { (-2 \pi i )^\ell \over \ell !} \, \delta ^{I_{r} \cdots I_{r+1 -\ell}} _L \DD^{I_1 \cdots I_{r-\ell}} (1,\cdots, n)
\no \\
\Delta _L^{(k)} \DD^{I_1 \cdots I_r} (1, \cdots, n) & = & 0
\no \\
\Delta _L^{(n)} \DD^{I_1 \cdots I_{r}} (1, \cdots, n) & = &
\sum _{\ell=1}^{r} { (2 \pi i )^\ell \over \ell !} \, \delta ^{I_1 \cdots I_\ell} _L \DD^{I_{\ell+1} \cdots I_r} (1,\cdots, n)
\eea
where the middle equation is absent for $n=2$.}

\sm

The expression for $\Delta _L ^{(1)} \DD^{I_1 \cdots I_r}(1)$  will be derived in compact form in terms of generating functions in section \ref{sec:3} below, and the proof of  Lemma \ref{2.lem:5} is relegated to appendix \ref{sec:B}.

\subsubsection{Proof of item $(a)$ of Theorem \ref{2.thm:3}}
\label{restprfa}

It remains to prove the induction step giving the derivation of the descent equations of (\ref{2.desc.2}) and (\ref{2.desc.3}) for $r=s$, assuming the validity of the descent equations for $0 \leq r \leq s-1$. To this end, we define the  combination $\tilde D_\delta^{I_1 \cdots I_s } (1, \cdots , n)$ for arbitrary $n \geq 3$ by, 
\bea
\label{2.tilde}
\tilde D_\delta^{I_1 \cdots I_s } (1, \cdots , n)  & = & 
\DD ^{I_1 \cdots I_s } (1, \cdots , n) 
+ \Big (   \chi^{I_s \cdots  I_1} (1,2) {-}   \chi^{I_s \cdots  I_1} (1,n) \Big ) C_\delta (2,\cdots, n)   
\no \\ && 
- \sum_{i=0}^{s-1} g^{I_s  \cdots I_{i+1}}{}_J(1,2) \DD ^{I_1 \cdots I_iJ} (2, \cdots , n)  
\eea
where we have chosen the right side to be the combination that enters the descent equation (\ref{2.desc.2}) for $r=s$ and the corresponding value of $n$. To prove the induction step, it will suffice to prove that $\tilde D_\delta^{I_1 \cdots I_s } (1, \cdots , n)$ is single-valued and holomorphic in $z_1$, so that it must a  linear combination of the holomorphic Abelian differentials $\om_J(1)$ and defines the corresponding coefficients $D_\delta^{I_1 \cdots I_s J} (2, \cdots , n)$ of rank $s+1$ by,
\bea
\tilde D_\delta^{I_1 \cdots I_s } (1, \cdots , n) = \om_J(1)  D_\delta^{I_1 \cdots I_s J} (2, \cdots , n)
\eea
To prove these properties of $\tilde D_\delta^{I_1 \cdots I_s } (1, \cdots , n)$, we make use of Lemmas \ref{2.lem:4} and \ref{2.lem:5}, as we shall now show in the remainder of this subsection.

\sm

To prove that $\tilde D_\delta^{I_1 \cdots I_s } (1, \cdots , n)$ is holomorphic in $z_1$ we use the fact that $0 \leq r \leq s-1$ implies that $1 \leq s$ so that the combination $\chi^{I_s \cdots  I_1} (1,2) {-}   \chi^{I_s \cdots  I_1} (1,n)$ is holomorphic inside the preferred fundamental domain $D$.  Furthermore, only the contribution from $i=s-1$ to the sum on the second line has a pole, so that,
\bea
\bar \p _1 \tilde D_\delta^{I_1 \cdots I_s } (1, \cdots , n)  & = & 
\bar \p_1 \DD ^{I_1 \cdots I_s } (1, \cdots , n) 
- \pi \delta (1,2)  \DD ^{I_1 \cdots I_s} (2, \cdots , n)  
\eea
In view of of the first line of (\ref{2.desc.8}) in Lemma \ref{2.lem:4}, which holds for $r=s$, the right side vanishes, so that $\tilde D_\delta^{I_1 \cdots I_s } (1, \cdots , n) $ is indeed holomorphic in $z_1$.

\sm 

To prove that $\tilde D_\delta^{I_1 \cdots I_s } (1, \cdots , n)$ is single-valued in $z_1$, we use the fact that its $\mA$ monodromy vanishes by construction while   its $\mB$ monodromy is given as follows, 
\bea
\Delta _L ^{(1)} \tilde D_\delta^{I_1 \cdots I_s } (1, \cdots , n)  & = & 
\Delta _L ^{(1)} \DD ^{I_1 \cdots I_s } (1, \cdots , n) 
- \sum_{i=0}^{s-1} \Delta _L ^{(1)} g^{I_s  \cdots I_{i+1}}{}_J(1,2) \DD ^{I_1 \cdots I_iJ} (2, \cdots , n) 
\no \\ &&
+  \Big ( \Delta _L ^{(1)}  \chi^{I_s \cdots  I_1} (1,2) {-}  \Delta _L ^{(1)} \chi^{I_s \cdots  I_1} (1,n) \Big ) C_\delta (2,\cdots, n)   
\eea
The first term on the right side may be evaluated using the first equation in (\ref{2.desc.8}) of Lemma \ref{2.lem:5}, while the monodromies of $g$ and $\chi$ are given by the first equations in (\ref{2.4.mon}) and (\ref{mclim.01}), respectively. Assembling all contributions and rearranging the double sum that emerges in the last line  above, we obtain,
\bea
\Delta _L ^{(1)} \tilde D_\delta^{I_1 \cdots I_s } (1, \cdots , n)  & = & 
\sum _{k=1}^s { (-2 \pi i)^k \over k!} \, \delta ^{I_s \cdots I_{s+1-k}}_L \bigg [
\DD^{I_1 \cdots I_{s-k}} (1, \cdots, n) 
\no \\ && \qquad
+ \Big ( \chi^{I_{s-k} \cdots  I_1} (1,2) -  \chi^{I_{s-k} \cdots  I_1} (1,n) \Big ) C_\delta (2,\cdots, n)   
\no \\ && \qquad
- \sum_{i=0}^{s-k} g^{I_{s-k}  \cdots I_{i+1}}{}_J(1,2) \DD ^{I_1 \cdots I_iJ} (2, \cdots , n)  \bigg ]
\eea
The combination inside the square brackets vanishes by the descent equation of (\ref{2.desc.2}) for $r=s-k \leq s-1$. 
This concludes the proof of item $(a)$ of Theorem \ref{2.thm:3}.

\subsubsection{Proof of item $(b)$ of Theorem \ref{2.thm:3}}
\label{restprfb}

In item $(b)$ we have $n=2$ and we define the corresponding combination, 
\bea
\tilde D_\delta^{I_1 \cdots I_s }(1,2) =
\DD ^{I_1 \cdots I_s }(1,2) 
- \partial_2 \chi^{I_s \cdots I_1}(1,2)
- \sum_{i=0}^{s-1}   g^{I_s \cdots I_{i+1} }{}_J(1,2) \DD^{ I_{1} \cdots I_i J}(2)
\eea
in (\ref{2.desc.3}) at $r=s$. Its holomorphicity in $z_1$ follows from the facts that $\partial_2 \chi^{I_s \cdots I_1}(1,2)$ is holomorphic in $z_1 \in D$ for $s \geq 1$ combined with the first line of (\ref{2.desc.6}) of Lemma \ref{2.lem:4}. The multiplet $\tilde D_\delta^{I_1 \cdots I_s }(1,2)$ is single-valued in $z_1$ since its $\mA$ monodromy vanishes by construction while its $\mB$ monodromy in $z_1$ may be evaluated using the first equation in (\ref{2.desc.8}) of Lemma \ref{2.lem:5}, while the monodromies of $g$ and $\chi$ are given by the first lines in (\ref{2.4.mon}) and (\ref{mclim.01}), respectively.  After some simplifications, we obtain, 
\bea
\Delta _L ^{(1)} \tilde D_\delta^{I_1 \cdots I_s } (1, 2)  & = & 
\sum _{k=1}^s { (-2 \pi i)^k \over k!} \, \delta ^{I_s \cdots I_{s+1-k}}_L \bigg [
\DD^{I_1 \cdots I_{s-k}} (1, 2)  + \p_2  \chi^{I_{s-k} \cdots  I_1} (1,2)   
\no \\ && \hskip 0.8in
- \sum_{i=0}^{s-k} g^{I_{s-k}  \cdots I_{i+1}}{}_J(1,2) \DD ^{I_1 \cdots I_iJ} (2)  \bigg ]
\eea
The combination inside the square brackets vanishes by the descent equation of (\ref{2.desc.2}) for $r=s-k \leq s-1$. Therefore, $\tilde D_\delta^{I_1 \cdots I_s } (1, 2)$ must be a linear combination of the holomorphic Abelian differentials $\om_J(1)$ and we shall define the corresponding coefficients $D_\delta^{I_1 \cdots I_s J} (2)$ of rank $s+1$ by,
\bea
\tilde D_\delta^{I_1 \cdots I_s }(1,2) & = & \om_J(1)  D_\delta^{I_1 \cdots I_s J}(2)
\eea
This completes the proof of item $(b)$ of Theorem \ref{2.thm:3}.

\newpage

\section{Generating functions}
\setcounter{equation}{0}
\label{sec:3}

In this section, the multiplets $\DD$ will be collected in a generating function $\DDD$ analogously to how the Enriquez kernels $g$ were collected in the  form $\KK_J$ given in (\ref{2.KJ}).  Both generating functions are valued in an infinite-dimensional  Lie algebra $\mg_b$ that is freely generated by $h$ elements $b=\{ b_1 , \cdots, b_h \}$. 
The recursion relations of Theorem~\ref{2.thm:3}, the differential equations of Lemma~\ref{2.lem:4}, and the monodromies of Lemma~\ref{2.lem:5}  will be compactly expressed in terms of these generating functions and used to provide the proof of items $(c)$ and $(d)$ of Theorem \ref{2.thm:3} in section \ref{sec:4}, thereby completing the proof of the theorem.

\subsection{Generating functions for the Enriquez kernels}

In this first subsection, we begin by reviewing and elaborating on the properties of the connection form $\KK_J$, viewed as a generating function for the Enriquez kernels $g^{I_1 \cdots I_r} {}_J$. Recall that the Taylor series expansion of the  form $ \KK_J(x,y;B)$ is given by equation (\ref{2.KJ}) which we repeat here for convenience, 
\bea
\KK_J(x,y;B) = \sum _{r=0}^\infty g^{I_1 \cdots I_r} {}_J (x,y) B_{I_1} \cdots B_{I_r}
\eea
with $g^\emptyset {}_J(x,y) = \om_J(x)$ as was stated already in footnote~\ref{foot:2} and where the $h$ elements $B_I$ generate the free Lie algebra $\mg_b$ by adjoint action $B_I X = [ b_I, X]$ for all $X \in \mg_b$.\footnote{Henceforth, the generators $a^1,\cdots,a^h$ will no longer play a role. The generators $b_I$ will enter only through their adjoint action via  $B_I$. The algebra freely generated by the $B_I$ is isomorphic to $\mg_b$.}
 It will be useful to introduce separate generating functions that
capture the decomposition (\ref{2.dec}) of Enriquez kernels $g^{I_1 \cdots I_r} {}_J (x,y) $
into their trace and traceless parts, given by $\chi ^{I_1 \cdots I_r}  (x,y) $ and $\varpi ^{I_1 \cdots I_r} {}_J (x)$, respectively,  as follows,
\begin{align}
\label{gfsKW} 
\XX (x,y;B) & =  \sum _{r=0}^\infty \chi ^{I_1 \cdots I_r}  (x,y) B_{I_1} \cdots B_{I_r}
\no \\ 
\OO _J(x;B) & =  \sum _{r=0}^\infty \varpi ^{I_1 \cdots I_r} {}_J (x) B_{I_1} \cdots B_{I_r}
\end{align}
with $\varpi ^\emptyset {}_J(x) = \om_J(x)$. The decomposition of (\ref{2.dec}) is then equivalent to,
\bea
\KK_J(x,y;B) & =  \OO_J(x;B) - \XX(x,y;B) B_J
\eea

The basic differential equations are stated in a \textit{preferred fundamental domain} $D$ for $\Sigma$ (see figure \ref{fig:1}) in which  $g^I{}_J(x,y)$ has a simple pole while $g^{I_1 \cdots I_r}{}_J(x,y)$ for $r=0$ and for $r \geq 2$ are non-singular.  For $x,y \in D$ we then have,
\begin{align}
\label{3.Kder}
\pbx \KK_J(x,y;B) & =  \pi \delta (x,y) B_J & \pbx \XX(x,y;B) & =  - \pi \delta (x,y) 
\no \\
\pby \KK_J(x,y;B) & = - \pi \delta (x,y) B_J & \pby \XX(x,y;B) & =   \pi \delta (x,y) 
\no \\
&& \pbx \OO_J(x;B) & = 0
\end{align}
Beyond the fundamental domain we will use the monodromy equations, to be derived below, to extend the above formulas to all of $\tilde \Sigma$ (see the discussion preceding figure \ref{fig:1}). As a result,  $\KK_J(x,y;B)$ and $\XX(x,y;B)$ will have simple poles at the points $\pi^{-1}(y)$ with residues that involve all powers of $B$ so that $g^{I_1 \cdots I_r}{}_J(x,y)$ will have poles for all $r \geq 1$.

\sm

The monodromy relations for $g$ in (\ref{2.4.mon}), for $\chi$ in (\ref{mclim.01}), and for $\varpi$ in (\ref{trcless}) translate into the following monodromy relations for $\KK_J$, 
\bea
\label{3.Kmon1}
\KK_J(\mB_L \cdot x, y;B) & = & e^{- 2 \pi i B_L} \, \KK_J (x,y;B)
\no \\
\KK_J(x, \mB_L \cdot y;B) & = & \KK_J (x,y;B) + \KK_L (x,y;B) \, { e^{ 2 \pi i B_L} -1 \over B_L} \, B_J
\eea
for the trace part $\XX$,
\bea
\label{3.Kmon2}
\XX(\mB_L \cdot x, y;B) & = & e^{-2 \pi i B_L} \, \XX(x,y;B) - { 1 \over h} \, { e^{-2 \pi i B_L} -1 \over B_L} \, \om_L(x)
\no \\
\XX(x, \mB_L \cdot y;B) & = & \XX(x,y;B) - \KK_L(x,y;B) \, { e^{ 2 \pi i B_L} -1 \over B_L}
\eea
and for the traceless part $\OO_J$, 
\bea
\label{3.Kmon3}
\OO_J(\mB_L \cdot x;B) = e^{-2 \pi i B_L} \OO_J(x;B) - { 1 \over h} \, { e^{-2 \pi i B_L} -1 \over B_L} \, B_J \om_L(x)
\eea
Finally, the integral relations for $g$ in (\ref{2.intg}) translate into the following integral of $\KK_J$,
\bea
\label{3.Kper}
\oint _{\mA^L} dt \, \KK_J(t,y;B) =  { - 2 \pi i B_L \over e^{-2 \pi i B_L} -1} \, \delta ^L_J
\eea
using the generating function for the Bernoulli numbers given in footnote \ref{foot:5}.

\subsection{Generating functions for  $\DD$}

The multiplets $\DD^{I_1 \cdots I_r}(1,\cdots, n)$ may be collected in a generating function that takes values in the Lie algebra $\mg_b$. Setting $\DD^\emptyset (1,\cdots, n) = C_\delta(1,\cdots, n)$ as in (\ref{2.desc.1}), we introduce the following closely related generating functions for arbitrary $n \geq 0$,\footnote{The reversal of the order in which the contractions with the generators $B_I$ are performed results from our choice for the order of the indices on $\DD$ and is purely a matter of (perhaps unfortunate) convention.} 
\bea
\label{4.gen}
\DDD(1,\cdots, n;B) & = & \sum_{r=0} ^\infty \DD^{I_1 \cdots I_r} (1,\cdots, n) B_{I_r} \cdots B_{I_1}
\no \\
\DDD^J(1,\cdots, n;B) & = & \sum_{r=0} ^\infty \DD^{I_1 \cdots I_r J} (1,\cdots, n) B_{I_r} \cdots B_{I_1}
\eea
These relations combined with (\ref{2.desc.1}) for $n \geq 2$ imply,
\bea
\label{10.DJD}
\DDD(1,\cdots, n;B)  & = & \CD(1,\cdots, n) + B_J \DDD^J(1,\cdots, n;B)   \hskip 0.6in n \geq 2
\eea
while for $n=0,1$ we have,
\begin{align}
\DD^\emptyset (\emptyset;B)  & = 0 & \DDD(\emptyset;B)  & =  B_J \DDD^J(\emptyset;B) 
\no \\
\DD^\emptyset (1;B)  & = 0 & \DDD(1;B)  & =  B_J \DDD^J(1;B)  
\end{align}
The generating function $\DDD^J$ may be deduced from $\DDD$  by the operation \textit{Iota}  on elements in the Lie algebra $\mg_b$, defined by, 
\bea
\label{10.iota}
\cI^J 1 = 0
\hskip 1in
\cI^J \big ( B_{I} X \big ) = \delta ^J_I \, X
\eea
for any $X \in \mg$. The relation in (\ref{10.DJD}) then gives,
\bea
\label{3.iotaD}
\DDD^J (1, \cdots, n)= \cI^J \DDD(1, \cdots, n)
\eea
Thus, the operation $\cI^J$ \textit{removes} the first factor of $B$ in any Taylor series in powers of $B$ to which it is being applied and generates the extra upper index $J$ as a result.

\subsection{Recursion  relations for $\DD$}

The recursion relations on the components $\DD^{I_1 \cdots I_r}(1,\cdots,n)$ for $n \geq 3$ in item $(a)$ and for $n=2$ in item $(b)$ of Theorem  \ref{2.thm:3} may be reformulated compactly  in terms of the generating functions $\DDD$ and  $\DDD^J$ in (\ref{4.gen}) via the following corollary.

{\cor
\label{3.cor:1}
The generating functions $\DDD$ and  $\DDD^J$ satisfy the following relations.\\
(a) For $n \geq 3$, 
\bea
\label{3.rec}
\DDD(1,\cdots, n;B) & = &  \KK_J(1,2;B)  \DDD^J(2,\cdots, n;B)  
\no \\ &&
-  \big ( \XX(1,2;B) - \XX(1,n;B) \big ) \CD(2,\cdots, n)
\eea
(b) For $n=2$, 
\bea
\DDD(1,2;B) =  \KK_J(1,2;B)  \DDD^J(2;B) + \p_2 \XX(1,2;B)
\label{dd12}
\eea
}

Parts $(a)$ and $(b)$ of the corollary are proven by reformulating items $(a)$ and $(b)$, respectively, of Theorem \ref{2.thm:3} into generating functions.

\subsection{Differential relations}

The system of differential equations of (\ref{2.desc.6}) and (\ref{2.desc.7}) of Lemma \ref{2.lem:4} on the components $\DD^{I_1 \cdots I_r}(1,\cdots,n)$ may be readily reformulated in terms of the differential equations for the generating functions given in the corollary below.  Equivalently, they may be derived directly from the relations (\ref{3.rec}) and (\ref{dd12}) on the generating functions together with the antiholomorphic derivatives (\ref{3.Kder}) of the generating series for Enriquez kernels.

{\cor
\label{3.cor:2}
The generating functions $\DDD$ satisfy the following differential relations. \\
(a) For $n \geq 3$ and $2 \leq k \leq n-1$, 
\bea
\label{genDdiffs}
\bar \p_1 \DDD(1,\cdots, n;B) & = & \pi \, \delta (1,2) \DDD(2,\cdots, n;B) - \pi \, \delta (1,n) C_\delta(2,\cdots, n)
 \\
\bar \p_k \DDD(1,\cdots, n;B) & = & \pi \Big ( \delta (k,k+1) - \delta (k, k-1) \big ) \DDD(1, \cdots, \hat k , \cdots, n;B)
\no \\
\bar \p_n \DDD(1,\cdots, n;B) & = & \pi \, \delta (n,1) C_\delta (1, \cdots, n-1) - \pi \, \delta (n,n-1) \DDD(1,\cdots, n-1;B)
\no 
\eea
(b) For $n=1,2$,
\bea
\bar \p_1 \DDD(1,2;B) & = & \pi \, \delta (1,2) \, \DDD(2;B) + \pi \, \p_1 \delta(1,2)
\no \\
\bar \p_2 \DDD(1,2;B) & = & - \pi \, \delta (1,2) \, \DDD(1;B) + \pi \, \p_2 \delta(1,2)
\no \\
\bar \p_1 \DDD(1;B) & = & 0
\eea}

The differential relations for the generating functions $\DDD^J$ may be easily obtained by applying the operator $\cI^J$ introduced in (\ref{10.iota}) to the relations of Corollary \ref{3.cor:2}. For $n \geq 2$ and $2 \leq k \leq n-1$ we obtain,
\bea
\bar \p_1 \DDD^J(1,\cdots, n;B) & = & \pi \, \delta (1,2) \DDD^J(2,\cdots, n;B) 
\no \\
\bar \p_k \DDD^J(1,\cdots, n;B) & = & \pi \big ( \delta (k,k+1) - \delta (k, k-1) \big ) \DDD^J(1, \cdots, \hat k , \cdots, n;B)
\no \\
\bar \p_n \DDD^J(1,\cdots, n;B) & = & - \pi \, \delta (n,n-1) \DDD^J(1,\cdots, n-1;B)
\eea
where the middle equation is absent for $n=2$,  while for $n=1$ we have,
\bea
\bar \p_1 \DDD^J(1;B) = 0
\eea

\subsection{Monodromy relations for $n \geq 2$}

The monodromy relations  of (\ref{2.desc.8}) in Lemma \ref{2.lem:5} on the components $\DD^{I_1 \cdots I_r}(1,\cdots,n)$ may be reformulated in terms of monodromy relations for the generating functions given in the corollary below.  Equivalently, they may be derived directly by evaluating the monodromy of the recursion relations (\ref{3.rec}) and (\ref{dd12}) on the generating functions through the monodromies (\ref{3.Kmon1}) and (\ref{3.Kmon2}) of the contributing series in Enriquez kernels.

{\cor
\label{3.cor:3}
The generating functions $\DDD$ satisfy the following monodromy relations for $n \geq 2$ and $2 \leq k \leq n-1$,
\bea
\label{3.cor.3}
\DDD(\mB_L \cdot 1,  \cdots, n;B) & = & e^{- 2 \pi i B_L} \DDD(1,\cdots, n;B)
\no \\
\DDD(1,  \cdots, \mB_L \cdot k, \cdots n;B) & = & \DDD(1,\cdots, n;B)
\no \\
\DDD( 1,  \cdots, \mB_L \cdot n;B) & = & \DDD(1,\cdots, n;B) \, e^{2 \pi i B_L} 
\eea
where the middle equation is absent when $n=2$.}

\sm

The monodromy relations for the generating functions $\DDD^J$ may be obtained by applying the operator $\cI^J$ of (\ref{10.iota}) to the monodromy relations of Corollary \ref{3.cor:3}. For $n \geq 2$ and $2 \leq k \leq n-1$ we obtain,
\bea
\DDD^J (\mB_L \cdot 1,  \cdots, n;B) & = & 
\DDD^J(1,\cdots , n;B) + \delta ^J_L \, { e^{- 2 \pi i B_L} -1 \over B_L} \, \DDD(1,\cdots, n;B)
\no \\
\DDD^J (1, \cdots ,\mB_L \cdot k, \cdots, n;B) & = &  \DDD^J (1,\cdots, n;B)
 \\
\DDD^J(1, \cdots ,  \mB_L \cdot n;B) & = &  
\DDD^J(1,\cdots, n;B) \, e^{2 \pi i B_L} + \delta ^J_L \, { e^{2 \pi i B_L} -1 \over B_L} C_\delta (1,\cdots ,n)
\no
\eea

\subsection{Monodromy relation for $\DDD(1;B)$}

The generating function $\DDD(2;B)$ is related to the generating function $\DDD(1,2;B)$ by equation (\ref{dd12}) with the help of the series $\KK_J(1,2)$ and $\XX(1,2)$ in Enriquez kenels. This relation allows us to evaluate the $\mB$ monodromy of $\DDD(2;B)$ in the variable $z_2$ in terms of the $\mB$ monodromy of $\DDD(1,2;B)$ in the same variable, which is given by the last equation of (\ref{3.cor.3}) of Corollary \ref{3.cor:3} and is stated in the lemma below. This lemma will serve as an intermediary step in the proof of items $(c)$ and $(d)$ of Theorem \ref{2.thm:3}.

{\lem
\label{3.lem:4}
The monodromy relations for the one-point functions are given by, 
\bea
\label{3.D1}
\DDD(\mB_L \cdot 1;B) & = & e^{-2 \pi i B_L} \, \DDD(1;B) \, e^{2 \pi i B_L}
\no \\
\DDD^J(\mB_L \cdot 1;B) & = & 
\DDD^J(1;B) \, e^{2 \pi i B_L} + \delta^J_L  \, { e^{- 2 \pi i B_L} -1 \over B_L} \,  \DDD(1;B) \, e^{2 \pi i B_L}
\eea}

\sm 

To prove the lemma, we apply  $\mB_L$ to the point $z_2$ in (\ref{dd12}), 
\bea
\DDD(1, \mB_L \cdot 2;B) =  \KK_J(1, \mB_L \cdot  2;B)  \DDD^J(\mB_L \cdot  2;B) + \p_2 \XX(1,\mB_L \cdot  2;B)
\eea
and then use the last equation of (\ref{3.cor.3}) for $n=2$ along with the monodromy relations given by the second equations of (\ref{3.Kmon1}) and (\ref{3.Kmon2}). The resulting identity,
\bea
  \KK_J(1,   2;B)  \DDD^J(  2;B) e^{2\pi i B_L}
   &=&   \KK_J(1,   2;B)  \DDD^J(\mB_L \cdot   2;B)  \\
   &&\quad
   +\KK_L(1,   2;B) \, \frac{e^{2\pi i B_L} - 1}{B_L} \, \DDD(\mB_L \cdot   2;B) 
   \notag
\eea
implies the first line of (\ref{3.D1}) by matching the residues of the pole in $z_1$ at $z_2$ where $\KK_J(1,   2;B) $ contributes via $B_J$ and the contractions $  \KK_J(1,   2;B)  \DDD^J(  i;B)$ thereby simplify to
$\DDD(  i;B)= B_J \DDD^J(i;B)$ (for $i=2$ or $i=\mB_L\cdot 2$).  The second equation of (\ref{3.D1}) then 
follows from applying the operator $\cI^J$ to the first equation.

\newpage

\section{Convolution periods}
\setcounter{equation}{0}
\label{sec:4}

To complete the proof of Theorem \ref{2.thm:3}, it remains to establish its items $(c)$ and $(d)$. We shall do so in this section with the help of the generating functions introduced in section~\ref{sec:3} and the use of $\mA$ periods of the multiplets $\DD^{I_1 \cdots I_r}(1, \cdots, n)$ which we shall introduce in this section. In the process we shall establish a recursion relation for the periods of the multiplets $\DD^{I_1 \cdots I_r}(1, \cdots, n)$; show that the operation of taking $\mA$ periods closes in the space of multiplets; and prove that the multiplets can be represented by multiple $\mA$ period integrals of cyclic products of Szeg\"o kernels.

\subsection{Recursion relation for $\mA$ periods}

The starting points for the recursion relations between the $\mA$ periods of the multiplets $\DD^{I_1 \cdots I_r}(1, \cdots, n)$ are the recursion relations (\ref{2.desc.2}) and (\ref{2.desc.3}) of items $(a)$ and $(b)$ of Theorem \ref{2.thm:3}, respectively, which were already proven in section \ref{sec:2.5}. To compute the $\mA$ periods,  we make use of the crucial property, shown in (\ref{2.intg}) and (\ref{3.Kper}), that the $\mA$ periods of $g^{I_1 \cdots I_r}{}_J(x,y)$ and $\KK_J(x,y)$ in the variable $x$ are independent of $y$. As a result, when taking the integrals of (\ref{2.desc.2}) and (\ref{2.desc.3}) in the variable $z_1$ over a cycle $\mA^L$,  the contributions from the terms $\chi^{I_r\cdots I_1}(1,2)-\chi^{I_r\cdots I_1}(1,n)$ and $\partial_2\chi^{I_r\cdots I_1}(1,2)$ vanish, and we obtain the following recursion relations for all $n \geq 2$, 
\bea
\label{4.DA.1}
\oint _{\mA^L} dt \, \DD ^{I_1 \cdots I_r } (t,2, \cdots , n) 
& = &
 \sum_{k=0}^{r} (-2 \pi i )^k \, {\Ber_k \over k!} \, \delta ^{I_r \cdots I_{r+1-k}}_L \, \DD^{I_1 \cdots I_{r-k} L} (2,\cdots, n)
\eea
where we recall that  $\Ber_k$ are the Bernoulli numbers (see footnote \ref{foot:5}). Reformulated in terms of generating functions, the relation (\ref{4.DA.1}) is equivalent to,
\bea
\label{4.DA.2}
\oint _{\mA^L}
dt \, \DDD (t,2, \cdots , n;B) = { - 2 \pi i B_L \over e^{-2 \pi i B_L} -1} \, \DDD^L(2,\cdots, n;B)
\eea
The prefactor on the right is invertible, since its  series expansion starts with the identity.  Its inverse gives the generating function for the multiplets $\DD^{I_1 \cdots I_r}(2,\cdots, n)$ in terms of an $\mA$ period of the generating function with an extra point $t$,
\bea
\label{4.DA.3}
\DDD^L(2,\cdots, n;B) = { e^{-2 \pi i B_L} -1 \over - 2 \pi i B_L} \oint _{\mA^L} dt \,  \DDD (t,2, \cdots , n;B) 
\eea
which translates into the following relation between the components,
\bea
\label{4.DA.4}
\DD^{I_1 \cdots I_r L} (2,\cdots, n)
= \sum_{\ell=0}^r  { (-2 \pi i)^{r-\ell} \over (r- \ell+1)!} \,  \delta ^{I_{\ell+1}  \cdots I_r}_L \oint _{\mA^L} 
dt \, \DD^{I_1 \cdots I_\ell} (t,2,\cdots, n)
\eea
For example, to low rank we have, 
\begin{align}
\label{4.DA.5}
\DD^L(2,\cdots,n) &= \oint _{\mA^L} dt \, \CD (t,2,\cdots,n)
 \no \\
\DD^{I L}(2,\cdots,n) &= 
\oint _{\mA^L} dt \, \bigg\{ \DD^{I}(t,2,\cdots,n)- i \pi \delta^{I}_L \CD (t,2,\cdots,n) \bigg\} 
\end{align}
One may now proceed by eliminating the multiplet $\DD^{I}(2,\cdots,n) $ between the first and second equations in (\ref{4.DA.5}) in order to obtain a formula for $\DD^{I L}(2,\cdots,n)$ in terms of $\mA$ periods of the cyclic product of Szeg\"o kernels only,
\bea
\label{4.DA.6}
\DD^{I L}(2,\cdots,n) = 
\oint _{\mA^L} \! \! dt_2 \,  \oint _{\mA^{I} } \! \! dt_1 \, \CD (t_1,t_2, 2,\cdots,n) 
- i \pi \delta^{I}_L \oint _{\mA^L} \! \! dt \, \CD (t,2,\cdots,n)  
\eea
Actually, the double integral on the right side is well-defined only when $L \not = I$. When  $L=I$ the pole of $C_\delta (t_1, t_2, 2, \cdots, n)$ in $t_2$ at $t_1$ sits on the contour of the integration in~$t_2$. A careful contour prescription is required for how the $t_2$ integration across this pole should proceed, which will be formulated in the next subsection.

\subsection{Prescription for multiple $\mA$ period integrations}
\label{sec:Acont}

The Enriquez kernels $g^{I_1\cdots I_r}{}_J(t,y)$ are defined for $t,y$ in the (open) interior $D^o$ of the fundamental domain $D$. To evaluate their $\mA^L$-periods in the variable $t$ using (\ref{2.intg}), we take their integral over a cycle $\mA^L _\ep$ that is homotopic to $\mA^L$ and is entirely contained in $D^o$ (barring the end points), as shown in the left panel of figure \ref{fig:eps},  and then take the limit of this integral as $\ep \to 0$ and $\mA^L_\ep \to \mA^L$ on the boundary of $D$, while leaving the point $y$ fixed in the interior $y \in D^o$. The small displacement $\mA^L _\ep$ of $\mA^L$ is understood to not cross the point $y \in D^o$ as we take $\ep \to 0$. This limiting procedure was implicit in our earlier manipulations of the $\mA$ period integrals where the $\mA$ cycles lie on the boundary of $D$ as shown in figure \ref{fig:1}. 

\begin{figure}[htb]
\begin{center}
\tikzpicture[scale=1.35]
\scope[xshift=6cm,yshift=2.2cm, scale=4.2, rotate=0]
\draw  [very  thick, color=red,  domain=179:198] plot ({cosh(0.6)*cos(20)+sinh(0.6)*cos(\x)}, {cosh(0.6)*sin(20)+sinh(0.6)*sin(\x)});
\draw  [very  thick, color=red,  dotted, domain=215:198] plot ({cosh(0.6)*cos(20)+sinh(0.6)*cos(\x)}, {cosh(0.6)*sin(20)+sinh(0.6)*sin(\x)});
\draw  [very  thick, color=red,  domain=359:340] plot ({cosh(0.65)*cos(155)+sinh(0.65)*cos(\x)}, {cosh(0.65)*sin(155)+sinh(0.65)*sin(\x)});
\draw  [very  thick, color=red,  dotted, domain=340:320] plot ({cosh(0.65)*cos(155)+sinh(0.65)*cos(\x)}, {cosh(0.65)*sin(155)+sinh(0.65)*sin(\x)});
\draw  [very  thick, color=blue,  domain=287:270] plot ({cosh(1)*cos(85)+sinh(1)*cos(\x)}, {cosh(1)*sin(85)+sinh(1)*sin(\x)});
\draw  [very  thick, color=blue, domain=270:242] plot ({cosh(1)*cos(85)+sinh(1)*cos(\x)}, {cosh(1)*sin(85)+sinh(1)*sin(\x)});
\draw  [very  thick,  domain=300:270] plot ({cosh(0.373)*cos(85)+sinh(0.7)*cos(\x)}, {cosh(0.373)*sin(85)+sinh(0.7)*sin(\x)});
\draw  [very  thick,  domain=270:229] plot ({cosh(0.373)*cos(85)+sinh(0.7)*cos(\x)}, {cosh(0.373)*sin(85)+sinh(0.7)*sin(\x)});
\draw  [very  thick,  domain=335:270] plot ({0.6*cos(85)+sinh(0.45)*cos(\x)}, {0.6*sin(85)+sinh(0.45)*sin(\x)});
\draw  [very  thick,  domain=270:195] plot ({0.6*cos(85)+sinh(0.45)*cos(\x)}, {0.6*sin(85)+sinh(0.45)*sin(\x)});
\draw (0.48,0.415) [fill=black] circle(0.015cm) ;
\draw (-0.41,0.495) [fill=black] circle(0.015cm) ;
\draw [color=red] (-0.55,0.3) node{\small  $\mB_L$};
\draw [color=blue] (0.1,0.45) node{\small $\mA^L$};
\draw  (0.1,0.25) node{\small $\mA^L_\ep$};
\draw  (0.35,0.15) node{\small $\mA^L_{2 \ep}$};
\draw  (-0.45,0.1) node{ \large $D$};
\draw (-0.29,0.1)node{$\bullet$}node[right]{$z_n$};
\draw (0,0.05)node{$\bullet$}node[left]{$z_2$};
\draw(-0.1,0.331)node{$\bullet$} node[below]{\small $t_1$};
\draw(0.2,0.155)node{$\bullet$} node[below]{\small $t_2$};
\endscope
\scope[xshift=12.5cm,yshift=2.2cm, scale=4.2, rotate=0]
\draw  [very  thick, color=red,  domain=179:198] plot ({cosh(0.6)*cos(20)+sinh(0.6)*cos(\x)}, {cosh(0.6)*sin(20)+sinh(0.6)*sin(\x)});
\draw  [very  thick, color=red,  dotted, domain=215:198] plot ({cosh(0.6)*cos(20)+sinh(0.6)*cos(\x)}, {cosh(0.6)*sin(20)+sinh(0.6)*sin(\x)});
\draw  [very  thick, color=red,  domain=359:340] plot ({cosh(0.65)*cos(155)+sinh(0.65)*cos(\x)}, {cosh(0.65)*sin(155)+sinh(0.65)*sin(\x)});
\draw  [very  thick, color=red,  dotted, domain=340:320] plot ({cosh(0.65)*cos(155)+sinh(0.65)*cos(\x)}, {cosh(0.65)*sin(155)+sinh(0.65)*sin(\x)});
\draw  [very  thick, color=blue,  domain=287:270] plot ({cosh(1)*cos(85)+sinh(1)*cos(\x)}, {cosh(1)*sin(85)+sinh(1)*sin(\x)});
\draw  [very  thick, color=blue, domain=270:242] plot ({cosh(1)*cos(85)+sinh(1)*cos(\x)}, {cosh(1)*sin(85)+sinh(1)*sin(\x)});
\draw  [very  thick,  domain=300:270] plot ({cosh(0.371)*cos(85)+sinh(0.7)*cos(\x)}, {cosh(0.371)*sin(85)+sinh(0.7)*sin(\x)});
\draw  [very  thick,  domain=270:229] plot ({cosh(0.371)*cos(85)+sinh(0.7)*cos(\x)}, {cosh(0.371)*sin(85)+sinh(0.7)*sin(\x)});
\draw(-0.1,0.331)node{$\bullet$} node[below]{\small $t_1$};
\draw(0.15,0.206)node{$\bullet$} node[above]{\small $t_{p-1}$};
\draw(0.2,0.155)node{$\bullet$} node[below]{\small $t_p$};
%
\draw(0.06,0.281)node[rotate = -5]{$\vdots$};
\draw  [very  thick,  domain=325:270] plot ({0.725*cos(85)+sinh(0.5)*cos(\x)}, {0.725*sin(85)+sinh(0.5)*sin(\x)});
\draw  [very  thick,  domain=270:205] plot ({0.725*cos(85)+sinh(0.5)*cos(\x)}, {0.725*sin(85)+sinh(0.5)*sin(\x)});
\draw  [very  thick,  domain=335:270] plot ({0.6*cos(85)+sinh(0.45)*cos(\x)}, {0.6*sin(85)+sinh(0.45)*sin(\x)});
\draw  [very  thick,  domain=270:195] plot ({0.6*cos(85)+sinh(0.45)*cos(\x)}, {0.6*sin(85)+sinh(0.45)*sin(\x)});
\draw (0.48,0.415) [fill=black] circle(0.015cm) ;
\draw (-0.41,0.495) [fill=black] circle(0.015cm) ;
\draw [color=blue] (0.1,0.41) node{\small $\mA^L$};
\draw (-0.005,0.26) node{\small $\mA^L_\ep $};
\draw (-0.29,0.1)node{$\bullet$}node[right]{$z_n$};
\draw (0,0.05)node{$\bullet$}node[left]{$z_2$};
\draw  (-0.45,0.1) node{ \large $D$};
\draw  (0.35,0.15) node{ \small $\mA^L_{p \ep}$};
\endscope
\endtikzpicture
\caption{The left panel depicts the cycles $\mA^L_\ep$ and $\mA^L_{2 \ep}$ for $\ep >0$ as a small homotopic deformation of $\mA^L$  contained in the  interior $D^o$ of $D$ (barring the end points). The right panel depicts the integration contours $\mA^L_{k \ep}$ for $k=1, \cdots , p$ for the multiple integrals in (\ref{4.DA.7}) and coincident indices $I_{i_1} = I_{i_2} =\cdots =I_{i_p} = L$,   with the associated integration variables $t_{i_1} , \cdots , t_{i_p}$. In both cases, any other arguments of the integrand, including $z_2 , \cdots ,z_n$, are assumed to be \textit{deeper inside} $D^o$ than any of the curves $\mA^L_{k \ep}$.
 \label{fig:eps}}
\end{center}
\end{figure}

\sm

When multiple integrations over $\mA$ cycles are considered, the above limiting procedure provides a unique prescription for how the nested integrations over multiple contours should be defined. Consider, for example, 
the double integral encountered in (\ref{4.DA.6}). When the indices take different values $L \not = I$, the integral is well-defined as it stands. When $L=I$, we introduce two curves  $\mA^L_\ep$ and $\mA^L_{2 \ep}$ which are homotopic to $\mA^L$, entirely contained in $D^o$,  such that $\mA^L_{2 \ep}$ lies \textit{deeper inside} $D^o$ than $\mA_\ep^L$ and all other arguments of the integrand, namely $z_2, \cdots , z_n$, lie deeper inside $D^o$ than the curve $\mA^L_{2 \ep}$. This set-up is shown in the left panel of figure \ref{fig:eps}. The double integral is then defined by,
\bea
\oint _{\mA^L} \! \! dt_2 \,  \oint _{\mA^{L} } \! \! dt_1 \, \CD (t_1,t_2, 2,\cdots,n) 
= \lim _{\ep \to 0} \oint _{\mA^L_{2 \ep} } \! \! dt_2 \,  \oint _{\mA^{L}_\ep } \! \! dt_1 \, \CD (t_1,t_2, 2,\cdots,n) 
\eea
To define the $r$-fold $\mA$ period integral of the product of Szeg\"o kernels, 
\bea
\label{4.DA.7}
\newH^{I_1 \cdots I_r} _\delta (1,\cdots, n) = 
\oint _{\mA^{I_r}} \!\! dt_r \cdots \oint _{\mA^{I_1}} \!\! dt_1 \, \CD(t_1, \cdots, t_r, 1,\cdots, n)
\eea
we proceed analogously.  When the indices $I_k$ and $I_\ell$ for $k \not= \ell$ take different values, the  integrals are well-defined as they stand. When $p$ of the  indices $I_{i_1} , \cdots , I_{i_p}$, with  $2 \leq p \leq r$,  are all equal to one another (and different from all other indices)  we shall denote this common value by $L$.  To define the integral over the variables $t_{i_1} , \cdots ,t_{i_p}$,  we introduce a sequence of contours $\mA^L _{k \ep}$ for $k=1, \cdots, p$ which are all homotopic to $\mA^L$, as shown in the right panel of  figure \ref{fig:eps}. Each contour is entirely contained in $D^o$; the contours are ordered so that $\mA^L _{k \ep}$ is \textit{deeper inside} $D^o$ than $\mA^L _{k' \ep}$ for all $k' < k$; and all other arguments of the integrand lie \textit{deeper inside} $D^o$ than $\mA^L_{p \ep}$. The $p$-fold integral is then defined by the following limit,
\bea
\lim _{\ep \to 0}  \oint _{\mA^L_{p \ep} } \!\! dt_{i_p}  \oint _{\mA^L_{(p-1) \ep} } \!\!\!\! dt_{i_{p-1}} \cdots 
\oint _{\mA^L_{ 2\ep} } \!\! dt_{i_2} \oint _{\mA^L_\ep } \!\! dt_{i_1} \, \CD(t_1, \cdots, t_r, 1,\cdots, n)
\label{contpres}
\eea
When several other indices in the multiple integral of (\ref{4.DA.7}) have the same value $K \not = L$, the above prescription is to be applied to their $\mA$ cycle integrations. We note that the same prescription to displace integration contours in iterated $\mA$ cycle convolutions was employed in the integral representations of Enriquez kernels in \cite{DHoker:2025dhv}.

\subsection{Expressing $\DDD$ in terms of multiple $\mA$ periods of $\CD$}
\label{sec:recdds}

With the integration contour prescription of section \ref{sec:Acont} in place, we are now equipped to recast the multiplets $D_\delta ^{I_1 \cdots I_r}(1,\cdots,n)$ with $n\geq 1$ in terms of the multiple $\mA$ periods of cyclic products $C_\delta$ defined in (\ref{4.DA.7}). To do so, we multiply both sides of (\ref{4.DA.3}) by $B_L$ and sum the result over all $L$,\footnote{Whenever the summation over repeated indices does not appear in the customary presentation, we shall explicitly include the corresponding summation sign in order to avoid any possible confusion. When convenient, we shall use the shorthand notation $\bz = (z_1, \cdots , z_n)$ for the ordered set of $n \geq 1 $ points.}
\beq
 B_L \DDD^L(\bz;B) =  \sum _L \beta _L  \oint_{\mA^L} dt \, \DDD(t,\bz;B) 
\label{4.beta.1}
\eeq
where $\b_L$ is given by the following series in $B_L$, 
\beq
\beta_L = \frac{e^{ -2\pi i B_L } - 1 }{-2\pi i} = B_L + \sum_{k=2}^{\infty} \frac{(-2\pi i)^{k-1}}{k!} (B_L)^k
\label{4.beta.2}
\eeq
Next, we use equation (\ref{10.DJD}) to express the left side  of (\ref{4.beta.1}) in terms of $\DDD(\bz;B)$ and $ C_\delta(\bz)$. Doing so, we obtain an integral equation that  relates generating functions $\DDD$ with different numbers of points to one another and $\CD(\bz)$,
\beq
\DDD(\bz;B) = \CD(\bz)+  \sum _L \beta _L  \oint_{\mA^L} dt \, \DDD(t,\bz;B) 
\label{4.beta.3}
\eeq
The first term on the right side is independent of the generators $B$ while the Taylor series expansion of $\beta$ starts at first order $B$, as shown in the second equality in (\ref{4.beta.2}). Therefore, by iterating equation (\ref{4.beta.3}) repeatedly, we obtain the Taylor expansion of $\DDD(\bz;B)$ in powers of $\beta$ with coefficients given by the multiple $\mA$ period integrals $\newH^{I_1\cdots I_r}_\delta(\bz)$ defined in (\ref{4.DA.7}) with the contour prescription (\ref{contpres}). For example, a single iteration of (\ref{4.beta.3}) gives,
\bea
\label{4.beta.4}
\DDD(\bz;B) & = & C_\delta(\bz) + \beta_{I_1}  \newH^{I_1} _\delta (\bz)
+ \sum_{I_1, I_2} \beta_{I_2} \beta _{I_1}  \oint_{\mA^{I_2}} dt_2 \oint_{\mA^{I_1}} dt_1 \, \DDD(t_1,t_2,\bz;B) 
\notag \\
&= & C_\delta(\bz) + \beta_{I_1}  \newH^{I_1} _\delta (\bz)
+ \beta_{I_2} \beta_{I_1}  \newH^{I_1 I_2}_\delta (\bz) + {\cal O}(\beta ^3) 
\eea
while the all-order expression for an arbitrary number $n\geq 1$ of points in $\bf z$ takes the form,
\bea
\label{4.beta.5}
\DDD(\bz;B) = C_\delta(\bz) + \sum_{r=1}^{\infty} \beta _{I_r} \cdots \beta_{I_1} \, \newH^{I_1 \cdots I_r} _\delta (\bz)  
\eea
By expanding each $\beta_{I_\ell}$ in powers of $B_{I_\ell}$ according to (\ref{4.beta.2}) and extracting the coefficient of $B_{I_r}\cdots B_{I_1}$, we obtain the  components $\DD^{I_1 \cdots I_r}(\bz)$, which at low rank are given  by,
\bea
\DD^{I_1}(\bz) & = & \newH^{I_1} _\delta(\bz) 
\label{4.beta.6} \\
\DD^{I_1 I_2}(\bz) & = & \newH_\delta ^{I_1I_2}(\bz) - i \pi \, \delta ^{I_1}_{I_2} \, \newH_\delta^{I_2}(\bz) 
\no \\
\DD^{I_1 I_2 I_3}(\bz) & = &  \newH_\delta ^{I_1 I_2 I_3}(\bz)  
- i \pi \, \Big [ \delta ^{I_1}_{I_2} \, \newH_\delta^{I_2 I_3}(\bz) 
 + \delta ^{I_2}_{I_3} \, \newH_\delta^{I_1I_3}(\bz) \Big]    
 -{2 \pi^2 \over 3}  \delta^{I_2 I_1 }_{I_3} \, \newH_\delta^{I_3}(\bz) 
\no
\eea
with no summation over any of the repeated indices.

\subsection{Construction  of the constant multiplets $\DD^{I_1 \cdots I_r}$}
\label{sec:recs.02}

We now return to completing the construction of the system of descent equations of Theorem \ref{2.thm:3} and to proving the remaining items $(c)$ and $(d)$, both of which involve constant multiplets denoted by $\DD^{I_1 \cdots I_r}$. In this subsection, we will  connect the definition (\ref{defcsts}) of these constant multiplets with the  $\mA$ periods of the previous subsections. This will pave the way for proving their cyclic symmetry property (\ref{2.desc.5cyc}) of item $(c)$ in section \ref{restthm} and then deriving the representation (\ref{2.desc.4}) of the one-point function
$D_\delta^{I_1\cdots I_r}(1)$ of item $(d)$ in section \ref{sec:4.6.1}.

\sm

To motivate the particular construction of the constant multiplets $\DD^{I_1 \cdots I_r}$
 to be given below, we begin by recasting the monodromy relation (\ref{3.D1}) for the one-point function of Lemma \ref{3.lem:4} in the following form, 
\bea
\label{4.D.1}
\DDD(\mB_L \cdot 1;B) = e^{- 2 \pi i \, \ad_{B_L}} \DDD(1;B)
\eea
To arrive at a relation of the form (\ref{2.desc.4}) in item $(d)$ of Theorem \ref{2.thm:3} we need a combination of constant multiplets and Enriquez kernels whose monodromy matches that of $\DDD(1;B)$ in (\ref{4.D.1}). Inspection of the monodromy relation for the generating function $\KK_J(x,y;B)$ of Enriquez kernels, given in (\ref{3.Kmon1}), shows that $\KK_J(1,p; \ad_{B})$, whose last argument is $\ad_B$ instead of $B$, fulfills this requirement. Although the dependence of $\KK_J(1,p; \ad_{B})$ on an extra point $p$ and the presence of a pole in $z_1$ at $p$, neither of which were present in $\DDD(1;B)$, may at first appear as a drawback, we shall soon show that consistency of the system of descent equations implies the absence of both. 

\sm

To proceed, we define the following combination,
\bea
\label{4.4.1}
\tilde {\bf D} _\delta (1,p;B) = \DDD(1;B) - \KK_J(1,p; \ad_{B}) \DDD^J(B) 
\eea
in terms of a generating function $\DDD^J(B) $ for as yet unknown constant multiplets $\DD^{I_1 \cdots I_r}$,
\bea
\label{4.4.2}
\DDD^J(B) = \sum _{r=1}^\infty   \DD^{I_1 \cdots I_r J } B_{I_r} \cdots B_{I_1}
\eea 
To determine $\DDD^J(B)$, we impose the vanishing of the $\mA$ periods in $t$ of $\tilde {\bf D} _\delta (t,p;B) $,
\bea
\label{4.4.3}
\oint _{\mA^L} dt \, \tilde {\bf D} _\delta (t,p;B) =0
\eea
To extract the condition this relation imposes on the constant multiplets, we use the fact that the $\mA$ periods of $\KK_J(1,p; \ad_{B})$ are independent of $p$ and may be read off from (\ref{3.Kper}),
\bea
\label{4.Kper.1}
\oint _{\mA^L} dt \, \KK_J(t,p; \ad_B) =   \delta ^L_J \, { - 2 \pi i \, \ad_{B_L} \over e^{-2 \pi i \, \ad_{B_L}} -1} 
= \delta ^L_J \sum_{k=0}^\infty ( - 2 \pi i)^k  { \Ber _k \over k!} ( \ad_{B_L})^k
\eea
The corresponding $h \times h$  matrix in the indices $J,L$ is diagonal and invertible. Therefore, the vanishing of the $\mA$ periods of $\tilde {\bf D} _\delta (1,p;B) $ implies the following expression for $\DDD^J(B)$, 
\bea
\label{4.Kper.2}
\DDD^J (B) =  { e^{- 2 \pi i \, \ad _{B_J} } -1 \over - 2 \pi i \, \ad_{B_J}} \ \oint _{\mA^J} dt \, \DDD(t;B)
\eea
whose coefficients of $B_{I_1}\cdots B_{I_r}$ feature the definition (\ref{defcsts}) for the
constant multiplets in item $(c)$ of Theorem \ref{2.thm:3}. This justifies the vanishing
$\mA$ periods of $\tilde {\bf D} _\delta (t,p;B)$ which were imposed in (\ref{4.4.3}).

\sm

The pole of $\tilde {\bf D} _\delta (1,p;B)$ in the variable $z_1$ at the point $p$ is given solely by the contribution of $g^{I}{}_J(1,p) \, \ad_{B_I}$ to $\KK_J(1,p; \ad_{B}) $ and its residue is,
\bea
\label{4.Kper.3}
\mathop{\mathrm{Res}}_{z_1=p} \, \tilde {\bf D} _\delta (1,p;B) = - \sum_J  \ad _{B_J} \DDD^J(B) = -  [B_J, \DDD^J(B)]
\eea
The commutator may be expressed in terms of the constant multiplets $\DD^{I_1 I_2 \cdots I_r}$
in the expansion (\ref{4.4.2}) and their cyclically permuted counterparts $\DD^{I_2 \cdots I_r I_1}$ as follows,
\bea
\label{4.Kper.4}
{} [B_J, \DDD^J(B)] = 
\sum_{r=2}^\infty \Big ( \DD^{I_1 I_2 \cdots I_r} - \DD^{I_2 \cdots I_r I_1} \Big ) B_{I_r} \cdots B_{I_1}
\eea
Independence of $\tilde {\bf D} _\delta (1,p;B)$ on the point $p$ will require, at the very least, that its residue in $z_1$ at $p$ cancels, namely that the commutator $[B_J, \DDD^J(B)]$ vanishes. We therefore proceed to showing the cyclicity of $\DD^{I_1 \cdots I_r} $ in item $(c)$ of Theorem \ref{2.thm:3}  which implies the vanishing of the commutator $[B_J, \DDD^J(B)]$ by (\ref{4.Kper.4}).

\subsection{Invariance of constant multiplets under cyclic permutations}
\label{restthm}

In this subsection, we shall prove item $(c)$ of Theorem \ref{2.thm:3}, namely the invariance of the constant multiplets $\DD^{I_1 \cdots I_r}$ under cyclic permutations of its indices. 

\sm

The starting point is the expression (\ref{4.Kper.2}) for the generating function $\DDD^L(B)$ of the constant multiplets $\DD^{I_1 \cdots I_r}$. Combining this formula with the integral of the expression
(\ref{4.beta.5}) for the one-point function, expanded in powers of the composite
letters $\beta_I$ in (\ref{4.beta.2}), we obtain,
\bea
\label{4.DD.1}
\DDD^L  (B) = {    e^{-2 \pi i \, \ad_{B_L} } -1 \over - 2 \pi i \, {\rm ad}_{B_L}  } \,
  \sum_{r=1}^{\infty} \beta _{I_r} \cdots \beta_{I_1}  \newH^{I_1 \cdots I_r L} _\delta 
\eea
Expanding the left side and the exponentials on the right side in Taylor series in powers of $B$ gives the constant multiplets $\DD^{I_1 \cdots I_r} $ in terms of the multiple $\mA$ periods $\newH^{J_1\cdots J_s} _\delta$ with $s\leq r$ in (\ref{4.DA.7}) where all points $t_i$ of the  cyclic product $C_\delta$ are integrated over. To low orders we obtain, for example,  
\bea
\DD^{I_1 I_2} & = & \newH_\delta ^{I_1I_2}
\label{6.3.g} \\
\DD^{I_1 I_2 I_3} & = &  \newH_\delta ^{I_1 I_2 I_3}  
- i \pi \, \big[ \delta ^{I_1}_{I_2} \, \newH_\delta^{I_2 I_3} 
 + \delta ^{I_2}_{I_3} \, \newH_\delta^{I_1I_3}  
 -  \delta ^{I_1}_{I_3} \, \newH_\delta^{I_2I_3}  \big]     \notag \\
\DD^{I_1 I_2 I_3 I_4} & = &  \newH_\delta ^{I_1 I_2 I_3 I_4}  
-i\pi \,
\big[  \delta ^{I_1}_{I_2} \, \newH_\delta^{I_2 I_3 I_4}
+  \delta ^{I_2}_{I_3} \, \newH_\delta^{I_1 I_3 I_4} 
+  \delta ^{I_3}_{I_4} \, \newH_\delta^{ I_1 I_2 I_4} 
- \delta ^{I_1}_{I_4} \, \newH_\delta^{ I_2 I_3 I_4}   \big]
\notag \\
&&\quad - \pi^2 \bigg[
 \delta ^{I_1}_{I_2} \, \delta ^{I_3}_{I_4} \, \newH_\delta^{I_2 I_4}
 + \frac{2}{3}  \delta ^{I_1 I_2}_{I_3} \,  \newH_\delta^{I_3 I_4} 
  + \frac{2}{3}  \delta ^{I_2 I_3}_{I_4} \,  \newH_\delta^{I_1 I_4}  \notag \\
 &&\quad\quad\ \
  -  \delta ^{I_1}_{I_4} \, \delta ^{I_2}_{I_3} \, \newH_\delta^{I_3 I_4}
    - \frac{4}{3}    \delta ^{I_1 I_3}_{I_4} \,  \newH_\delta^{I_2 I_4} 
 + \frac{2}{3}     \delta ^{I_1 I_2}_{I_4} \,  \newH_\delta^{I_3 I_4} 
\bigg]
\notag 
\eea
with no summation over repeated indices implied. The generating series identity (\ref{4.DD.1}) for constant multiplets lends itself particularly well to the evaluation of the commutator in (\ref{4.Kper.3}) and (\ref{4.Kper.4}) and we obtain,
\bea
\label{4.5.1}
{} [B_J, \DDD^J(B)] = { i \over 2 \pi} \sum_L \Big (  e^{-2 \pi i \, \ad_{B_L} } -1 \Big )  \,
  \sum_{r=1}^{\infty} \beta _{I_r} \cdots \beta_{I_1}  \newH^{I_1 \cdots I_r L} _\delta 
\eea
This formula expresses the commutator $[B_J, \DDD^J(B)] $ and therefore the transformation of the constant multiplets under cyclic permutations in its components in terms of multiple $\mA$ periods $\newH^{I_1 \cdots I_r}_\delta$. As we will see soon, the vanishing of the right side of (\ref{4.5.1}) through interrelations of the summands for different values of $r$ hinges on the transformation properties of $\newH^{I_1 \cdots I_r}_\delta$ under cyclic permutations that   we shall derive in the next subsection.

\subsubsection{Transformation of $\newH^{I_1 \cdots I_r}_\delta$ under cyclic permutations}
\label{sec:concross}

A cyclic permutation of the indices $\newH_\delta^{I_1 I_2 \cdots I_r } \rightarrow \newH_\delta^{I_2 \cdots I_r I_1}$ modifies the order in which the curves $\mA^{I_1},\cdots,\mA^{I_r}$ are extended into the interior $D^o$ of the fundamental domain $D$ by the prescription for multiple $\mA$ periods given in section \ref{sec:Acont}.  In particular, the outermost curve $\mA^{I_1}_\ep $ of $\newH_\delta^{I_1 I_2 \cdots I_r }$ is moved to the innermost position $\mA^{I_1}_{r \varepsilon}$ under the cyclic permutation,  as illustrated in figure \ref{fig:noncyc}. By doing so, the curve $\mA^{I_1}_\ep $ crosses all of those curves $\mA^{I_2}_{2 \ep}, \cdots, \mA^{I_r}_{r \ep}$ with $I_j = I_1$ when comparing the prescriptions  for their relative ordering in $\newH_\delta^{I_1 I_2 \cdots I_r }$ and 
 $\newH_\delta^{I_2 \cdots I_r I_1}$.  The quantitative relation is given by  the following lemma.

{\lem
\label{4.lem:10}
The multiple $\mA$ period $\newH^{I_1 \cdots I_r}_\delta$ of the cyclic product of Szeg\"o kernels defined in  (\ref{4.DA.7}) transforms as follows under cyclic permutations of its indices,
\bea
\label{4.lem.10}
\newH_\delta^{I_2 \cdots I_r I_1} = \newH_\delta^{I_1 I_2 \cdots I_r} 
+ 2\pi i \big ( \delta^{I_1}_{I_r} - \delta^{I_1}_{I_2} \big ) \newH_\delta^{I_2 \cdots I_r } 
\eea
}

\begin{figure}[htb]
\begin{center}
\tikzpicture[scale=0.7]
\draw[very thick] (-2,0) node[left]{$\partial D$} -- (2,0);
\draw[thick] (-2,-0.5) -- (2,-0.5);
\draw[thick] (-2,-1) -- (2,-1);
\draw (-1.3,-1.325) node{$\vdots$};
\draw (0,-1.325) node{$\vdots$};
\draw (1.3,-1.325) node{$\vdots$};
\draw[thick] (-2,-2) -- (2,-2);
\draw[thick] (-2,-2.5) -- (2,-2.5);
\draw(2.4,-0.5)node{$t_2$};
\draw(2.4,-1.1)node{$t_3$};
\draw(2.5,-1.9)node{$t_r$};
\draw(2.4,-2.5)node{$t_1$};
\draw[arrows={-Stealth[width=1.8mm, length=2.1mm]}](0,-0.5)--(-0.1,-0.5);
\draw[arrows={-Stealth[width=1.8mm, length=2.1mm]}](0,-1)--(-0.1,-1);
\draw[arrows={-Stealth[width=1.8mm, length=2.1mm]}](0,-2)--(-0.1,-2);
\draw[arrows={-Stealth[width=1.8mm, length=2.1mm]}](0,-2.5)--(-0.1,-2.5);
\draw(3.5,-1)node{$=$};
\scope[xshift=7cm]
\draw[very thick] (-2,0)node[left]{$\partial D$} -- (2,0);
\draw[thick] (-2,-0.5) -- (2,-0.5);
\draw[thick] (-2,-1) -- (2,-1);
\draw (-1.3,-1.325) node{$\vdots$};
\draw (0,-1.325) node{$\vdots$};
\draw (1.3,-1.325) node{$\vdots$};
\draw[thick] (-2,-2) -- (2,-2);
\draw(2.4,0.1)node{$t_1$};
\draw(2.4,-0.5)node{$t_2$};
\draw(2.4,-1.1)node{$t_3$};
\draw(2.5,-2)node{$t_r$};
\draw[arrows={-Stealth[width=1.8mm, length=2.1mm]}](0,-0.5)--(-0.1,-0.5);
\draw[arrows={-Stealth[width=1.8mm, length=2.1mm]}](0,-1)--(-0.1,-1);
\draw[arrows={-Stealth[width=1.8mm, length=2.1mm]}](0,-2)--(-0.1,-2);
\draw[arrows={-Stealth[width=1.8mm, length=2.1mm]}](0,-0)--(-0.1,-0);
\endscope
\draw(10.75,-1)node{$-\ \delta^{I_1}_{I_2}$};
\scope[xshift=14.5cm]
\draw[very thick] (-2,0)node[left]{$\partial D$} -- (2,0);
\draw[thick] (-2,-0.5) -- (2,-0.5);
\draw[thick] (-2,-1) -- (2,-1);
\draw (-1.3,-1.325) node{$\vdots$};
\draw (0,-1.325) node{$\vdots$};
\draw (1.3,-1.325) node{$\vdots$};
\draw[thick] (-2,-2) -- (2,-2);
\draw(2.4,-0.5)node{$t_2$};
\draw(2.4,-1.1)node{$t_3$};
\draw(2.5,-2)node{$t_r$};
\draw[red](0.6,-0.25)node{\footnotesize $t_1$};
\draw[arrows={-Stealth[width=1.8mm, length=2.1mm]}](1,-0.5)--(0.9,-0.5);
\draw[arrows={-Stealth[width=1.8mm, length=2.1mm]}](-1,-0.5)--(-1.1,-0.5);
\draw(0,-0.5)node{$\bullet$};
\draw[arrows={-Stealth[width=1.8mm, length=2.1mm]}](0,-1)--(-0.1,-1);
\draw[arrows={-Stealth[width=1.8mm, length=2.1mm]}](0,-2)--(-0.1,-2);
\draw[red, thick] (0,-0.5) circle (0.35);
\draw[red,arrows={-Stealth[width=1.8mm, length=2.1mm]}](-0.35*0.707, -0.5+0.35*0.707)--(-0.35*0.707-0.1, -0.5+0.35*0.707-0.1);
\endscope
%
\draw(1.25,-5)node{$-\ \delta^{I_1}_{I_3}$};
\scope[xshift=5cm, yshift=-4cm]
\draw[very thick] (-2,0)node[left]{$\partial D$} -- (2,0);
\draw[thick] (-2,-0.5) -- (2,-0.5);
\draw[thick] (-2,-1) -- (2,-1);
\draw (-1.3,-1.325) node{$\vdots$};
\draw (0,-1.325) node{$\vdots$};
\draw (1.3,-1.325) node{$\vdots$};
\draw[thick] (-2,-2) -- (2,-2);
\draw(2.4,-0.5)node{$t_2$};
\draw(2.4,-1.1)node{$t_3$};
\draw(2.5,-2)node{$t_r$};
\draw[arrows={-Stealth[width=1.8mm, length=2.1mm]}](0,-0.5)--(-0.1,-0.5);
\draw[red](0.6,-0.75)node{\footnotesize $t_1$};
\draw[arrows={-Stealth[width=1.8mm, length=2.1mm]}](-1,-1)--(-1.1,-1);
\draw[arrows={-Stealth[width=1.8mm, length=2.1mm]}](1,-1)--(0.9,-1);
\draw(0,-1)node{$\bullet$};
\draw[arrows={-Stealth[width=1.8mm, length=2.1mm]}](0,-2)--(-0.1,-2);
\draw[red, thick] (0,-1) circle (0.35);
\draw[red,arrows={-Stealth[width=1.8mm, length=2.1mm]}](-0.35*0.707, -1+0.35*0.707)--(-0.35*0.707-0.1, -1+0.35*0.707-0.1);
\endscope
\draw(9.1,-5)node{$-\ \cdots$};
\draw(10.75,-5)node{$-\ \delta^{I_1}_{I_n}$};
\scope[xshift=14.5cm, yshift=-4cm]
\draw[very thick] (-2,0)node[left]{$\partial D$} -- (2,0);
\draw[thick] (-2,-0.5) -- (2,-0.5);
\draw[thick] (-2,-1) -- (2,-1);
\draw (-1.3,-1.325) node{$\vdots$};
\draw (0,-1.325) node{$\vdots$};
\draw (1.3,-1.325) node{$\vdots$};
\draw[thick] (-2,-2) -- (2,-2);
\draw(2.4,-0.5)node{$t_2$};
\draw(2.4,-1.1)node{$t_3$};
\draw(2.5,-2)node{$t_r$};
\draw[arrows={-Stealth[width=1.8mm, length=2.1mm]}](0,-0.5)--(-0.1,-0.5);
\draw[arrows={-Stealth[width=1.8mm, length=2.1mm]}](0,-1)--(-0.1,-1);
\draw[red](0.6,-1.75)node{\footnotesize $t_1$};
\draw[arrows={-Stealth[width=1.8mm, length=2.1mm]}](-1,-2)--(-1.1,-2);
\draw[arrows={-Stealth[width=1.8mm, length=2.1mm]}](1,-2)--(0.9,-2);
\draw(0,-2)node{$\bullet$};
\draw[red, thick] (0,-2) circle (0.35);
\draw[red,arrows={-Stealth[width=1.8mm, length=2.1mm]}](-0.35*0.707, -2+0.35*0.707)--(-0.35*0.707-0.1, -2+0.35*0.707-0.1);
\endscope
\endtikzpicture
\caption{Contour deformation that brings the cyclically permuted ordering of contour displacements with $t_1$ in the innermost position back to the original order of (\ref{4.DA.7})  with $t_1$ in the outermost position. The crossing of the $t_1$ contour with those of $t_2,\cdots,t_r$ through the contour deformation is homotopic to
infinitesimal circles around $t_j$ drawn in red which only arise if $I_1 = I_j$. Moreover, the
residue structure (\ref{rescdelta}) of the integrand implies that the circles around $t_3,\cdots,t_{r-1}$ 
in the second line of the figure do not contribute to (\ref{4.lem.10}).  \label{fig:noncyc}}
\end{center}
\end{figure}

\sm
  
To prove the lemma, we trace the effects of $\mA^{I_1}_\ep$ crossing $\mA^{I_2}_{2 \ep} ,\cdots, \mA^{I_r}_{r \ep}$.  
As shown in figure \ref{fig:noncyc}, each time $\mA^{I_1}_\ep$ crosses a curve $\mA^{I_k}_{k \ep}$ with $I_k=I_1$
the pole in $t_1$ at $t_k$ produces an extra contribution given by the integral in $t_1$ around a small circle centered at $t_k$. These integrals may be evaluated using the residues,  
\beq
\mathop{\mathrm{Res}}_{~t_k = t_{k \pm 1}} \, C_\delta(t_1,\cdots,t_r) = \pm C_\delta(t_1,\cdots,\hat{t}_k ,\cdots,t_r)
\label{rescdelta}
\eeq
of the cyclic products $C_\delta$  for nearest neighbor points. No contributions arise for $I_k \not = I_1$ or for non-nearest neighbor points.  Thus, the only non-trivial residues arise for the circles in $t_1$ around the points $t_2$ and $t_r$ provided $I_1=I_2$ and/or $I_1=I_r$, while the contributions around $t_3,\cdots,t_{r-1}$ vanish. Taking into account the residue structure of (\ref{rescdelta}) proves formula (\ref{4.lem.10}) and thus Lemma \ref{4.lem:10}. A similar relation for the exchange symmetry between $\mA$ periods due to simple poles in the integrand was pointed out in \cite{DHoker:2025dhv}.

\subsubsection{Lemma \ref{4.lem:10} implies item $(c)$ of Theorem \ref{2.thm:3}}
\label{sec:cycDs.two}

With the cyclic transformation law in (\ref{4.lem.10}) established in Lemma \ref{4.lem:10}, we shall now prove the invariance of the constant multiplets $D_\delta^{I_1\cdots I_r}$ under cyclic permutations of its indices and thereby provide a proof of item $(c)$ of Theorem \ref{2.thm:3}. 

\sm

The starting point is the expression in (\ref{4.5.1}) for the commutator, whose vanishing is equivalent to the cyclic symmetry of all constant multiplets in view of (\ref{4.Kper.4}). The expression (\ref{4.5.1}) is rendered more explicit upon the use of the following relation,
\bea
\big (  e^{-2 \pi i \, \ad_{B_L} } -1 \big ) X =  e^{-2 \pi i B_L } X e^{ 2 \pi i B_L} - X
=
- 2 \pi i \big [ \beta _L , X \big ] \big ( 1- 2 \pi i \b_L \big )^{-1}
\eea
for arbitrary combinations $X$ of words in $B_I$. Here, we have used the definition of $\b_L$ given in (\ref{4.beta.2}) to express the exponential as follows $e^{-2 \pi i B_L} = 1- 2 \pi i \b_L$ and to pass from the first equality to the second. Applying this formula to the series for the commutator in (\ref{4.DD.1})  gives, 
\bea
{} \big [ B_L, \DDD^L(B) \big ] 
=  \sum_{r=2}^{\infty} \big [ \QQ_{I_r} , \QQ_{I_{r-1}}\cdots  \QQ_{I_{1}} \big ] \, (1- 2\pi i \QQ_{I_r})^{-1} \, \newH_\delta^{I_1 \cdots I_r } 
\eea
Writing out the commutator of the $\beta$ in the summand on the right and using a suitable relabelling of the indices, we obtain, 
\bea
{} \big [ B_L , \DDD^L(B) \big ] 
&= &  \sum_{r=2}^{\infty} \QQ_{I_r}  \cdots \QQ_{I_{2}}  \QQ_{I_{1}} \,
\Big\{
(1- 2\pi i \QQ_{I_r})^{-1} \, \newH_\delta^{I_1 I_2 \cdots I_r } 
 \\ && \hskip 1.2in
 - (1- 2\pi i \QQ_{I_1})^{-1} \,  \newH_\delta^{I_2 \cdots I_r I_1 }   \Big\} 
\no
\eea
Substituting the expression for the cyclic permute $\newH_\delta^{I_2 \cdots I_r I_1 }$ from (\ref{4.lem.10}) of Lemma \ref{4.lem:10}, the above expression becomes, 
\bea
\label{cycprf.05}
{} \big [ B_L , \DDD^L(B) \big ] 
 &=& \sum_{r=2}^{\infty} \QQ_{I_r}  \cdots \QQ_{I_{2}}  \QQ_{I_{1}} \,
\Big\{ (1- 2\pi i \QQ_{I_r})^{-1} \,  \newH_\delta^{I_1 I_2 \cdots I_r }  
 \\
&&\qquad
 - (1- 2\pi i \QQ_{I_1})^{-1} \,
\big[ \newH_\delta^{I_1 I_2 \cdots I_r } 
+ 2\pi i \big ( \delta^{I_1}_{I_r} - \delta^{I_1}_{I_2} \big ) \, \newH_\delta^{I_2 \cdots I_r } 
\big] \Big\}
\no
\eea
The correction terms proportional to $  2\pi i (\delta^{I_1}_{I_r} - \delta^{I_1}_{I_2})$ on the second line cancel for $r=2$ and only contribute starting at $r=3$. After shifting the summation variable $r$ by one in the summation of these terms to restore the original range $r\geq 2$, and a suitable relabeling of the indices, we obtain,
\bea
\label{cycprf.06}
{} \big [ B_L , \DDD^L(B) \big ] 
 &= & \sum_{r=2}^{\infty} \QQ_{I_r}  \cdots \QQ_{I_{2}}  \QQ_{I_{1}} \,
 \newH_\delta^{I_1 I_2 \cdots I_r }  \Big\{
(1- 2\pi i \QQ_{I_r})^{-1} \,   - (1- 2\pi i \QQ_{I_1})^{-1} 
 \\
&&\qquad\qquad
 -  2\pi i \QQ_{I_r} \, (1- 2\pi i \QQ_{I_r})^{-1} 
 + 2\pi i  \QQ_{I_1} \, (1- 2\pi i \QQ_{I_1})^{-1}  
\big] \Big\} 
\no
 \eea
The first and third terms inside the braces add up to 1 while the second and fourth terms add up to $-1$ so that the above sum cancels term by term and we conclude that,
\bea
{} \big [ B_L , \DDD^L(B) \big ]  =0
\label{finallyzero}
\eea
which by (\ref{4.Kper.4}) proves the invariance of the constant multiplets $\DD^{I_1 \cdots I_r}$ under cyclic permutations of its indices, and thereby proves item $(c)$ of Theorem \ref{2.thm:3}.

\subsection{Completing the proof of Theorem \ref{2.thm:3}: $(c)$  implies~$(d)$}
\label{sec:4.6.1}

Having established cyclic symmetry of the constant multiplets $\DD^{I_1 \cdots I_r}$ in the previous subsection, it follows that the combination $\tilde {\bf D} _\delta (1,p;B)$ in (\ref{4.4.1}) simplifies to
\bea
\tilde {\bf D} _\delta (1,p;B) &=& \DDD(1;B) - {\bf W}_J(1; \ad_{B}) \DDD^J(B)  + {\bf X}(1,p; \ad_{B}) 
[B_J, \DDD^J(B)] \notag \\
&=& \DDD(1;B) - {\bf W}_J(1; \ad_{B}) \DDD^J(B) 
\eea
which exposes holomorphicity in $z_1$ and independence on $p$. Moreover, the discussion 
around (\ref{4.4.3}) to (\ref{4.Kper.2}) revealed that $\tilde {\bf D} _\delta (1,p;B)$
has vanishing $\mA$ periods in $z_1$ (as an equivalent of the definition (\ref{defcsts})
of $D_\delta^{I_1\cdots I_r}$),  and (\ref{4.4.1}) together with (\ref{3.D1}) imply that $\tilde {\bf D} _\delta (1,p;B)$
has the following monodromy,
\bea
\label{4.D.1alt}
\tilde {\bf D} _\delta (\mB_L \cdot 1,p;B) = e^{- 2 \pi i B_L} \tilde {\bf D} _\delta (1,p;B) e^{ 2 \pi i B_L}
\eea
Taylor expanding in powers of $B$, the components $\tilde D _\delta ^{I_1 \cdots I_r} (1,p)$ defined by,\footnote{Given that the series expansion of both ingredients $\DDD(1;B)$ and $ \DDD^J(B) $ of $\tilde {\bf D} _\delta (1,p;B)$
starts at the first order in $B_I$, terms at the zeroth order in $B_I$ are guaranteed to be absent from (\ref{taytd}).}
\bea
\tilde {\bf D} _\delta (1,p;B) = \sum _{r=1}^\infty \tilde D _\delta ^{I_1 \cdots I_r} (1,p) B_{I_r} \cdots B_{I_1}
\label{taytd}
\eea
have the following monodromy in $z_1$, 
\bea
\tilde D _\delta^{I_1 \cdots I_r}  ( \mB _L \cdot 1,p) =
\sum _{k=0}^r \sum _{\ell =0}^{r-k} { (-)^k (2 \pi i)^{k+\ell} \over k ! \, \ell ! } 
\, \delta ^{I_1 \cdots I_\ell}_L \,
 \tilde D _\delta^{ I_{\ell +1} \cdots I_{r-k}   }  ( 1,p) 
\, \delta ^{ I_{r-k+1}  \cdots I_r } _L
\eea
The above properties are essential to prove the following proposition that relates 
the one-point function to the generating series of the constant multiplet. 
{\prop
\label{prop:td}
The combination $\tilde {\bf D} _\delta (1,p;B)$ defined by (\ref{4.4.1}) vanishes identically,
\bea
\tilde {\bf D} _\delta (1,p;B) =0
\label{proptd.1}
\eea
which implies the following expression for the one-point function
$ \DDD(1;B)$ in terms of the constant multiplets in $\DDD^J(B) $,
\bea
 \DDD(1;B) =\KK_J(1,p; \ad_{B}) \DDD^J(B) 
=  {\bf W}_J(1; \ad_{B}) \DDD^J(B) 
\label{proptd.2}
\eea}

\sm

The first statement (\ref{proptd.1}) of the proposition is proven by contradiction. Let us assume that the lowest order term in the Taylor expansion (\ref{taytd}) of $\tilde {\bf D} _\delta (1,p;B)$ in $B$ is non-zero. In view of the monodromy relation (\ref{4.D.1alt}), this lowest order term  must have vanishing $\mB$ monodromy. Since it is a $(1,0)$ form holomorphic in $z_1$, it must be a linear combination of the Abelian differentials $\om_J(1)$. But since its $\mA$ periods vanish in view of (\ref{4.4.3}), this lowest order term must vanish, in contradiction to our initial hypothesis that the lowest order term is non-vanishing. This implies the vanishing of $\tilde {\bf D} _\delta (1,p;B)$ in (\ref{proptd.1}).

\sm

The expression (\ref{proptd.2}) for the one-point function in terms of $\KK_J(1,p; \ad_{B})$ is an immediate consequence of (\ref{proptd.1}) and the expression (\ref{4.4.1}) for $\tilde {\bf D} _\delta (1,p;B)$. The second equality in (\ref{proptd.2}) follows from (\ref{finallyzero}), concluding the proof of Proposition \ref{prop:td}.

\sm

Expanding the last statement $ \DDD(1;B) =  {\bf W}_J(1; \ad_{B}) \DDD^J(B) $ of 
Proposition \ref{prop:td} in components according to (\ref{gfsKW}) and (\ref{4.gen})
results in (\ref{2.desc.4}) and thereby concludes the proof of item $(d)$ of Theorem \ref{2.thm:3}.

\newpage

\section{Decomposing cyclic products into Enriquez kernels}
\setcounter{equation}{0}
\label{sec:5}

In this section, we describe the simplified representations of cyclic products $C_\delta$ 
in terms of Enriquez kernels and constant multiplets $D_\delta^{I_1\cdots I_r}$ 
that result from the meromorphic descent in Theorem \ref{2.thm:3} and list several properties of
the main constituents. 

\subsection{Meromorphic decomposition of cycles of Szeg\"o kernels}
\label{sec:2.6}

The meromorphic descent described in the previous sections leads to the decomposition of the cyclic product $C_\delta(1,\cdots,n)$ of Szeg\"o kernels in the following theorem, which was already announced in (\ref{intro.03}) and will be referred to as the \textit{meromorphic decomposition}.\footnote{Similar decomposition formulas for $C_\delta(1,\cdots,n)$ were proposed in \cite{Tsuchiya:2017joo, Tsuchiya:2022lqv}, but it remains unclear how to reconcile those claims with the expressions in this work beyond genus one.}

\sm

{\thm
\label{2.thm:cdec} 
The spin-structure dependence of $C_\delta$ can be fully separated from the dependence on the points $z_1,\cdots,z_n$ for $n \geq 2$ through the meromorphic decomposition,
\begin{align}
C_\delta (1,\cdots, n) &=
  {\cal W}(1,\cdots,n)  +  \sum_{r=2}^{n}
  {\cal W}_{I_1  \cdots I_{r}}(1,\cdots,n)  \, D_\delta^{I_1 \cdots I_{r}}
\label{again.03}
\end{align}
The multiplets ${\cal W}_{I_1 \cdots I_{r}}(1,\cdots,n)$ are independent of the spin structure $\delta$, meromorphic 
and single-valued $(1,0)$ forms in $z_1,\cdots,z_n \in \Sigma$, cyclically symmetric in the indices $I_1\cdots I_r$ and expressible in terms of  (products, index contractions and derivatives of) Enriquez kernels.  In particular, the $r=n$ term is given by a cyclically symmetrized product of holomorphic Abelian differentials, 
\bea
{\cal W}_{I_1 \cdots I_n}(1,\cdots,n) = 
\frac{1}{n} \, \omega_{I_1}(1) \, \cdots \,  \omega_{I_n}(n) + {\rm cycl}(I_1,\cdots,I_n)
\label{eq2.6.1}
\eea
The coefficients $D_\delta^{I_1 \cdots I_{r}}$ in (\ref{again.03}) are those produced in the descent procedure of Theorem \ref{2.thm:3} for $r=2,\cdots,n$; they are independent of the points  $z_1, \cdots, z_n $; cyclically symmetric; locally holomorphic in the moduli of $\Sigma$ and carry all the dependence of $C_\delta$ on $\delta$.}

\sm

Note that, while the multiplets $ {\cal W}_{I_1  \cdots I_{r}}(1,\cdots,n) $ at fixed rank $r \leq n$ are distinct for different values of $n$, their coefficients $D_\delta^{I_1 \cdots I_{r}}$ are independent of  $n$ and thus universal.

\subsubsection{Proof of Theorem \ref{2.thm:cdec}}
\label{sec:2.6.0}

We present a constructive proof of Theorem \ref{2.thm:cdec} which proceeds by rearranging the descent equations (\ref{2.desc.2}) as  follows with $n \geq 3$ and $r \geq 0$,  
\begin{align}
\label{rewdesc.1}
\DD ^{I_1 \cdots I_r } (1, \cdots , n) 
& = \om_J(1) \DD ^{I_1 \cdots I_r J} (2, \cdots , n) 
+ \sum_{i=0}^{r-1} g^{I_r  \cdots I_{i+1}}{}_J(1,2) \DD ^{I_1 \cdots I_iJ} (2, \cdots , n) 
\no \\ 
&\quad  + \Big (  \chi^{I_r \cdots  I_1} (1,n)   {-} \chi^{I_r \cdots  I_1} (1,2) \Big ) C_\delta (2,\cdots, n)   
\end{align}
where $\DD^\emptyset (1,\cdots, n) = C_\delta(1,\cdots, n)$. We also rearrange the descent equations  (\ref{2.desc.3}) and (\ref{2.desc.4}),  for $r \geq 0$ in the first line below and $r \geq 1$ in the second line below, as follows, 
\begin{align}
\DD ^{I_1 \cdots I_r }(1,2) 
& =  \om_J(1) \DD^{ I_{1} \cdots I_r J}(2)
+ \sum_{i=0}^{r-1}   g^{I_r \cdots I_{i+1} }{}_J(1,2) \DD^{ I_{1} \cdots I_i J}(2)
+ \partial_2 \chi^{I_r \cdots I_1}(1,2)
\no \\
 \DD^{I_1 \cdots I_r }(1) 
& = 
\om_J(1) \DD ^{I_1 \cdots I_r J} 
+ \sum_{{0 \leq i < j \leq r \atop (i,j) \not= ( 0,r)} } (-1)^i 
\varpi ^{ I_1 \cdots I_i \, \shuffle \, I_r  \cdots  I_{j+1}  }{}_J(1) \DD^{I_{i+1} \cdots I_j J} 
\label{rewdesc.2}
\end{align}
The meromorphic decomposition (\ref{again.03}) along with the explicit form of the multiplets $ {\cal W}_{I_1 \cdots I_{r}}(1,\cdots,n)$ may now be obtained by eliminating from this system of $n$ equations (and their cyclic permutations in the points) all the intermediate multiplets $\DD^{I_1 \cdots I_s}(r,\cdots)$, that depend on at least one point and at most $n-1$ points, in favor of the constant coefficients $\DD^{I_1 \cdots I_s}$. This may be done recursively by increasing the rank $r$ or, equivalently, by decreasing the number of points  via iterative use of (\ref{rewdesc.1}) and (\ref{rewdesc.2}).  The last step of this recursive procedure gives rise to the holomorphic $r=n$ term
 $\omega_{I_1}(1) \cdots  \omega_{I_n}(n) D_\delta^{I_1 \cdots I_{n}}$ in (\ref{again.03})
 and (\ref{eq2.6.1}) by the first term on the right side of (\ref{rewdesc.2}). The Enriquez-kernel representation of $ {\cal W}_{I_1  \cdots I_{r}}(1,\cdots,n)$ with $r\leq n-1$ follows by tracking the appearance of the constant $D_\delta^{I_1 \cdots I_{r}}$ in each of the $n$ steps of the meromorphic descent, where
$ {\cal W}(1,\cdots,n)$ without any indices is obtained from the contributions without any
accompanying factor of $D_\delta^{I_1 \cdots I_{r}}$. This procedure leads to the $n$ terms
in the meromorphic decomposition~(\ref{again.03}).
 
\sm 

The cyclic symmetry of $\DD^{I_1 \cdots I_r}$ in its indices $I_1, \cdots, I_r$, guaranteed by item $(c)$ of Theorem~\ref{2.thm:3}, implies that we may choose $ {\cal W}_{I_1  \cdots I_{r}}(1,\cdots,n)$ to be cyclically invariant in its indices, 
\beq
{\cal W}_{I_1 I_2 \cdots I_{r}}(1,\cdots,n) = {\cal W}_{I_2  \cdots I_{r} I_1}(1,\cdots,n)
\label{cycW}
\eeq
and we shall do so throughout. The single-valuedness of $ {\cal W}_{I_1  \cdots I_{r}}(1,\cdots,n)$  emerges through the particular combinations of Enriquez kernels that arise in the descent procedure, but this property is not manifest  term by term. To prove it, one may use the fact that  Theorem~\ref{2.thm:3} involves only the cyclic invariance of $\DD^{I_1 \cdots I_r}$ with $2\leq r \leq n$ without relying on any relation between $\DD^{I_1 \cdots I_r}$ of different rank. Since $C_\delta(1,\cdots, n)$ is single-valued it follows that the individual terms $ {\cal W}_{I_1  \cdots I_{r}}(1,\cdots,n) \DD^{I_1 \cdots I_r}$ must also be single-valued. Our choice (\ref{cycW}) then ensures that already the individual coefficients $ {\cal W}_{I_1  \cdots I_{r}}(1,\cdots,n)$ are single-valued, irrespective on their contraction with the cyclic $\DD^{I_1 \cdots I_r}$. This completes the proof of the single-valuedness of $ {\cal W}_{I_1  \cdots I_{r}}(1,\cdots,n)$ in all of $z_1,\cdots,z_n$ and of Theorem \ref{2.thm:cdec}.

\subsubsection{Examples of single-valued combinations of Enriquez kernels}
\label{sec:2.ex}

We shall now spell out the explicit forms of the meromorphic and single-valued multiplets
$ {\cal W}_{I_1 \cdots I_{r}}(1,\cdots,n)$ for $n=2,3$ points that follow
from the constructive procedure outlined in the above proof. Their $n=4$ point
counterparts can be found in appendix \ref{sec:2.6.2}. There is no need to
spell out the cases with $r=n$ in view of (\ref{eq2.6.1}).

\sm

For $n=2$ points, matching the expression (\ref{2.1.a}) for the two-cycle of 
Szeg\"o kernels with the general form (\ref{again.03}) of the decomposition readily allows us to identify,
\bea
 {\cal W}(1,2) & = \p_2 \chi(1,2) 
   \label{eq2.6.4}
\eea
For $n=3$ points, eliminating all of $C_\delta(2,3), D_\delta^{I}(2,3), D_\delta^{K}(3),D_\delta^{IJ}(3)$ from the 
descent equations (\ref{2.thm.1}) with the help of the two-point identities (\ref{2.1.a}) and (\ref{2.1.b})
casts $C_\delta(1,2,3)$ into the form of (\ref{again.03}). By isolating the coefficient of $D_\delta^{JK}$
and the terms without any accompanying $D_\delta^{JK}, D_\delta^{IJK}$ in all steps of the
three-point descent, we are led to the following expressions for the
forms ${\cal W}_{JK}(1,2,3) $ and $ {\cal W}(1,2,3) $, respectively,
\begin{align}
{\cal W}_{JK}(1,2,3) & =  
\frac{1}{2} \Big[ \big ( \chi(1,3) - \chi(1,2) \big )  \om_J(2) \om_K(3) 
 + \om_I(1) g^I{}_{J} (2,3) \om_K(3) \notag \\
 &\quad\quad +  \big( \om_J(1) \om_I(2) - \om_I(1) \om_J(2) \big) \varpi ^I{}_{K}(3) + (J \leftrightarrow K) \Big] \no \\
 {\cal W}(1,2,3) & =    \big (  \chi(1,3) - \chi(1,2)  \big )  \p_3 \chi(2,3)  
  + \om_I(1)  \p_3 \chi^I(2,3) 
  \label{eq2.6.5}
\end{align}
The prescription to add the image under $J\leftrightarrow K$ applies to both
lines of the expression for ${\cal W}_{JK}(1,2,3)$ and implements the
cyclic symmetry according to our choice (\ref{cycW}).

\subsection{Further properties of   $D_\delta^{I_1\cdots I_r}$ and $ {\cal W}_{I_1 \cdots I_{r}}(1,\cdots,n) $}
\label{sec:rmks}

The propositions below collect further properties of the  multiplets in the meromorphic decomposition of
cyclic products of Szeg\"o kernels in Theorem \ref{2.thm:cdec}, including a reflection symmetry of $D_\delta^{I_1\cdots I_r}$ and various further properties of  $ {\cal W}_{I_1 \cdots I_{r}}(1,\cdots,n) $.

{\prop 
\label{5.reflect}
The multiplets $\DD^{I_1  \cdots I_r}$ are invariant under alternating reflection symmetry for any $r\geq 2$, 
\beq
\DD^{I_1  I_2 \cdots I_{r-1} I_r} = (-1)^r \DD^{I_r I_{r-1} \cdots  I_2 I_1}
\label{reflD}
\eeq}
 The proof is given in appendix \ref{sec:moreprf}.
 
 \sm

{\prop
\label{mdec.props}
The multiplets $ {\cal W}_{I_1 \cdots I_{r}}(1,\cdots,n) $ exhibit the following further properties.
\begin{itemize}
\itemsep=-0.03in
\item[$(a)$] For $n\geq 3$, $r\leq n-1$ and $2\leq k \leq n-1$ they have simple poles at adjacent points,
\bea
\bar \partial_1 {\cal W}_{I_1  \cdots I_{r}}(1,2,\cdots,n)
&=&  \pi \, \big ( \delta (1,2) - \delta (1,n) \big ) \,
{\cal W}_{I_1  \cdots I_{r}}(2,\cdots, n) \notag \\
\bar \partial_k {\cal W}_{I_1  \cdots I_{r}}(1,2,\cdots,n)
&=&  \pi \, \big ( \delta (k,k+1) - \delta (k,k-1) \big ) \,
{\cal W}_{I_1  \cdots I_{r}}(1,\cdots,\hat k ,\cdots, n) \notag \\
\bar \partial_n {\cal W}_{I_1  \cdots I_{r}}(1,2,\cdots,n)
&=&  \pi \, \big ( \delta (n,1) - \delta (n,n-1) \big ) \,
{\cal W}_{I_1  \cdots I_{r}}(1,\cdots, n-1)
 \label{wdec.03}
\eea
and are holomorphic in all points if $r=n$.
\item[$(b)$] For $n\geq 2$ and $0\leq r \leq n$ and for a fixed choice of their indices $I_1,\cdots,I_r$, 
they are invariant under cyclic permutations of the points,
\beq
{\cal W}_{I_1  \cdots I_{r}}(1,2,\cdots,n) = {\cal W}_{I_1  \cdots I_{r} }(2,\cdots,n,1)
\label{cycWlegmero}
\eeq
\item[$(c)$] They are invariant under the simultaneous reflection of the indices and the points,
\beq
 {\cal W}_{I_1 I_2 \cdots I_{r}}(1,2,\cdots,n)  = (-1)^{n+r}  {\cal W}_{I_r  \cdots  I_2 I_{1}}(n,\cdots,2,1) 
 \label{reflWW}
\eeq
\end{itemize}
}

A direct proof of item $(a)$ uses the fact that the proof of Theorem~\ref{2.thm:3} does not involve any properties of  $\DD^{I_1 \cdots I_r}$ with $2\leq r \leq n$ other than their cyclic invariance, so that they may be treated effectively as linearly independent of one another. Substituting the relation (\ref{again.03}) into (\ref{2.desc.5pol}) and identifying the coefficients of the various multiplets $\DD^{I_1 \cdots I_r}$ then proves (\ref{wdec.03}). In the process, one makes use of the fact that the multiplet $ {\cal W}_{I_1  \cdots I_{r}}(1,\cdots,n)$ for $r=n$ in (\ref{eq2.6.1}) is holomorphic in all points.

\sm

The proofs of items $(b)$, $(c)$ are relegated to appendix \ref{sec:4.cor}. They rely on the modular counterpart of the meromorphic decomposition (\ref{again.03}) to be described in section \ref{sec:4.dec}. 

\sm

Note that the expressions for the multiplets $ {\cal W}_{I_1 I_2 \cdots I_{r}}(1,\cdots,n)$
with $r\geq 3$ obtained from the meromorphic descent equations as described below (\ref{rewdesc.2}) 
do not match the images $(-1)^r {\cal W}_{I_r  \cdots I_{2} I_1}(1,\cdots,n)$ of
reflecting only the indices and not the points. More precisely, the multiplets $D_\delta^{I_1\cdots I_r}$
in the second line of the descent equation (\ref{rewdesc.2}) and the decomposition (\ref{again.03})
are understood as cyclically invariant but otherwise unspecified symbols for the purpose of
defining the $ {\cal W}_{I_1 \cdots I_{r}}(1,\cdots,n)$ as coefficients, 
without any reference to their reflection property (\ref{reflD}).
This approach is possible since the meromorphic descent does not rely on (\ref{reflD})
and allows us to construct a larger class of single-valued and meromorphic 
$ {\cal W}_{I_1 \cdots I_{r}}(1,\cdots,n)$ (expected to be relevant to future work) than the combinations 
realized in (\ref{again.03}) upon contraction with $D_\delta^{I_1\cdots I_r}$.

\newpage

\section{Descent in terms of DHS kernels}
\setcounter{equation}{0}
\label{sec:6}

The descent relations of the cyclic product of Szeg\"o kernels $\CD(1,\cdots, n)$ in terms of the single-valued modular invariant but non-meromorphic DHS kernels introduced in \cite{DHS:2023} were presented in \cite{DHoker:2023khh}, where a proof was given for low rank only. Here, we shall offer a convenient reformulation of these descent relations and provide a full proof.

\subsection{Definition and properties of DHS kernels}
\label{sec:4.1}

The DHS kernels $f^{I_1 \cdots I_r}{}_J(x,y)$ for $r \geq 0$ and $I_1, \cdots, I_r, J \in \{ 1, \cdots, h\}$ were defined in \cite{DHS:2023} in terms of the Arakelov Green function $\cG(x,y)$ (see appendix \ref{sec:A.3} for its definition and properties), holomorphic Abelian differentials $\om_I(x)$ and their complex conjugates. The DHS kernel for $r=1$ is defined by the  integral,
\bea
f^I{}_J(x,y) & = & 
 \int _\Sigma d^2 t \, \p_x \cG(x,t) \Big (  \bar \om^I(t)  \om_J(t) - \delta ^I_J \, \delta (t,y) \Big )
 \hskip 0.6in 
 d^2 t = { i \over 2} dt \wedge d \bar t
 \label{recDHSpre} 
\eea
while for $r \geq 2$ the kernels are defined recursively in the rank  by,
\bea
f^{I_1 \cdots I_r}{}_J (x,y) & = & \int _\Sigma d^2 t \,  \p_x \cG(x,t) \, \bar \om^{I_1} (t) f^{I_2 \cdots I_r}{}_J(t,y) 
\label{recDHS}
\eea
The DHS kernels satisfy a Massey system of differential relations, as may be established using the differential equations (\ref{A.arak1}) and (\ref{A.arak2}) for the Arakelov Green function,
\bea
\bar \p_x  f^{I_1} {}_J(x,y) & = & - \pi \, \bar \om ^{I_1} (x) \, \om_J(x) + \pi \, \delta ^{I_1}_J \, \delta (x,y) 
\no \\
\bar \p_y f^{I_1} {}_J(x,y) & = & \pi \, \delta ^{I_1}_J \,  \om_K(x) \, \bar \om^K(y) - \pi \, \delta ^{I_1}_J \, \delta(x,y)
\label{massey.1}
\eea
and for $r \geq 2$,
\bea
\bar \p_x f^{I_1 \cdots I_r}{}_J(x,y) & = & - \pi \, \bar \om^{I_1} (x) \, f^{I_2 \cdots I_r}{}_J(x,y)
\no \\
\bar \p_y f^{I_1 \cdots I_r}{}_J(x,y) & = &  \pi \, \delta ^{I_r}_J \, f^{I_1 \cdots I_{r-1} }{}_K(x,y) \,  \bar \om^{K} (y)
\label{massey.2}
\eea
Finally, $f^{I_1 \cdots I_r}{}_J(x,y)$ admits a decomposition into a traceless and trace part, given by,
\bea
\label{4.PhiG}
f^{I_1 \cdots I_r}{}_J(x,y) = \p_x \Phi  ^{I_1 \cdots I_r}{}_J(x) - \delta ^{I_r}_J\, \p_x \cG^{I_1 \cdots I_{r-1}} (x,y) 
\eea
where $\Phi  ^{I_1 \cdots I_{r-1} J}{}_J(x)=0$ and, similar to (\ref{2.dec}), the dependence on $y$ is concentrated in the trace $\delta^{I_r}_J$ with respect to the rightmost indices. Using this relation for $r=1$, one may recast the recursion relation (\ref{recDHS}) entirely in terms of DHS kernels $f$, 
\bea
f^{I_1 \cdots I_r}{}_J (x,y) =  - \int _\Sigma  d^2 t \, f^{I_1}{}_K(x,t)  \, \bar \om^{K} (t)  f^{I_2 \cdots I_r}{}_J(t,y) 
\eea
The DHS kernel $f^{I_1 \cdots I_r}{}_J(x,y)$ is a single-valued $(1,0)$ form in $x \in \Sigma$ and a single-valued $(0,0)$ form in $y \in \Sigma$ and transforms as a modular tensor  under  $\Sp(2h,\ZZ)$ \cite{vdG3, Kawazumi:lecture, DHoker:2020uid}. The form $f^{I_1 \cdots I_r}{}_J(x,y)$ is real analytic for  $x ,y  \in \Sigma $ with $x \not= y$ for all $r\geq 0$, just as  ${\cal G}(x,y)$ is. In view of (\ref{A.arakas}), the  $r=1$ DHS kernel exhibits a single simple pole at $x=y$,
\bea
f^I{}_J(x,y) = { \delta ^I_J  \over x-y} + \hbox{regular}
\eea
While the form $f^{I_1I_2}{}_J(x,y)$ for $r=2$ does not have a limit as $y \to x$, the  combination $f^{I_1I_2}{}_J(x,y) +\delta^{I_2}_J \frac{\pi }{x-y} \int^x_y \bar \omega^{I_1}$ does, see Lemma 8.1 of \cite{DHoker:2024ozn}. For $r \geq 3$ the form $f^{I_1 \cdots I_r}{}_J(x,y)$ has a  limit as $y \rightarrow x$.

\subsection{Formulation of the descent in terms of DHS kernels}

The descent of the cyclic product of Szeg\"o kernels in terms of DHS kernels was presented in equations (33-35) of \cite{DHoker:2023khh} and may be restated as follows. 

\sm

The functions $\CD ^{I_1 \cdots I_r}(1,\cdots,n)$ are defined for $r \geq 0$ to be differential $(1,0)$ forms in the variables $z_1, \cdots,  z_n$ for $n \geq 1$ and constants on $\Sigma$ for $n=0$. For $r=0$ and $n \geq 2$ they are defined by the cyclic product of $n$ Szeg\"o kernels introduced in (\ref{intro.01}), namely, 
\beq
\CD^\emptyset (1,\cdots, n) = \CD (1,\cdots, n)
\eeq
and to vanish for $n=0,1$. For $r\geq 1$ and $n \geq 0$, they  are defined recursively by,
\bea
\label{4.5}
\CD^{I_1 \cdots I_r} (1, \cdots , n)
= \int _\Sigma d^2 t_r \,  \bar \om^{I_r}(t_r) \, \CD^{I_1 \cdots I_{r-1}} (t_r, 1, \cdots, n) 
\eea
and vanish for $(r,n)=(1,0)$. These integrals are absolutely convergent for $n+r \geq 3$ and conditionally convergent for $n+r=2$. In particular,  the constant modular tensors are obtained from (\ref{4.5}) for $n=0$. For $r \geq 3$ they are given by,
\bea
\CD^{I_1 \cdots I_r} 
= \int _\Sigma d^2 t_1 \, \bar \om^{I_1} (t_1) \cdots \int _\Sigma d^2 t_r \, \bar \om^{I_r} (t_r) \,
\CD(t_1, \cdots , t_r)
\label{mdec.01}
\eea
for $r=2$ by $C_\delta ^{IJ} = D^{IJ} _\delta - \pi Y^{IJ}$ with $D_\delta^{IJ} $ given by (\ref{2.1.b}) and vanish for $r=0,1$.

{\prop
\label{6.prop:10}
The functions $\CD^{I_1 \cdots I_r} (1,\cdots, n)$ satisfy the following properties \cite{DHoker:2023khh}. \begin{description}
\itemsep =0in 
\item (a) They are modular tensors under the congruence subgroup $\Gamma_h(2)$ of ${\rm Sp}(2h,\ZZ)$, 
\bea
\Gamma_h(2) = \big\{ M \in {\rm Sp}(2h,\mathbb Z)\, | \, M = I \ {\rm mod} \ 2 \big\}
\label{defgam}
\eea
that leaves each spin structure $\delta$ and associated Szeg\"o kernel $S_\delta(x,y)$ invariant. 
\item (b) The constant tensors $\CD^{I_1 \cdots I_r} $ are invariant under cyclic permutations of its indices,
\bea
C_\delta^{I_1 I_2 \cdots I_r} = C_\delta^{I_2 \cdots I_r I_1}
\label{cyccdelta}
\eea
\item (c) and transform as follows under reflection of its indices,
\bea
\label{6.refl}
\CD^{I_1 I_2 \cdots I_{r-1} I_r} & = & (-)^r \CD ^{I_r I_{r-1} \cdots I_2 I_1}
\eea
\end{description}
}

The proof of the proposition is readily obtained by inspection of the definition of $\CD^{I_1 \cdots I_r} (1, \cdots , n)$. In particular item $(a)$  follows from the invariance of the cyclic product $C_\delta$ under $\Gamma_h(2)$ (see (\ref{trfdelta}) for the ${\rm Sp}(2h,\mathbb Z)$ transformation of the spin structures) together with the modular transformation of the differentials $\bar \omega^{I_r}(t_r)$ in the recursion (\ref{4.5}), while items $(b)$ and $(c)$ follow from the cyclic and reflection properties of $C_\delta(t_1,\cdots,t_r)$ in the integrand of (\ref{mdec.01}). 

\sm

The theorem below presents a constructive proof of the \textit{modular descent} in terms of DHS kernels,  
already presented without proof  in equations (33-35) of \cite{DHoker:2023khh}.

{\thm
\label{4.thm:1}
The family of modular tensors $\CD^{I_1 \cdots I_r} (1,\cdots, n)$ solves the following system of descent equations.  For $n \geq 3$ and $r \geq 0$, we have, 
\bea
\CD ^{I_1 \cdots I_r } (1, \cdots , n) 
& = & 
\omega_J(1) \CD ^{I_1 \cdots I_r J} (2, \cdots , n)   
+  \sum_{i=0}^{r-1} f^{I_{r}  \cdots I_{i+1} }{}_J(1,2) \CD ^{I_1 \cdots I_i J} (2, \cdots , n)    
\no \\ &&
 +   \Big (   \p_1 \cG ^{I_{r} \cdots  I_1} (1,n){-} \p_1 \cG ^{I_{r} \cdots  I_1} (1,2) \Big ) C_\delta (2,\cdots, n) 
\label{4.1}
\eea
For $n=2$ with $r \geq 0$ and for $n=1$ with $r \geq 1$, we have,
\bea
\CD ^{I_1 \cdots I_r} (1,2) & = &
\omega_J(1)\CD^{I_1 \cdots I_r J}(2)
+  \sum_{i=0}^{r-1} f^{I_r \cdots I_{i+1}}{}_J(1,2) \CD^{I_1 \cdots I_i J}(2)
+ \p_1 \p_2 \cG^{I_r \cdots I_1} (1,2)
\notag \\
\CD^{I_1 \cdots I_r}(1) & = & \om_J(1) \CD^{I_1 \cdots I_r J} 
+ \sum_{{0 \leq i < j \leq r \atop (i,j) \not= (0,r)}} (-1)^i 
\p_1 \Phi ^{I_1 \cdots I_i \shuffle I_r \cdots I_{j+1}}{}_J (1) \CD^{I_{i+1} \cdots I_j J}
\label{4.2}
\eea
where we use the notation $f^\emptyset {}_J(1,2) = \om_J(1)$ while  $\p \cG$ and $\p \Phi$ are introduced in (\ref{4.PhiG}).}

\subsection{Proof of Theorem \ref{4.thm:1}}

To prove Theorem \ref{4.thm:1}, we begin by proving the following lemma. 

{\lem
\label{4.lem:1}
The functions $\CD^{I_1 \cdots I_r}(1,\cdots, n)$, defined in (\ref{4.5}), satisfy the following system of differential equations.  For $r=0$ and $n \geq 3$ with $1 \leq k \leq n$, 
\bea
\label{4.3}
\bar \p_k C_\delta (1,\cdots, n) 
= \pi \, \big ( \delta (k,k+1) - \delta (k,k-1) \big ) \, \CD (1, \cdots, \hat k, \cdots, n)
\eea
For $r \geq 1$ and $n \geq 2$ with $2 \leq k \leq n-1$, we have,
\bea
\label{4.4}
\bar \p_1 \CD^{I_1 \cdots I_r} (1, \cdots, n) & = & \pi \, \delta (1,2) \CD^{I_1 \cdots I_r} (2, \cdots , n)
- \pi \, \bar \om^{I_r}(1) \, \CD^{I_1 \cdots I_{r-1}} (1, \cdots, n) 
 \\
\bar \p _k \CD^{I_1 \cdots I_r} (1, \cdots, n) 
& = & \pi \, \big ( \delta (k,k+1)-\delta(k,k-1) \big ) \, \CD^{I_1 \cdots I_r} (1, \cdots, \hat k, \cdots, n) 
\no \\
\bar \p _n \CD^{I_1 \cdots I_r} (1, \cdots, n) 
& = & 
\pi \bar \om ^{I_1} (n) \, \CD^{I_2 \cdots I_r}(1,\cdots, n)
- \pi \, \delta(n,n-1) \, \CD^{I_1 \cdots I_r} (1, \cdots, n-1) 
 \no
\eea
For $n=1$ and $r \geq 2$, we have,
\bea
\label{4.4b}
\bar \p_1 \CD^{I_1 \cdots I_r} (1) = \pi \bar \om^{I_1}(1) \CD^{I_2 \cdots I_r }(1)
- \pi \bar \om^{I_r}(1) \CD^{I_1 \cdots I_{r-1}} (1) 
\eea}
The proof of the lemma proceeds by induction in $r$ for all values of $n$. For $r=0$ and all values of $n \geq 3$, equation (\ref{4.3}) is an immediate consequence of the $\bar \p$ derivative of the Szeg\"o kernel.  For $r=1$ and $n \geq 2$ equation (\ref{4.5}) reduces to,
\bea
\CD^I(1,\cdots, n) = \int _\Sigma d^2 t \, \bar \om^I(t) \, \CD(t,1,\cdots,n)
\eea
Its $\bar \p_1$ and $\bar \p_n$ derivatives may be evaluated using (\ref{4.3}) on $\CD(t,1,\cdots,n)$ and readily produce the first and last equations in (\ref{4.4}) while its $\bar \p_k$ derivative gives the middle equations for $2 \leq k \leq n-1$. Assuming now that the system of equations (\ref{4.4}) holds for all $r \leq s-1$ we shall show that it also holds for $r=s$. Indeed, from the definition of $\CD^{I_1 \cdots I_s} (1, \cdots , n)$  in (\ref{4.5}), we have for all $1 \leq j \leq n$,
\bea
\p_j \CD^{I_1 \cdots I_s} (1, \cdots , n)
= \int _\Sigma d^2 t \,  \bar \om^{I_s}(t) \, \bar \p_j \CD^{I_1 \cdots I_{s-1}} (t, 1, \cdots, n)  
\eea
For each value of $j$ we use the corresponding equation in (\ref{4.4}) for $r=s-1$ to evaluate the $\bar \p_j$ derivative and readily verify that the result is the corresponding equality of (\ref{4.4}) for $r=s$, which completes the proof of the lemma.

\subsubsection{Completing the proof of Theorem \ref{4.thm:1}}

To prove Theorem \ref{4.thm:1} we introduce the following combination for $n \geq 3$, 
\bea
\label{4.pr1}
\tilde C_\delta^{I_1 \cdots I_r}(1,\cdots, n) 
& = & 
\CD ^{I_1 \cdots I_r } (1, \cdots , n) 
- \sum_{i=0}^{r} f^{I_{r}  \cdots I_{i+1} }{}_J(1,2) \CD ^{I_1 \cdots I_i J} (2, \cdots , n)  
\no \\ &&
+  \Big (  \p_1 \cG ^{I_{r} \cdots  I_1} (1,2) {-}  \p_1 \cG ^{I_{r} \cdots  I_1} (1,n) \Big ) C_\delta (2,\cdots, n)   
\eea
while for $n=2$ and $n=1$, we define, 
\bea
\tilde C_\delta^{I_1 \cdots I_r} (1,2) & = & \CD ^{I_1 \cdots I_r} (1,2)  - \sum_{i=0}^{r} f^{I_r \cdots I_{i+1}}{}_J(1,2) \CD^{I_1 \cdots I_i J}(2)
- \p_1 \p_2 \cG^{I_r \cdots I_1} (1,2)
 \label{other4.pr1} \\
\tilde C_\delta^{I_1 \cdots I_r}(1) & = &  \CD^{I_1 \cdots I_r}(1) - \om_J(1) \CD^{I_1 \cdots I_r J} 
- \!\! \sum_{{0 \leq i < j \leq r \atop (i,j) \not= (0,r)}} (-1)^i 
\p_1 \Phi ^{I_1 \cdots I_i \shuffle I_r \cdots I_{j+1}}{}_J (1) \CD^{I_{i+1} \cdots I_j J}
\no 
\eea
Manifestly, the vanishing of these quantities is equivalent to  equation (\ref{4.1}) for $n \geq 3$ and equation (\ref{4.2}) for $n=2,1$. By Lemma \ref{4.lem:1} and the Massey system (\ref{massey.1}), (\ref{massey.2}) of DHS kernels, their $\bar \p_1$ derivatives for $n \geq 2$ and $r \geq 1$ evaluate as follows,
\bea
\label{4.pr2}
\bar \p_1  \tilde \CD (1,\cdots, n)  & = & 0
\no \\
\bar \p_1  \tilde C_\delta^{I_1 \cdots I_r}(1,\cdots, n)  & = & - \pi \bar \om^{I_r} (1) \tilde C_\delta^{I_1 \cdots I_{r-1}}(1,\cdots, n) 
\eea
while for $n=1$ we have,
\bea
\bar \p _1 \tilde C_\delta^{I_1 \cdots I_r} (1) = 
\pi \bar \om^{I_1}(1) \tilde C_\delta^{I_2 \cdots I_r}(1) - \pi \tilde C_\delta^{I_1 \cdots I_{r-1} }(1) \bar \om^{I_r}  (1)
\label{n1lim}
\eea
The derivation of (\ref{n1lim}) makes use of the cyclic symmetry of the constants of (\ref{cyccdelta}).
Furthermore, the integral against $\bar \om^{I_{r+1}}(1)$ over $z_1 \in \Sigma$ may be evaluated with the help of (\ref{4.5}) and the fact that the integral of $\bar \om^{I_r}(1)$ against $f^{I_{r}  \cdots I_{i+1} }{}_J(1,2)$ and $\p_1 \cG ^{I_{r} \cdots  I_1} (1,k)$ vanishes for all $ i \not = r$ and all $k$  and we find, 
\bea
\label{4.pr3}
\int _\Sigma d^2 t \, \bar \om^{I_{r+1}}(t) \, \tilde C_\delta^{I_1 \cdots I_r}(t,2,\cdots, n) 
=0
\eea
We now proceed to a proof by induction in $r$. For $r=0$ the first equation in (\ref{4.pr2}) tells us that $\tilde \CD (1,\cdots, n) $ is holomorphic in $z_1$, so that it must be a linear combination of the holomorphic Abelian differentials $\om_J(1)$ with $z_1$-independent coefficients. Then  (\ref{4.pr3}) implies that $\tilde \CD (1,\cdots, n) =0$. Let us now assume that $\tilde C_\delta^{I_1 \cdots I_r}(1,\cdots, n)=0$ for all $r  \leq s-1$. Therefore, the second equation of (\ref{4.pr2}) and (\ref{n1lim}) imply that  $\tilde C_\delta^{I_1 \cdots I_s}(1,\cdots, n)$ is holomorphic in $z_1$ for all $n\geq 1$ so that it must be a linear combination of the holomorphic Abelian differentials $\om_J(1)$ with $z_1$-independent coefficients. Then  (\ref{4.pr3}) implies that $\tilde C_\delta^{I_1 \cdots I_s}(1,\cdots, n) = 0$. Thus, we conclude that $\tilde C_\delta^{I_1 \cdots I_r}(1,\cdots, n)=0$ for all $n$ and $r$, which completes the proof of Theorem \ref{4.thm:1}.

\subsection{Modular decomposition of cyclic products of Szeg\"o kernels}
\label{sec:4.dec}

The modular descent described in the previous subsections leads to the decomposition of the cyclic product $C_\delta(1,\cdots,n)$ of Szeg\"o kernels in the following theorem, which was already announced in (\ref{intro.04}) and will be referred to as the \textit{modular decomposition},

{\thm
\label{4.thm:cdec} 
The spin-structure dependence of $C_\delta$ can be fully separated from the dependence on the points $z_1,\cdots,z_n$ for $n \geq 2$ through the modular decomposition,
\begin{align}
C_\delta (1,\cdots, n) &=
  {\cal V}(1,\cdots,n)  +  \sum_{r=2}^{n}
  {\cal V}_{I_1 \cdots I_{r}}(1,\cdots,n) \, C_\delta^{I_1 \cdots I_{r}}
\label{again.02}
\end{align}
The multiplets ${\cal V}_{I_1 \cdots I_{r}}(1,\cdots,n)$ are independent of the spin structure, meromorphic and single-valued $(1,0)$ forms in $z_1,\cdots,z_n \in \Sigma$,  modular tensors under the full modular group ${\rm Sp}(2h,\ZZ)$, cyclically symmetric in the indices $I_1\cdots I_r$ and expressible in terms of  (products, index contractions and derivatives of) DHS kernels.  In particular, the term with $r=n$ is given by cyclically symmetrized products  of holomorphic Abelian differentials, 
\bea
\cV_{I_1 \cdots I_n} (1,\cdots, n) = 
\frac{1}{n} \, \omega_{I_1}(1) \, \cdots \,  \omega_{I_n}(n) + {\rm cycl}(I_1,\cdots,I_n)
\label{w0calv}
\eea
}

The proof of the theorem largely follows the ideas in the proof of Theorem \ref{2.thm:cdec}.
The modular decomposition (\ref{again.02}) and the explicit form of 
${\cal V}_{I_1 \cdots I_{r}}(1,\cdots,n)$ in terms of DHS kernels follow iteratively from the modular descent in 
(\ref{4.1}) and (\ref{4.2}) in the same way as the meromorphic decomposition (\ref{again.03}) is
obtained from the meromorphic descent in (\ref{rewdesc.1}) and (\ref{rewdesc.2}). 
The resulting ${\cal V}_{I_1 \cdots I_{r}}(1,\cdots,n)$ are term by term modular tensors 
under the full modular group $\Sp(2h,\mathbb Z)$ since this is already the case for the 
composing DHS kernels \cite{DHS:2023} and Abelian differentials.
Meromorphicity of the ${\cal V}_{I_1 \cdots I_{r}}(1,\cdots,n)$ with $r<n$ in $z_1,\cdots,z_n$ is not manifest
term by term in their DHS-kernel representation but guaranteed by the
meromorphicity (\ref{4.3}) of the cyclic product $C_\delta$ and the
analysis of antiholomorphic derivatives in the modular descent. Finally, by the cyclic symmetry of $\CD^{I_1 \cdots I_r}$, we may choose to define the $\cV_{I_1 \cdots I_r}(1,\cdots, n)$ to be cyclically symmetric in their indices,
\beq
{\cal V}_{I_1 I_2 \cdots I_{r}}(1,\cdots,n) = {\cal V}_{I_2  \cdots I_{r} I_1}(1,\cdots,n)
\label{cycV}
\eeq
and we shall do so throughout. This completes the proof of Theorem \ref{4.thm:cdec}.

\sm

Note that, while the modular tensors $ {\cal V}_{I_1  \cdots I_{r}}(1,\cdots,n) $ at fixed rank $r \leq n$ 
are distinct for different values of $n$, their coefficients $C_\delta^{I_1 \cdots I_{r}}$
are independent of $n$ and thus universal.

\sm

Following our choice for their meromorphic counterparts explained
at the end of section \ref{sec:rmks}, the modular tensors $\cV_{I_1I_2 \cdots I_r}(1,\cdots, n)$
with $r\geq 3$ are not taken to match their reflection images $(-1)^r\cV_{I_r \cdots I_2I_1}(1,\cdots, n)$.
This is again possible by considering the descent equations (\ref{4.2}) and decomposition (\ref{again.02})
for cyclically symmetric but otherwise unspecified  $\CD^{I_1 \cdots I_r}$ and has the advantage of introducing
a larger class of single-valued and meromorphic modular tensors $\cV_{I_1 \cdots I_r}(1,\cdots, n)$
to be used in future work.

\subsubsection{Examples for $n=2,3$ points}
\label{sec:exampledhs}

For $n=2,3$ points, the modular descent in (\ref{4.1}) and (\ref{4.2})
leads to the following examples of ${\cal V}_{I_1  \cdots I_{r}}(1,2,\cdots,n)$
at non-maximal rank $r<n$ (see (\ref{w0calv}) for the $n=r$ case),
\begin{align}
{\cal V}(1,2) &= \p_1 \p_2 \cG(1,2) 
 \label{mdec.06} \no \\
{\cal V}_{JK}(1,2,3) & =   \frac{1}{2} \, \Big[
\big (  \p_1 \cG(1,3) - \p_1 \cG(1,2) \big )  \om_J(2) \om_K(3) 
 + \om_I(1) f^I{}_{J} (2,3) \om_K(3) \notag \\
 &\quad +  \big( \om_J(1) \om_I(2) - \om_I(1) \om_J(2) \big) \p_3 \Phi^I{}_{K}(3) 
 + (J\leftrightarrow K) \Big] \no \\
 {\cal V}(1,2,3) & =    \big ( \p_1 \cG(1,3) -\p_1 \cG(1,2)  \big )  \p_2 \p_3 \cG(2,3)  
  + \om_I(1) \p_2 \p_3 \cG^I(2,3)  
\end{align}
The symmetrization $J\leftrightarrow K$ applies to both lines of the expression for ${\cal V}_{JK}(1,2,3)$
and implements our choice (\ref{cycV}).

\sm

Note that the relation $\p_1 \p_2  \cG(1,2) = \p_2 \chi(1,2) + \pi Y^{IJ}\omega_I(1) \omega_J(2)$ 
between the $z_i$-dependent parts ${\cal W}(1,2)$ and ${\cal V}(1,2)$ in the
decompositions of $C_\delta(1,2)$ implies the relation,
\beq
C_\delta^{I J}  = D_\delta^{I J}  - \pi Y^{I J} 
\label{catrk2}
\eeq
with $D_\delta^{IJ}$ given by (\ref{2.1.b}).
The term involving the inverse imaginary part $Y^{IJ}$ of the period matrix
exemplifies the non-meromorphicity of the constants $C_\delta^{I_1\cdots I_r} $
in the moduli of $\Sigma$ and is needed for $C_\delta^{I J}$ to transform
as a modular tensor of $\Gamma_h(2)$.

\subsubsection{Further properties of $ {\cal V}_{I_1 \cdots I_{r}}(1,\cdots,n) $}
\label{sec:dhsrem}

Similar to the meromorphic case in section \ref{sec:rmks}, we gather several properties of the
constituents of the modular decomposition (\ref{again.02}) in the following proposition.

{\prop
\label{prop:mdec}
The modular tensors $ {\cal V}_{I_1 \cdots I_{r}}(1,\cdots,n) $  in the modular decomposition (\ref{again.02})
of cyclic products exhibit the following properties.
\begin{itemize}
\item[$(a)$] For  $n\geq 3$, $r\leq n-1$ and $2\leq k \leq n-1$ they have simple poles in adjacent points, 
\bea
\bar \partial_1 {\cal V}_{I_1  \cdots I_{r}}(1,2,\cdots,n)
&=&  \pi \, \big ( \delta (1,2) - \delta (1,n) \big ) \,
{\cal V}_{I_1  \cdots I_{r}}(2,\cdots, n) \notag \\
\bar \partial_k {\cal V}_{I_1  \cdots I_{r}}(1,2,\cdots,n)
&=&  \pi \, \big ( \delta (k,k+1) - \delta (k,k-1) \big ) \,
{\cal V}_{I_1  \cdots I_{r}}(1,\cdots,\hat k ,\cdots, n) \notag \\
\bar \partial_n {\cal V}_{I_1  \cdots I_{r}}(1,2,\cdots,n)
&=&  \pi \, \big ( \delta (n,1) - \delta (n,n-1) \big ) \,
{\cal V}_{I_1  \cdots I_{r}}(1,\cdots, n-1)
 \label{mdec.03}
\eea
and are holomorphic in all points at $r=n$.
\item[$(b)$] For $n\geq 2$ and $0\leq r \leq n$ and for a fixed choice of their indices $I_1,\cdots,I_r$, they are invariant under cyclic permutations of the points,
\beq
{\cal V}_{I_1  \cdots I_{r}}(1,2,\cdots,n) = {\cal V}_{I_1  \cdots I_{r} }(2,\cdots,n,1)
\label{cycWleg}
\eeq
\item[$(c)$] They are  invariant under  simultaneous reflection of the indices and the points,
\beq
 {\cal V}_{I_1 I_2 \cdots I_{r}}(1,2,\cdots,n)  = (-1)^{n+r}  {\cal V}_{I_r  \cdots I_2 I_{1}}(n,\cdots,2,1) 
 \label{reflVV}
\eeq
\end{itemize}
}

\sm

The proof of item $(a)$ follows the same logic used to prove item $(a)$ of Proposition \ref{mdec.props} where the tensor $ {\cal V}_{I_1  \cdots I_{r}}(1,\cdots,n)$ replaces the multiplet $ {\cal W}_{I_1  \cdots I_{r}}(1,\cdots,n)$ and the tensor $C_\delta^{I_1\cdots I_r}$ replaces  the multiplet $D_\delta^{I_1\cdots I_r}$.

\sm

Items $(b)$ and $(c)$ are equivalent to the vanishing of the following combinations,
\bea
 {\cal P}_{I_1 \cdots I_{r}}(1,2,\cdots,n) &=&
 {\cal V}_{I_1 \cdots I_{r}}(1,2,\cdots,n)  -  {\cal V}_{I_1 \cdots I_{r}}(2,\cdots,n,1) 
 \label{defPQ} \\
 {\cal Q}_{I_1 \cdots I_{r}}(1,2,\cdots,n) &=&
 {\cal V}_{I_1 I_2 \cdots I_{r}}(1,2,\cdots,n) -  (-1)^{n+r} {\cal V}_{I_r \cdots I_2 I_{1}}(n,\cdots,2,1)
 \notag
 \eea
which we shall prove by induction in $n-r$. For $r=n$, 
the vanishing of $ {\cal P}_{I_1 \cdots I_{n}}(1,\cdots,n)$ and 
$ {\cal Q}_{I_1 \cdots I_{n}}(1,\cdots,n) $ immediately follows from
the expression (\ref{w0calv}) for ${\cal V}_{I_1 \cdots I_{n}}(1,\cdots,n)$
in terms of cyclically symmetrized Abelian differentials.
As an inductive step, let us assume that $ {\cal P}_{I_1 \cdots I_{r}}(1,\cdots,n) $
and $ {\cal Q}_{I_1 \cdots I_{r}}(1,\cdots,n) $ in (\ref{defPQ}) vanish for $n-r=s$,
then we will show that this implies their vanishing at $n-r=s+1$. For this purpose,
we note that the antiholomorphic derivatives (\ref{mdec.03}) of $ {\cal V}_{I_1 \cdots I_{r}}(1,\cdots,n) $
established as item $(a)$ imply
\bea
\bar \partial_k {\cal P}_{I_1  \cdots I_{r}}(1,2,\cdots,n)
&=&  \pi \, \big ( \delta (k,k{+}1) - \delta (k,k{-}1) \big ) \,
{\cal P}_{I_1  \cdots I_{r}}(1,\cdots,\hat k ,\cdots, n) 
\label{aderpq}
\eea
for $k=2,\cdots,n$ and $\bar \partial_1 {\cal P}_{I_1  \cdots I_{r}}(1,\cdots,n) = 0$.
As a result, $ {\cal P}_{I_1 \cdots I_{r}}(1,\cdots,n) $ at $n-r=s+1$ is holomorphic 
since its counterparts at $n-r=s$ on the right side of (\ref{aderpq}) vanish by the 
inductive assumption. So it can at best be a combination of Abelian differentials
\bea
{\cal P}_{I_1  \cdots I_{r}}(1,2,\cdots,n) \, \big|_{n-r=s+1}
= \omega_{J_1}(1) \cdots \omega_{J_n}(n) \, \mathfrak{P}^{J_1\cdots J_n}_{I_1  \cdots I_{r}}
\label{aderpqc}
\eea
with modular tensors $\mathfrak{P}^{J_1\cdots J_n}_{I_1  \cdots I_{r}}$ independent
on the points. The constant tensors $\mathfrak{P}^{J_1\cdots J_n}_{I_1  \cdots I_{r}}$ are obtained
 by integrating $ {\cal P}_{I_1 \cdots I_{r}}(1,\cdots,n) $
over $n$ copies of $\Sigma$ against $\bar \omega^{J_1}(1) \cdots \bar \omega^{J_n}(n)$.
These surface integrals vanish since each $ {\cal V}_{I_1 \cdots I_{r}}(1,\cdots,n) $ in the
modular decomposition (\ref{again.02}) with $r\leq n-1$ integrates to zero.
This follows from the fact that each contribution from the modular descent
equations (\ref{4.1}), (\ref{4.2}) is a total derivative of a single-valued function in at least one
of the points. The same reasoning applies to $ {\cal Q}_{I_1 \cdots I_{r}}(1,\cdots,n) $
in (\ref{defPQ}) where the $\bar \partial_k$ derivatives take the form of (\ref{aderpq})
with ${\cal P}\rightarrow {\cal Q}$ on both sides for all of $k=1,\cdots,n$.
This completes our inductive proof of items $(b)$ and $(c)$.

\sm

We also note that the modular tensors ${\cal V}_{I_1  \cdots I_{r}}(1,\cdots,n)$ and $\CD^{I_1  \cdots I_r} $ are generally not locally holomorphic in the complex-structure moduli of $\Sigma$, as will be detailed in a forthcoming paper \cite{DHoker:2025comp}. Still, the antiholomorphic moduli variations of $ C_\delta^{I_1 \cdots I_{r}}$ in (38-41) of \cite{DHoker:2023khh} identify the totally symmetrized components $\CD^{(I_1  \cdots I_r)} $ as locally holomorphic.

\subsection{Meromorphic versus modular descent and decomposition}
\label{sec:mevsmod}

As was already advertised in the Introduction, there exists a remarkable correspondence between the modular decomposition (\ref{again.02}) and  the meromorphic decomposition (\ref{again.03}) of the cyclic product of Szeg\"o kernels, which follows from the same correspondence between the respective systems of descent equations. 
Indeed, the meromorphic descent equations in (\ref{rewdesc.1}) and (\ref{rewdesc.2}) are mapped, term by term, to the modular descent equations in (\ref{4.1}) and (\ref{4.2}) by simultaneously swapping their elements as follows,
\bea
\DD ^{I_1 \cdots I_r } (1, \cdots , n)   & \quad \longleftrightarrow \quad  & \CD ^{I_1 \cdots I_r } (1, \cdots , n) 
\no \\
g^{I_1 \cdots I_r}{}_J(x,y) & \longleftrightarrow & f^{I_1 \cdots I_r}{}_J(x,y)
\no \\
\chi^{I_1 \cdots I_s}(x,y)  & \longleftrightarrow & \partial_x \cG^{I_1 \cdots I_s}(x,y)  
\no \\
\varpi^{I_1 \cdots I_r}{}_J(x)  & \longleftrightarrow & \partial_x \Phi^{I_1 \cdots I_r}{}_J(x) 
\label{subgtof}
\eea
This correspondence of the modular and meromorphic descent equations under the map (\ref{subgtof}) implies that the  constituents  ${\cal V}_{I_1  \cdots I_{r}}(1,\cdots,n)$ and ${\cal W}_{I_1  \cdots I_{r}}(1,\cdots,n)$ are mapped into one another, 
\bea
{\cal V}_{I_1  \cdots I_{r}}(1,2,\cdots,n) = {\cal W}_{I_1  \cdots I_{r}}(1,2,\cdots,n) \, \big|_{(\ref{subgtof})}
 \label{mdec.05}
\eea
Concretely, ${\cal V}_{I_1  \cdots I_{r}}(1,\cdots,n)$ is obtained from $ {\cal W}_{I_1  \cdots I_{r}}(1,\cdots,n) $ by  simultaneously mapping their constituents term-by-term according to (\ref{subgtof}). This reproduces the modular $n=2,3$ examples in section \ref{sec:exampledhs} from the meromorphic ones in section \ref{sec:2.ex}, and the  expressions for  ${\cal V}_{I_1  \cdots I_{r}}(1,2,3,4)$ with $r=0,2,3$ may be read off term-by-term from those of ${\cal W}_{I_1  \cdots I_{r}}(1,2,3,4)$ in (\ref{eq2.6.6}).

\subsubsection{Correspondence of $\mA$ periods and surface integrals}
\label{sec:4.qual}

At a more formal level, the correspondence (\ref{subgtof}) extends to swapping the $\mA$ periods for the meromorphic descent with surface integrals for the modular descent,
\bea
\oint_{\mA^L} dt \quad  \longleftrightarrow \quad \int_\Sigma d^2t \, \bar \omega^L(t)
\label{AvsSigma}
\eea
and the first line of (\ref{subgtof}). However, the $\mA$ integral representations (\ref{6.3.g})
of the constant multiplets $D_\delta^{I_1\cdots I_r}$ with $r\geq 3$ exhibit a tail of lower-rank convolutions with
(rational multiples of) powers of $2\pi i$ as coefficients which do not have any modular counterpart in
the surface-integral representation (\ref{mdec.01}) of $C^{I_1\cdots I_r}_\delta$. 
Similar tails of lower-rank convolutions on $\mA$ cycles also occur for the
multiplets $D^{I_1\cdots I_r}_\delta(1,\cdots,n)$ at $n\geq 1$ in (\ref{4.beta.6}) which 
are absent from the modular $C^{I_1\cdots I_r}_\delta(1,\cdots,n)$ in (\ref{4.5}).
Still, the leading order $\beta_I = B_I + \cO(2\pi i B^2)$ in the generating series
(\ref{4.beta.5}) implies that the terms in the $\mA$-integral representation of 
$D^{I_1\cdots I_r}_\delta(1,\cdots,n)$ without reference to $2\pi i$ correspond
to the expressions for $C^{I_1\cdots I_r}_\delta(1,\cdots,n)$ as surface integrals 
(\ref{4.5}) under (\ref{AvsSigma}) for all $n,r\geq 1$ (also see (\ref{4.DD.1}) for
the cases with $n=0$ and $r\geq 2$). An analogous correspondence via (\ref{AvsSigma}) was observed in \cite{DHoker:2025dhv} between the surface integral representation (\ref{recDHSpre})  and (\ref{recDHS}) of DHS kernels and the $\mA$-integral representation of Enriquez kernels, including a tail of simpler $\mA$ convolutions with powers of $2\pi i$ as prefactors in the meromorphic case.

\subsubsection{An informal argument:  derivatives versus monodromies}
\label{sec:4.vs}

The combinations ${\cal V}_{I_1  \cdots I_{r}}(1,\cdots,n)$ in (\ref{again.02}) and ${\cal W}_{I_1  \cdots I_{r}}(1,\cdots,n)$
in (\ref{again.03}) are both single-valued and meromorphic in the points $z_1, \cdots, z_n \in \Sigma$, even if  these properties are realized in different ways on $\cV$ and $\cW$.  While  ${\cal V}_{I_1  \cdots I_{r}}(1,\cdots,n)$ is manifestly single-valued, its meromorphicity relies on the cancellation of the $\bar \om^I(z_i)$ from their derivatives  $\bar \p_k$ that result from applying  (\ref{massey.1}) and (\ref{massey.2}) term-by-term and eventually conspire to (\ref{mdec.03}). Conversely, while ${\cal W}_{I_1  \cdots I_{r}}(1,\cdots,n)$ is manifestly meromorphic, its single-valuedness relies on cancellations between individual $\mB$-monodromies (\ref{2.4.mon}) of the Enriquez kernels.

\sm

The combinatorial mechanisms for the cancellations of $\bar \p_k$ derivatives in ${\cal V}_{I_1  \cdots I_{r}}(1,\cdots,n)$ and $\mB_L$ monodromies in  ${\cal W}_{I_1  \cdots I_{r}}(1,\cdots,n)$ turn out to be closely related.
A direct link may be established by \textit{formally truncating} the $\mB_L$ monodromies $\Delta^{(k)}_L $ of the Enriquez kernels in (\ref{2.4.mon}) to the first order in $2\pi i$, and denoting the corresponding operation by $\delta_L^{(k)}$,
\begin{align}
\delta^{( x )}_L g^{I_1 \cdots I_r}{}_J(x,y) & =  - 2 \pi i \,  \delta^{I_1}_L \, g^{I_2 \cdots I_r}{}_J(x,y)
\no \\
\delta^{( y )}_L g^{I_1 \cdots I_r}{}_J(x,y) & =  2 \pi i \, \delta ^{I_r}_J \, g^{I_1 \cdots I_{r-1} }{}_L(x,y) 
 \label{mdec.07}
\end{align}
These \textit{differential monodromies} $\delta^{(k)}_L$ are in one-to-one correspondence with the coefficients of 
$\bar \omega^L(z_k)$ in the $\bar \p_k$ derivatives of the DHS kernels. More specifically, the Massey system (\ref{massey.1}) and  (\ref{massey.2}) is in formal correspondence with (\ref{mdec.07}) through the substitution rules,
\bea
g^{I_1 \cdots I_r}{}_J(x,y) &\quad \longleftrightarrow \quad & f^{I_1 \cdots I_r}{}_J(x,y)
 \label{mdec.08} 
\no \\
2 \pi i \, \delta^I_L\ {\rm in} \ \delta^{(k)}_L &\longleftrightarrow & \pi \bar \omega^I(k)\ {\rm in} \ \bar \partial_k
\eea
By this link between antiholomorphic derivatives of DHS kernels and 
differential monodromies of Enriquez kernels, the meromorphicity
(\ref{mdec.03}) of ${\cal V}_{I_1  \cdots I_{r}}(1,\cdots,n)$ 
can be viewed as a consequence of the vanishing
$\mB_L$ monodromies of ${\cal W}_{I_1  \cdots I_{r}}(1,\cdots,n)$ (as
established in Theorem \ref{2.thm:cdec}), truncated to the first order in $2\pi i$.
More generally, the differential $\mB$ monodromies
of the statements in Theorem \ref{2.thm:3} on the meromorphic descent
offer an alternative, if informal, proof of the modular descent.

\sm

It would be interesting to investigate whether the contributions with $\geq 2$ powers of
$2\pi i$ in the $\mB_L$ monodromies (\ref{2.4.mon}) of Enriquez kernels have
an echo in terms of DHS kernels.

\newpage

\section{Descent of linear chain products of Szeg\"o kernels}
\setcounter{equation}{0}
\label{sec:ch}

Cyclic products of Szeg\"o kernels for even spin structure $\delta$ at generic moduli, discussed earlier in sections \ref{sec:5} and \ref{sec:6}, constitute an important ingredient in the evaluation of superstring amplitudes in the RNS formulation. However, other ingredients are required as well, such as worldsheet fermion correlators involving the worldsheet  supercurrent and stress tensor \cite{DHoker:2002hof,DHoker:2005dys}. More specifically, contributions of worldsheet fermions with an even spin structure $\delta$ to $N$-point genus-two amplitudes with massless  external NS states may all be reduced to cyclic products of Szeg\"o  kernels, linear chain products of Szeg\"o  kernels defined by,
\bea
\LL(x; z_1, \cdots, z_n; y) = S_\delta(x,z_1) S_\delta (z_1, z_2) \cdots S_\delta (z_{n-1} , z_n) S_\delta (z_n, y)
\label{defLdelta}
\eea
or products of cyclic products and linear chain products anchored at different points on the surface.
For example, the insertion of the stress tensor $T_{ww}$ into a cyclic product of Szeg\"o kernels  is achieved by taking the following limit of linear chain products,
\bea
\lim _{x,y \to w} \left ( \half \p_x \LL(x; z_1, \cdots, z_n; y) - \half \p_y \LL(x; z_1, \cdots, z_n; y)  \right )
\eea
The cyclic product $\CD(z_1, \cdots , z_{n+1})$ itself may be obtained trivially from the linear chain product 
$\LL(x; z_1, \cdots, z_n; y)$ by \textit{closing the chain},
\bea
\CD(z_1, \cdots, z_{n+1}) = \lim _{x,y \to z_{n+1}} \LL(x; z_1, \cdots, z_n; y)
\eea
 The functions $\LL(x; z_1, \cdots, z_n; y) $ are meromorphic in all their arguments; single-valued $(1,0)$ forms in the \textit{internal points} $z_1, \cdots, z_n$; and $(\half, 0)$ forms in the \textit{end points} $x,y$ in which they inherit the monodromies associated with the spin structure $\delta$. Following earlier notation, $\LL(x; z_1, \cdots, z_n; y)$ will often be abbreviated $\LL(x;1, \cdots, n; y)$.

\sm

In this section, we shall extend the descent procedures and resulting decomposition formulae of sections \ref{sec:2}, \ref{sec:5} and \ref{sec:6} to the case of linear chain products. The goal of the descent procedure here, as it was in the case of the cyclic product of Szeg\"o kernels, is to reduce the spin structure dependence to elements that are as simple as possible, and in particular  independent of the internal points $z_1, \cdots, z_n$. Since $\LL(x; z_1, \cdots, z_n; y) $ is a $(\half,0)$ form in the end points $x,y$, the simple spin structure dependent elements we seek for linear chain products must inevitably retain the dependence on~$x,y$.

\subsection{The case $n=1$}

A simple example is provided by the case $n=1$, where the Fay~trisecant  identity
(see footnote \ref{foot:6}) may be used to obtain the decomposition of $\LL(x;z;y) \!=\! S_\delta(x,z) S_\delta(z,y)$,
\bea
\LL(x;z;y) =  \p_z \ln {E(z,y) \over E(z,x)} \, S_\delta(x,y) - \om_I(z) \, { \p^I \tet [\delta](x-y) \over \tet[\delta](0) \, E(x,y)}
\label{dec2chain}
\eea
We refer to (\ref{defvth}) and (\ref{A.Szego2}) of appendix \ref{sec:A}  for the definition and properties of Riemann $\vartheta$-functions and their  appearance in the Szeg\"o kernel. Throughout, the argument $x-y$ of the $\tet$ function stands for the Abel map so that  $ \tet [\delta](x-y) $ is a shorthand for~$ \tet [\delta](\int^x_y \omega ) $.

\sm

In each term on the right side of (\ref{dec2chain}), the left factor is $\delta$-independent and contains all the dependence of the term on the internal point $z$;  the right factor is $z$-independent and contains all the  $\delta$-dependence of the term; and both factors may depend on the end points $x,y$. Using the second relation in (\ref{2.chi0}), the $z$-dependence of both terms may be expressed in terms of Enriquez kernels,
\bea
\LL(x;z;y) =  \big ( \chi(z,x) - \chi(z,y) \big ) M_\delta(x,y) + \om_I(z) M_\delta^I(x,y) 
\label{Lmot.1}
\eea
where the entire spin-structure dependence is carried by,
\begin{align}
M_\delta(x,y)&= S_\delta(x,y) \label{Lmot.2} \\
M_\delta^I(x,y) &= - { \p^I \tet [\delta](x-y) \over \tet[\delta](0) \, E(x,y)}
=   - { \p^I \tet [\delta](x-y) \over \tet[\delta](x-y) } S_\delta(x,y) 
\notag
\end{align} 
Alternatively, the decomposition may be formulated in terms of DHS kernels,
\bea
\LL(x;z;y) =  \big ( \p_z \cG(z,x) - \p_z \cG(z,y) \big ) \LL(x,y) + \om_I(z) L_\delta^I(x,y) 
\label{Lmot.3}
\eea
where the spin-structure dependence is now carried by,
\begin{align}
\LL(x,y)&= S_\delta(x,y)
\label{Lmot.4} \\
L_\delta^I(x,y) &=   \bigg( {-} { \p^I \tet [\delta](x-y) \over \tet[\delta](x-y) } 
+ 2 \pi i \, \Im \int ^y _x \om^I \bigg) S_\delta(x,y) 
\notag
\end{align}
The advantage of the decomposition (\ref{Lmot.3})  is that its terms are individually modular 
tensors under the congruence subgroup $\Gamma_h(2)$ of (\ref{defgam}), at 
the cost of introducing non-meromorphic dependence 
on the end points $x,y$ into $L_\delta^I(x,y)$ which is compensated by the Arakelov Green 
functions in (\ref{Lmot.3}).  

\sm

In the remainder of this section, we shall generalize both the manifestly meromorphic  decomposition (\ref{Lmot.1}) of the linear chain $L_\delta(x;z;y)$ and its modular  counterpart (\ref{Lmot.3}) to linear chains (\ref{defLdelta}) with an arbitrary number $n\geq 2$ of internal points. The modular descent of Theorem \ref{4.thm:1} and the decomposition formula in (\ref{again.02}) of the cyclic products $C_\delta(1,\cdots,n)$ admit  integral representations (\ref{mdec.01}) that are somewhat simpler than those of the meromorphic formulation in Theorem \ref{2.thm:cdec}, see section \ref{sec:4.qual}. Therefore, we shall start by presenting a modular decomposition of the linear chains (\ref{defLdelta}) in section \ref{sec:L.mod}, \ref{sec:modlv} and then proceed to its meromorphic counterpart in section \ref{sec:L.mero}.

\subsection{Descent for chains of arbitrary length via DHS kernels}
\label{sec:L.mod}

The descent procedure for linear chains $\LL(x;1, \cdots, n;y)$ with an arbitrary number $n$ of internal points in terms of DHS kernels $f$  may be formulated analogously to the descent procedure  for the case of cyclic products of Szeg\"o kernels in Theorem~\ref{4.thm:1}. 

\sm

The functions $L_\delta^{I_1 \cdots I_r} (x;1, \cdots, n;y)$ are differential $(1,0)$ forms in the variables $z_1, \cdots, z_n$ and $(\half,0)$ forms in $x$ and $y$. For $r=0$ they are defined  by,
\bea
L_\delta^\emptyset (x;1, \cdots, n;y)= \LL(x;1, \cdots, n; y)
 \label{Lmot.31} 
\eea
while for $r \geq 1$ they are defined recursively as follows,
\bea
L_\delta^{I_1 \cdots I_r}(x;1, \cdots, n;y) = \int _\Sigma d^2 t_r \, \bar \om^{I_r} (t_r) \, L_\delta^{I_1 \cdots I_{r-1}} (x; t_r, 1, \cdots, n; y)
 \label{Lmot.32} 
\eea  
In particular, for $n=0$ the repeated  iteration of (\ref{Lmot.32}) gives, 
\bea
L_\delta^{I_1 \cdots I_r}(x,y) = 
\int_\Sigma d^2 t_1 \, \bar \om^{I_1}(t_1) \cdots \int_\Sigma d^2 t_r \, \bar \om^{I_r}(t_r)
\LL(x;t_1, \cdots , t_r;y)
\label{Ldelint}
\eea

{\prop
\label{6.prop:12}
The functions $L_\delta^{I_1 \cdots I_r} (x;1, \cdots, n;y)$ satisfy the following properties.
\begin{description}
\itemsep=0in
\item (a) For arbitrary $n$, they are tensors under the congruence subgroup $\Gamma_h(2)$ of (\ref{defgam}).
\item (b) For $n=0$, the spinors $L_\delta^{I_1 \cdots I_r}(x,y) $  transform as follows under reflections,
\beq
 L_\delta^{I_1 I_2\cdots I_r}(y,x)  = (-1)^{r+1}  L_\delta^{I_r\cdots I_2 I_1}(x,y) 
\label{Lmot.27}
\eeq
\item (c) their antiholomorphic derivatives for $r\geq 1$ are given by,
\begin{align}
\bar \p_x L_\delta^{I_1\cdots I_r}(x,y) &= \pi \bar \om^{I_1}(x) L_\delta^{I_2\cdots I_r}(x,y) 
\label{aholoLd}\\
\bar \p_y L_\delta^{I_1\cdots I_r}(x,y) &= - \pi \bar \om^{I_r}(y) L_\delta^{I_1\cdots I_{r-1}}(x,y) 
\notag
\end{align}
\item (d) and their monodromies are given by, 
\begin{align}
L_\delta^{I_1\cdots I_r}(\mA^K\! \cdot \! x,y) &= e^{2\pi i \delta'_K} L_\delta^{I_1\cdots I_r}(x,y) 
\no \\ 
L_\delta^{I_1\cdots I_r}(\mB_K \! \cdot \! x,y) &= e^{2\pi i \delta_K''} L_\delta^{I_1\cdots I_r}(x,y)
\no \\
L_\delta^{I_1\cdots I_r}(x,\mA^K \! \cdot \! y) &= e^{2\pi i \delta'_K} L_\delta^{I_1\cdots I_r}(x,y)
\no \\ 
L_\delta^{I_1\cdots I_r}(x,\mB_K \! \cdot \! y) &= e^{2\pi i \delta_K''} L_\delta^{I_1\cdots I_r}(x,y)
\label{almostsv}
\end{align}
where we use the standard decomposition  $\delta = \delta'' + \Omega \delta'$ with $2 \delta ', 2 \delta '' \in \ZZ_2^h$. 
\end{description}
}

The following theorem shows how the functions $L_\delta^{I_1 \cdots I_r} (x;1, \cdots, n;y)$ solve the descent equations for the case of linear chain products of Szeg\"o kernels.

{\thm
\label{4.thm:4}
For $n \geq 2$ the functions $L_\delta^{I_1 \cdots I_r} (x;1, \cdots, n;y)$ satisfy the following system of descent equations, 
\bea
L_\delta^{I_1 \cdots I_r} (x;1, \cdots, n;y) & = &
\sum_{i=0}^{r} f^{I_r \cdots I_{i+1}} {}_J (1,2) \, L_\delta^{I_1 \cdots I_i J} (x;2, \cdots, n;y)
 \label{Ldesc1} \\ &&
- \Big ( \p_1 \cG^{I_r \cdots I_1} (1,2) - \p_1 \cG^{I_r \cdots I_1} (1,x) \Big ) L_\delta(x;2, \cdots, n;y)
\no
\eea
while for $n=1$ they obey,
\bea
\L_\delta^{I_1 \cdots I_r} (x;z;y) &=& 
\sum_{i=0}^{r} f^{I_r \cdots I_{i+1}}{}_J(z,y) \, L_\delta^{I_1 \cdots I_i J} (x,y)
 \label{Ldesc2}   \\ &&
- \Big ( \p_z \cG^{I_r \cdots I_1} (z,y) - \p_z \cG^{I_r \cdots I_1} (z,x) \Big )L_\delta(x,y)
\no
\eea}

\sm

To prove the theorem, we need the result analogous  to Lemma \ref{4.lem:1} giving the differential equations. For $r=0$ with $1 \leq k \leq n$ and setting $z_0=x$ and $z_{n+1}=y$ we have, 
\bea
\bar \p _k \LL(x;1, \cdots, n;y) & = & 
\pi \big ( \delta (k, k+1) - \delta (k,k-1) \big ) \LL(x; 1, \cdots, \hat k, \cdots , n;y) 
 \label{Lmot.33} 
\eea
For $ r \geq 1$ with $k=1$,  $2 \leq k \leq n-1$ and $k=n$ we have respectively,
\bea
\bar \p _1 L_\delta^{I_1 \cdots I_r} (x;1, \cdots, n;y) & = & \pi  \delta (1,2) \, L_\delta^{I_1 \cdots I_r} (x;2, \cdots, n;y)  
 \label{Lmot.34}  \\ &&
- \pi \bar \om^{I_r} (z_1) L_\delta^{I_1 \cdots I_{r-1}} (x;2, \cdots, n;y)
\no \\
\bar \p _k L_\delta^{I_1 \cdots I_r} (x;1, \cdots, n;y) & = & 
\pi \big ( \delta (k, k+1) - \delta (k,k-1) \big )
\no \\ && \quad \times  
L_\delta^{I_1 \cdots I_r}  (x; 1, \cdots, \hat k, \cdots , n;y)
\no \\
\bar \p _n  L_\delta^{I_1 \cdots I_r} (x;1, \cdots, n;y) & = & 
\pi \big ( \delta (n, y) - \delta (n,n-1) \big ) L_\delta^{I_1 \cdots I_r}  (x; 1, \cdots , n;y)
\no
\eea
The proof of the theorem proceeds along the same lines as the proof of Theorem \ref{4.thm:1}.

\subsection{Modular decomposition of linear chain products}
\label{sec:modlv}

As a result of iterating the modular descent in (\ref{Ldesc1}) and (\ref{Ldesc2}), linear chains (\ref{defLdelta}) of Szeg\"o kernels with an arbitrary number $n$  of internal points can be decomposed as,
\bea
L_\delta(x;1,\cdots,n;y) &= & \sum_{r=0}^{n} \cV_{I_1\cdots I_r}(x;1,\cdots,n;y) L_\delta^{I_1\cdots I_r}(x,y)
\label{moddecL}
\eea
Similar to the modular decomposition (\ref{again.02}) of the cyclic products $C_\delta$ of Szeg\"o kernels, all the dependence on the internal points $z_1,\cdots,z_n$ is carried by DHS kernels, namely by the  tensors $f$ and $\partial \cG$ in (\ref{Ldesc1}) and (\ref{Ldesc2}). The DHS kernels are grouped into  combinations $ \cV_{I_1\cdots I_r}(x;1,\cdots,n;y)$ meromorphic in the internal points and single-valued in all points according to the indices of the accompanying spinors
$L^{I_1\cdots I_r}_\delta(x,y)$ in (\ref{Ldelint}) that capture the entire spin-structure dependence of (\ref{moddecL}). By the composition of $\cV_{I_1\cdots I_r}(x;1,\cdots,n;y)$ from  DHS kernels and the integral representation
(\ref{Ldelint}) of $L^{I_1\cdots I_r}_\delta(x,y)$, they are modular tensors
of $\Sp(2h,\mathbb Z)$ and $\Gamma_h(2)$, respectively, and we
refer to (\ref{moddecL}) as the \textit{modular decomposition} of linear chain products. In particular, for $r=n$ one readily establishes,
\bea
\label{moddecL.2}
\cV _{I_1 \cdots I_n} (x;1,\cdots, n;y) & = & \omega_{I_1}(1) \cdots \omega_{I_n}(n)
\eea
A significant difference between the decompositions of cyclic products and linear chain products of Szeg\"o kernels is that the extra dependence on the end points in the latter case enters both the $ \cV_{I_1\cdots I_r}(x;1,\cdots,n;y)$ and the  $L^{I_1\cdots I_r}_\delta(x,y)$ on the right side of (\ref{moddecL}).

\subsubsection{Examples of the modular tensors $\cV_{I_1\cdots I_r}(x;1,\cdots,n;y)$}
\label{sec:exv12}

The simplest non-trivial example of the modular tensors
$ \cV_{I_1\cdots I_r}(x;1,\cdots,n;y) $ in (\ref{moddecL}) involving
one internal point can be read off from (\ref{Lmot.3}),
\beq
\cV(x;1;y) =  \p_1 \cG(1,x) - \p_1 \cG(1,y) 
 \label{Lmot.36} 
\eeq
With two internal points, the non-trivial $ \cV$ tensors 
in (\ref{moddecL}) are given by,
\begin{align}
\cV_J(x;1,2;y) &= \p_1 \big( \cG(1,x) -\cG(1,2) \big) \om_J(2) + \om_I(1) f^I{}_J(2,y)
 \label{Lmot.37} \\
\cV(x;1,2;y) &=  \p_1\big( \cG(1,x) -\cG(1,2) \big) \p_2 \big( \cG(2,x) -\cG(2,y) \big) 
\no \\ & \qquad 
+ \om_I(1) \p_2 \big( \cG^I(2,x) -\cG^I(2,y) \big)  \notag
\end{align}
Their analogues with three internal points and rank $\leq 2$
can be found in appendix \ref{sec:vwith3}. The cases at 
highest rank $r=n$ follow the simple formula in (\ref{moddecL.2}).

\subsubsection{Properties of the modular tensors $\cV_{I_1\cdots I_r}(x;1,\cdots,n;y)$}
\label{sec:manyp}

We gather several properties of the  modular tensors $\cV_{I_1\cdots I_r}(x;1,\cdots,n;y)$ entering the modular decomposition (\ref{moddecL}) of linear chains  in the following~proposition.

{\prop
\label{ldec.props}
The modular tensors $\cV_{I_1\cdots I_r}(x;1,\cdots,n;y)$ satisfy the properties.
\begin{itemize}
\itemsep=0in
\item[$(a)$] The antiholomorphic derivatives of $\cV_{I_1\cdots I_r}(x;1,\cdots,n;y)$
with $r\leq n-1$ and $k=1,\cdots,n$ are given by (setting $z_0 = x$ and $z_{n+1} = y$),
\bea
\bar \p_k  \cV_{I_1\cdots I_r}(x;1,\cdots,n;y) &= & \pi 
 \big ( \delta (k, k+1) - \delta (k,k-1) \big )  
\cV_{I_1 \cdots I_r}  (x; 1, \cdots, \hat k, \cdots , n;y)
\no \\
\bar \p_x  \cV_{I_1\cdots I_r}(x;1,\cdots,n;y) &= & 
- \pi \bar \omega^K(x) \cV_{ K I_1\cdots I_r}(x;1,\cdots,n;y)
\no \\ &&
+ \pi \delta(1,x) \cV_{ I_1\cdots I_r}(x;2,\cdots,n;y)
\no \\
\bar\p_y \cV_{I_1\cdots I_r}(x;1,\cdots,n;y) &=  & 
\pi \cV_{  I_1\cdots I_r K }(x;1,\cdots,n;y) \bar \om^K(y)
\no \\ &&
 - \pi \delta(n,y)  \cV_{ I_1\cdots I_r}(x;1,\cdots,n{-}1;y)
 \label{Lmot.30}
\eea
while the cases with $r=n$ in (\ref{moddecL.2}) are holomorphic.
\item[$(b)$] The modular tensors $\cV_{I_1\cdots I_r}(x;1,\cdots,n;y)$ exhibit the 
following alternating parity under simultaneous reflection
of the indices and the points,
\beq
 \cV_{I_1 I_2\cdots I_r}(y;n,\cdots,2,1;x) = (-1)^{n-r}  \cV_{I_r\cdots I_2 I_1}(x;1,2,\cdots,n;y)
  \label{Lmot.29} 
\eeq
\end{itemize}
}

We note that the cyclic symmetry of $\cV _{I_1 \cdots I_r} (1,\cdots, n)$, established in (\ref{cycV}),
has no counterpart for $\cV_{I_1\cdots I_r}(x;1,\cdots,n;y)$.

\sm

To prove item $(a)$ of Proposition \ref{ldec.props}, we proceed as follows. In the antiholomorphic derivatives (\ref{Lmot.30}) of item $(a)$, the $\delta$-functions follow from the pole structure (\ref{Lmot.33}) of the linear chain
product. The modular descent is designed to preserve this pole structure without referring to any relations between spinors $L_\delta^{I_1\cdots I_r}(x,y)$ of different rank, so it must hold for each term in the decomposition (\ref{moddecL}), except for the non-singular term of rank $r=n$ since $L_\delta^{I_1\cdots I_n}(x,y)$ does not enter
linear chains with $n-1$ internal points. The contributions of $\bar \om^K(x),\bar \om^K(y)$ to the antiholomorphic derivatives (\ref{Lmot.30}) of item $(a)$ can be inferred by imposing meromorphicity of the modular decomposition.
They follow by evaluating $\bar\partial_x,\bar\partial_y$ of (\ref{moddecL}), using the
antiholomorphic derivatives of $L_\delta^{I_1\cdots I_r}(x,y)$ in item $(a)$
and imposing the coefficients of the resulting $L_\delta^{J_1\cdots J_s}(x,y)$
to be free of $\bar \om^K(x),\bar \om^K(y)$. This can be imposed separately for each value
of $s = 0, 1,\cdots,n-1$ since the modular descent is designed to reflect the meromorphicity of
linear chain products without relying on any relations between
spinors $L_\delta^{I_1\cdots I_r}(x,y)$ of different rank.

\sm

Item $(b)$ is equivalent to the vanishing of,
\bea
 \cR_{I_1 \cdots I_r}(x;1,\cdots,n;y) = 
 \cV_{I_1 \cdots I_r}(x;1,\cdots,n;y) - (-1)^{n-r}  \cV_{I_r\cdots I_1}( y;n,\cdots,1;x)
 \label{defRdiff}
\eea
and proven by induction in $n-r$ similar to the proof of items $(b)$ and $(c)$
of Proposition \ref{prop:mdec}. The base case of $r=n$ in
(\ref{moddecL.2}) evidently gives rise to a vanishing $\cR$. The inductive step
consists of relating (\ref{defRdiff}) with
$n-r=s$ and $n-r=s+1$ by antiholomorphic derivatives in $z_1,\cdots,z_n$
and noting that all instances of $\cR$ with $n-r\geq 1$ vanish upon integrating
$z_1,\cdots,z_n$ over $n$ copies of the surface against $\bar \omega^{J_1}(1)
\cdots \bar \omega^{J_n}(n)$.

\sm

Note that both of $ \cV_{I_1\cdots I_r}(x;1,\cdots,n;y)$ at $r\leq n-1$ 
and $ L_\delta^{I_1 \cdots I_r}(x,y)$ at $r\geq 1$ individually depend
non-meromorphically on the complex-structure moduli of $\Sigma$. Still, their 
combination in (\ref{moddecL}) resulting in a linear chain of Szeg\"o kernels
is guaranteed to yield a meromorphic function of the moduli.

\subsection{Descent for chains of arbitrary length via Enriquez kernels}
\label{sec:L.mero}

The meromorphic descent procedure given in Theorem \ref{2.thm:3} and the modular descent procedure given in Theorem \ref{4.thm:1} for cyclic products of Sze\"o kernels were found to be related by the correspondence of (\ref{subgtof}) converting Enriquez kernels into DHS kernels and vice-versa. The same type of correspondence relates the modular descent for linear chain products of Szeg\"o kernels described in the previous sections to their meromorphic counterpart.

\sm

The multiplets $M_\delta^{I_1 \cdots I_r} (x;1, \cdots, n;y)$ are differential $(1,0)$ forms in the 
internal points $z_1, \cdots, z_n$ and $(\half,0)$ forms in the end points $x$ and $y$. For $r=0$, they are defined  by,
\bea
M_\delta^\emptyset (x;1, \cdots, n;y)= \LL(x;1, \cdots, n; y)
 \label{Lmot.51} 
\eea
and for $r \geq 1$ they are defined recursively in the rank $r$ as follows, 
\bea
 \label{Lmot.52} 
M_\delta^{I_1 \cdots I_{r}}(x;1, \cdots, n;y) & = & 
\oint _{\mA^{I_{r}}} dt \, M_\delta ^{I_1 \cdots I_{r-1}} (x;t,1,\cdots, n;y)
 \\ &&
-\sum_{\ell=1}^{r-1} (-2 \pi i )^{r  - \ell} { {\rm Ber} _{r -\ell} \over (r  -\ell)!} \, 
\delta ^{I_{r} \cdots I_{\ell}} _J \, M_\delta ^{I_1 \cdots I_{\ell-1} J} (x;1,\cdots, n;y)
\no
\eea
The following theorem shows how the functions $M_\delta^{I_1 \cdots I_{r}}(x;1, \cdots, n;y) $ solve the system of descent equations for linear chain products of Szeg\"o kernels in terms of Enriquez kernels.

{\thm
\label{thm.meroL}
For $n \geq 2$ the functions $M_\delta^{I_1 \cdots I_{r}}(x;1, \cdots, n;y) $ satisfy the following system of descent equations, 
\bea
M_\delta^{I_1 \cdots I_r} (x;1, \cdots, n;y) & = &
\sum_{i=0}^{r} g^{I_r \cdots I_{i+1}} {}_J (1,2) \, M_\delta^{I_1 \cdots I_i J} (x;2, \cdots, n;y)
 \label{Lmot.53} \\ &&
- \Big (  \chi^{I_r \cdots I_1} (1,2) - \chi^{I_r \cdots I_1} (1,x) \Big ) L_\delta(x;2, \cdots, n;y)
\no
\eea
while for $n=1$ we have,
\bea
M_\delta^{I_1 \cdots I_r} (x;z;y) &=& 
\sum_{i=0}^{r} g^{I_r \cdots I_{i+1}}{}_J(z,y) \, M_\delta^{I_1 \cdots I_i J} (x,y)
 \label{Lmot.54}   \\ &&
- \Big (\chi^{I_r \cdots I_1} (z,y) - \chi^{I_r \cdots I_1} (z,x) \Big ) S_\delta(x,y)
\no
\eea
Iterating the recursion relations reduces all the spin structure dependence of a general function $M_\delta^{I_1 \cdots I_s}(x;1,\cdots, n;y)$ to a linear combination of the basic multiplets $M_\delta^{I_1 \cdots I_r}(x,y)$ with $\delta$-independent coefficients. These basic multiplets are expressible through multiple $\mA$ convolutions of linear chain products,
\bea
\MM_\delta^{I_1 \cdots I_r}(x,y) = \oint_{\mA^{I_r}} dt_r  \cdots  \oint_{\mA^{I_1}} dt_1 \, L_\delta(x;t_1,\cdots,t_r;y)
\label{defcms}
\eea
according to the generating series,
\bea
\sum_{r=1}^\infty B_{I_r}\cdots B_{I_1} M_\delta^{I_1 \cdots I_r}(x,y) = 
\sum_{r=1}^\infty \QQ_{I_r}\cdots \QQ_{I_1} \MM_\delta^{I_1 \cdots I_r}(x,y)
\label{Lmot.55}
\eea
see (\ref{4.beta.2}) for the  series expansion of $ \QQ_{I} $ in terms of $B_I$ and section \ref{sec:Acont} for the integration contour prescription that defines the right side of (\ref{defcms}).}

\sm

The proof of the theorem proceeds as for Theorem \ref{2.thm:3} and is left to the reader. A few remarks are as follows. First,  the recursion (\ref{Lmot.53}) in the rank implies the recursion (\ref{Lmot.52}) in the number of points by integrating in $z_1$ over $\mA^{I_{r}}$, using the $\mA$ period (\ref{2.intg}) and the vanishing $\mA$  periods of $\chi^{I_r \cdots I_1} (1,2) - \chi^{I_r \cdots I_1} (1,x)$ in $z_1$. Second, the relation of (\ref{Lmot.55})  follows from iterations of (\ref{Lmot.52}), which are conveniently collected in a generating series akin to (\ref{4.gen}) by closely  following the derivation in section \ref{sec:recdds}  for cyclic products. Note that for linear chain products, the edge
case $n=0$ does not require a separate treatment in contrast to the case for cyclic products.

\subsubsection{Meromorphic decomposition of linear chains}

By iterating the meromorphic descent (\ref{Lmot.53}) and (\ref{Lmot.54}), we find the following 
alternative decomposition of the linear chain product (\ref{defLdelta}):
\bea
\label{merodecL}
L_\delta(x;1,\cdots,n;y) &= & 
 \sum_{r=0}^{n} \cW_{I_1\cdots I_r}(x;1,\cdots,n;y) M_\delta^{I_1\cdots I_r}(x,y)
\eea
Just like in the term-by-term modular decomposition (\ref{moddecL}), the dependence on the internal points $z_1,\cdots,z_n$ of the linear chain via $\cW_{I_1\cdots I_r}(x;1,\cdots,n;y)$ is fully separated from that on
the even spin structure $\delta$ via $M_\delta^{I_1\cdots I_r}(x,y)$. In contrast to (\ref{moddecL}), the decomposition (\ref{merodecL}) is meromorphic  term by term in both the points $x,y,z_1,\cdots,z_n$ and the moduli of $\Sigma$. However, the single-valuedness of the composing Szeg\"o kernels in the internal points is realized through cancellations of $\mB$ monodromies between individual contributions to $\cW_{I_1\cdots I_r}(x;1,\cdots,n;y)$ with $r\leq n-1$. Their counterparts at $r=n$ in turn are holomorphic and single-valued,
\bea
\cW_{I_1\cdots I_n}(x;1,\cdots,n;y) & = & \omega_{I_1}(1) \cdots \omega_{I_n}(n) 
\label{merodecL.2}
\eea
As will be detailed in Proposition \ref{lmdec.props} below, the monodromies of (\ref{merodecL})
in the end points $x,y$ according to the Szeg\"o kernels in (\ref{defLdelta}) additionally rely on the 
interplay between the multiplets $\cW_{I_1\cdots I_r}(x;1,\cdots,n;y)$ and $M_\delta^{I_1\cdots I_r}(x,y)$.

\subsubsection{Correspondence with the modular descent}

The meromorphic descent (\ref{Lmot.53}) and (\ref{Lmot.54}) may be obtained from the modular one in (\ref{Ldesc1}) and (\ref{Ldesc2}) by converting DHS kernels to Enriquez kernels according to, 
\bea
\label{subftog}
L_\delta^{I_1 \cdots I_r} (x;1, \cdots, n;y)  &\quad \longleftrightarrow \quad & M_\delta^{I_1 \cdots I_r} (x;1, \cdots, n;y)
\no \\
  f^{I_1 \cdots I_r}{}_J(x,y) &\longleftrightarrow & g^{I_1 \cdots I_r}{}_J(x,y)
\no \\
  \partial_x \cG^{I_1 \cdots I_s}(x,y)  &\longleftrightarrow & \chi^{I_1 \cdots I_s}(x,y)  \no \\
\partial_x \Phi^{I_1 \cdots I_r}{}_J(x)  &\longleftrightarrow & \varpi^{I_1 \cdots I_r}{}_J(x) 
\eea
which adapt the analogous substitutions (\ref{subgtof})
for the cyclic products to linear chain products. Hence, the combinations of Enriquez  kernels
$\cW_{I_1\cdots I_r}(x;1,\cdots,n;y)$ that carry the entire dependence on
the internal points in (\ref{merodecL}) are simply obtained by applying the substitution (\ref{subftog})
to the DHS kernels within the $\cV_{I_1\cdots I_r}(x;1,\cdots,n;y)$ in the modular decomposition,
\beq
\cW_{I_1\cdots I_r}(x;1,\cdots,n;y) = \cV_{I_1\cdots I_r}(x;1,\cdots,n;y) \, \Big|_{(\ref{subftog})}
\label{Lmot.61}
\eeq
The same correspondence $g^{J_1\cdots J_s}{}_K(x,y)\leftrightarrow f^{J_1\cdots J_s}{}_K(x,y)$ was found in (\ref{mdec.05}) to relate $\cW_{I_1\cdots I_r}(1,\cdots,n) \leftrightarrow \cV_{I_1\cdots I_r}(1,\cdots,n) $ in the decompositions of cyclic products.

\subsubsection{Examples of the multiplets $\cW_{I_1\cdots I_r}(x;1,\cdots,n;y)$ and $M_\delta^{I_1\cdots I_r}(x,y)$}

The simplest examples of the multiplets
$\cW_{I_1\cdots I_r}(x;1,\cdots,n;y)$ with $r\leq n-1$ can be obtained
by applying the correspondence (\ref{Lmot.61})
to the expressions for the modular tensors $\cV_{I_1\cdots I_r}(x;1,\cdots,n;y)$
with $n\leq 3$ in (\ref{Lmot.36}), (\ref{Lmot.37}) and (\ref{Lmot.38}).

\sm

The simplest examples of $M_\delta^{I_1 \cdots I_r}(x,y)$, written as
multiple $\mA$ convolution integrals (\ref{defcms}), are obtained by expanding
the letters $\beta_I$ on the right side of (\ref{Lmot.55})
via (\ref{4.beta.2}) and isolating the coefficients of $B_{I_r}\cdots B_{I_1}$,
\begin{align}
M_\delta^{I_1}(x,y) & =  \MM^{I_1} _\delta(x,y) 
\label{exofms} \\
M_\delta^{I_1 I_2 }(x,y) & =  \MM_\delta ^{I_1I_2}(x,y) - i \pi \, \delta ^{I_1}_{I_2} \, \MM_\delta^{I_2}(x,y) 
\no \\
M_\delta^{I_1 I_2 I_3}(x,y) & =  \MM_\delta ^{I_1 I_2 I_3}(x,y)  
- i \pi \, \big[ \delta ^{I_1}_{I_2} \, \MM_\delta^{I_2 I_3}(x,y) 
 + \delta ^{I_2}_{I_3} \, \MM_\delta^{I_1I_3}(x,y) \big]    -{2 \pi^2 \over 3}  \delta^{I_2 I_1 }_{I_3} \, \MM_\delta^{I_3}(x,y)
 \notag
 \end{align}
The composition of indices of the terms of lower rank is in
one-to-one correspondence with the analogous representations of the 
multiplets $D_\delta^{I_1\cdots I_r}(1,\cdots,n)$ for cyclic products in
(\ref{4.beta.6}) since the respective generating series match.
However, the expressions for the constant multiplets $D_\delta^{I_1\cdots I_r}$
in terms of similar convolutions in (\ref{6.3.g}) feature extra terms
whose index structure does not have any counterparts in (\ref{exofms}).

\subsubsection{Properties of the multiplets $\cW_{I_1\cdots I_r}(x;1,\cdots,n;y)$ and $M_\delta^{I_1\cdots I_r}(x,y)$}
\label{proswms}

We gather several properties of the 
multiplets $\cW_{I_1\cdots I_r}(x;1,\cdots,n;y)$ and $M_\delta^{I_1\cdots I_r}(x,y)$
entering the meromorphic decomposition (\ref{merodecL}) of linear chains 
in the following~proposition:

{\prop
\label{lmdec.props}
The multiplets $\cW_{I_1\cdots I_r}(x;1,\cdots,n;y)$
and $M_\delta^{I_1\cdots I_r}(x,y) $ exhibit the following properties:
\begin{itemize}
\itemsep=-0.03in
\item[$(a)$] The simple poles of the multiplets $\cW_{I_1\cdots I_r}(x;1,\cdots,n;y)$
with $r\leq n-1$ as well as $k=1,\cdots,n$ are determined by (again setting $z_0 = x$ and $z_{n+1} = y$),
\bea
\bar \p_k  \cW_{I_1\cdots I_r}(x;1,\cdots,n;y) &= & \pi 
 \big ( \delta (k, k+1) - \delta (k,k-1) \big )  
\cW_{I_1 \cdots I_r}  (x; 1, \cdots, \hat k, \cdots , n;y)
\no \\
\bar \p_x  \cW_{I_1\cdots I_r}(x;1,\cdots,n;y) &= & 
 \pi \delta(1,x) \cW_{ I_1\cdots I_r}(x;2,\cdots,n;y)
\no \\
\bar\p_y \cW_{I_1\cdots I_r}(x;1,\cdots,n;y) &=  & 
 - \pi \delta(n,y)  \cW_{ I_1\cdots I_r}(x;1,\cdots,n{-}1;y)
 \label{Lmot.30w}
\eea
while the cases with $r=n$ in (\ref{merodecL.2}) are holomorphic.
\item[$(b)$] The multiplets $\cW_{I_1\cdots I_r}(x;1,\cdots,n;y)$ are single-valued
in the internal points. Their monodromies in the end points $x,y$ are
trivial for $\mA$ cycles and take the following form for $\mB$ cycles,
\begin{align}
\Delta_L^{(x)}  \cW_{I_1\cdots I_r}(x;1,\cdots,n;y) &=
 \sum_{k=1}^{n-r} \frac{(-2\pi i)^k}{k!} \cW_{  \vec{L}_k I_1\cdots I_r}(x;1,\cdots,n;y)
\label{Lmot.60}\\
\Delta_L^{(y)} \cW_{I_1\cdots I_r}(x;1,\cdots,n;y) &=  \sum_{k=1}^{n-r} \frac{(2\pi i)^k}{k!}
\cW_{  I_1\cdots I_r  \vec{L}_k  }(x;1,\cdots,n;y)
\notag
\end{align}
with the shorthand $\vec{L}_k$ for $k$ consecutive indices $LL\cdots L$.
\item[$(c)$] The $\mA$ monodromies of the spinors
$M_\delta^{I_1\cdots I_r}(x,y) $ are identical to those of the 
individual Szeg\"o kernels and their modular counterparts $L_\delta^{I_1\cdots I_r}(x,y) $
in (\ref{almostsv}). Their $\mB$ monodromies in turn are given by
\begin{align}
M_\delta^{I_1\cdots I_r}(\mB_L\cdot x,y) &= 
e^{2\pi i \delta_L''}  \sum_{k=0}^r \frac{(2\pi i)^k}{k!} \delta^{I_1 \cdots I_k}_L  M_\delta^{I_{k+1}\cdots I_r}(x,y) 
\label{Lmot.70} \\
M_\delta^{I_1\cdots I_r}(x,\mB_L\cdot y) &= 
e^{2\pi i \delta_L''}  \sum_{k=0}^r \frac{(-2\pi i)^k}{k!} \delta^{I_{r-k+1} \cdots I_r}_L  M_\delta^{I_{1}\cdots I_{r-k}}(x,y)
\notag
\end{align}
\item[$(d)$] The multiplets $\cW_{I_1\cdots I_r}(x;1,\cdots,n;y)$ exhibit the 
following alternating parity property under simultaneous reflection
of the indices and the points,
\beq
 \cW_{I_1 I_2\cdots I_r}(y;n,\cdots,2,1;x) = (-1)^{n-r}  \cW_{I_r\cdots I_2 I_1}(x;1,2,\cdots,n;y)
  \label{Wmot.29} 
\eeq
\item[$(e)$] The analogous reflection properties of the spinors
$M_\delta^{I_1\cdots I_r}(x,y) $ are,
\beq
 M_\delta^{I_1 I_2\cdots I_r}(y,x)  = (-1)^{r+1}  M_\delta^{I_r\cdots I_2 I_1}(x,y) 
\label{Lmot.71}
\eeq
\end{itemize}
}

\sm

Item $(a)$ is proven through the same arguments that give rise to the
poles in item $(b)$ of Proposition \ref{ldec.props}, i.e.\ the delta distrubutions in
(\ref{Lmot.30}) while ignoring the additional terms involving $\bar \omega^K$.

\sm

Items $(b)$ and $(c)$ can be understood from the notion of {\it differential $\mB$ monodromies}
introduced in section \ref{sec:4.vs}. The first order in $2\pi i$ on the right side of (\ref{Lmot.60})
follows from (\ref{Lmot.61}) and the terms $\bar \omega^K(x), \bar \omega^K(y)$ in (\ref{Lmot.30}) through the correspondence between the differential 
monodromies of Enriquez kernels
and antiholomorphic derivatives of DHS kernels in (\ref{mdec.08}). The differential
monodromies in the $k=1$ terms of (\ref{Lmot.70}) are then a consequence of imposing
single-valuedness on the meromorphic decomposition (\ref{merodecL})
of linear chain products to first order in $2\pi i$ in the term-by-term
monodromies using linear independence of
$ \cW_{I_1\cdots I_r}(x;1,\cdots,n;y) $ at different rank.

\sm

The full proof of items $(b)$ and $(c)$ including higher orders in $2\pi i$
is most conveniently carried out by organizing the Enriquez kernels and 
spinors $ M_\delta^{I_1 \cdots I_r}(x,y) $ into generating functions similar 
to those in section \ref{sec:3}.
The differential monodromies derived in the previous paragraphs imply the
exact statements (\ref{Lmot.60}) and (\ref{Lmot.70}) of items $(b)$ and $(c)$
once the powers of $2\pi i$ in the monodromies of these generating functions
are shown to exponentiate as it is the case in (\ref{3.Kmon1}).

\sm

The proof of items $(d)$ and $(e)$ again exploits the correspondence (\ref{subftog})
between the constituents of the meromorphic and modular descents for linear
chain products, following the logic of the proof of
Proposition \ref{5.reflect} and items $(b)$, $(c)$ of Proposition \ref{mdec.props}. 
\begin{itemize} 
\item We prove (\ref{Wmot.29}) in item $(d)$ by adapting the arguments
in appendix \ref{sec:4.cor}: checking the 
reflection properties (\ref{Lmot.29}) of individual $ \cV_{I_1\cdots I_r}(x;1,\cdots,n;y)$
solely requires the interchange lemma and Fay identities of DHS kernels which
hold in identical form for Enriquez kernels \cite{DHoker:2024ozn} and thereby
reduce the verifications of (\ref{Wmot.29}) to those of (\ref{Lmot.29}) which are proven in section
\ref{sec:manyp}. Note that the decompositions of linear chain products in this section do not involve 
any derivatives of DHS or Enriquez kernels, so the derivation of (\ref{Wmot.29})
does not require the relations (\ref{mdec.11}) and (\ref{mdec.13}) among derivatives.
\item The statement (\ref{Lmot.71}) of item $(e)$ can be inferred from the
reflection parity $(-1)^{n+1}$ of linear chain products $L_\delta(x;1,\cdots,n;y)$, 
imposed at the level of the meromorphic decomposition (\ref{merodecL}). The proof is most
conveniently carried out by induction in the rank $r$ of $M_\delta ^{I_1 \cdots I_r  }(x,y)$ as done
in appendix \ref{sec:moreprf} in the context of cyclic products, using the
reflection properties (\ref{Wmot.29}) of the multiplets $ \cW_{I_1\cdots I_r}(x;1,\cdots,n;y)$ in item $(d)$.
At rank $r \leq 3$, a direct proof of the reflection properties based on the representations
(\ref{exofms}) and contour deformation techniques can be found in appendix~\ref{sec:refmm}.
\end{itemize}

\subsection{Coincident limits of $M_\delta ^{I_1 \cdots I_r }(x,y)$ and $L_\delta ^{I_1 \cdots I_r  }(x,y)$}
\label{sec:L.coin}

In this subsection, we relate the coincident limit $y \rightarrow x$ of the multiplets $M_\delta ^{I_1 \cdots I_r  }(x,y)$ and the tensors $L_\delta ^{I_1  \cdots I_r }(x,y)$ to the constant multiplets $D_\delta^{I_1  \cdots I_r }$ and the constant tensors $C_\delta^{I_1  \cdots I_r }$, respectively. These relations derive  from the fact that the cyclic case is given by the closure $y\rightarrow x$ of the linear chain case,
\beq
\lim_{y\rightarrow x} L_\delta(x;1,\cdots,n;y) = C_\delta(1,\cdots,n,x)
\label{L.coin.01}
\eeq
We shall treat the meromorphic and modular cases separately below as the structure of their limit differs to some degree.

\subsubsection{The limit of $M_\delta ^{I_1 \cdots I_r }(x,y)$}
\label{sec:L.coin.1}

Imposing the meromorphic decompositions of (\ref{again.03}) and (\ref{merodecL}) on the relation (\ref{L.coin.01}) leads to recursion relations for the coincident limits $\lim_{y\rightarrow x} M_\delta^{I_1 \cdots I_r }(x,y)$. These limits exist and are finite for $r\geq 1$. For example, to the lowest few ranks we obtain, 
\bea
 \label{L.coin.07}
M_\delta^{I }(x,x) & = & D_\delta^{IJ} \omega_J(x)
 \\
M_\delta^{IJ}(x,x) & =  & D_\delta^{IJK} \omega_K(x)
+  D_\delta^{IK }  \varpi^J{}_K(x) -   D_\delta^{JK }  \varpi^I{}_K(x)
\no \\
M_\delta^{IJK}(x,x) &= & D_\delta^{IJKL} \omega_L(x)
+  D_\delta^{IJL }  \varpi^K{}_L(x) -   D_\delta^{JK L }  \varpi^I{}_L(x) +  D_\delta^{I L }  \varpi^{KJ}{}_L(x)
\no  \\ && 
  +   D_\delta^{K L}  \varpi^{ IJ }{}_L(x)
 - D_\delta^{J L} \big(\varpi^{IK }{}_L(x) + \varpi^{KI}{}_L(x) \big)
\no
\eea
The combinatorial structure of these expressions for $M_\delta^{I_1 \cdots I_r}(x,x) $ at $r=1,2,3$ suggests the following generalization to arbitrary rank.

{\prop
\label{6:conj.1}
The coincident limit $y \rightarrow x$ of the spinors $M_\delta^{I_1 \cdots I_r}(x,y) $ in the meromorphic decomposition
(\ref{merodecL}) of linear chain products is given as follows in terms of Enriquez kernels and  constant multiplets $D_\delta^{J_1\cdots J_s}$,
\begin{align}
&M_\delta^{I_1 \cdots I_r}(x,x )  = D_\delta^{I_1 \cdots I_r K} \omega_K(x)
 + \! \! \sum_{0 \leq i < j \atop{(i,j)\neq (0,r)} }^r \! \!  (-1)^{i} D_\delta^{I_{i+1}\cdots I_j K}
\varpi^{I_1 \cdots I_{i} \shuffle I_{r} \cdots I_{j+1}}{}_K(x)
\label{L.coin.08}
\end{align}
}
The proof relates $M_\delta^{I_1 \cdots I_r}(x,x )$ to the one-point function $\DD^{I_1 \cdots I_r}(x )$, which may be expressed in terms of the constant multiplets using item $(d)$ of Theorem \ref{2.thm:3}. We note the formal
similarity of (\ref{L.coin.07}) with the formulae for the coincident limits of $\chi^I(x,y), \chi^{IJ}(x,y)$ and $\chi^{IJK}(x,y)$ in section 9.4 of \cite{DHoker:2024ozn}. More specifically, Conjecture 9.8 of \cite{DHoker:2024ozn}
relates $\chi^{I_1\cdots I_r}(x,x)$ to combinations of $\varpi^{J_1\cdots J_s}{}_K(x)$ and certain constants $\mN^{I_p\cdots I_q K}$ in the place of $D_\delta^{I_p\cdots I_q K}$ which mirror the double sums and
shuffle products of (\ref{L.coin.08}).

\subsubsection{The limit of $L_\delta ^{I_1 \cdots I_r}(x,y)$}
\label{sec:L.coin.3}

The correspondence of the meromorphic and modular decompositions
of linear chain products in (\ref{subftog}) does not always extend on a term-by-term
basis to coincident limits: the offset between the rank-one spinors in (\ref{Lmot.4}),
\beq
L_\delta^I(x,y) = M_\delta^I(x,y) + 2\pi i \, S_\delta(x,y) \, \Im \int^y_x \omega^I
\label{L.coin.09}
\eeq
does not have a well-defined limit $y \rightarrow x$ in view of the direction dependence of the limit of the last term above. Instead, the proper limit should be taken as follows,
\beq
\lim_{y\rightarrow x} \bigg( L_\delta^I(x,y) + \frac{\pi}{x-y} \int^y_x \bar \omega^I \bigg)
= M_\delta^I(x,x) - \pi \omega^I(x) = C_\delta^{IJ} \omega_J(x)
\label{L.coin.13}
\eeq
The case for arbitrary rank $r \geq 2$ follows its meromorphic counterpart
(\ref{L.coin.08}) under the correspondence $(M_\delta,D_\delta ,\varpi) \leftrightarrow 
(L_\delta,C_\delta ,\partial \Phi) $ and is given by the following proposition.

{\prop
\label{6:conj.2}
The coincident limit $y \rightarrow x$ of the spinors $L_\delta^{I_1 \cdots I_r}(x,y) $ for rank $r\geq 2$ exists and is given in terms of DHS kernels and the constant tensors $C_\delta^{J_1\cdots J_s}$ as follows,
\begin{align}
&L_\delta^{I_1 \cdots I_r}(x,x )  = C_\delta^{I_1 \cdots I_r K} \omega_K(x)
 + \! \! \sum_{0 \leq i < j \atop{(i,j)\neq (0,r)} }^r \! \!  (-1)^{i} C_\delta^{I_{i+1}\cdots I_j K}
\partial_x \Phi^{I_1 \cdots I_{i} \shuffle I_{r} \cdots I_{j+1}}{}_K(x)
\label{L.coin.17}
\end{align}
}
The proof proceeds by relating $L_\delta^{I_1 \cdots I_r}(x,x ) $ to the one-point tensor $\CD^{I_1 \cdots I_r}(x)$ and using the last item in Theorem \ref{4.thm:1} to express the result in terms of the constant tensors $\CD^{I_1 \cdots I_r}$.

\newpage

\section{Specializing to genus one}
\setcounter{equation}{0}
\label{sec:h1}

In this section, we shall examine the decompositions of cyclic products of Szeg\"o kernels in (\ref{again.03}), (\ref{again.02}) and of linear chain products of Szeg\"o kernels in (\ref{moddecL}), (\ref{merodecL}) for the special case of genus one, namely $h=1$. The torus $\Sigma$ will be represented as the quotient $\Sigma = \CC / \Lambda$ where the lattice $\Lambda$ is generated by the periods 1 and $\tau$ with $\Im \tau >0$. In the formulas below, the torus $\Sigma$ will be considered fixed and the dependence on $\tau$ will generally not be exhibited.  As will be shown below, the spin-structure dependence can be separated from the dependence on the points of the Szeg\"o kernels in terms of Jacobi $\vartheta_{\kappa}$ functions for $\kappa=1,2,3,4$ and coefficients of the Kronecker-Eisenstein series that furnish the integration kernels of elliptic polylogarithms  \cite{Levin:2007, BrownLevin, Broedel:2014vla, Broedel:2017kkb, Enriquez:2023}. In particular, the integral representations of the constants  $D_\delta^{I_1\cdots I_r}$ and $C_\delta^{I_1\cdots I_r}$ in section \ref{sec:recs.02} and (\ref{mdec.01}), respectively, can be made fully explicit.

\subsection{Enriquez and DHS kernels}
\label{sec:h1.1}

We begin by reviewing the restriction of the Enriquez and DHS kernels to genus one and expressing them in terms of expansion coefficients of Kronecker-Eisenstein series. More specifically, we will encounter two closely related variants of the Kronecker-Eisenstein series that are given in terms of the unique odd Jacobi theta function $\vartheta_1$ by, 
\beq
F(z,\eta) = \frac{\vartheta_1'(0) \vartheta_1(z+\eta) }{\vartheta_1(z)  \vartheta_1(\eta) }
\hskip 1in
\Omega(z,\eta) = \exp \bigg(2\pi i \eta \, \frac{\Im z} {\Im \tau} \bigg) F(z,\eta) 
\label{h1eq.01}
\eeq
The function $F(z,\eta)$ is meromorphic and multiple-valued in $z\in \Sigma$, while  $\Omega(z,\eta)$ is single-valued but non-meromorphic in $z \in \Sigma$. The $\mA$ monodromy $z \to z+1$ of $F(z,\eta)$ is trivial while its $\mB$ monodromy $z \to z+\tau$ is given by,
\bea
F(z{+}\tau,\eta)   = e^{-2\pi i \eta} F(z,\eta) 
\eea
Both $F(z,\eta)$ and $\Omega(z,\eta)$ are meromorphic and multiple-valued  in $\eta$ and their monodromies may be  obtained from those in $z$ by using the relation $F(z,\eta) = F(\eta, z)$.  The Kronecker-Eisenstein integration kernels $g^{(r)} $ and $f^{(r)}$ are obtained as the coefficients of the Laurent expansions of $F(z,\eta)$ and $\Omega(z,\eta)$  in powers of $\eta$, respectively,
\beq
F(z,\eta) = \frac{1}{\eta} + \sum_{r=1}^\infty \eta^{r-1} g^{(r)}(z)
\hskip 1in
\Omega(z,\eta) =  \frac{1}{\eta} + \sum_{r=1}^\infty \eta^{r-1} f^{(r)}(z)
\label{h1eq.02}
\eeq
The Enriquez and DHS kernels reduce to $g^{(r)}(z)$ and $f^{(r)}(z)$ for $r \geq 1$, respectively \cite{Enriquez:2011, DHS:2023}, 
\beq
g^{I_1\cdots I_r}{}_J(x,y) \, \big|_{h=1} = g^{(r)}(x{-}y)
\hskip 1in
f^{I_1\cdots I_r}{}_J(x,y) \, \big|_{h=1} = f^{(r)}(x{-}y)
\label{h1eq.03}
\eeq
where all the indices on the left sides are restricted to take the value $1$. In particular, the $y$-independent traceless parts in the decompositions (\ref{2.dec}) and (\ref{4.PhiG}) vanish, so that,
\begin{align}
\varpi^{I_1\cdots I_r}{}_J(x) \, \big|_{h=1} &= 0 
&\chi^{I_1\cdots I_s}(x,y) \, \big|_{h=1} &= - g^{(s+1)}(x{-}y)
\notag \\
\partial_x \Phi^{I_1\cdots I_r}{}_J(x) \, \big|_{h=1} &= 0
&\partial_x \cG^{I_1\cdots I_s}(x,y) \, \big|_{h=1} &= - f^{(s+1)}(x{-}y)
\label{h1eq.04}
\end{align}
For $r\geq 3$, the  coincident limit of $ g^{(r)}(x{-}y)$  produces the holomorphic Eisenstein series~$ {\rm G}_r$, 
\beq
\lim _{y \to x} g^{(r)}(x-y) = - {\rm G}_r 
= - \sum_{m,n \in \mathbb Z \atop (m,n) \neq (0,0)} \frac{1}{(m\tau {+} n)^r}
\label{h1eq.05}
\eeq
which are modular forms of weight $(r,0)$ under ${\rm SL}(2,\ZZ)$  and vanish for odd $r$.
For $r=2$, the double sum in (\ref{h1eq.05}) is conditionally convergent and may be defined by using 
the Eisenstein summation prescription  \cite{123modular, DHoker:2024cup} which results in the 
quasi-modular form ${\rm G}_2$. A modular-covariant counterpart of $ {\rm G}_2$ is given by the 
almost holomorphic modular form,
\beq
\widehat {\rm G}_2 = {\rm G}_2- \frac{\pi}{\Im \tau}
\label{h1eq.06}
\eeq

\subsection{Szeg\"o kernels for even spin structures}
\label{sec:h1.1a}

For genus one, the Szeg\"o kernel $S_\delta(x,y)$ for an even spin structure $\delta$ may be expressed in terms of the even Jacobi theta functions $\vartheta_\delta$ with $\delta = 2,3,4$,
\beq
S_\delta(z)  = \frac{\tet_1'(0) \tet_\delta(z)}{\tet_1(z) \tet_\delta(0)}
\label{h1eq.07}
\eeq
where translation invariance on the torus implies that $S_\delta(x,y) |_{h=1} = S_\delta(z)$ only depends on the difference of the points $z=x-y$. The Szeg\"o kernel $S_\delta(z)$ of  (\ref{h1eq.07}) is closely related to  the Kronecker-Eisenstein series $F(z,\eta)$ of  (\ref{h1eq.01}) when $\eta$ is set to the corresponding half-period $\hp_\delta$ given as follows,
\bea
\label{h1eq.08}
\om_\delta = u_\delta \tau + v_\delta 
\hskip 0.8in 
(u_2,v_2) = (0, \thalf), ~ (u_3,v_3) = (\thalf, \thalf), ~(u_4,v_4) = (\thalf, 0)
\eea
and we have \cite{Tsuchiya:2017joo}, 
\bea
\label{h1eq.09}
S_\delta(z) =  e^{ 2 \pi i  z u_\delta }  F(z,\hp_\delta) = F(z,\hp_\delta) \, \times\, \begin{cases}
\ \ \ 1 &: \ \delta=2 \\
e^{ i \pi z} &: \ \delta=3,4
\end{cases} 
\eea
Alternatively, we may express $S_\delta(z)$ in terms of $\Omega(z,\omega_\delta)$ of (\ref{h1eq.01}), which is conveniently done with the help of real \textit{co-moving coordinates} $u,v \in \RR/\ZZ$ related to $z$ by $z = u \tau + v$, 
\bea
S_\delta(z) =  \Omega(\hp_\delta,z) =  e^{ 2 \pi i  (v \, u_\delta  - u \, v_\delta )}  \,  \Omega(z, \hp_\delta)
\label{h1eq.09a}
\eea
The Fourier expansion \cite{Broedel:2018iwv, Gerken:2018},
\beq
\Omega(z,\eta) =  \sum_{m,n \in \mathbb Z}  \frac{ e^{ 2\pi i (mv - nu)} }{m\tau + n + \eta}
\label{h1eq.11}
\eeq
of the doubly-periodic Kronecker-Eisenstein series $\Omega(z,\eta)$ combined with (\ref{h1eq.09a})
gives the Fourier expansion of the Szeg\"o kernel \cite{Stieberger:2002wk}, 
\beq
S_\delta(z) =  e^{ 2 \pi i  (v \, u_\delta  - u \, v_\delta )}
\sum_{m,n \in \mathbb Z}  \frac{ e^{ 2\pi i (mv - nu)} }{m\tau + n + \hp_\delta}
\label{h1eq.12}
\eeq
which exposes the signs $e^{2\pi i u_\delta}$ and  $e^{-2\pi i v_\delta}$ produced by
the $\mA$ monodromies $v \rightarrow v+1$ and $\mB$ monodromies $u \rightarrow u+1$
of the Szeg\"o kernel, respectively.

\subsection{Cyclic products of Szeg\"o kernels}
\label{sec:h1.2}

The cyclic product of $n$ Szeg\"o kernels on the torus with even spin structure $\delta$ is given by,
\bea
C_\delta(1,2,\cdots, n) = S_\delta (z_{12}) S_\delta(z_{23})  \cdots S_\delta(z_{n1})
\eea
where we use the notation $z_{ij} = z_i - z_j$. In view of the relations (\ref{h1eq.09}) and (\ref{h1eq.09a}), the cyclic product may be expressed alternatively as follows, 
\begin{align}
C_\delta(1,2,\cdots,n) &=  F(z_{12},\hp_\delta) F(z_{23},\hp_\delta) \cdots F(z_{n1},\hp_\delta) \notag \\
&=  \Omega(z_{12},\hp_\delta) \Omega(z_{23},\hp_\delta) \cdots \Omega(z_{n1},\hp_\delta)
\label{h1eq.21}
\end{align}
In these products, all the non-trivial monodromy of the individual $F$ factors and all the non-meromorphicity of the individual $\Omega$ factors cancels so that $C_\delta(1,\cdots, n)$ is meromorphic and single-valued in $z_1,\cdots,z_n \in \Sigma$. The decomposition of $C_\delta(1,\cdots, n)$ into a sum of $\delta$-dependent constants with $\delta$ independent coefficients is given by the following proposition, which summarizes and clarifies some of the results obtained earlier  in \cite{Tsuchiya:2012nf, Tsuchiya:2017joo}. 
{\prop 
\label{8.thm:1}
The cyclic product $C_\delta(1,\cdots, n)$ of Szeg\"o kernels on the torus with spin structure $\delta$ may be decomposed as follows, 
\bea
\label{8.dec}
C_\delta (1,\cdots, n) = V_n (1,\cdots, n) + \sum _{k=1} ^{[n/2]} R_{2k}(e _\delta) V_{n - 2 k} (1,\cdots, n)
\eea
where $R_{2k}(e _\delta)$ is constant on $\Sigma$ and a modular form of weight $(2k,0)$ under the congruence subgroup $\Gamma(2) \subset {\rm SL}(2,\ZZ)$, while $V_r(1,\cdots, n)$ are $\delta$-independent elliptic (i.e.\ meromorphic doubly periodic) functions in the points $z_i$. The functions $V_r(1,\cdots, n)$ may be expressed either in terms of the meromorphic Kronecker-Eisenstein coefficients,
\bea
V_r(1,\cdots,n) = \sum_{s_1,s_2,\cdots,s_n \geq 0
\atop{ s_1+ s_2+\cdots + s_n = r}} g^{(s_1)}(z_{12})g^{(s_2)}(z_{23})\cdots g^{(s_n)}(z_{n1})
 \label{h1eq.34}  
\eea
or in terms of their single-valued counterparts, 
\bea
 \label{h1eq.34a} 
V_r(1,\cdots,n) = \sum_{s_1,s_2,\cdots,s_n \geq 0
\atop{ s_1+ s_2+\cdots + s_n = r}} f^{(s_1)}(z_{12})f^{(s_2)}(z_{23})\cdots f^{(s_n)}(z_{n1})
\eea
with $g^{(0)}(z) =  f^{(0)}(z)=1$ and thus $V_0(1,\cdots,n)=1$. The modular forms $R_{2k}(e _\delta)$ under $\Gamma (2)$ may be expressed in terms of the Weierstrass function $\wp(z)$ and its derivatives, 
evaluated at the half period $\om_\delta$ in (\ref{h1eq.08}) corresponding to the spin structure $\delta$,
\bea
\label{8.Dk}
R_{2}(e _\delta) = \wp (\om_\delta) = e_\delta
\hskip 0.8in 
R_{2k}(e _\delta) = \frac{ \wp^{(2k-2)}(\hp_\delta) }{(2k{-}1)!} - {\rm G}_{2k} \quad \hbox{ for } k \geq 2
\eea
or may alternatively be written as degree-two polynomials in $e_\delta = \wp(\om_\delta)$,
\bea
\label{8.Dk1}
R_{2k}(e _\delta) = a_{2k-4} e_\delta ^2 + b_{2k-2} e_\delta + c_{2k}
\eea
where $a_{2k-4}, b_{2k-2} $ and $c_{2k}$ are modular forms under ${\rm SL}(2,\ZZ)$ of weight $2k-4$, $2k-2$ and $2k$, respectively, for $k \geq 2$, with $a_0=b_0=1$ and $a_2=b_2=0$.}

\subsubsection{Proof of Proposition \ref{8.thm:1}}
\label{sec:h1.prf}

To prove the proposition we use the relation between $C_\delta(1,\cdots, n)$ and the generating functions $F$ and $\Omega$ evaluated at the half periods, as spelled out in (\ref{h1eq.21}). Considering the product of $F$ or $\Omega$ factors with arbitrary values of the parameter $\eta \in \CC$ instead, the Laurent expansions of (\ref{h1eq.02}) imply the following Laurent expansion \cite{Dolan:2007eh, Tsuchiya:2012nf}, 
\begin{align}
\label{h1eq.22}
F(z_{12},\eta)F(z_{23},\eta)  \cdots F(z_{n1},\eta) =  \sum_{r=0}^\infty \eta^{r-n} V_r(1,2,\cdots,n)
\end{align}
The expressions (\ref{h1eq.34}) and (\ref{h1eq.34a}) for $V_r(1,\cdots,n)$ in terms of $g^{(s)}$ or $f^{(s)}$ kernels
readily follow from inserting the expansion (\ref{h1eq.02}) of the individual Kronecker-Eisenstein
series into (\ref{h1eq.22}). It remains to show that, when $\eta$ is set to the half period $\om_\delta$, then its dependence may be assembled into the above modular forms of $\Gamma(2)$.

\sm 

To do so, we use the fact that $F(z,\eta)$ satisfies $F(z, \eta+1) = F(z,\eta)$ and $F(z, \eta + \tau) = e^{- 2 \pi i z} F(z,\eta)$ to verify that the combination of (\ref{h1eq.22}) is an elliptic (i.e.\ meromorphic and doubly periodic) function in $\eta$, all of whose poles in $\eta  \in \Sigma$ are located at $\eta =0$ and are of maximum order $n$. Therefore, the combination of (\ref{h1eq.22}) is a linear combination of a finite number of derivatives $\wp^{(\ell)} (\eta)$ of $\wp (\eta)$ with $\ell+2 \leq n$. The precise form of the coefficients is obtained by using the following expansion, 
\bea
\wp (\eta) = { 1 \over \eta^2} + \cO(\eta) 
\hskip 0.8in 
{ \wp ^{(\ell)} (\eta) \over (\ell+1)!} - {\rm G}_{\ell+2} = { (-)^\ell \over \eta ^{\ell+2}} + \cO(\eta)
\quad \hbox{ for } \ell \geq 1
\eea
Matching the poles gives, 
\bea
F(z_{12},\eta)  \cdots F(z_{n1},\eta) & = & V_n (1,\cdots, n) +   \wp(\eta ) V_{n-2} (1,\cdots, n) 
 \\ &&
+ \sum_{\ell=1}^{n-2} (-)^\ell \left ( { \wp ^{(\ell)} (\eta) \over (\ell+1)!} - {\rm G}_{\ell+2} \right ) 
V_{n-\ell-2} (1,\cdots,n) + \cO(\eta) 
\no
\eea
Since the left side is an elliptic function in $\eta$ whose poles, which are all at 
$\eta=0$ and translates by $\mathbb Z {+} \tau \mathbb Z$, are matched by the poles on the right side,  the terms $\cO(\eta)$ on the right side vanish by Liouville's theorem. Furthermore, the derivatives $\wp ^{(\ell)} (\eta)$ evaluated at the half periods $\eta = \om_\delta$  vanish for all odd values of $\ell$, so that we recover (\ref{8.dec}) with $R_{2k}(e _\delta)$ given by (\ref{8.Dk}) upon setting $\ell{+}2=2k$. Finally, using the defining equation of~$\wp(\eta)$,
\bea
\wp'(\eta)^2 = 4 \wp(\eta)^3 - 60 {\rm G}_4 \wp (\eta) - 140 {\rm G}_6 
\label{defwp}
\eea
one readily expresses the function $\wp^{(2k-2)}(\eta)$ in terms of a polynomial in $\wp(\eta)$ of degree $k$. 
Evaluating the derivatives at $\eta=\om_\delta$ and using the cubic equation satisfied by $e_\delta$,
\bea
\label{8.edelta}
e_\delta ^3 - 15 {\rm G}_4 e_\delta - 35 {\rm G}_6=0
\eea
one iteratively reduces $\wp^{(2k-2)}(\hp_\delta)$ to a degree-two polynomial
in $\wp(\hp_\delta)$ and thereby demonstrates the validity of  (\ref{8.Dk1}). For the lowest values of $k$
the $\delta$-independent coefficients in these polynomials are given as follows, 
\begin{align}
a_0 &= 1
&b_2 &= 0
&c_4 &= 6 {\rm G}_4  \label{exabc}\\
a_2 &= 0
&b_4 &= 6 {\rm G}_4
&c_6 &= 20 {\rm G}_6 \notag \\
a_4 &= 3 {\rm G}_4
&b_6 &= 15 {\rm G}_6
&c_8 &= 14 {\rm G}_8 \notag \\
a_6 &= 10 {\rm G}_6
&b_8 &= 70 {\rm G}_8
&c_{10} &= 120 {\rm G}_{10}  \notag
\end{align}
Note that, similar to the equation (\ref{8.edelta}) for the roots at genus one, cyclic products of Szeg\"o kernels at genus two can be decomposed into degree-two polynomials in the entries $\mL_\delta^{11},\mL_\delta^{12},\mL_\delta^{22}$ thanks to the system of trilinear equations these objects satisfy \cite{DHoker:2022xxg}, also see \cite{DHoker:2023khh} for a representation of $\mL_\delta^{IJ}$ at $I,J \in \{1,2\}$ in terms of Riemann theta functions.

\subsubsection{Extracting the constants $C^{I_1\cdots I_r}_\delta$ and $D^{I_1\cdots I_r}_\delta$}
\label{sec:h1.2.3}

As a consequence of Proposition \ref{8.thm:1}, we obtain simple expressions for the reduction of the constants $C^{I_1\cdots I_r}_\delta$ and $D^{I_1\cdots I_r}_\delta$ to the case of genus one, as spelled out in following proposition.

{\prop
\label{cor:cgenus1}
For genus one, the constants $C^{I_1\cdots I_r}_\delta$  and $D^{I_1\cdots I_r}_\delta$ in the modular and meromorphic descents  are given by,
\beq
C_\delta^{I_1 \cdots I_r} \, \big|_{h=1} = \begin{cases} 
\ \ \ e_\delta + \widehat {\rm G}_2 &: \ r =2   \\
R_r(e_\delta) +  {\rm G}_n &: \ r \geq 4  \ {\rm even} \\
\ \ \ \ \ \ \ \, 0 &: \ r \geq 3  \ {\rm odd} \end{cases}
 \label{h1eq.41c}
\eeq
and
\beq
D_\delta^{I_1 \cdots I_r} \, \big|_{h=1} = \begin{cases} 
\ \ \  e_\delta +  {\rm G}_2 &: \ r =2   \\
R_r(e_\delta) +  {\rm G}_n &: \ r \geq 4  \ {\rm even} \\
\ \ \ \ \ \ \ \, 0 &: \ r \geq 3  \ {\rm odd} \end{cases}
 \label{h1eq.41d}
\eeq
respectively. The polynomials $R_r(e_\delta)$ in $e_\delta= \wp (\om_\delta) $ are obtained from (\ref{8.Dk}) and (\ref{defwp}).
}

\sm

The proof of the proposition can be found in appendix \ref{app:D.1}. We note the examples
\begin{align}
C_\delta^{I_1 \cdots I_4}  \, \big|_{h=1} &= e_\delta^2 - 5 {\rm G}_4 
&C_\delta^{I_1 \cdots I_8}  \, \big|_{h=1} &= 3 {\rm G}_4 e_\delta^2 + 15 {\rm G}_6 e_\delta + 15 {\rm G}_8
 \label{h1eq.42} \\
 C_\delta^{I_1 \cdots I_6}  \, \big|_{h=1} &=  6 {\rm G}_4 e_\delta+21 {\rm G}_6
&C_\delta^{I_1 \cdots I_{10}}  \, \big|_{h=1} &= 10 {\rm G}_6 e_\delta^2 + 70  {\rm G}_8 e_\delta
+ 121 {\rm G}_{10}
 \notag
\end{align}
and the following general relation between the constants $C^{I_1\cdots I_r}_\delta$ 
and $D^{I_1\cdots I_r}_\delta$ at genus one which can be read off from 
(\ref{h1eq.41c}) and (\ref{h1eq.41d}):
\beq
D_\delta^{I_1 \cdots I_r} \, \big|_{h=1} = \begin{cases} 
\ \, \Big. C_\delta^{I_1 I_2} \, \big|_{h=1}  + \frac{\pi}{\Im \tau} \Big. &: \ r =2   \\
\ \ \ \ \  \,  \Big. C_\delta^{I_1 \cdots I_r} \, \big|_{h=1}  \Big. &: \ r \geq 3   \end{cases}
 \label{cvsdath1}
\eeq
The genus one instances of $ C_\delta^{I_1 \cdots I_r} $ at $r\leq 8$ were reported 
without proof in \cite{DHoker:2023khh}, and the expressions for $ C_\delta^{I_1 \cdots I_6}  \, \big|_{h=1} $ 
and $ C_\delta^{I_1 \cdots I_8}  \, \big|_{h=1} $ experienced corrections in the most recent arXiv version of the reference.

\subsection{Linear chain products at genus one}
\label{sec:h1.3}

For linear chain products (\ref{defLdelta}) at genus one, the dependence on both the marked points
and on the even spin structure $\delta$ can be made fully explicit as done in Proposition
\ref{8.thm:1} for cyclic products. The results have not appeared in the literature prior to this work
and are summarized in the following theorem.

{\thm
\label{thm:linh1}
The genus one instances of linear chain products (\ref{defLdelta}) of Szeg\"o kernels can be
decomposed in two different ways,
\begin{align}
L_\delta(x;1,\cdots,n ;y) &=
\sum_{r=0}^{n} M_\delta^{I_1 \cdots I_r}(x,y) \, \big|_{h=1} 
W_{n-r}( x;1,\cdots,n ;y) 
\label{ldech1.01} \\
&= \sum_{r=0}^{n} L_\delta^{I_1 \cdots I_r}(x,y) \, \big|_{h=1} 
V_{n-r}( x;1,\cdots,n ;y) 
\notag
\end{align}
where the first line is term-by-term meromorphic in all variables and the second
line exposes the modular properties. In both cases, the dependence on the marked points is carried
by elliptic functions of the internal points $z_1,\cdots,z_n$
\begin{align}
W_r(x;1,2,\cdots,n;y) &=  \! \! \! \sum_{s_1,\cdots,s_{n+1}\geq 0 \atop{s_1+\cdots+s_{n+1}= r}} \! \! \! g^{(s_1)}(x{-}z_1) g^{(s_2)}(z_{12})
\cdots g^{(s_n)}(z_{n-1,n}) g^{(s_{n+1})}(z_{n}{-}y)
 \label{h1eq.56} \\
 V_r(x;1,2,\cdots,n;y) &=  \! \! \! \sum_{s_1,\cdots,s_{n+1}\geq 0 \atop{s_1+\cdots+s_{n+1}= r}} \! \! \! f^{(s_1)}(x{-}z_1) f^{(s_2)}(z_{12})
\cdots f^{(s_n)}(z_{n-1,n}) f^{(s_{n+1})}(z_{n}{-}y)
\notag
\end{align}
and the accompanying $\delta$ dependent spinors in $x$ and $y$ are expressible
in terms of Kronecker-Eisenstein derivatives and lattice sums (see (\ref{h1eq.08}) for
the co-moving coordinates $u_\delta,v_\delta$ and recall that $z=x{-}y = u \tau {+}v$ with $u,v \in \RR / \ZZ$)
\begin{align}
M_\delta^{I_1 \cdots I_r}(x,y) \, \big|_{h=1}  &=
 \frac{(-1)^r}{r!} \, e^{2\pi i u_\delta (x{-}y) } 
\, \partial^r_\eta F(x{-}y ,\eta) \, \big|_{\eta = \hp_\delta} 
 \label{h1eq.60} \\
 L_\delta^{I_1 \cdots I_r}(x,y) \, \big|_{h=1}   &= 
e^{ 2 \pi i  (v u_\delta  - u v_\delta )} \sum_{m,n \in \mathbb Z}  \frac{ e^{ 2\pi i (mv - nu)} }{(m\tau + n + \hp_\delta)^{r+1}}
 \notag
\end{align}
}

The proof of Theorem \ref{thm:linh1} may be found in appendix \ref{app:D.3}.
The expressions (\ref{h1eq.60}) make the genus one instances of the integral representations 
(\ref{Ldelint}) and (\ref{Lmot.55}) of the quantities $L_\delta^{I_1\cdots I_r}(x,y)$ and 
$M_\delta^{I_1\cdots I_r}(x,y)$ in the 
decompositions (\ref{moddecL}) and (\ref{merodecL}) fully explicit.
In combination with the relation (\ref{L.coin.17}) between the coincident limits
of $L_\delta^{I_1\cdots I_r}(x,x)$ and the constant tensors $C_\delta^{I_1\cdots I_r}$
at $r\geq 3$, we are led to the following corollary.

{\cor
\label{latsumcdel}
The constants  $C_\delta^{I_1\cdots I_r}$ and  $D_\delta^{I_1\cdots I_r}$ at 
genus one in (\ref{h1eq.41c}) can be alternatively
written in terms of the following lattice sums
\bea
C_\delta^{I_1\cdots I_r} \, \big|_{h=1}
= D_\delta^{I_1\cdots I_r} \, \big|_{h=1}
= \sum_{m,n \in \mathbb Z} \frac{1}{(m\tau + n + \hp_\delta)^r} \, , \ \ \ \ r\geq 3
\eea
The analogous expressions of the $r=2$ cases can be read off from (\ref{h1eq.41c}) and \ref{h1eq.41d})  using the lattice-sum representation of $e_\delta= \wp(\hp_\delta)$.
}

\newpage

\section{Outlook}
\setcounter{equation}{0}
\label{sec:out}

In this work, we have established that the integration kernels for polylogarithms on a Riemann surface $\Sigma$ of arbitrary genus provide a natural space of functions in terms of which the dependence of fermion correlators on points in $\Sigma$  may be expressed. For both cyclic products and linear chain products of Szeg\"o kernels, the descent procedures in Theorems \ref{2.thm:3} and \ref{4.thm:1} systematically give their $\Sigma$-dependence in terms of Enriquez or DHS kernels. In this way, their dependence on the spin structure is concentrated in constants
on $\Sigma$ in the case of cyclic products or spinors on $\Sigma$ 
depending solely on the end points of linear chain products. The spin-structure
dependence is expressed in terms of multiple convolution integrals, either
over homology cycles or over the surface, which extend the analogous convolution
representations of Enriquez kernels \cite{DHoker:2025dhv} and DHS kernels \cite{DHS:2023}.

\sm

The decomposition formulae (\ref{again.03}), (\ref{again.02}), (\ref{moddecL})
and (\ref{merodecL}) for products of Szeg\"o kernels obtained
from our descent procedure are expected to substantially simplify the evaluation
of superstring amplitudes: First, the spin structure sums for arbitrary chiral measures
can be performed at the level of constants or spinors depending only on the end points
of linear chain products instead of functions of the other points. Second, the link to
higher-genus polylogarithms offers
a growing arsenal of techniques for the integration over the points of the fermion
correlators in a low-energy expansion of string amplitudes.

\sm

More generally, the integration kernels produced in our decompositions of the cyclic products
and linear chains are believed to span the function space needed to express the complete
moduli-space integrands of string amplitudes. Under this assumption,
our results should feed into bootstrap constructions of higher-point and higher-genus amplitudes beyond today's reach of direct calculations.  In particular, the meromorphic function  space of Enriquez kernels \cite{Enriquez:2011} is compatible with the chiral-splitting formulation of string amplitudes
\cite{Verlinde:1987sd, DHoker:1988pdl, DHoker:1989cxq} and expected to fruitfully combine with the methods
of \cite{Geyer:2021oox, Geyer:2024oeu} to incorporate the information
about the supergravity amplitudes in the field-theory limit.

\sm

Our results raise several follow-up questions in a broader mathematical context
and suggest concrete steps towards their string theory applications including,
\begin{itemize}
\itemsep=-0.03in
\item converting the integral representations of the constants
$D_\delta^{I_1\cdots I_r}$ and $C_\delta^{I_1\cdots I_r}$ in the
decompositions (\ref{again.03}), (\ref{again.02}) of cyclic products
into expansion formulae around boundaries of moduli space;
\item relating $D_\delta^{I_1\cdots I_r} \leftrightarrow C_\delta^{I_1\cdots I_r}$
and the forms $\cW_{I_1\cdots I_r}(1,\cdots,n) \leftrightarrow \cV_{I_1\cdots I_r}(1,\cdots,n)$
that capture the dependence of cyclic products on the points,
using the gauge transformation and Lie-algebra automorphism
relating the Enriquez and DHS connections~\cite{DHoker:2025szl};
\item expressing higher-genus correlation functions of the current algebra
of heterotic strings in terms of Enriquez kernels
and DHS kernels as done at genus one \cite{Dolan:2007eh}.
\end{itemize}

\newpage

\appendix

\section{Function theory on a Riemann surface}
\setcounter{equation}{0}
\label{sec:A}

Throughout, $\Sigma$ will be a compact Riemann surface of arbitrary genus $h \geq 1$ with simply connected covering space $\tilde \Sigma$ and associated projection by $\pi : \tilde \Sigma \to \Sigma $. The first homology group $H_1(\Sigma, \ZZ)$ is endowed with an intersection pairing $\mJ$. A canonical basis for $H_1(\Sigma, \ZZ)$ is given by cycles $\mA^I$ and $\mB_I$ that obeys $\mJ(\mA^I, \mA^J) = \mJ (\mB_I, \mB_J)=0$ and $\mJ(\mA^I, \mB_J)=\delta ^I_J$ for $I,J \in \{ 1, \cdots, h\}$. Choosing the cycles $\mA^I$ and $\mB_J$ to share a~common base point $q$ promotes them into a set of generators of the first homotopy group $\pi_1(\Sigma, q)$ of $\Sigma$, as illustrated in figure \ref{fig:1} for a Riemann surface of genus two.

\sm

A basis for the Dolbeault cohomology group $H_1^{(1,0)}(\Sigma) $ is given by the holomorphic Abelian differentials $\om_J = \om_J(x) dx$, whose integrals on $\mA^I$ cycles are normalized and whose integrals on $\mB_I$ cycles give the entries of the  period matrix $\Omega$ of the surface $\Sigma$,
\bea
\label{A.abel}
\oint _{\mA^I} \om_J = \delta ^I_J \hskip 0.7in \oint _{\mB_I} \om_J = \Omega _{IJ}
\eea
The Riemann relations imply that the period matrix $\Omega$ is symmetric $\Omega ^t = \Omega$ and that its imaginary part $Y = \Im(\Omega)$ is a positive definite matrix.

\subsection{The prime form}
\label{sec:A.1}

A key ingredient in the theory of functions and differential forms on $\Sigma$ is provided by the prime form $E(x,y)$, defined for $x, y \in \tilde \Sigma$ to be a differential form of conformal weight $(-\half, 0)$ in both $x$ and $y$, anti-symmetric $E(y,x)=-E(x,y)$ with a single zero for $x,y $ in a given fundamental domain $D$, normalized by $E(x,y) = (x-y) dx^{-\half} dy ^{-\half} + \cO(x-y)^3$ in a system of local coordinates. The Riemann $\tet$-function of rank $h$ with characteristics $\kappa = (\kappa', \kappa'')$ with $\kappa ', \kappa''   \in \CC^{h}$ for period matrix $\Omega$ and $\zeta \in \CC^h$ is defined by,\footnote{The dependence on $\Omega$ will be suppressed throughout.} 
\bea
\tet [\kappa] (\zeta ) = 
\sum_{n \in \ZZ^h} \exp \big \{ \pi i (n+\kappa' )^t \Omega (n+\kappa') + 2 \pi i (n+\kappa')^t (\zeta + \kappa '') \big \}
\label{defvth}
\eea
An explicit formula for the prime form is given in \cite{Fay:1973} by,
\bea
E(x,y) = { \tet[\nu] (\int _y^x \omega ) \over h_\nu(x) \, h_\nu (y)} 
\hskip 0.8in 
h_\nu(x)^2 = \om_I(x) { \p \over \p \zeta _I}  \tet [\nu](\zeta ) \Big |_{\zeta=0}
\eea
where $\nu$ is an odd half-integer characteristic, or \textit{spin structure on $\Sigma$}  (for which $2 \nu', 2 \nu'' \in \ZZ_2^h$ and the integer $4 (\nu')^t \nu''$ is odd-valued) and $h_\nu(x)$ is the holomorphic $(\half, 0)$ form whose square is given by the second equation above. The above ratio defining the prime form is actually independent of the choice of odd half-integer characteristic $\nu$. The periodicity properties of the Riemann theta function with respect to $\zeta$ lead to the following monodromies of the prime form
\bea
E(\mA^I \cdot x, y) & = & - E(x,y)
\no \\
E(\mB_I \cdot x, y) & = & - E(x,y) \, \exp \Big \{-i \pi \Omega _{II} - 2 \pi i \int ^x _y \om_I \Big \}
\eea
As a consequence, $\p_x \ln  E(x,y)$ has no $\mA^I$ monodromies in $x,y$, and its
$\mB_I$ monodromies are given by
\bea
\label{Emon}
\p_x \ln E(\mB_I \cdot x, y) & = &  \p_x \ln  E(x,y)  - 2 \pi i  \om_I (x)
\no \\
\p_x \ln E(x, \mB_I \cdot y) & = &  \p_x \ln  E(x,y)  + 2 \pi i  \om_I (x)
\eea
This can be used to reproduce the monodromies (\ref{2.4.mon}) of Enriquez kernels from their
representations as $\mA$-cycle convolutions of prime forms and Abelian differentials \cite{DHoker:2025dhv}.

\subsection{The Szeg\"o kernel}
\label{sec:A.2}

The Szeg\"o kernel $S_\delta(x,y)$ for even spin structure $\delta$ and a \textit{generic} period matrix $\Omega$ (namely, for which $\tet[\delta](0 ) \not=0$)  is a meromorphic $(\half, 0)$ form in $x,y \in \Sigma$ defined by,
\bea
\label{A.Szego1}
S_\delta(x,y) = { \tet [\delta](\int ^x _y \om) \over \tet [\delta] (0) \, E(x,y)}
\eea
It follows that $S_\delta(x,y)= - S_\delta(y,x)$ has a single pole in $x$ at $y$ with unit residue so that,
\bea
\label{A.Szego2}
\bar \partial_x S_\delta (x,y) = \pi \delta(x,y)
\eea
For even spin structures and non-generic moduli for which $\tet[\delta](0 ) =0$ and for all odd spin structures, the Cauchy-Riemann operator $\bar \partial_x $ acting on $(\half, 0)$ forms has zero modes so that the definition of the Szeg\"o kernel requires making choices. Here, we shall only consider the generic case.

\subsection{The Arakelov Green function}
\label{sec:A.3}

The \textit{Arakelov Green function} $\cG(x,y) = \cG(y,x)$ \cite{Faltings} (see also \cite{Alvarez-Gaume:1986nqf, DHoker:2017pvk} for its use in physics) is a real single-valued  scalar in $x,y \in \Sigma$ uniquely defined by the following  equations,
\bea
\label{A.arak1}
\bar \p_x \p_x \cG(x,y)= - \pi \delta(x,y) + \pi \kappa (x) 
\hskip 1in 
\int _\Sigma d^2 t \, \kappa(t) \cG(t,y)=0
\eea
where $\kappa$ is the canonical volume form on $\Sigma$, defined by,
\bea
\kappa (x)  = { 1 \over  h} \om_I(x) \, \bar \om^I(x) 
\eea
An explicit formula for the Arakelov Green function may be obtained in terms of the \textit{string Green function} $G(x,y)
$ which is given by,
\bea
G(x,y) = - \ln |E(x,y)|^2 + 2 \pi  \left ( \Im \int ^x _y \om_I \right )  \left (\Im \int ^x _y  \om^I \right ) 
\eea
as follows,
\bea
\cG(x,y) = G(x,y) - \gamma (x) - \gamma(y) + \int _\Sigma d^2t \, \kappa(t) \gamma (t)
\eea
where
\bea 
\gamma(x) = \int _\Sigma d^2 t \, \kappa (t) G(x,t) 
\eea
The defining equations (\ref{A.arak1}) and the canonical volume form $\kappa$  being conformal and modular invariant, it follows that the Arakelov Green function $\cG(x,y)$ is a modular invariant conformal scalar (while the string Green function is not conformally invariant). The function $\cG(x,y)$ is real-analytic for $x,y \in \Sigma$ away from $x= y$, where it has the following asymptotic behavior as $x \to y$,
\bea
\label{A.arakas}
\cG(x,y) = - \ln |x-y|^2 + \hbox{regular} 
\eea
It will also be useful to have the following mixed differential equation,
\bea
\label{A.arak2}
\bar \p_y \p_x \cG(x,y) = \pi \delta(x,y) - \pi \om_I(x) \bar \om^I(y)
\eea

\subsection{Enriquez kernels}
\label{sec:A.4}

The basic definition of the Enriquez kernel $g^{I_1 \cdots I_r}{}_J(x,y)$, for $r \geq 0$ and $I_1, \cdots, I_r, J \in \{ 1, \cdots, h\}$ was presented in section \ref{sec:2.1} as a meromorphic $(1,0)$ form in  $x \in \tilde \Sigma$ and $(0,0)$ form in $y \in \tilde \Sigma$.
The Enriquez kernel $g^{I_1 \cdots I_r}{}_J(x,y)$ is holomorphic in the interior $x,y \in D^o$
of a preferred fundamental domain for $\Sigma$ for $r \geq 2$,
has a single simple pole in $x$ at $y$ with residue $\delta^{I_1}_J$ for $r=1$,  and is given by $g^\emptyset {}_J(x,y) = \om_J(x)$ for $r=0$. Its monodromies in $x$ and $y$  around $\mA$ cycles are trivial, while its monodromies around $\mB$ cycles are given by (\ref{2.4.mon}) using the notations of (\ref{2.mondef}) and (\ref{2.kron}). The forms $g^{I_1 \cdots I_r}{}_J(x,y)$ may have poles in $x$ at $\pi^{-1}(y)$ for all $r \geq 1$, as mandated by the monodromy relations.  

\sm

In the sequel, it will be useful to have the monodromies around $\mB$ cycles of the traceless and trace parts defined by (\ref{2.dec}) separately at our disposal. They may be obtained by decomposing the monodromy relations of $g$ given in (\ref{2.4.mon}) into their traceless and trace parts, and we find (see (\ref{2.mondef}) for the $\Delta _L^{(x)}$ notation), 
\bea
\label{trcless}
\Delta _L^{(x)} \varpi^{I_1\cdots I_r}{}_J(x) = 
 \sum_{k=1}^{r} \frac{(-2\pi i)^k}{k!} \delta^{I_1\cdots I_k}_L   \varpi^{I_{k+1}\cdots I_r}{}_J(x) 
- \frac{(-2\pi i)^r}{r! \, h}  \delta^{I_r}_J \delta^{I_1\cdots I_{r-1}}_L  \omega_L(x) 
\qquad
\eea
whereas the $\mB$ monodromies of $\chi^{I_1 \ldots I_s}(x,y) $ with $s\geq 0$  are given by,
\bea
\Delta _L^{(x)}  \chi^{I_1\cdots I_{s}}(x,y) &= & 
 \sum_{k=1}^s \frac{(-2\pi i)^k}{k!} \delta^{I_1 \cdots I_k}_L \chi^{I_{k+1} \cdots I_s}(x,y)
-\frac{(-2\pi i)^{s+1}}{(s{+}1)! h } \delta^{I_1\cdots I_s}_L \omega_L(x)
\notag \\
\Delta _L^{(y)}  \chi^{I_1\cdots I_{s}}(x,y) &= & 
- \sum_{k=0}^s \frac{(2\pi i)^{s-k+1}}{(s{-}k{+}1)!} \, g^{I_1 \cdots I_k}{}_L(x,y) \delta_L^{I_{k+1}\cdots I_s}
\label{mclim.01}
\eea
The periods around $\mA^L$ cycles on the boundary of the fundamental domain $D$ in figure~\ref{fig:1} are given in terms of Bernoulli numbers $\Ber_r$ by (\ref{2.intg}) and its decomposition into traceless and trace parts was given by (\ref{2.dec}). Thus, we have, 
\bea
\label{A.intchi}
\oint _{\mA^L} dt \, \chi^{I_1 \cdots I_r}(t,y) &=& -  \frac{1}{h}  ( -2 \pi i )^{r+1} { \Ber _{r+1} \over (r+1)!} \,  \delta ^{I_1 \cdots I_r}_L
\no \\
\oint _{\mA^L} dt \, \varpi^{I_1 \cdots I_r}{}_J (t) &=&  ( -2 \pi i )^{r} { \Ber _{r} \over r!} \,  
\bigg ( \delta ^{I_1 \cdots I_r L}_J -  \frac{1}{h} \delta^{I_r}_J  \delta ^{I_1 \cdots I_{r-1}} _L \bigg)
\eea

\subsubsection{Low order monodromy formulas}

It will be useful to dispose of some low order formulas for the $\mB$ monodromies of $g$, $\varpi$ and $\chi$ in all their arguments. The $\mB$ monodromies of $g$ are given by,  
\bea
\label{A.g.1}
\Delta _L^{(x)} g^I{}_J(x, y) & = & - 2 \pi i \, \delta ^I_L \, \om_J(x)
\no \\
\Delta _L^{(y)} g^I{}_J(x, y) & = &   2 \pi i \, \delta ^I_J \, \om_L(x)
\no \\
\Delta _L^{(x)} g^{I_1 I_2} {}_J(x, y) & = &  - 2 \pi i \, \delta ^{I_1} _L \, g^{I_2} {}_J(x,y)  - 2 \pi^2 \, \delta^{I_1 I_2}_L \, \om_J(x)
\no \\
\Delta ^{(y)}_L g^{I_1 I_2} {}_J(x, y) & = &  2 \pi i \, \delta ^{I_2} _J  \, g^{I_1} {}_L(x,y)  - 2 \pi^2 \, \delta^{I_2}_J \, \delta^{I_1}_L \, \om_L(x)
\eea
those of $\chi$ are given by, 
\bea
\label{A.chi}
\Delta _L^{(x)}  \chi(x,y) & = &  2 \pi i  \,  \om_L(x)/h
\no \\
\Delta ^{(y)}_L \chi(x, y) & = & - 2 \pi i \, \om_L(x)
\no \\
\Delta _L^{(x)}  \chi^I (x,y) & = & - 2 \pi i \, \delta ^I_L \, \chi(x,y) +  2 \pi^2  \, \delta^I_L  \, \om_L (x)/h
\no \\
\Delta ^{(y)}_L \chi^I (x, y) & = &  - 2 \pi i \, g^I{}_L(x,y) + 2 \pi^2 \, \delta ^I_L \, \om_L(x) 
\eea
and those of $\varpi$ are given by,
\bea
\label{A.varpi}
\Delta _L^{(x)} \varpi ^I{}_J(x) & = & 
- 2 \pi i  \delta ^I_L \, \om_J(x) + 2 \pi i   \delta ^I_J \, \om_L(x) /h
\no \\
\Delta _L^{(x)} \varpi ^{I_1 I_2} {}_J(x) & = & 
- 2 \pi i \, \delta ^{I_1} _L \, \varpi ^{I_2} {}_J(x) 
- 2 \pi^2 \, \delta ^{I_1 I_2 }_L  \om_J(x) 
+ 2 \pi^2 \delta ^{I_1}_L  \delta ^{I_2}_J \, \om_L(x) /h
\eea

\newpage

\section{Proof of Lemma \ref{2.lem:4}}
\setcounter{equation}{0}
\label{sec:Ba}

To prove Lemma \ref{2.lem:4}, we note that (\ref{2.desc.5pol}) follows from (\ref{2.desc.1}) and the known poles of the Szeg\"o kernels in the cyclic product $\CD(1,\cdots, n)$, while the relations (\ref{2.desc.6}) may be verified directly for the case $r=1$ using the first line in (\ref{2.desc.0}).  To establish the relations (\ref{2.desc.6}) for $r \geq 2$, we proceed by induction in $r$. To do so we reformulate the system of equations in (\ref{2.desc.6}) in terms of the following combinations for $n \geq 2$ and $r \geq 1$, 
\bea
\label{2.Sdk}
S_{\delta \, k} ^{I_1 \cdots I_r} (1, \cdots, n)  &=&  \bar \p_k \DD^{I_1 \cdots I_r} (1,\cdots, n)  \\
 &&\quad - \pi \Big ( a_k \, \delta (k, k+1) - b_k \, \delta (k, k-1) \Big ) \DD^{I_1 \cdots I_r} (1, \cdots, \hat k , \cdots, n)
\notag
\eea
where the constants $a_k$ and $b_k$ are given as follows, 
\bea
\begin{cases} a_1 =1 \cr\; \!  b_1=0 \end{cases}
\hskip 0.5in 
\begin{cases} a_n =0 \cr\, b_n=1 \end{cases}
\hskip 0.5in 
\begin{cases} a_k =1 \cr\; \! b_k=1 \end{cases} \quad \hbox{for} \quad 2 \leq k \leq n-1
\eea
With this notation, the system of equations in (\ref{2.desc.6}) is equivalent to $S_{\delta \, k} ^{I_1 \cdots I_r} (1, \cdots, n) =0$ for all values of $k$ in the range $1 \leq k \leq n$. Next, by differentiating (\ref{2.desc.2}) with respect to $z_k$ for $2 \leq k \leq n$, eliminating the terms proportional to $\chi^{I_r \cdots  I_1} (1,2) {-}   \chi^{I_r \cdots  I_1} (1,n)$ in terms of $\DD$ functions whose argument $z_k$ is missing, we find that the entire remaining relation may be recast in terms of the combinations $S_{\delta \, k} $ in (\ref{2.Sdk}) for $1 \leq k \leq n$ as follows, 
\bea
\label{2.desc.9}
\om_{J} (1) S_{\delta \, k} ^{I_1 \cdots I_r J} (2, \cdots, n) & = &
S_{\delta \, k} ^{I_1 \cdots I_r } (1, \cdots, n) 
 \\ &&
- \sum _{i=0}^{r-1} g^{I_r \cdots I_{i+1}}{}_J(1,2) \, S_{\delta \, k} ^{I_1 \cdots I_i J} (2, \cdots, n)
\no
\eea
Recall that these relations are derived for $1 \leq r \leq s-1$ in view of the assumptions made in the formulation of  Lemma \ref{2.lem:4}. Using the fact that the rank on the left is $r+1$ but the rank on the right is at most $r$, and that $S_{\delta \, k} ^{I_1 \cdots I_r } (1, \cdots, n) =0$ for $r=1$, it follows by induction in $r$ that $S_{\delta \, k} ^{I_1 \cdots I_r } (1, \cdots, n) =0$ for all $r \leq s$. This completes the proof of  the system of equations (\ref{2.desc.6}). 

\sm

To prove (\ref{2.desc.7}), it suffices to define $S_{\delta \, k} ^{I_1 \cdots I_r } (1, \cdots, n) $ for the case $n=1$ by setting $a_k=b_k=0$ in (\ref{2.Sdk}).  The equations of (\ref{2.desc.9}) are then satisfied for $n=2$. Using (\ref{2.desc.9}) for $n=2$ and the fact that the case $r=1$ holds true by Proposition \ref{2.thm:1} then readily leads to a proof of (\ref{2.desc.7}) by induction in $r$ and completes the proof of Lemma \ref{2.lem:4}.

\newpage

\section{Proof of Lemma \ref{2.lem:5}}
\setcounter{equation}{0}
\label{sec:B}

In this appendix, we shall prove the monodromy relations in (\ref{2.desc.8}) of Lemma \ref{2.lem:5} for the multiplets $ \DD^{I_1 \cdots I_r} (1,\cdots, n)$ up to rank $r \leq s$ under the assumption of the lemma that the descent equations (\ref{2.desc.2})  hold true for all $r \leq s-1$ and all $n \geq 3$. Recall that all $\mA$ cycle monodromies manifestly vanish term by term. The $\mB$ cycle monodromies in the points $z_k$ of $ \DD^{I_1 \cdots I_r} (2,\cdots, n)$ will be evaluated separately for the three cases distinguished in (\ref{2.desc.8}), starting with the middle line for which $3 \leq k \leq n-1$, which turns out to be the simplest of the three and, in turn,  will be used to prove the case $k=2$.

\subsection{Monodromy of $\DD ^{I_1 \cdots I_r} (2, \cdots , n) $ in $z_k$ for $3 \leq k \leq n-1$}
\label{sec:B.1}

Applying the monodromy operator $\Delta _L ^{(k)} $ for $3 \leq k \leq n-1$ to both sides of (\ref{2.desc.2}) gives, 
\bea
\label{B.k}
\om_J(1)  \Delta _L ^{(k)} \DD ^{I_1 \cdots I_r J} (2, \cdots , n) 
& = &
 \Delta _L ^{(k)} \DD ^{I_1 \cdots I_r } (1, \cdots , n) 
 \\ &&
- \sum_{i=0}^{r-1} g^{I_{r}  \cdots I_{i+1}}{}_J(1,2) \Delta _L ^{(k)} \DD ^{I_1 \cdots I_i J} (2, \cdots , n)  
 \no
\eea
Evaluating the monodromy of the first equation in (\ref{2.desc.0}), we readily establish that $\Delta _L ^{(k)} \DD ^{J} (1, \cdots , n) =0$ for all $n \geq 1$, which confirms equation (\ref{B.k}) for the case $r=0$. For $r \geq 1$ we proceed by using (\ref{B.k}) by induction in $r$. Since the rank on the left side is one higher than the maximum rank on the right side, the induction is straightforward and establishes the middle equation of (\ref{2.desc.8}).

\subsection{Monodromy of $\DD ^{I_1 \cdots I_r} (2, \cdots , n) $ in $z_2$ for $3 \leq n$}
\label{sec:B.2}

In terms of the following combination,
\bea
\label{B.T}
T^{I_1 \cdots I_r}_\delta  (1,\cdots, n) 
& = &
\Delta ^{(1)} _L \DD^{I_1 \cdots I_r} (1,\cdots, n) 
 \\ &&
- \sum_{\ell=1}^r { (-2 \pi i)^\ell \over \ell! } \delta ^{I_r \cdots I_{r+1-\ell}} _L 
\DD^{I_1 \cdots I_{r -\ell} } (1,\cdots, n)
\no
\eea
the first equation of (\ref{2.desc.8}) is equivalent to $T^{I_1 \cdots I_r}_\delta (1,\cdots, n) =0$. Next, we apply the operator $\Delta ^{(2)}_L$ to (\ref{2.desc.2}),  and take care of including the double monodromy given by the sum on the last line below, 
\bea
\label{B.J}
\om_J(1) \, \Delta ^{(2)}_L \DD ^{I_1 \cdots I_r J} (2, \cdots , n) 
& = &
\Delta ^{(2)}_L \, \DD ^{I_1 \cdots I_r } (1, \cdots , n) 
+ \Delta ^{(2)}_L   \chi^{I_r \cdots  I_1} (1,2) \, C_\delta (2,\cdots, n)   
\no \\ &&
- \sum_{i=0}^{r-1} \Delta ^{(2)}_L  g^{I_r  \cdots I_{i+1}}{}_J(1,2) \DD ^{I_1 \cdots I_iJ} (2, \cdots , n)  
\no \\ &&
- \sum_{i=0}^{r-1}  g^{I_r  \cdots I_{i+1}}{}_J(1,2) \, \Delta ^{(2)}_L  \DD ^{I_1 \cdots I_iJ} (2, \cdots , n)  
\no \\ &&
- \sum_{i=0}^{r-1} \Delta ^{(2)}_L  g^{I_r  \cdots I_{i+1}}{}_J(1,2) \, \Delta ^{(2)}_L  \DD ^{I_1 \cdots I_iJ} (2, \cdots , n)  
\eea
Using the results established in section \ref{sec:B.1}, the first term on the first line on the right side of (\ref{B.J}) vanishes, $\Delta ^{(2)}_L \DD^{I_1 \cdots I_r}(1,\cdots, n)=0$. To compute the monodromies involving the Enriquez kernels, we use the second equation of (\ref{2.4.mon}) for $g$ and the second equation of (\ref{mclim.01}) for $\chi$, and suitably rearrange the indices as follows, 
\bea
\Delta _L^{(2)} g^{I_r \cdots I_i}{}_J(1,2) & = &
\delta ^{I_i}_J \sum_{k=1}^{r+1-i}  {(2 \pi i)^k \over k!} g^{I_r \cdots I_{k+i}}{}_L(1,2) \, \delta ^{I_{k+i-1} \cdots I_{i+1}} _L
\no \\
\Delta _L^{(2)} \chi^{I_r \cdots I_1} (1,2) & = &
- \sum_{k=1}^{r+1} {(2 \pi i)^k \over k!} g^{I_r \cdots I_k}{}_L (1,2) \, \delta ^{I_{k-1} \cdots I_1} _L
\eea
After some regrouping of sums, the result may be written as $L_1+L_2+L_3=0$ where,
\bea
L_1 & = & 
\sum_{i=0}^r g^{I_r \cdots I_{i+1} }{}_J(1,2) \, \Delta _L ^{(2)} \DD^{I_1 \cdots I_i J} (2,\cdots , n) 
\no \\
L_2 & = & 
\sum_{i=0}^r \sum_{k=1}^{r+1-i} { (2 \pi i)^k \over k!} g^{I_r \cdots I_{I+k}}{}_L(1,2) \, 
\delta ^{I_{i+k-1} \cdots I_{i+1}} _L \DD^{I_1 \cdots I_i} (2, \cdots, n)
\no \\
L_3 & = &
\sum_{i=1}^r \sum_{k=1}^{r+1-i} { (2 \pi i)^k \over k!} g^{I_r \cdots I_{I+k}}{}_L(1,2) \,
\delta ^{I_{i+k-1} \cdots I_{i+1}} _L \Delta ^{(2)}_L \DD^{I_1 \cdots I_i} (2, \cdots, n)
\eea
Next, we eliminate the $\Delta _L^{(2)} \DD$ terms in favor of the combination $T_\delta$ that was introduced in (\ref{B.T}). The contributions involving $T_\delta$ are given by,
\bea
L_1^T & = & 
\sum_{i=0}^r g^{I_r \cdots I_{i+1} }{}_J(1,2) \, T_\delta^{I_1 \cdots I_i J} (2,\cdots , n) 
\no \\
L_3^T & = &
\sum_{i=1}^r \sum_{k=1}^{r+1-i} { (2 \pi i)^k \over k!} g^{I_r \cdots I_{i+k}}{}_L(1,2) \,
\delta ^{I_{i+k-1} \cdots I_{i+1}} _L T_\delta^{I_1 \cdots I_i} (2, \cdots, n)
\eea
In terms of $L_1^T$ and $L_3^T$, the contributions $L_1$ and $L_3$ take the following form, 
\bea
L_1 & = & L_1^T - \sum_{i=0}^r \sum _{\ell=1}^{i+1} { (- 2 \pi i )^\ell \over \ell !} g^{I_r \cdots I_{i+1}} {}_L(1,2) \, 
\delta ^{ I_i \cdots I_{i+2-\ell}}_L \DD^{I_1 \cdots I_{i+1-\ell}}(2,\cdots, n)
 \\
L_3 & = & L_3^T + \sum _{i=1}^r \sum_{k=1}^{r+1-i} \sum_{\ell=1}^i (-)^\ell {  (2 \pi i)^{k+\ell}  \over k! \, \ell !}
g^{I_r \cdots I_{i+k}}{}_L(1,2) \, \delta ^{I_{i+k-1} \cdots I_{i-\ell+1}}_L \DD^{I_1 \cdots I_{i-\ell}} (2,\cdots, n)
\no
\eea
where we have used the identity, 
\bea
\delta ^{I_{k+i-1} \cdots I_{i+1}} _L \, \delta ^{I_i \cdots I_{i+1-\ell}} _L = 
\delta ^{I_{k+i-1} \cdots  I_{i+1-\ell}} _L 
\eea
to simplify the summand of the triple sum in $L_3$. As a result of using this identity, the factors $g$ and $\DD$ in the summand depend only on the combinations $k+i$ and $i-\ell$. Hence, it is possible to carry out one of the three summations explicitly. To do so, we perform a double change of variables in the triple sum for $L_3$, with the following ranges,
\begin{align} 
(k,i) & \to  (p,i) & p&=k+i  & 1 & \leq i \leq p \leq r +1
\no \\
(\ell, i) & \to (q,i) & q &=i-\ell & 0& \leq q \leq i-1 \leq p-1
\end{align}
so that the three summations in $L_3$ may be rearranged as follows,
\bea
\sum_{i=1}^{r} \sum_{k=1}^{r+1-i} \sum_{\ell=1}^i 
= \sum_{p=2}^{r+1} \sum_{q=0}^{p-2} \sum_{i=q+1}^{p-1}
\eea
As a result, $L_3$ may be expressed as follows, 
\bea
L_3 & = & L_3^T + \sum_{p=2}^{r+1} \sum_{q=0}^{p-2} \sum_{i=q+1}^{p-1} 
{ (-)^{i+q}  (2 \pi i)^{p-q}  \over (p-i)! \, (i-q) !}
g^{I_r \cdots I_p}{}_L(1,2) \, \delta ^{I_{p-1} \cdots I_{q+1}}_L \DD^{I_1 \cdots I_q} (2,\cdots, n)
\qquad
\eea
The sum over $i$ may be carried out exactly,
\bea
\sum_{i=q+1}^{p-1} {(-)^{i+q} \over (p-i)! \, (i-q)!} = - { 1 + (-)^{p-q} \over (p-q)!}
\eea
leaving the following double sum,
\bea
L_3 = L_3^T - \sum_{p=2}^{r+1} \sum_{q=0}^{p-2} \big ( 1+(-)^{p-q} \big )
{ (2 \pi i)^{p-q}  \over (p-q)! }
g^{I_r \cdots I_p}{}_L(1,2) \, \delta ^{I_{p-1} \cdots I_{q+1}}_L \DD^{I_1 \cdots I_q} (2,\cdots, n)
\qquad
\eea
In view of the factor $ ( 1+(-)^{p-q}  )$ in the summand, we are free to extend the summation over $q$ to include $q=p-1$ and, once that has been done, we are also free to extend the summation range of $p$ to include $p=1$ since the second sum then forces $q=0$ which again vanishes in view of the factor $ ( 1+(-)^{p-q}  )$. Separating now the contributions from the two terms in the factor $ ( 1+(-)^{p-q}  )$, we have, 
\bea
L_3 = L_3^T && - \sum_{p=1}^{r+1} \sum_{q=0}^{p-1} 
{ (2 \pi i)^{p-q}  \over (p-q)! }
g^{I_r \cdots I_p}{}_L(1,2) \, \delta ^{I_{p-1} \cdots I_{q+1}}_L \DD^{I_1 \cdots I_q} (2,\cdots, n)
\no \\ && 
- \sum_{p=1}^{r+1} \sum_{q=0}^{p-1} 
{ (- 2 \pi i)^{p-q}  \over (p-q)! }
g^{I_r \cdots I_p}{}_L(1,2) \, \delta ^{I_{p-1} \cdots I_{q+1}}_L \DD^{I_1 \cdots I_q} (2,\cdots, n)
\eea
Changing summation variables in $L_1- L_1^T$ from $i$ to $p=i+1$ and $\ell$ to $q = p -\ell$, and in $L_2$ from $k$ to $p=i+k$ and from $i$ to $q$, we obtain,
\bea
L_1 & = & L_1^T + \sum  _{p=1}^{r+1} \sum_{q=0}^{p-1} { (-2 \pi i )^{p-q} \over (p-q)!} g^{I_r \cdots I_p} {}_L(1,2) 
\delta ^{I_{p-1} \cdots I_{q+1}} _L \DD^{I_1 \cdots I_q} (2, \cdots, n)
\no \\
L_2 & = & 
\sum_{p=1}^{r+1} \sum_{q=0}^{p-1} { (2 \pi i)^{p-q} \over (p-q)!} g^{I_r \cdots I_p}{}_L(1,2) \, 
\delta ^{I_{p-1} \cdots I_{q+1}} _L \DD^{I_1 \cdots I_q} (2, \cdots, n)
\eea
We see that, in the sum $L_1+L_2+L_3$, the sum of $L_1-L_1^T$ cancels the sum of the second line of $L_3 - L_3^T$ while $L_2$ cancels the sum on the first line of $L_3-L_3^T$. Therefore, the relation $L_1+L_2+L_3=0$ reduces to $L_1^T+L_3^T=0$ or more explicitly, 
\bea
\label{B.TT}
\om_J(1) T_\delta^{I_1 \cdots I_r J}(2,\cdots, n)  & = & - \sum_{i=0}^{r-1} g^{I_r \cdots I_{i+1} }{}_J(1,2) \, T_\delta^{I_1 \cdots I_i J} (2,\cdots , n) 
 \\ &&
- \sum_{i=1}^r \sum_{k=1}^{r+1-i} { (2 \pi i)^k \over k!} g^{I_r \cdots I_{i+k}}{}_L(1,2) \,
\delta ^{I_{i+k-1} \cdots I_{i+1}} _L T_\delta^{I_1 \cdots I_i} (2, \cdots, n)
\no
\eea
where the term on the left corresponds to the $i=r$ contribution to $L_1^T$. 

\sm

We are now ready to complete the proof by induction. For $r=1$ we have,
\bea
\label{B.Tnew}
T^{I}_\delta  (2,\cdots, n) 
= \Delta ^{(1)} _L \DD^{I} (2,\cdots, n)  + 2 \pi i \, \delta ^{I} _L  \CD(2,\cdots, n)
\eea
which vanishes in view of the expression for $\DD^{I_1} (2,\cdots, n) $ in the first line of (\ref{2.desc.0}) and the $\mB$ monodromy transformation of $\chi(1,2)$ in the point $z_2$, given in  (\ref{A.chi}). Since the left side of equation (\ref{B.TT}) is of rank~$r+1$ while the maximum rank on the right side is~$r$, it follows by induction in $r$ that we have $T_\delta^{I_1 \cdots I_r }(2,\cdots, n) =0$ for all $r$ and all $n \geq 2$. This completes the proof of the monodromy formula of (\ref{2.desc.8}) for $k=2$.

\subsection{Monodromy of $\DD ^{I_1 \cdots I_r} (2, \cdots , n) $ in $z_n$ for $3 \leq n$}
\label{sec:B.3}

The final element in the proof of Lemma \ref{2.lem:5} consists of proving the last equation in (\ref{2.desc.8}). To do this we introduce the following combination, 
\bea
U^{I_1 \cdots I_r}_\delta (1,\cdots, n) 
= \Delta ^{(n)} _L \DD^{I_1 \cdots I_r} (1,\cdots, n)
- \sum_{\ell=1}^r { (2 \pi i)^\ell \over \ell !} \delta ^{I_1 \cdots I_\ell}_L \DD^{I_{\ell+1} \cdots I_r} (1,\cdots, n)
\qquad
\eea
in terms of which the last equation in (\ref{2.desc.8}) is equivalent to $U^{I_1 \cdots I_r}_\delta (1,\cdots, n) =0$. Applying $\Delta ^{(n)}_L$ to equation (\ref{2.desc.2}) gives, 
\bea
0 & = & \Delta ^{(n)} _L \DD^{I_1 \cdots I_r} (1,\cdots, n) - \Delta ^{(n)} _L \chi ^{I_r \cdots I_1} (1,n) \, \CD(2,\cdots, n)
\no \\ &&
- \sum_{i=0}^r g^{I_r \cdots I_{i+1}} {}_J(1,2) \Delta ^{(n)} _L \DD^{I_1 \cdots I_i J } (2,\cdots, n) 
\eea
Eliminating $\Delta ^{(n)} _L \DD$ in favor of $U_\delta$, we may organize the result as $L_4^T+L_4=0$ where,
\bea
L^T_4 = U^{I_1 \cdots I_r}(1,\cdots, n) 
- \sum_{i=0}^r g^{I_r \cdots I_{i+1}} {}_J(1,2) U^{I_1 \cdots I_i J} _\delta (2,\cdots , n)
\label{l4teq}
\eea
and $L_4$ simplifies to give,
\bea
L_4 & = & \sum _{\ell=1}^r { (2 \pi i )^\ell \over \ell !} \delta ^{I_1 \cdots I_\ell}_L 
\bigg [ \DD^{I_{\ell+1} \cdots I_r} (1, \cdots , n) 
\no \\ && \hskip 1.2in 
+ \big ( \chi^{I_r \cdots I_{\ell+1}} (1,2) - \chi^{I_r \cdots I_{\ell+1}} (1,n) \big ) C_\delta (2,\cdots, n) 
\no \\ && \hskip 1.2in 
- \sum _{k=\ell+1}^{r+1} g^{I_r \cdots I_k}{}_J(1,2) \DD^{I_{\ell+1} \cdots I_{k-1} J} (2, \cdots,n) \bigg ]
\eea
The rank of the terms on the right is at most $r-1$. The expression coincides with equation (\ref{2.desc.2}) and therefore vanishes. The remaining expression for $L^T_4$ in (\ref{l4teq}) involves only $U_\delta$ and may again be analyzed by induction in $r$, starting with the vanishing of $U^I_\delta (2,\cdots,n)$ which follows from the monodromy in $z_n$ of the first equation in (\ref{2.desc.2}). This concludes the proof of the $z_n$ monodromy and therefore of the entire Lemma \ref{2.lem:5}.

\newpage

\section{Proving properties of meromorphic multiplets}
\setcounter{equation}{0}
\label{app:E}

This appendix gathers proofs of several symmetry properties of the meromorphic multiplets 
$D_\delta^{I_1\cdots I_r}$, $\cW_{I_1\cdots I_r}(1,\cdots,n)$,  and $M_\delta^{I_1\cdots I_r}(x,y)$ in the decompositions
 (\ref{again.03}) and (\ref{merodecL}) of cyclic products and linear chain products, respectively.

\subsection{Reflection property of $D_\delta^{I_1  \cdots I_r} $ from contour deformations}
\label{sec:refDs}

The contour deformation techniques of figure \ref{fig:noncyc}, which were the key to
proving the cyclic invariance of the constant multiplets $D_\delta^{I_1\cdots I_r}$ 
in section \ref{restthm}, can also be used to establish their reflection property at fixed rank $r$ stated in
Proposition \ref{5.reflect}. Before presenting an indirect proof of the proposition for arbitrary rank in appendix \ref{sec:moreprf}, we shall provide some explicit calculations to rank $r \leq 5$.

\sm

For this purpose, we shall evaluate $\newH_\delta^{I_1 I_r \cdots I_2}$ in terms of $ \newH_\delta^{I_1 I_2\cdots I_r} $ which will give the departures  from reflection symmetry of the convolution integrals in (\ref{4.DA.7}) for $n=0$ by contour deformations similar to those of  figure \ref{fig:noncyc}. \footnote{Note that the coefficient of $\delta^{I_4}_{I_3} $ in the last line can be rewritten as
$ \newH_\delta^{I_1 I_2 I_3 I_5 } 
- 2\pi i \delta^{I_2}_{I_3}  \newH_\delta^{I_1 I_2 I_5  } 
=  \newH_\delta^{I_1 I_5 I_3 I_2 } 
- 2\pi i \delta^{I_5}_{I_3}  \newH_\delta^{I_1 I_5 I_2  } $
by the rank-four identity in (\ref{refls34}) which establishes the
symmetry of the combination $\newH_\delta^{I_1 I_2 I_3 I_4  I_5}
+\newH_\delta^{I_1 I_5 I_4 I_3  I_2}$ under the simultaneous
swap $I_2 \leftrightarrow I_5$ and $I_3 \leftrightarrow I_4$.}
\begin{align}
\newH_\delta^{I_1 I_3  I_2} &= -  \newH_\delta^{I_1 I_2  I_3} + 2\pi i \delta^{I_3}_{I_2}  \newH_\delta^{I_1 I_2  }
\label{refls34} 
\no \\
\newH_\delta^{I_1 I_4 I_3  I_2} &=  \newH_\delta^{I_1 I_2  I_3 I_4} 
+ 2\pi i \delta^{I_4}_{I_3}  \newH_\delta^{I_1 I_3 I_2  }
- 2\pi i \delta^{I_3}_{I_2}  \newH_\delta^{I_1 I_2 I_4 } \notag \\
\newH_\delta^{I_1 I_5 I_4 I_3  I_2} &=  - \newH_\delta^{I_1 I_2  I_3 I_4 I_5} 
+ 2\pi i \delta^{I_5}_{I_4}  \newH_\delta^{I_1 I_4 I_3 I_2  }
+ 2\pi i \delta^{I_3}_{I_2}  \newH_\delta^{I_1 I_2 I_4 I_5 }  
\notag \\
&\quad + 2\pi i \delta^{I_4}_{I_3} \big[  \newH_\delta^{I_1 I_2 I_3 I_5 } 
- 2\pi i \delta^{I_2}_{I_3}  \newH_\delta^{I_1 I_2 I_5  } \big]  
\end{align}
These expressions are derived by deforming the
cascade of displaced integration contours for $t_1,\cdots,t_r$ associated with
$ \newH_\delta^{ I_1 I_r \cdots I_2 }$ to the simpler
arrangement of contours on the right side
of figure \ref{fig:noncyc} that corresponds to $ \newH_\delta^{ I_1 I_2 \cdots I_r }$,
with $t_1$ at the outermost, $t_2$ at the next-to-outermost and $t_r$ at the innermost placement in 
the interior $D^o$ of the fundamental domain.
The contours for $t_2,\cdots,t_r$ of the starting point
$ \newH_\delta^{ I_1 I_r \cdots I_2 }$ are ordered in the opposite way, with $t_2$
at the innermost placement in $D^o$ and $t_r$ at the outermost one besides $t_1$.
One proceeds by deforming the contour for $t_2$ past all of those for $t_3,\cdots,t_r$,
followed by deforming all further contours for $t_j$ past those for $t_{j+1},\cdots,t_r$
in the order of increasing $j=3,\cdots,r-1$. All crossings of contours for
$t_j$ with those of $t_i$ contribute via integrals over infinitesimal circles 
of $t_j$ around $t_i$ if $I_i = I_j$ which we evaluate via
residues as in the proof of Lemma \ref{4.lem:10}. 
Again, the pole structure of the integrand $C_\delta(t_1,\cdots,t_r)$ implies
that the only non-vanishing residues (\ref{rescdelta}) arise for circles of $t_k$
around $t_{k\pm 1}$.

\sm

The alternating reflection parity (\ref{reflD}) at rank $r\leq 5$ then follows by applying
the properties (\ref{refls34}) and (\ref{4.lem.10}) of the $\mA$ convolutions to the expressions for 
$D_\delta^{I_1  \cdots I_r} $  in terms of
$ \newH_\delta^{ J_1\cdots J_s }$. At rank $r\leq 4$, these expressions 
can be found in (\ref{6.3.g}), and the $r=5$ case is obtained from the 
components of (\ref{4.Kper.2}), where the integrand 
$ \DDD(t;B)$ on the right side is expanded via (\ref{4.beta.4}).

\subsection{Proof of items $(b)$ and $(c)$ of Proposition \ref{mdec.props}}
\label{sec:4.cor}

We shall here prove the symmetry properties (\ref{cycWlegmero}), (\ref{reflWW})
of the ${\cal W}_{I_1   \cdots I_{r}}(1,\cdots,n) $ in
items $(b)$, $(c)$ of Proposition \ref{mdec.props}, using the fact
that the same properties are already established for their modular
counterparts $\cW \leftrightarrow \cV$ in items $(b)$, $(c)$ of Proposition \ref{prop:mdec}.

\sm

At fixed $n$ and $r$, the proven symmetry properties (\ref{cycWleg}) and (\ref{reflVV})
 of ${\cal V}_{I_1   \cdots I_{r}}(1,\cdots,n) $ can alternatively be checked by means of the 
 Fay identities and interchange lemmas \cite{DHoker:2024ozn} of the DHS kernels.
For instance, the weight-one interchange lemma $ \omega_I(x) f^I{}_J(y,x) + \omega_I(y) f^I{}_J(x,y)  = 0$ \cite{DHoker:2020uid}
 together with the alternating symmetry of double derivatives
 \beq
 \partial_x \partial_y\cG^{I_1\cdots I_r}(x,y) = (-1)^r \partial_x \partial_y \cG^{I_r\cdots I_1}(y,x) 
  \label{mdec.11} 
 \eeq
 suffice to check that the simplest instances of $ {\cal V}_{I_1  \cdots I_{r}}(1,\cdots,n) $ 
 in (\ref{mdec.06}) obey
\bea
\label{mdec.12} 
{\cal V}(1,2) & = &   {\cal V}(2,1) 
\no \\
  {\cal V}_{JK}(1,2,3) & = & - {\cal V}_{KJ}(2,1,3) 
  \no \\
{\cal V}(1,2,3) & = & -  {\cal V}(2,1,3)
  \eea
The key idea of this proof is to transfer these symmetry checks from the
DHS kernels in ${\cal V}_{I_1   \cdots I_{r}}(1,\cdots,n) $ to the Enriquez kernels
in ${\cal W}_{I_1   \cdots I_{r}}(1,\cdots,n) $. This can be done since the 
explicit form of ${\cal W}_{I_1   \cdots I_{r}}(1,\cdots,n) $ is obtained
from ${\cal V}_{I_1   \cdots I_{r}}(1,\cdots,n) $ by substituting $f^{J_1\cdots J_s}{}_K(x,y)
\rightarrow g^{J_1\cdots J_s}{}_K(x,y)$ term by term, see (\ref{mdec.05}).

\sm
  
 As detailed in section 9 of \cite{DHoker:2024ozn}, the interchange lemmas
 and Fay identities of Enriquez kernels take the same form as those of
 the DHS kernels (see \cite{DHoker:2024ozn} for a proof of the DHS-kernel identities and
 the meromorphic interchange lemma and \cite{Baune:2024ber} for a proof of meromorphic Fay identities). 
 Similarly, the alternating symmetry (\ref{mdec.11}) has
 a direct counterpart \cite{DHoker:2024ozn}
\beq
\partial_y \chi^{I_1\cdots I_r}(x,y) = (-1)^r \partial_x  \chi^{I_r\cdots I_1}(y,x) 
  \label{mdec.13} 
 \eeq
at the level of Enriquez kernels. Accordingly, the identities (\ref{mdec.12})
among modular tensors at $n\leq 3$ points propagate to
\bea
 {\cal W}(1,2) & = &   {\cal W}(2,1)
 \no \\
  {\cal W}_{JK}(1,2,3) & = & - {\cal W}_{KJ}(2,1,3) 
  \no \\
   {\cal W}(1,2,3) & = & -  {\cal W}(2,1,3)
  \label{mdec.14} 
\eea
since the underlying manipulations of Enriquez kernels take the same
form as the DHS-kernel identities required for the derivation of (\ref{mdec.12}).

\sm 

More generally, the one-to-one correspondence between linear and quadratic
relations among Enriquez kernels and those of DHS kernels implies that the
Fay identities and interchange lemmas needed to establish the properties (\ref{cycWleg}) and (\ref{reflVV})
of $ {\cal V}_{I_1  \cdots I_{r}}(1,\cdots,n) $ are preserved under
$f^{J_1\cdots J_s}{}_K(x,y) \rightarrow g^{J_1\cdots J_s}{}_K(x,y)$. 
This relies on the uniqueness results on the relations among
Enriquez kernels in Lemma 11 and Theorem 12 of \cite{Baune:2024ber}.
Hence, the images $  {\cal W}_{I_1  \cdots I_{r}}(1,\cdots,n) $ of 
$ {\cal V}_{I_1  \cdots I_{r}}(1,\cdots,n) $ under (\ref{subgtof}) will obey
the formal images of the properties (\ref{cycWleg}) and (\ref{reflVV}) under
$\cV \rightarrow \cW$. These images are the statements in items $(b)$ and $(c)$
of Proposition \ref{mdec.props} which thereby completes their proof.

\subsection{Proof of Proposition \ref{5.reflect}}
\label{sec:moreprf}

We prove the alternating reflection parity $\DD^{I_1 I_2  \cdots I_r} = (-1)^r \DD^{I_r  \cdots I_2 I_1} $
in Proposition \ref{5.reflect} by induction in the rank $r$. As a base case at $r=2$, the parity
$\DD^{I_1 I_2} = \DD^{I_2 I_1}$ is clear both from the cyclicity of $\DD^{I_1  \cdots I_r} $ 
in Theorem~\ref{2.thm:3} and from the explicit form of $\DD^{I_1 I_2} $ in (\ref{2.1.b}).

\sm

Assuming that $\DD^{I_1 I_2  \cdots I_s} = (-1)^s \DD^{I_s  \cdots I_2 I_1} $
hold for $s\leq r-1$, we will now infer the $s=r$ case from
item $(c)$ of Proposition \ref{mdec.props}
and the alternating parity 
\bea
C_\delta(1,2,\cdots,r) = (-1)^r C_\delta(r,\cdots,2,1) 
\label{prfrd.01}
\eea
of cyclic products which follows from the antisymmetry $S_\delta(x,y) = - S_\delta(y,x)$
of Szeg\"o kernels. For this purpose, the alternating parity (\ref{prfrd.01}) will be
imposed at the level of the meromorphic decomposition (\ref{again.03})
which we split into the form of
\bea
C_\delta (1,\cdots, r) &=& \check C_\delta(1,\cdots,r)+ \omega_{I_1}(1) \cdots \omega_{I_r}(r) D_\delta^{I_1 \cdots I_r} \label{prfrd.02}\\
\check C_\delta(1,\cdots,r) &=&
  {\cal W}(1,\cdots,r)  +  \sum_{s=2}^{r-1}
  {\cal W}_{I_1  \cdots I_{s}}(1,\cdots,r)  \, D_\delta^{I_1 \cdots I_{s}}
\notag
\eea
The alternating parity of $ {\cal W}_{I_1  \cdots I_{s}}(1,\cdots,r)$ established
in item $(c)$ of Proposition \ref{mdec.props} together with the
inductive assumption imply that (\ref{prfrd.01}) holds separately for the 
part $\check C_\delta(1,\cdots,r) $ in (\ref{prfrd.02}),
\bea
\check C_\delta(1,\cdots,r) &=&
 (-1)^r {\cal W}(r,\cdots,1)  +  \sum_{s=2}^{r-1} (-1)^{r-s}
  {\cal W}_{I_s  \cdots I_{1}}(r,\cdots,1)  \, D_\delta^{I_1 \cdots I_{s}}
\notag\\
 &=&
(-1)^r \, \bigg\{
{\cal W}(r,\cdots,1)  +  \sum_{s=2}^{r-1}  
  {\cal W}_{I_s  \cdots I_{1}}(r,\cdots,1)  \, D_\delta^{I_s \cdots I_{1}}
\bigg\}
\notag \\
&=& (-1)^r \, \check C_\delta(r,\cdots,1)
\label{prfrd.03}
\eea
where we emphasize that $D_\delta^{I_1 \cdots I_{s}} = (-1)^s D_\delta^{I_s \cdots I_{1}}$
has only been used for $s\leq r-1$. As a consequence of (\ref{prfrd.03}), the contributions
from $\check C_\delta(1,\cdots,r)$ drop out in the following rewriting of (\ref{prfrd.01}):
\bea
0 &=& C_\delta(1,2,\cdots,r) - (-1)^r C_\delta(r,\cdots,2,1) 
\label{prfrd.04} \\ 
&=&  \omega_{I_1}(1) \cdots \omega_{I_r}(r) D_\delta^{I_1 \cdots I_r} 
- (-1)^r \omega_{I_1}(r) \cdots \omega_{I_r}(1) D_\delta^{I_1 \cdots I_r}  \notag \\
&=& \omega_{I_1}(1) \cdots \omega_{I_r}(r) \big( D_\delta^{I_1 \cdots I_r} 
- (-1)^r  D_\delta^{I_r \cdots I_1} \big)
\notag
\eea
which completes the inductive step and therefore the proof of
Proposition \ref{5.reflect} for arbitrary rank $r$
(see appendix \ref{sec:refDs} for an alternative proof at rank $r\leq 5$).

\subsection{Reflection properties of $M_\delta^{I_1  \cdots I_r}$ from contour deformations}
\label{sec:refmm}

The reflection property (\ref{Lmot.27}) of the spinors $ L_\delta^{I_1 \cdots I_r}(x,y)$
in the modular decomposition (\ref{moddecL}) of linear chain products is
manifest from their surface-integral representation (\ref{Ldelint}). However,
this does not immediately carry over to the reflection properties (\ref{Lmot.71})
of their meromorphic counterparts $M_\delta^{I_1  \cdots I_r}(x,y)$ when expressed
in terms of the $\mA$ convolutions generated by (\ref{Lmot.55}). As an alternative
to the indirect proof in section \ref{proswms} that is valid for arbitrary rank, we shall here
provide a direct proof of (\ref{Lmot.71}) at rank $r\leq 3$, 
using the contour deformations of section \ref{sec:concross}
which may also be extended to $r\geq 4$.
 
\sm

The salient point is that the $\mA$ convolutions $\MM_\delta ^{I_1 \cdots I_r}(x,y)$ at $r\geq 2$ 
fail to exhibit any reflection parity $\pm 1$ since the transition to $\MM_\delta ^{I_r \cdots I_1}(y,x)$ 
requires the crossing of certain integration contours $\mA^{I_k}$ in (\ref{defcms}) which results in 
residue contributions similar to those in section \ref{sec:concross}.
By following the integration contour deformation techniques illustrated in
figure \ref{fig:noncyc}, one can derive examples such as,
\begin{align}
\MM_\delta ^{I_1 }(y,x) &= \MM_\delta ^{ I_1 }(x,y) 
\label{reflmdel}\\
\MM_\delta ^{I_1 I_2 }(y,x) &= - \MM_\delta ^{ I_2 I_1}(x,y) 
 + 2\pi i  \delta^{I_2}_{I_1}  \MM_\delta ^{I_1}(x,y)\notag \\
\MM_\delta ^{I_1 I_2 I_3}(y,x) &= \MM_\delta ^{I_3 I_2 I_1}(x,y) 
 - 2\pi i \big[ \delta^{I_3}_{I_2}  \MM_\delta ^{I_2 I_1}(x,y)
 + \delta^{I_2}_{I_1}  \MM_\delta ^{I_3 I_1}(x,y) \big]
 \no \\ & \qquad
 + ( 2\pi i )^2 \delta^{I_1 I_2}_{I_3}  \MM_\delta ^{I_3}(x,y)
 \notag
\end{align}
see (\ref{refls34}) for the analogous reflection formulae
for $\mA$ convolutions of cyclic products. Just as it was the
case for the constants (\ref{6.3.g}) in the cyclic products, the
correction terms in (\ref{exofms}) by convolutions of shorter linear
chain products compensate for the lack of simple reflection parities
$\pm 1$ in (\ref{reflmdel}).

\sm

\newpage

\section{Low rank examples}
\setcounter{equation}{0}
\label{app:F}

This appendix gathers further examples of the meromorphic
and single-valued forms ${\cal W}$ and ${\cal V}$ in the reduction
of products of Szeg\"o kernels. We follow the presentation of 
sections \ref{sec:2.6}, \ref{sec:modlv} and spell out the explicit
form of ${\cal W}_{I_1\cdots I_r}(1,2,3,4)$ for cyclic products
in terms of Enriquez kernels and ${\cal V}_{I_1\cdots I_r}(x;1,2,3;y)$
for linear chains in terms of DHS kernels. Their counterparts
${\cal V}_{I_1\cdots I_r}(1,2,3,4)$ and ${\cal W}_{I_1\cdots I_r}(x;1,2,3;y)$
involving the opposite type of higher-genus kernels can be straightforwardly
obtained from the subsequent expressions via $g^{I_1\cdots I_r}{}_J(x,y) \leftrightarrow f^{I_1\cdots I_r}{}_J(x,y)$, i.e.\ through the correspondences
(\ref{mdec.05}) and (\ref{Lmot.61}).

\subsection{Meromorphic ${\cal W}$ for cyclic products of four Szeg\"o kernels}
\label{sec:2.6.2}

The single-valued combinations ${\cal W}_{I_1\cdots I_r}(1,\cdots,n)$ of Enriquez
kernels at $n=2,3$ points can be found in section \ref{sec:2.ex}.
At $n=4$, the expressions below for
${\cal W}_{I_1\cdots I_r}(1,2,3,4) $ at $r=0,2,3$ are derived by matching the results 
of the descent (\ref{5.h}) with the meromorphic decomposition (\ref{again.03}).
Similar to the $n=3$ case, one eliminates $C_\delta(2,3,4), C_\delta(3,4)$
and all the multiplets $D_\delta^{I_1\cdots I_s}(r,\cdots)$ that depend on at least one
point $z_r,\cdots$ from (\ref{5.h}). The resulting coefficients of $D_\delta^{JKL}$,
$D_\delta^{JK}$ and $\delta$-independent terms are given by,
\begin{align}
{\cal W}_{JKL}(1,2,3,4) & = \frac{1}{3} \Big[ \big ( \chi(1,4) - \chi(1,2) \big )  \om_J(2) \om_K(3)  \om_L(4)
\label{eq2.6.6} \\
&\quad\quad
+ \om_I(1) g^I{}_{J} (2,3) \om_K(3) \om_L(4)
+ \om_J(1)  \om_I(2) g^I{}_{K} (3,4)  \om_L(4) \notag \\
&\quad\quad
+  \big( \om_J(1) \om_K(2)\om_I(3) - \om_I(1) \om_J(2)  \om_K(3)\big) \varpi ^I{}_{L}(4)
+ {\rm cycl}(J,K,L) \Big]
\no \\
{\cal W}_{JK}(1,2,3,4) & = \frac{1}{2} \Big[
\big ( \chi(1,4) - \chi(1,2) \big )  \big ( \chi(2,4) - \chi(2,3) \big )  \om_J(3) \om_K(4)
\notag \\
&\quad\quad +   \big ( \chi(1,4) - \chi(1,2) \big )  \om_I(2) g^I{}_J(3,4) \om_K(4) \notag \\
&\quad\quad +  \big ( \chi(1,4) - \chi(1,2) \big )
\big( \om_J(2) \om_I(3) - \om_I(2) \om_J(3) \big) \varpi ^I{}_{K}(4) \notag \\
&\quad\quad +\omega_I(1) \big ( \chi^I(2,4) - \chi^I(2,3) \big )  \om_J(3) \om_K(4) \notag \\
&\quad\quad + \omega_I(1) g^I{}_M(2,3) g^M{}_J(3,4) \omega_K(4)
+ \omega_I(1) \omega_M(2) g^{MI}{}_J(3,4) \omega_K(4) \notag \\
&\quad \quad +  \omega_I(1)\big( g^I{}_J(2,3) \omega_M(3) -  g^I{}_M(2,3) \omega_J(3) \big) \varpi^M{}_{K}(4)  \notag \\
&\quad \quad +  \omega_I(2)  \big( \omega_J(1)g^I{}_M(3,4) -   \omega_M(1)  g^I{}_J(3,4) \big) \varpi^M{}_{K}(4)  \notag \\
&\quad \quad +   \big( \om_J(1) \om_I(2)\om_M(3) - \om_I(1) \om_J(2)  \om_M(3)
\notag \\
&\quad\quad\quad -  \om_M(1) \om_J(2)\om_I(3)+  \om_M(1) \om_I(2)\om_J(3)
\big) \varpi^{MI}{}_{K}(4) + (J \leftrightarrow K) \Big]
\no \\
{\cal W}(1,2,3,4) & =  \big( \chi(1,4) - \chi(1,2) \big) \big( \chi(2,4) - \chi(2,3) \big) \partial_4 \chi(3,4) \notag \\
&\quad + \big( \chi(1,4) - \chi(1,2) \big) \omega_I(2) \partial_4 \chi^I(3,4) + \omega_I(1) g^I{}_J(2,3)  \partial_4 \chi^J(3,4) \notag \\
&\quad + \omega_I(1) \big( \chi^I(2,4) - \chi^I(2,3) \big)
\partial_4 \chi(3,4) + \omega_I(1) \omega_J(2) \partial_4 \chi^{JI}(3,4)
\notag
\end{align}
The prescriptions $+ {\rm cycl}(J,K,L) $ and $+ (J \leftrightarrow K) $ refer to 
all lines of the expressions for ${\cal W}_{JKL}(1,2,3,4)$ and ${\cal W}_{JK}(1,2,3,4) $,
respectively, and again implement the cyclic symmetrization in the indices
according to the choice (\ref{cycW}). The rank-four case of ${\cal W}_{IJKL}(1,2,3,4)
= \frac{1}{4} \om_I(1)\om_J(2)\om_K(3)\om_L(4)+ {\rm cycl}(I,J,K,L)$
lines up with the general formula (\ref{eq2.6.1}) for maximal rank $r=n$.

\subsection{Modular ${\cal V}$ for linear chains with three internal points}
\label{sec:vwith3}

The combinations $\cV_{I_1\cdots I_r}(x;1,\cdots ,n;y)$ of DHS kernels that are
meromorphic in the internal points $z_1,\cdots,z_n$ of linear chain products are spelt out
for $n=1,2$ in section \ref{sec:exv12}.
Their three-point analogues following from the modular descent equations
of Theorem \ref{4.thm:4} are given by
\begin{align}
\cV_{JK}(x;1,2,3;y) &=  \p_1 \big( \cG(1,x) -\cG(1,2) \big) \om_J(2) \om_K(3)   \label{Lmot.38}  \\
&\quad + \om_I(1) f^I{}_J(2,3) \om_K(3) + \om_J(1)  \om_I(2)  f^I{}_K(3,y)   \notag\\
\cV_J(x;1,2,3;y) &=  \p_1 \big( \cG(1,x) -\cG(1,2) \big) \p_2 \big( \cG(2,x) -\cG(2,3) \big)  \om_J(3) \notag \\
&\quad+ \p_1\big( \cG(1,x) -\cG(1,2) \big) \om_I(2) f^I{}_J(3,y) \notag \\
&\quad
+ \om_I(1) \p_2\big( \cG^I(2,x) - \cG^I(2,3) \big)  \om_J(3) \notag \\
&\quad
+ \om_I(1) \om_K(2) f^{KI}{}_J(3,y) + \om_I(1) f^I{}_K(2,3) f^K{}_J(3,y)  \notag \\
\cV(x;1,2,3;y) &=  \p_1\big( \cG(1,x) -\cG(1,2) \big)
\p_2 \big( \cG(2,x) -\cG(2,3) \big)
\p_3 \big( \cG(3,x) -\cG(3,y) \big) \notag \\
&\quad + \p_1 \big( \cG(1,x) -\cG(1,2) \big) \om_K(2) \p_3 \big( \cG^K(3,x) -\cG^K(3,y) \big) \notag \\
&\quad + \om_K(1)\p_2 \big( \cG^K(2,x) -\cG^K(2,3) \big)\p_3 \big( \cG(3,x) -\cG(3,y) \big)  \notag \\
&\quad + \om_I(1) \om_K(2)\p_3 \big( \cG^{KI}(3,x) -\cG^{KI}(3,y) \big) \notag \\
&\quad + \om_I(1) f^I{}_K(2,3)  \p_3\big( \cG^{K}(3,x) -\cG^{K}(3,y) \big) \notag
\end{align}
The rank-three case of $\cV_{IJK}(x;1,2,3;y)
=  \om_I(1)\om_J(2)\om_K(3)$
is covered by the all-multiplicity formula in the
second line of (\ref{moddecL}).

\newpage

\section{Proofs at genus one}
\setcounter{equation}{0}
\label{app:D}

In this last appendix we gather the proofs of  Proposition \ref{cor:cgenus1} and Theorem \ref{thm:linh1}
on products of Szeg\"o kernels at genus one.

\subsection{Proof of Proposition \ref{cor:cgenus1}}
\label{app:D.1}

We shall derive the expressions (\ref{h1eq.41c}) and (\ref{h1eq.41d})  in Proposition \ref{cor:cgenus1} for 
the specializations of the constants $C^{I_1\cdots I_r}_\delta$ and $D^{I_1\cdots I_r}_\delta$ 
to genus one.

\sm

The first step is to derive the expressions (\ref{h1eq.41c}) 
for $C_\delta^{I_1 \cdots I_n} |_{h=1}$ at $n\geq 3$ by integrating all the $n$
points $z_1,\cdots,z_n$ in (\ref{8.dec}) over the genus one surface against
$\prod_{j=1}^n d^2 z_j/(\Im \tau)$.
The left side integrates to $C_\delta^{I_1 \cdots I_n} |_{h=1}$ by the integral
representation (\ref{mdec.01}) of $C_\delta^{I_1 \cdots I_n}$ at arbitrary genus. 
Integrating the right side of (\ref{8.dec}) over $n$
copies of the torus produces ${\rm G}_n$ from the
singular term $f^{(1)}(z_{12}) f^{(1)}(z_{23}) \cdots f^{(1)}(z_{n1})$ of $V_n(1,\cdots,n)$ 
whereas all $V_k(1,\cdots,n)$
at $1\leq k \leq n{-}1$ integrate to zero. This follows from the fact that
$\int_\Sigma d^2 z \,f^{(k)}(z{-}x)= 0$ for any $k\geq 1$ and $x \in \mathbb C$ since
the Fourier zero mode $m=n=0$ of (\ref{h1eq.11}) only features the singular
term $1/\eta$. Finally, the summand in (\ref{8.dec}) at $k = \lfloor  n/2\rfloor$
integrates to $R_n(e_\delta)$ if $n$ is even since $V_0(1,\cdots,n)=1$ and to zero if $n$ is odd.

\sm

The next step is to prove the $n=2$ instances of both (\ref{h1eq.41d}) 
and (\ref{h1eq.41c}) by specializing the explicit results (\ref{2.1.b}) and
(\ref{catrk2}) for $D_\delta^{IJ}$ and $C_\delta^{IJ}$ to genus $h=1$.
The double-derivatives of theta functions in (\ref{2.1.b}) reduce to those
in the expressions,
\bea
e_\delta = - \partial_\eta^2 \ln \vartheta_\delta(\eta) \,\big|_{\eta = 0} - {\rm G}_2
\eea
as a consequence of $\wp(\eta) = - \partial_\eta^2 \ln \vartheta_1(\eta) - {\rm G}_2$ 
at $\eta = \hp_\delta$. In this way, we define the regularization of
the conditionally convergent integral of $C_\delta(1,2)$ over its two points.

\sm 

The third and most involved part of the proof of Proposition \ref{cor:cgenus1} concerns
the equality of $C_\delta^{I_1 \cdots I_n}  |_{h=1}$ and $D_\delta^{I_1 \cdots I_n} |_{h=1}$ 
for $n\geq 3$. The key idea is to compare the expressions (\ref{h1eq.34a}) and
(\ref{h1eq.34}) for the elliptic functions $V_r(1,\cdots,n)$ of the points $z_1,\cdots,z_n$ 
with the genus one instances of the modular tensors 
${\cal V}_{I_1 \cdots I_{r}}(1,\cdots,n)  $
in the modular decomposition (\ref{again.02}) 
and their counterparts ${\cal W}_{I_1 \cdots I_{r}}(1,\cdots,n)  $
in the meromorphic decomposition (\ref{again.03}).

\sm 

By (\ref{h1eq.03}) and (\ref{h1eq.04}), all instances 
of ${\cal W}_{I_1 \cdots I_{r}}(1,\cdots,n)  $ and
${\cal V}_{I_1 \cdots I_{r}}(1,\cdots,n)  $ at non-zero rank $r\geq 2$ reduce to
products of undifferentiated Kronecker-Eisenstein coefficients $g^{(s)}(z_{ij})$ and
$f^{(s)}(z_{ij})$ akin to (\ref{h1eq.34}) and (\ref{h1eq.34a}), respectively. The 
expressions are gathered in the following lemma to be proven in section \ref{app:lemp.1} below:

{\lem
\label{VWnonzero}
The genus one instances of the meromorphic and single-valued functions
${\cal V}_{I_1\cdots I_r}(1,\cdots,n) $ 
and ${\cal W}_{I_1\cdots I_r}(1,\cdots,n)$ of $z_1,\cdots,z_n$ at rank $r\geq 2$
in the meromorphic and modular decompositions (\ref{again.03}) and (\ref{again.02})
can be expressed in terms of the elliptic functions $V_s(1,\cdots,n)$ in (\ref{h1eq.34a}) via
\beq
{\cal V}_{I_1 \cdots I_{r}}(1,\cdots,n)  \, \big|_{h=1} = V_{n-r}(1,\cdots,n) \, , \ \ \ \
2\leq r\leq n
 \label{h1eq.35mod}
\eeq
which coincides with the meromorphic counterparts at genus one:
\beq
{\cal W}_{I_1 \cdots I_{r}}(1,\cdots,n)  \, \big|_{h=1} = V_{n-r}(1,\cdots,n) \, , \ \ \ \
2\leq r\leq n
 \label{h1eq.35mero}
\eeq}

\sm

The comparison of the scalars  ${\cal V}(1,\cdots,n) $ and ${\cal W}(1,\cdots,n) $ at
genus one with combinations of the elliptic $V_{r}(1,\cdots,n)$ requires a refined analysis
since their construction from the respective descent procedures
introduces derivatives of Enriquez or DHS kernels,
see the examples in (\ref{eq2.6.5}), (\ref{mdec.06}) and (\ref{eq2.6.6}).
The appearance of these derivatives at genus one is captured by the following lemma
to be proven in section \ref{app:lemp.2} below.

{\lem
\label{derivV}
The spin structure independent terms ${\cal V}(1,\cdots,n) $ and ${\cal W}(1,\cdots,n) $
in the meromorphic and modular decompositions (\ref{again.03}) and (\ref{again.02}) at
genus one are given by the coefficients of $1/\eta$ in the generating functions,
\beq
{\cal V}(1,\cdots,n) \, \big|_{h=1} =
 \mathop{\mathrm{Res}}_{\eta = 0}  \Omega(z_{n1},\eta) \Omega(z_{12},\eta) \cdots
\partial_{n-1} \Omega(z_{n-1,n},\eta) 
 \label{pfs.13} 
\eeq
as well as,
\beq
{\cal W}(1,\cdots,n) \, \big|_{h=1} =
 \mathop{\mathrm{Res}}_{\eta = 0}  F(z_{n1},\eta) F(z_{12},\eta) \cdots
\partial_{n-1} F(z_{n-1,n},\eta) 
 \label{pfs.21} 
\eeq
}

\sm
The expressions in (\ref{pfs.13}) and (\ref{pfs.21}) will be shown in  section \ref{app:lemp.3} to imply the
alternative representations in the following lemma which completely bypass derivatives
of Kronecker-Eisenstein coefficients.

{\lem
\label{lem:calvg1}
The spin structure independent terms ${\cal V}(1,\cdots,n) $ and ${\cal W}(1,\cdots,n) $
in the meromorphic and modular decompositions (\ref{again.03}) and (\ref{again.02}) at
genus one can be expressed in terms of the elliptic functions $V_r(1,\cdots,n)$
in (\ref{h1eq.34a}) and (almost) holomorphic Eisenstein series via
\beq
{\cal V}(1,\cdots,n) \, \big|_{h = 1} = 
V_n(1,\cdots,n) 
- \widehat {\rm G}_{2} V_{n-2}(1,\cdots,n) 
- \sum_{k=2}^{ \lfloor n/2 \rfloor } {\rm G}_{2k} V_{n-2k}(1,\cdots,n) 
\label{eq2.6.3}
\eeq
as well as,
\begin{align}
{\cal W}(1,\cdots,n) \, \big|_{h = 1} &= 
V_n(1,\cdots,n) 
-  {\rm G}_{2} V_{n-2}(1,\cdots,n) 
- \sum_{k=2}^{ \lfloor n/2 \rfloor } {\rm G}_{2k} V_{n-2k}(1,\cdots,n) 
\label{hoeq2.6.3}
\end{align}
}

\sm

Note that the right side of (\ref{hoeq2.6.3}) only differs from that of  
(\ref{eq2.6.3}) by the appearance of the meromorphic ${\rm G}_{2} $ in the place of
the modular $\widehat {\rm G}_{2} $.

\sm

On the basis of Lemmas \ref{VWnonzero} and \ref{lem:calvg1}, we can complete the proof of 
Proposition \ref{cor:cgenus1}: by equating the meromorphic
and modular decompositions (\ref{again.03}), (\ref{again.02}) at genus $h=1$ and inserting
all of (\ref{h1eq.35mod}), (\ref{h1eq.35mero}), (\ref{eq2.6.3}), (\ref{hoeq2.6.3})
for the dependence on the marked points, we can solve for $D_\delta^{I_1\cdots I_n}  |_{h=1}$
in terms of $C_\delta^{I_1\cdots I_n}  |_{h=1}$ and arrive at (\ref{cvsdath1}). This can be
done separately for $n=2,3,4,\cdots$ which leads to a single undetermined 
instance of $D_\delta^{I_1 \cdots I_n}  |_{h=1}$ in each case. This concludes our proof of
(\ref{cvsdath1}) which, in combination with (\ref{h1eq.41c}) established in earlier
steps, implies the second main statement (\ref{h1eq.41d}) of Proposition~\ref{cor:cgenus1}.

\subsection{Proof of Lemmas \ref{VWnonzero} to \ref{lem:calvg1}}

In this section we shall prove Lemmas \ref{VWnonzero}, \ref{derivV} and \ref{lem:calvg1}
which were used in the previous section to prove Proposition \ref{cor:cgenus1}.

\subsubsection{Proof of Lemma \ref{VWnonzero}}
\label{app:lemp.1}

The first statement (\ref{h1eq.35mod}) of Lemma \ref{VWnonzero} follows from the fact that the pole structure (\ref{mdec.03}) of the left side matches that of the right side in all variables,
\beq
\bar \partial_k V_{s}(1,\cdots,n)
=  \pi \, \big ( \delta (k,k+1) - \delta (k,k-1) \big ) \,
V_{s-1}(1,\cdots,\hat k ,\cdots, n)
 \label{h1mdec.03}
\eeq
with $s = n-r \geq 1$ and $k=1,\cdots,n$, so that the difference between the left and right sides must be independent of $z_1, \cdots, z_n$ and the fact that the multiple integral over all the points $z_1,\cdots,z_n$ over the torus vanishes on both sides. The two simple poles in $z_k$ at $z_{k\pm 1}$ follow from the generating function (\ref{h1eq.22}) of
the elliptic functions $V_{s}(1,\cdots,n)$ and the singular behavior
$F(z,\eta) = \frac{1}{z}+ {\cal O}(z^0)$. In particular, one
clearly reproduces $V_{0}(1,\cdots,n)=1$ from the simple formula
(\ref{w0calv}) for ${\cal V}_{I_1 \cdots I_{r}}(1,\cdots,n) $ at rank $r=n$.

\sm

The meromorphic counterpart (\ref{h1eq.35mero})
of (\ref{h1eq.35mod}) follows from the facts that 
\begin{enumerate}
\itemsep=-0.03in
\item[$(a)$] the expressions for 
${\cal V}_{I_1 \cdots I_{r}}(1,\cdots,n)$ and ${\cal W}_{I_1 \cdots I_{r}}(1,\cdots,n)$
are related by swapping Enriquez kernels and DHS kernels as in (\ref{mdec.05});
\item[$(b)$] their $h=1$ instances are related by swapping $g^{(s)}(z_{ij})
\leftrightarrow f^{(s)}(z_{ij})$ by virtue of~(\ref{h1eq.03});
\item[$(c)$] undifferentiated $h=1$ kernels $g^{(s)}(z_{ij})$ and
$f^{(s)}(z_{ij})$ obey the same Fay identities~\cite{Broedel:2014vla}.
\end{enumerate}
The outcome ${\cal V}_{I_1 \cdots I_{r}}(1,\cdots,n) |_{h=1}$
of the modular descent at genus one is related
the manifestly cyclic expressions for $V_{n-r}(1,\cdots,n)$ in (\ref{h1eq.34a})
by a sequence of Fay identities among $f^{(s)}(z_{ij})$.
By item $(a)$ and $(b)$, the expressions for 
${\cal W}_{I_1 \cdots I_{r}}(1,\cdots,n) |_{h=1}$ resulting from the
meromorphic descent at genus one is obtained from ${\cal V}_{I_1 \cdots I_{r}}(1,\cdots,n) |_{h=1}$
via $f^{(s)}(z_{ij}) \rightarrow g^{(s)}(z_{ij})$. Given that $g^{(s)}(z_{ij})$ and
$f^{(s)}(z_{ij})$ obey the same Fay identities, see item $(c)$, one can attain manifestly cyclic rewritings
of ${\cal W}_{I_1 \cdots I_{r}}(1,\cdots,n) |_{h=1}$ from those of 
${\cal V}_{I_1 \cdots I_{r}}(1,\cdots,n) |_{h=1}$ via $f^{(s)}(z_{ij}) \rightarrow g^{(s)}(z_{ij})$.
Since the elliptic $V_{n-r}(1,\cdots,n)$ in the earlier result (\ref{h1eq.35mod}) for
${\cal V}_{I_1 \cdots I_{r}}(1,\cdots,n) |_{h=1}$ are invariant under
$g^{(s)}(z_{ij}) \leftrightarrow f^{(s)}(z_{ij})$, see (\ref{h1eq.34})
and (\ref{h1eq.34a}), we are led to the second statement
(\ref{h1eq.35mero}) of Lemma \ref{VWnonzero}.

\subsubsection{Proof of Lemma \ref{derivV}}
\label{app:lemp.2}

The first statement (\ref{pfs.13}) of Lemma \ref{derivV} is readily checked for
the cases of $n=2,3$ through the Laurent expansion (\ref{h1eq.02}) of $\Omega(z,\eta)$
in $\eta$ and the specializations of (\ref{mdec.06}) to genus one,
\begin{align}
{\cal V}(1,2) \, \big|_{h=1} &= \partial_1 f^{(1)}_{12}
\label{pfs.01} \\
{\cal V}(1,2,3) \, \big |_{h=1} &= (f^{(1)}_{12} - f^{(1)}_{13})
 \partial_2 f^{(1)}_{23}
+ \partial_2 f^{(2)}_{23}
\notag 
\end{align}
using the shorthands
\beq
f^{(s)}_{ij} = f^{(s)}(z_{i}-z_j) 
 \label{pfs.00} 
\eeq
Our proof of (\ref{pfs.13}) at general $n$ proceeds in three steps by showing
that its right side
\begin{enumerate}
\itemsep=-0.03in
\item[$(a)$] is cyclic in $z_1,z_2,\cdots,z_n$;
\item[$(b)$] has the same recursive simple-pole structure in the number of points
as the left side with ${\rm Res}_{z_n = z_1}{\cal V}(1,\cdots,n) |_{h=1} = {\cal V}(1,\cdots,n-1) |_{h=1} $
at $n\geq 3$;
\item[$(c)$] vanishes upon integrating all of $z_1,\cdots,z_n$ over the torus.
\end{enumerate}
Item $(a)$ follows from the fact that the cyclic product
$\Omega(z_{12},\eta)  \Omega(z_{23},\eta) \cdots \Omega(z_{n1},\eta) $
without the $z_{n-1}$ derivative in (\ref{pfs.13}) is an elliptic function of $\eta$ and
therefore does not have a residue, 
\beq
V_{n-1}(1,\cdots,n) = \mathop{\mathrm{Res}}_{\eta=0}
 \Omega(z_{12},\eta)  \Omega(z_{23},\eta) \cdots \Omega(z_{n1},\eta)  = 0 
 \label{pfs.03} 
\eeq
Taking derivatives of (\ref{pfs.03}) in the $z_i$ then leads to differences such as
\begin{align}
0 &= \partial_{n}   \mathop{\mathrm{Res}}_{\eta = 0} \Omega(z_{12},\eta) \cdots
\Omega(z_{n-1,n},\eta)  \Omega(z_{n1},\eta)   \label{pfs.14} \\
&=  \mathop{\mathrm{Res}}_{\eta = 0} \Omega(z_{12},\eta) \cdots
\partial_{n} \big[\Omega(z_{n-1,n},\eta)  \Omega(z_{n1},\eta)  \big] \notag \\
&=  \mathop{\mathrm{Res}}_{\eta = 0} \big[ \Omega(z_{12},\eta) \cdots
\Omega(z_{n-1,n},\eta)  \partial_{n}  \Omega(z_{n1},\eta) 
- \Omega(z_{n1},\eta) 
 \Omega(z_{12},\eta) \cdots
  \partial_{n-1} \Omega(z_{n-1,n},\eta)  \big]
 \notag  
   \end{align}
 using translation invariance $(\p_i + \p_j) \Omega(z_{ij},\eta) =0$ in the last step.
We conclude from (\ref{pfs.14}) that the right side of (\ref{pfs.13}) is
cyclically invariant under $(z_1,z_2,\cdots,z_n) \rightarrow  (z_2,\cdots,z_n,z_1)$.

  \sm 
  
Item $(b)$ amounts to showing that the right side of (\ref{pfs.13}) with $n\geq 3$
has a simple pole in $z_{n1}$ with residue given by 
$\mathop{\mathrm{Res}}_{\eta=0}  \Omega(z_{12},\eta)\Omega(z_{23},\eta) \cdots
  \partial_{n-1} \Omega(z_{n-1,1},\eta) $. This immediately follows from 
  $ \Omega(z_{ij},\eta) = 1/z_{ij}+ \cO(z_{ij}^0)$ and by item $(a)$ implies that
 the right side of (\ref{pfs.13}) has simple poles in each pair $z_{i,i+1}$ of
 consecutive points, with an $(n-1)$-point instance
 of the same expression as a residue. Hence, the difference of
 the left and right side of (\ref{pfs.13}) at $n$ points is holomorphic if its
 $(n-1)$-point instance vanishes.
 
 \sm
 
 Item $(c)$ follows from the fact that both $f^{(s)}_{ij}$ and $\p_i f^{(s)}_{ij}$
 with $s\geq 1$ vanish upon integrating either $z_i$ or $z_j$ over the torus.
Accordingly, each term on the right side of
\beq
\mathop{\mathrm{Res}}_{\eta=0} \Omega(z_{n1},\eta) \Omega(z_{12},\eta) \cdots
\partial_{n-1} \Omega(z_{n-1,n},\eta) 
=  \! \! \! \! \! \! \sum_{s_1,s_2,\cdots,s_n \geq 0
\atop{ s_1+ s_2+\cdots + s_n = n-1}} \! \! \! \! \! \!
 f^{(s_1)}_{n1} f^{(s_2)}_{12} \cdots \p_{n-1} f^{(s_n)}_{n-1,n}
 \label{pfs.16}
\eeq
integrates to zero over one of $z_1,\cdots,z_n$ since
$s_1+\cdots + s_n = n-1$ is incompatible with having
all $s_j >0$. So there are at most $n-1$ factors of 
$f^{(s)}_{ij}$ or $\p_i f^{(s)}_{ij}$ with $s>0$ per summand, and there is at least
one point which only enters one of the factors.
Since also the left side of (\ref{pfs.13}) integrates to zero over $z_i$
and in fact for $\cV(1,\cdots,n)$ at arbitrary genus $h\geq 1$,
the difference of the left and right sides of
(\ref{pfs.13}) integrates to zero at all~$n\geq 2$. 

\sm

We can now show the equality (\ref{pfs.13}) by induction in $n$.
The base cases at $n=2,3$ are already checked in (\ref{pfs.01}).
Item $(b)$ implied that the difference of the left and right side of (\ref{pfs.13})
is holomorphic at $n$ points, assuming that it vanishes
at $\leq n-1$ points and $n\geq 3$. As a holomorphic function of $z_1,\cdots,z_n$, the
difference must be constant, and by the result of item $(c)$, this
constant vanishes.

\sm

Since ${\cal W}(1,\cdots,n)$ are obtained from ${\cal V}(1,\cdots,n)$ by the conversion
(\ref{mdec.05}) of DHS kernels into Enriquez kernels, their genus one instances
are related by $f^{(s)}_{ij} \leftrightarrow g^{(s)}_{ij}$. Hence, (\ref{pfs.13}) 
implies the second statement (\ref{pfs.21}) of the lemma and thereby concludes its proof.

\subsubsection{Proof of Lemma \ref{lem:calvg1}}
\label{app:lemp.3}

With the representations (\ref{pfs.13}) and (\ref{pfs.21}) of ${\cal V}(1,\cdots,n) |_{h=1}$
and ${\cal W}(1,\cdots,n) |_{h=1}$
at hand, we can now prove the statements (\ref{eq2.6.3})
and (\ref{hoeq2.6.3}) of Lemma \ref{lem:calvg1} by means of the identities
\begin{align}
\p_zF(z,\eta) &=\p_\eta F(z,\eta)  + 
\bigg(
g^{(1)}(\eta)   - g^{(1)}(z)
\bigg) F(z,\eta) 
 \label{pfs.17} \\
\p_z \Omega(z,\eta) &=\p_\eta \Omega(z,\eta)  + 
\bigg(
g^{(1)}(\eta) + \frac{\pi \eta}{\Im \tau} - f^{(1)}(z)
\bigg)\Omega(z,\eta) 
\notag
\end{align}
among Kronecker-Eisenstein series. The first line is a simple consequence of the 
theta function representations in (\ref{h1eq.01}) and the second line follows from
the first one via $ F(z,\eta) = e^{-2\pi i \eta \Im z/\Im \tau} \Omega(z,\eta) $.

\sm

By inserting the second line of (\ref{pfs.17}) into (\ref{pfs.13}), we find the alternative form
\begin{align}
{\cal V}(1,\cdots,n) \, \big|_{h=1} &=
\mathop{\mathrm{Res}}_{\eta=0} \Omega(z_{n1},\eta) \Omega(z_{12},\eta) \cdots
\Omega(z_{n-2,n-1},\eta) 
 \label{pfs.18} \\
 &\quad \times \bigg( g^{(1)}(\eta) + \frac{\pi \eta}{\Im \tau} - f^{(1)}(z) + \partial_\eta \bigg) \Omega(z_{n-1,n},\eta)\notag
 \end{align}
However, the last two terms of the second line give rise to an elliptic function in
$\eta$ whose residue at $\eta=0$ vanishes, see (\ref{pfs.03}).
Hence, only the first two terms in the parenthesis in the second line of
(\ref{pfs.18}) contribute to ${\cal V}(1,\cdots,n) |_{h=1}$. By inserting their Laurent expansion
\beq
g^{(1)}(\eta) + \frac{\pi \eta}{\Im \tau} = \frac{1}{\eta} - \widehat {\rm G}_2 \, \eta
- \sum_{k=4}^\infty {\rm G}_k \, \eta^{k-1}
 \label{pfs.19}
 \eeq
 into (\ref{pfs.18}), we arrive at the generating series of the
right side of (\ref{eq2.6.3}),
\begin{align}
&\bigg( g^{(1)}(\eta) + \frac{\pi \eta}{\Im \tau}  \bigg)\Omega(z_{12},\eta) \Omega(z_{23},\eta) \cdots \Omega(z_{n1},\eta)  \label{pfs.20} \\
&=  \bigg( \frac{1}{\eta} - \widehat {\rm G}_2 \, \eta
- \sum_{k=4}^\infty {\rm G}_k \, \eta^{k-1}  \bigg)  \, \frac{1}{\eta^n }\bigg( 1 + \sum^{\infty}_{r=1} \eta^r V_r(1,\cdots,n) \bigg) \notag
\end{align}
such that taking the residue of its simple pole
at $\eta=0$ implies the first statement (\ref{eq2.6.3}) of the lemma.

\sm

Similarly, after eliminating $\partial_{n-1} F(z_{n-1,n},\eta) $ from (\ref{pfs.21})
 through the first line of (\ref{pfs.17}) and 
discarding the elliptic functions of $\eta$ due to $ \partial_\eta   F(z_{n-1,n},\eta)$
and $ g^{(1)}(z) F(z_{n-1,n},\eta)$,
the residue in (\ref{pfs.21}) can be rewritten as
\beq
{\cal W}(1,\cdots,n) \, \big|_{h=1} =
\mathop{\mathrm{Res}}_{\eta=0} \,
g^{(1)}(\eta) F(z_{12},\eta) F(z_{23},\eta) \cdots F(z_{n1},\eta) 
 \label{pfs.22} 
\eeq
without the extra term $\frac{\pi \eta}{\Im \tau} $ which accompanied
$g^{(1)}(\eta) $ in the doubly-periodic case (\ref{pfs.17}). As a result,
the Laurent expansion $g^{(1)}(\eta) = \frac{1}{\eta} -  {\rm G}_2  \eta
- \sum_{k=4}^\infty {\rm G}_k  \eta^{k-1}$ involves the holomorphic but quasi-modular
Eisenstein series $ {\rm G}_2 $ in the place of the modular but almost holomorphic
$ \widehat {\rm G}_2 $ in (\ref{pfs.19}). The generating series
\begin{align}
&  g^{(1)}(\eta) F(z_{12},\eta) F(z_{23},\eta) \cdots  F(z_{n1},\eta)  \label{pfs.23} \\
&=  \bigg( \frac{1}{\eta} - {\rm G}_2 \, \eta
- \sum_{k=4}^\infty {\rm G}_k \, \eta^{k-1}  \bigg)  \, \frac{1}{\eta^n }\bigg( 1 + \sum^{\infty}_{r=1} \eta^r V_r(1,\cdots,n) \bigg) \notag
\end{align}
is therefore identical to (\ref{pfs.20}) up to ${\rm G}_2 \leftrightarrow
\widehat {\rm G}_2$, and taking the residue at $\eta = 0$
reproduces the second statement (\ref{hoeq2.6.3}) of the lemma and
thereby concludes its proof.

\subsection{Proof of Theorem \ref{thm:linh1}}
\label{app:D.3}

The proof of Theorem \ref{thm:linh1}
will be organized into multiple steps.

\subsubsection{Matching elliptic functions of $z_1,\cdots,z_n$}
\label{prfpart1}

The first step in proving Theorem \ref{thm:linh1} is to match the elliptic functions
$W_{n-r},V_{n-r}$ of $z_1,\cdots,z_n$ in (\ref{h1eq.56}) with the $h=1$ instance of the combinations
${\cal W}_{I_1 \cdots I_{r}}(x;1,\cdots,n;y)$ and ${\cal V}_{I_1 \cdots I_{r}}(x;1,\cdots,n;y)$
of Enriquez kernels and DHS kernels defined by the descents in sections \ref{sec:L.mod} 
and \ref{sec:L.mero}. Similar to the
elliptic functions (\ref{h1eq.35mod}) and (\ref{h1eq.35mero}) in the cyclic products,
we will now derive the dictionary
\begin{align}
{\cal V}_{I_1   \cdots I_{r}}(x;1,\cdots,n;y)  \, \big|_{h=1} &=  V_{n-r}(x;1,\cdots,n;y)  
\label{h1eq.59}\\
{\cal W}_{I_1  \cdots I_{r}}(x;1,\cdots,n;y)  \, \big|_{h=1} &=  W_{n-r}(x;1,\cdots,n;y)  
\notag
\end{align}
valid for $r=0,1,\cdots,n$.

\sm

For the modular case in the first line of (\ref{h1eq.59}), we observe that both sides are meromorphic
in the internal points and by (\ref{Lmot.30}) have the same simple
poles in adjacent points of the chain products with residues $\pm 1$. Moreover, both sides vanish
upon integrating $z_1,\cdots,z_n$ over the genus one surface since
$\int_\Sigma d^2 z \,f^{(k)}(z{-}x)= 0$ for any $k\geq 1$ and each term
in the second line of (\ref{h1eq.56}) at $r\leq n$ integrates to zero on these grounds.

\sm
The second line of (\ref{h1eq.59}) again follows from the identical Fay identities
among the Kronecker-Eisenstein kernels $g^{(s)}(z_{ij})$ and
$f^{(s)}(z_{ij})$: in the first place, the genus one expressions for 
${\cal V}_{I_1   \cdots I_{r}}(x;1,\cdots,n;y)$ and
${\cal W}_{I_1   \cdots I_{r}}(x;1,\cdots,n;y)$ that follow from the modular and
merormorphic descents of Theorems \ref{4.thm:4} and \ref{thm.meroL} 
via (\ref{h1eq.03}) do not line up with $V_{n-r}$ and $W_{n-r}$ in (\ref{h1eq.56}).
At fixed $n$, the first line of (\ref{h1eq.59}) which was established
on general grounds in the previous paragraph can be explicitly verified via
repeated use of the Fay identities among the $f^{(s)}(z_{ij})$.
The meromorphic counterparts ${\cal W}_{I_1  \cdots I_{r}}(x;1,\cdots,n;y)$
produced by the descent are obtained by converting the DHS kernels
of  ${\cal V}_{I_1  \cdots I_{r}}(x;1,\cdots,n;y)$ into Enriquez kernels, see (\ref{Lmot.61}),
so the respective genus one instances are related by 
$f^{(s)}(z_{ij}) \leftrightarrow g^{(s)}(z_{ij})$. The Fay identities among $f^{(s)}(z_{ij})$ that produce the 
expressions (\ref{h1eq.56}) for $V_{n-r}$ from the outcome of the modular descent apply in identical
form to the $g^{(s)}(z_{ij})$ in the outcome of the meromorphic descent.
Hence, the ${\cal W}_{I_1  \cdots I_{r}}(x;1,\cdots,n;y) $ at genus one admit alternative
representations obtained from substituting $f^{(s)}(z_{ij}) \rightarrow g^{(s)}(z_{ij})$ 
in any expression for ${\cal V}_{I_1  \cdots I_{r}}(x;1,\cdots,n;y) |_{h=1}$. Applying this
substitution $f^{(s)}(z_{ij}) \rightarrow g^{(s)}(z_{ij})$
to the expressions for $V_{n-r}$ in the first line of (\ref{h1eq.56}) casts
${\cal W}_{I_1  \cdots I_{r}}(x;1,\cdots,n;y) |_{h=1}$ into the form of
$W_{n-r}$ in the second line of (\ref{h1eq.56}). This concludes the derivation of
the second line in (\ref{h1eq.59}) from the first line.

\sm

Note that the reasoning of the previous paragraph relies on the absence of differentiated 
Enriquez kernels and DHS kernels in the ${\cal W}_{I_1  \cdots I_{r}}(x;1,\cdots,n;y)$
and ${\cal V}_{I_1  \cdots I_{r}}(x;1,\cdots,n;y)$ produced by the descents for linear chain
products. As a consequence, we do not encounter any Kronecker-Eisenstein derivatives 
$\partial_{z_i}g^{(s)}(z_{ij})$ or $\partial_{z_i}f^{(s)}(z_{ij})$ in the genus one limit,
and the Fay identities used to connect the two lines of (\ref{h1eq.59}) always involve three
different points. In particular, we do not encounter the different coefficients  ${\rm G}_2$ vs.\ 
$\widehat {\rm G}_2$ in the $\eta$ expansion of (\ref{pfs.17}) when explicitly 
verifying (\ref{h1eq.59}) at fixed multiplicity.

\subsubsection{Deriving the meromorphic decomposition at genus one}

We shall next prove the first line of (\ref{ldech1.01}) using the
Kronecker-Eisenstein representation of linear chain products
\begin{align}
L_\delta(x;1,\cdots,n ;y) &= S_\delta(x{-}z_1)  S_\delta(z_{12})\cdots S_\delta(z_{n-1,n}) S_\delta(z_n{-}y)
 \label{h1eq.51}\\
&= e^{2\pi i u_\delta (x{-}y)} F(x{-}z_1,\hp_\delta)  F(z_{12},\hp_\delta)\cdots 
F(z_{n-1,n},\hp_\delta) F(z_n{-}y,\hp_\delta) 
\notag
\end{align}
The second line follows from the expression (\ref{h1eq.09}) for the genus one Szeg\"o kernel
and will be simplified by means of the auxiliary function
\beq
P_\delta(\eta | x;1,\cdots,n ;y) =  F(x{-}y,\hp_\delta{-}\eta)
F(x{-}z_1,\eta)  F(z_{12},\eta)\cdots 
F(z_{n-1,n},\eta) F(z_n{-}y,\eta) 
 \label{h1eq.52}
\eeq
As a meromorphic function of $\eta$ at fixed $x,y,z_1,\cdots,z_n$, the right side of (\ref{h1eq.52})
is doubly periodic (the individual phases of $F(z,\eta{+}\tau) = e^{-2\pi i z} F(z,\eta)$ 
cancel from the product) and only has poles at the two points $\eta= 0$ and $\eta = \hp_\delta$ 
in a fundamental domain of the torus. Hence, (\ref{h1eq.52}) as an elliptic function of $\eta$
has vanishing total residue,
\beq
\mathop{\mathrm{Res}}_{\eta=0} P_\delta(\eta | x;1,\cdots,n ;y) 
+  \mathop{\mathrm{Res}}_{ \eta=\hp_\delta} P_\delta(\eta | x;1,\cdots,n ;y)  = 0
 \label{h1eq.53}
\eeq
By the Laurent expansion $F(z,\eta) = \frac{1}{\eta}+{\cal O}(\eta^0)$,
the second residue in (\ref{h1eq.53}) is given by
\begin{align}
\mathop{\mathrm{Res}}_{\eta=\hp_\delta} P_\delta(\eta | x;1,\cdots,n ;y)  &= 
{-} F(x{-}z_1,\hp_\delta)  F(z_{12},\hp_\delta)\cdots 
F(z_{n-1,n},\hp_\delta) F(z_n{-}y,\hp_\delta) 
\notag\\
&= {-}  e^{ 2\pi i u_\delta (y{-}x)} L_\delta(x;1,\cdots,n ;y) 
\label{ressum} 
\end{align}
where the linear chain product $L_\delta(x;1,\cdots,n ;y) $ has been identified using the 
second line of (\ref{h1eq.51}). The residue of the auxiliary function (\ref{h1eq.52}) at $\eta=0$ 
in turn follows from combining the Taylor expansion of $ F(x{-}y,\hp_\delta{-}\eta)
=  F(x{-}y,\hp_\delta) \! - \! \eta \big( \partial_\eta  F(x{-}y,\eta) |_{\eta = \hp_\delta} \big)
\! + \! {\cal O}(\eta^2)$ around $\eta = \hp_\delta$ with the Laurent expansion
\beq
F(x{-}z_1,\eta)  F(z_{12},\eta)\cdots 
F(z_{n-1,n},\eta) F(z_n{-}y,\eta) = \frac{1}{\eta^{n+1}} \bigg\{ 
1 + \sum_{r=1}^{\infty} \eta^r W_r(x;1,2,\cdots,n;y)
\bigg\}
 \label{h1eq.55}
\eeq
which generates the elliptic functions $W_r(x;1,\cdots,n;y)$ of the internal points $z_1,\cdots,z_n$
in the first line of (\ref{h1eq.56}) and leads to the representation
\beq
\mathop{\mathrm{Res}}_{\eta=0} P_\delta(\eta | x;1,\cdots,n ;y) 
= \sum_{r=0}^{n} \frac{(-1)^r}{r!} \, \partial^r_\eta F(x{-}y ,\eta) \, \big|_{\eta = \hp_\delta} 
W_{n-r}( x;1,\cdots,n ;y) 
 \label{h1eq.57}
\eeq
of the first residue in (\ref{h1eq.53}). By equating the right side of (\ref{h1eq.57}) 
with minus the expression (\ref{ressum}) for the second residue of $P_\delta(\eta | x;1,\cdots,n ;y) $,
we arrive at the following equivalent of the meromorphic decomposition 
in the first lines of (\ref{ldech1.01}) to (\ref{h1eq.60}):
\begin{align}
L_\delta(x;1,\cdots,n ;y) &=
- e^{2\pi i u_\delta (x{-}y) }
\mathop{\mathrm{Res}}_{\eta=\hp_\delta} P_\delta(\eta | x;1,\cdots,n ;y) 
 \label{h1eq.57alt} \\
&=  e^{ 2\pi i u_\delta (x{-}y) }
\sum_{r=0}^{n} \frac{(-1)^r}{r!} \, \partial^r_\eta F(x{-}y ,\eta) \, \big|_{\eta = \hp_\delta} 
W_{n-r}( x;1,\cdots,n ;y) 
\notag
\end{align}
In particular, the $W_{n-r}( x;1,\cdots,n ;y) $ were shown in section \ref{prfpart1} to line up with the
$h=1$ instance of the individual ${\cal W}_{I_1  \cdots I_{r}}(x;1,\cdots,n;y) $
of the meromorphic decomposition (\ref{merodecL}). By the linear independence of 
$W_{s}( x;1,\cdots,n ;y) $ at different values of $s=0,1,\cdots,n$, this implies
that the expressions for $M_\delta^{I_1\cdots I_r}(x,y) |_{h=1} $ in the first line 
of (\ref{h1eq.60}) can indeed be read off by comparing (\ref{h1eq.57alt}) with (\ref{ldech1.01}).

\sm

Note that, by the $\vartheta$-function representation (\ref{h1eq.01}) of the
Kronecker-Eisenstein series, all instances of $M_\delta^{I_1\cdots I_r}(x,y) |_{h=1} $
in (\ref{h1eq.60}) can be represented via $\vartheta_1(x-y+\hp_\delta)$,
$\vartheta_1(\hp_\delta)$ and their derivatives in the first argument. 
The rank $r=1$ case can be further
simplified to $M_\delta^{I}(x,y) |_{h=1} = - S_\delta(x{-}y) 
\frac{ \p_x \vartheta_\delta(x-y)}{ \vartheta_\delta(x-y) }$,
consistently with (\ref{Lmot.2}) at genus $h=1$.

\subsubsection{Matching with the modular decomposition}
\label{sec:h1.3.2}

The meromorphic decomposition (\ref{h1eq.57alt}) of linear chain products at genus one can
be reformulated in terms of the doubly-periodic generating series $\Omega(z,\eta)$ in 
(\ref{h1eq.01}). The dependence on the internal points then occurs through the
coefficients $V_r$ of
\beq
\Omega(x{-}z_1,\eta)  \Omega(z_{12},\eta)\cdots 
\Omega(z_{n-1,n},\eta) \Omega(z_n{-}y,\eta) = \frac{1}{\eta^{n+1}} \bigg\{ 
1 + \sum_{r=1}^{\infty} \eta^r V_r(x;1,2,\cdots,n;y)
\bigg\}
 \label{h1eq.62}
\eeq
in the second line of (\ref{h1eq.56}). Comparing with the
meromorphic generating function in (\ref{h1eq.55}) exposes that
\beq
W_r(x;1,\cdots,n;y) = \sum_{\ell=0}^r \frac{(-2\pi i u)^\ell}{\ell!} \, V_{r-\ell}(x;1,\cdots,n;y)
 \label{h1eq.63}
\eeq
with the co-moving coordinates $u,v \in \RR/\mathbb Z$ defined by $x-y=z=u\tau+v$.
In order to deduce the second line of (\ref{ldech1.01}) from the first line
(demonstrated in the previous section),
it therefore remains to show that the spinors 
$L^{I_1\cdots I_r}_\delta(x,y) |_{h=1} $ in the modular decomposition (\ref{moddecL})
of linear chains are related by
\beq
M_\delta^{I_1 \cdots I_r}(x,y) \, \big|_{h=1}  =
 \sum_{\ell=0}^r \frac{(-2\pi i u)^{r-\ell}}{(r-\ell )!} \,  L_\delta^{I_1\cdots I_{\ell}}(x,y) \, \big|_{h=1} 
 \label{h1eq.64}
\eeq
In the first place, their representation (\ref{Ldelint}) as convolutions
together with the Fourier expansion (\ref{h1eq.12}) of the Szeg\"o kernel implies
the alternative representation
\beq
L_\delta^{I_1\cdots I_r}(x,y) \, \big|_{h=1}  = e^{ 2 \pi i  (v u_\delta  - u v_\delta )} \sum_{m,n \in \mathbb Z}  \frac{ e^{ 2\pi i (mv - nu)} }{(m\tau + n + \hp_\delta)^{r+1}}
 \label{h1eq.65}
\eeq
Consistency with the expression in (\ref{h1eq.64}) and 
$M_\delta^{I_1 \cdots I_{r}}(x,y) |_{h=1} $ given by the first line of (\ref{h1eq.60}) 
can be seen from
\begin{align}
\partial_\eta^r F(z,\eta) &= 
\partial_\eta^r   e^{-2\pi i \eta u} 
 \sum_{m,n \in \mathbb Z} \frac{ e^{ 2\pi i (mv - nu)} }{ (m \tau + n + \eta)} 
 \label{h1eq.71} 
 \\
 &= e^{-2\pi i \eta u}  \sum_{\ell=0}^r  {r \choose \ell}
(-2\pi i u)^{r-\ell} \sum_{m,n \in \mathbb Z} \frac{ (-1)^\ell \, \ell! \, e^{ 2\pi i (mv - nu)} }{ (m \tau + n + \eta)^{\ell+1}} \notag
\\
&= (-1)^r \,  e^{-2\pi i \eta u}  \sum_{\ell=0}^r  \frac{r!}{(r-\ell)!} (2\pi i u)^{r-\ell}
\sum_{m,n \in \mathbb Z} \frac{  e^{ 2\pi i (mv - nu)} }{ (m \tau + n + \eta)^{\ell+1}} 
 \notag
\end{align}
where we have used the Fourier expansion (\ref{h1eq.11}) of $\Omega(z,\eta)$
in the first line. Setting $\eta = \hp_\delta$, inserting into the first line of
(\ref{h1eq.60}) and identifying the
Fourier expansion of $L_\delta^{I_1 \cdots  I_\ell}(x,y)|_{h=1} $ in the parenthesis of
\beq
M_\delta^{I_1 \cdots I_r}(x,y) \, \big|_{h=1}  = 
 \sum_{\ell=0}^r \frac{(-2\pi i u)^{r-\ell}}{(r-\ell )!} \,  \bigg( 
 e^{2 \pi i  (v u_\delta  - u v_\delta )} \sum_{m,n \in \mathbb Z}  \frac{ e^{ 2\pi i (mv - nu)} }{(m\tau + n + \hp_\delta)^{\ell+1}} \bigg)
 \label{h1eq.72}
\eeq
then reproduces the form (\ref{h1eq.64}) of $M_\delta^{I_1 \cdots I_r}(x,y) |_{h=1}$
mandated by (\ref{ldech1.01}). This concludes the proof of Theorem \ref{thm:linh1}.



\newpage

\begin{thebibliography}{99}
\itemsep=-0.02in

\bibitem{Berkovits:2022ivl}
N.~Berkovits, E.~D'Hoker, M.~B.~Green, H.~Johansson and O.~Schlotterer,
``Snowmass White Paper: String Perturbation Theory,''
[arXiv:2203.09099].

\bibitem{DHoker:2024cup}
E.~D'Hoker and J.~Kaidi,
``Modular Forms and String Theory,'' Cambridge University Press (2024),
ISBN: 9781009457538.

\bibitem{Bourjaily:2022bwx}
J.~L.~Bourjaily, J.~Broedel, E.~Chaubey, C.~Duhr, H.~Frellesvig, M.~Hidding, R.~Marzucca, A.~J.~McLeod, M.~Spradlin and L.~Tancredi, \textit{et al.}
``Functions Beyond Multiple Polylogarithms for Precision Collider Physics,''
[arXiv:2203.07088].

\bibitem{Abreu:2022mfk}
S.~Abreu, R.~Britto and C.~Duhr,
``The SAGEX review on scattering amplitudes Chapter 3: Mathematical structures in Feynman integrals,''
J. Phys. A \textbf{55}, no.44, 443004 (2022),
doi:10.1088/1751-8121/ac87de
[arXiv:2203.13014].

\bibitem{Blumlein:2022qci}
J.~Bl{\"u}mlein and C.~Schneider,
``The SAGEX review on scattering amplitudes Chapter 4: Multi-loop Feynman integrals,''
J. Phys. A \textbf{55}, no.44, 443005 (2022),
doi:10.1088/1751-8121/ac8086
[arXiv:2203.13015].

\bibitem{Dorigoni:2022iem}
D.~Dorigoni, M.~B.~Green and C.~Wen,
``The SAGEX review on scattering amplitudes Chapter 10: Selected topics on modular covariance of type IIB string amplitudes and their~~supersymmetric Yang{\textendash}Mills duals,''
J. Phys. A \textbf{55}, no.44, 443011 (2022),
doi:10.1088/1751-8121/ac9263
[arXiv:2203.13021].

\bibitem{Mafra:2022wml}
C.~R.~Mafra and O.~Schlotterer,
``Tree-level amplitudes from the pure spinor superstring,''
Phys. Rept. \textbf{1020} (2023), 1-162,
doi:10.1016/j.physrep.2023.04.001
[arXiv:2210.14241].

\bibitem{Brown:2011ik}
F.~Brown, ``On the decomposition of motivic multiple zeta values,''
  in \textit{Galois-{T}eichm\"uller theory and arithmetic geometry}, vol.~63 of
 Adv. Stud. Pure Math., pp.~31--58, Math. Soc. Japan, Tokyo, 2012,
 [arXiv:1102.1310].

\bibitem{Brown:2019wna}
F.~Brown and C.~Dupont,
``Single-valued integration and superstring amplitudes in genus zero,''
Commun. Math. Phys. \textbf{382} (2021) no.2, 815-874,
doi:10.1007/s00220-021-03969-4
[arXiv:1910.01107].

\bibitem{Mizera:2017cqs}
S.~Mizera,
``Combinatorics and Topology of Kawai-Lewellen-Tye Relations,''
JHEP \textbf{08} (2017), 097,
doi:10.1007/JHEP08(2017)097
[arXiv:1706.08527].

\bibitem{Mizera:2017rqa}
S.~Mizera,
``Scattering Amplitudes from Intersection Theory,''
Phys. Rev. Lett. \textbf{120}, no.14, 141602 (2018),
doi:10.1103/PhysRevLett.120.141602
[arXiv:1711.00469].

\bibitem{DHoker:2015wxz}
E.~D'Hoker, M.~B.~Green, \"O.~G\"urdogan and P.~Vanhove,
``Modular Graph Functions,''
Commun. Num. Theor. Phys. \textbf{11} (2017), 165-218,
doi:10.4310/CNTP.2017.v11.n1.a4
[arXiv:1512.06779].

\bibitem{DHoker:2016mwo}
E.~D'Hoker and M.~B.~Green,
``Identities between Modular Graph Forms,''
J. Number Theor. \textbf{189} (2018), 25-80,
doi:10.1016/j.jnt.2017.11.015
[arXiv:1603.00839].

\bibitem{Brown:2017qwo}
F.~Brown, ``A class of non-holomorphic modular forms I,''
Res. Math. Sci. {\bf 5} (2018)  5:7,
doi:10.1007/s40687-018-0130-8
[arXiv:1707.01230].

\bibitem{Brown:2017qwo2}
F.~Brown, ``A class of non-holomorphic modular forms II : equivariant iterated
  Eisenstein integrals,'' 
  Forum~of~Mathematics,~Sigma {\bf 8} (2020)  1,
  doi:10.1017/fms.2020.24
  [arXiv:1708.03354].
  
 \bibitem{BrownLevin}
F.~Brown and A.~Levin, ``Multiple Elliptic Polylogarithms,'' [arXiv:1110.6917].

\bibitem{Enriquez:Emzv}
B.~Enriquez,
``Analogues elliptiques des nombres multiz{\'e}tas,''
Bull. Soc. Math. Fr. \textbf{144}, no.3, 395-427 (2016),
doi:10.24033/bsmf.2718
[arXiv:1301.3042].
  
\bibitem{Brown:2014mmv}
 F.~Brown, ``Multiple modular values and the relative completion of the fundamental group of ${\cal M}_{1,1}$,'' [arXiv:1407.5167].
  
\bibitem{Broedel:2015hia}
J.~Broedel, N.~Matthes and O.~Schlotterer,
``Relations between elliptic multiple zeta values and a special derivation algebra,''
J. Phys. A \textbf{49} (2016) no.15, 155203,
doi:10.1088/1751-8113/49/15/155203
[arXiv:1507.02254].

  \bibitem{kawazumi2008johnson}
N.~Kawazumi,
``Johnson's homomorphisms and the Arakelov-Green function,''
[arXiv:0801.4218].

\bibitem{zhang2010gross}
S.~Zhang,
``Gross--Schoen cycles and dualising sheaves,''
Inventiones mathematicae \textbf{179} (2010),
doi:10.1007/s00222-009-0209-3
[arXiv:0812.0371].

\bibitem{DHoker:2013fcx}
E.~D'Hoker and M.~B.~Green,
``Zhang-Kawazumi Invariants and Superstring Amplitudes,''
J. Number Theor. \textbf{144} (2014), 111-150,
doi:10.1016/j.jnt.2014.03.021
[arXiv:1308.4597].

\bibitem{DHoker:2014oxd} 
  E.~D'Hoker, M.~B.~Green, B.~Pioline and R.~Russo,
  ``Matching the $D^{6}R^{4}$ interaction at two-loops,''
  JHEP {\bf 1501}, 031 (2015),
  doi:10.1007/JHEP01(2015)031
  [arXiv:1405.6226].

\bibitem{DHoker:2017pvk}
E.~D'Hoker, M.~B.~Green and B.~Pioline,
``Higher genus modular graph functions, string invariants, and their exact asymptotics,''
Commun. Math. Phys. \textbf{366} (2019) 927-979,
doi:10.1007/s00220-018-3244-3
[arXiv:1712.06135].

\bibitem{DHoker:2020uid}
E.~D'Hoker and O.~Schlotterer,
``Identities among higher genus modular graph tensors,''
Commun. Num. Theor. Phys. \textbf{16} (2022) no.1, 35-74,
doi:10.4310/CNTP.2022.v16.n1.a2
[arXiv:2010.00924].

\bibitem{Tsuchiya:1988va}
A.~Tsuchiya,
``More on One Loop Massless Amplitudes of Superstring Theories,''
Phys. Rev. D \textbf{39} (1989), 1626,
doi:10.1103/PhysRevD.39.1626.

\bibitem{Stieberger:2002wk}
S.~Stieberger and T.~R.~Taylor,
``NonAbelian Born-Infeld action and type 1. - heterotic duality 2: Non-renormalization theorems,''
Nucl. Phys. B \textbf{648} (2003), 3-34,
doi:10.1016/S0550-3213(02)00979-3
[arXiv:hep-th/0209064].

\bibitem{Bianchi:2006nf}
M.~Bianchi and A.~V.~Santini,
``String predictions for near future colliders from one-loop scattering amplitudes around D-brane worlds,''
JHEP \textbf{12} (2006), 010,
doi:10.1088/1126-6708/2006/12/010
[arXiv:hep-th/0607224].

\bibitem{Tsuchiya:2012nf}
A.~G.~Tsuchiya,
``On the pole structures of the disconnected part of hyper elliptic g loop M point super string amplitudes,''
[arXiv:1209.6117].

\bibitem{Atick:1986rs}
J.~J.~Atick and A.~Sen,
``Covariant One Loop Fermion Emission Amplitudes in Closed String Theories,''
Nucl. Phys. B \textbf{293} (1987), 317-347,
doi:10.1016/0550-3213(87)90075-7.

\bibitem{Lin:1988xb}
Z.~H.~Lin,
``One Loop Closed String five Particle Fermion Amplitudes in the covariant Formulation,''
Int. J. Mod. Phys. A \textbf{5} (1990), 299,
doi:10.1142/S0217751X90000131.

\bibitem{Lee:2017ujn}
S.~Lee and O.~Schlotterer,
``Fermionic one-loop amplitudes of the RNS superstring,''
JHEP \textbf{03} (2018), 190,
doi:10.1007/JHEP03(2018)190
[arXiv:1710.07353]. 


\bibitem{DHoker:2005vch}
E.~D'Hoker and D.~H.~Phong,
``Two-loop superstrings VI: Non-renormalization theorems and the 4-point function,''
Nucl. Phys. B \textbf{715} (2005), 3-90,
doi:10.1016/j.nuclphysb.2005.02.043
[arXiv:hep-th/0501197].

\bibitem{DHoker:2021kks}
E.~D'Hoker and O.~Schlotterer,
``Two-loop superstring five-point amplitudes. Part III. Construction via the RNS formulation: even spin structures,''
JHEP \textbf{12} (2021), 063,
doi:10.1007/JHEP12(2021)063
[arXiv:2108.01104].

\bibitem{Berkovits:2005ng}
N.~Berkovits and C.~R.~Mafra,
``Equivalence of two-loop superstring amplitudes in the pure spinor and RNS formalisms,''
Phys. Rev. Lett. \textbf{96} (2006), 011602,
doi:10.1103/PhysRevLett.96.011602
[arXiv:hep-th/0509234].

\bibitem{DHoker:2020prr}
E.~D'Hoker, C.~R.~Mafra, B.~Pioline and O.~Schlotterer,
``Two-loop superstring five-point amplitudes. Part I. Construction via chiral splitting and pure spinors,''
JHEP \textbf{08} (2020), 135,
doi:10.1007/JHEP08(2020)135
[arXiv:2006.05270].


\bibitem{Friedan:1985ge}
D.~Friedan, E.~J.~Martinec and S.~H.~Shenker,
``Conformal Invariance, Supersymmetry and String Theory,''
Nucl. Phys. B \textbf{271}, 93-165 (1986),
doi:10.1016/S0550-3213(86)80006-2.

\bibitem{DHoker:1988pdl} 
  E.~D'Hoker and D.~H.~Phong,
  ``The Geometry of String Perturbation Theory,''
  Rev.\ Mod.\ Phys.\  {\bf 60}, 917 (1988),
  doi:10.1103/RevModPhys.60.917.

\bibitem{DHoker:2005dys}
E.~D'Hoker and D.~H.~Phong,
``Two-loop superstrings. V. Gauge slice independence of the N-point function,''
Nucl. Phys. B \textbf{715}, 91-119 (2005),
doi:10.1016/j.nuclphysb.2005.02.042
[arXiv:hep-th/0501196].

\bibitem{Broedel:2014vla}
J.~Broedel, C.~R.~Mafra, N.~Matthes and O.~Schlotterer,
``Elliptic multiple zeta values and one-loop superstring amplitudes,''
JHEP \textbf{07} (2015), 112,
doi:10.1007/JHEP07(2015)112
[arXiv:1412.5535].

\bibitem{Tsuchiya:2017joo}
A.~G.~Tsuchiya,
``On new theta identities of fermion correlation functions on genus g Riemann surfaces,''
[arXiv:1710.00206].

\bibitem{Levin:2007}
A.~Levin and G.~Racinet, ``Towards multiple elliptic polylogarithms," [arXiv:math/0703237].

\bibitem{Broedel:2017kkb}
J.~Broedel, C.~Duhr, F.~Dulat and L.~Tancredi,
``Elliptic polylogarithms and iterated integrals on elliptic curves. Part I: general formalism,''
JHEP \textbf{05} (2018), 093,
doi:10.1007/JHEP05(2018)093
[arXiv:1712.07089].

\bibitem{Dolan:2007eh}
L.~Dolan and P.~Goddard,
``Current Algebra on the Torus,''
Commun. Math. Phys. \textbf{285} (2009), 219-264,
doi:10.1007/s00220-008-0542-1
[arXiv:0710.3743].

\bibitem{Gerken:2018}
J.~E. Gerken, A.~Kleinschmidt and O.~Schlotterer,
``Heterotic-string amplitudes at one loop: modular graph forms and relations to open strings,''
JHEP {\bf 01} (2019), 052,
doi:10.1007/JHEP01(2019)052
[arXiv:1811.02548].

\bibitem{Berg:2016wux}
M.~Berg, I.~Buchberger and O.~Schlotterer,
``From maximal to minimal supersymmetry in string loop amplitudes,''
JHEP \textbf{04} (2017), 163,
doi:10.1007/JHEP04(2017)163
[arXiv:1603.05262].

\bibitem{Mafra:2016nwr}
C.~R.~Mafra and O.~Schlotterer,
``One-loop superstring six-point amplitudes and anomalies in pure spinor superspace,''
JHEP \textbf{04} (2016), 148
doi:10.1007/JHEP04(2016)148
[arXiv:1603.04790].

\bibitem{Mafra:2018qqe}
C.~R.~Mafra and O.~Schlotterer,
``Towards the n-point one-loop superstring amplitude. Part III. One-loop correlators and their double-copy structure,''
JHEP \textbf{08} (2019), 092,
doi:10.1007/JHEP08(2019)092
[arXiv:1812.10971].

\bibitem{Gerken:2019cxz}
J.~E.~Gerken, A.~Kleinschmidt and O.~Schlotterer,
``All-order differential equations for one-loop closed-string integrals and modular graph forms,''
JHEP \textbf{01} (2020), 064,
doi:10.1007/JHEP01(2020)064
[arXiv:1911.03476].

\bibitem{Balli:2024wje}
F.~M.~Balli, A.~Edison and O.~Schlotterer,
``Pinching rules in the chiral-splitting description of one-loop string amplitudes,''
JHEP \textbf{05} (2025), 101,
doi:10.1007/JHEP05(2025)101
[arXiv:2410.19641].

\bibitem{Adamo:2013tsa}
T.~Adamo, E.~Casali and D.~Skinner,
``Ambitwistor strings and the scattering equations at one loop,''
JHEP \textbf{04} (2014), 104,
doi:10.1007/JHEP04(2014)104
[arXiv:1312.3828].

\bibitem{Geyer:2015jch}
Y.~Geyer, L.~Mason, R.~Monteiro and P.~Tourkine,
``One-loop amplitudes on the Riemann sphere,''
JHEP \textbf{03} (2016), 114,
doi:10.1007/JHEP03(2016)114
[arXiv:1511.06315].

\bibitem{He:2017spx}
S.~He, O.~Schlotterer and Y.~Zhang,
``New BCJ representations for one-loop amplitudes in gauge theories and gravity,''
Nucl. Phys. B \textbf{930} (2018), 328-383,
doi:10.1016/j.nuclphysb.2018.03.003
[arXiv:1706.00640].

\bibitem{DHoker:2022xxg}
E.~D'Hoker, M.~Hidding and O.~Schlotterer,
``Cyclic products of Szeg\"o kernels and spin structure sums. Part I. Hyper-elliptic formulation,''
JHEP \textbf{05} (2023), 073,
doi:10.1007/JHEP05(2023)073
[arXiv:2211.09069].

\bibitem{DHoker:2001qqx}
E.~D'Hoker and D.~H.~Phong,
``Two loop superstrings. 2. The Chiral measure on moduli space,''
Nucl. Phys. B \textbf{636}, 3-60 (2002),
doi:10.1016/S0550-3213(02)00431-5
[arXiv:hep-th/0110283].


\bibitem{DHoker:2025szl}
E.~D'Hoker, B.~Enriquez, O.~Schlotterer and F.~Zerbini,
``Relating flat connections and polylogarithms on higher genus Riemann surfaces,''
[arXiv:2501.07640].

\bibitem{Enriquez:2011}
B.~Enriquez, 
``Flat connections on configuration spaces and braid groups of surfaces,"
Advances in Mathematics {\bf 252} (2014), 204--226,
doi:10.1016/j.aim.2013.10.025
[arXiv:1112.0864].

\bibitem{DHoker:2025dhv}
E.~D'Hoker and O.~Schlotterer,
``Meromorphic higher-genus integration kernels via convolution over homology cycles,''
J. Phys. A \textbf{58} (2025) no.33, 33LT01,
doi:10.1088/1751-8121/adf789
[arXiv:2502.14769].

\bibitem{Baune:2024biq}
K.~Baune, J.~Broedel, E.~Im, A.~Lisitsyn and F.~Zerbini,
``Schottky\textendash{}Kronecker forms and hyperelliptic polylogarithms,''
J. Phys. A \textbf{57} (2024) no.44, 445202,
doi:10.1088/1751-8121/ad8197
[arXiv:2406.10051].

\bibitem{DHS:2023}
E.~D'Hoker, M.~Hidding and O.~Schlotterer,
``Constructing polylogarithms on higher-genus Riemann surfaces,''
Commun. Num. Theor. Phys. {\bf 19} (2025) no.2, 355-413,
doi:10.4310/cntp.250531031558
[arXiv:2306.08644].

 \bibitem{DHoker:2023khh}
E.~D'Hoker, M.~Hidding and O.~Schlotterer,
``Cyclic Products of Higher-Genus Szeg\"o Kernels, Modular Tensors, and Polylogarithms,''
Phys. Rev. Lett. \textbf{133} (2024) no.2, 021602,
doi:10.1103/PhysRevLett.133.021602
[arXiv:2308.05044].

\bibitem{DHoker:2024ozn}
E.~D'Hoker and O.~Schlotterer,
``Fay identities for polylogarithms on higher-genus Riemann surfaces,''
[arXiv:2407.11476].

\bibitem{Enriquez:2023}
B.~Enriquez and F.~Zerbini,
``Elliptic hyperlogarithms,''
Canad. J. Math. (2025), 1-36,
doi:10.4153/S0008414X24001068
[arXiv:2307.01833]. 

\bibitem{DHoker:2007csw}
E.~D'Hoker and D.~H.~Phong,
``Two-Loop Superstrings. VII. Cohomology of Chiral Amplitudes,''
Nucl. Phys. B \textbf{804}, 421-506 (2008),\\
doi:10.1016/j.nuclphysb.2008.04.030
[arXiv:0711.4314].


\bibitem{Enriquez:next}
B.~Enriquez and F.~Zerbini, 
``Higher-genus polylogarithms from multivalued Maurer-Cartan elements,''
work in progress.

\bibitem{Enriquez:2021}
B.~Enriquez and F.~Zerbini,
``Construction of Maurer-Cartan elements over configuration spaces of curves,''
[arXiv:2110.09341].

\bibitem{Enriquez:2022}
B.~Enriquez and F.~Zerbini,
``Analogues of hyperlogarithm functions on affine complex curves,''
To appear in Publ. Res. Inst. Math. Sci.
[arXiv:2212.03119].
  
\bibitem{Fay:1973}
J.~D.~Fay, ``Theta Functions on Riemann Surfaces,'' Lecture Notes in Math. {\bf 352} (1973),
doi:10.1007/BFb0060090.

\bibitem{Tsuchiya:2022lqv}
A.~G.~Tsuchiya,
``On a formula of spin sums, Eisenstein-Kronecker series in higher genus Riemann surfaces,''
Nucl. Phys. B \textbf{997} (2023), 116383,
doi:10.1016/j.nuclphysb.2023.116383
[arXiv:2209.14633].

  \bibitem{vdG3}
F.~Cl\'ery and G.~van der Geer,
 ``Constructing vector-valued Siegel modular forms from scalar-valued Siegel modular forms,"
[arXiv:1409.7176].

\bibitem{Kawazumi:lecture}
N.~Kawazumi, 
``Some tensor field on the Teichm\"uller space,''
\href{http://www.ms.u-tokyo.ac.jp/~kawazumi/OIST1610_v1.pdf}{Lecture at MCM2016}, OIST (2016).

\bibitem{DHoker:2025comp}
E.~D'Hoker and O.~Schlotterer, work in progress. 

\bibitem{DHoker:2002hof}
E.~D'Hoker and D.~H.~Phong,
``Lectures on two loop superstrings,''
Conf. Proc. C \textbf{0208124} (2002), 85-123
[arXiv:hep-th/0211111].

\bibitem{123modular}
J.~H.~Bruinier, G.~van der Geer, G.~Harder and D.~Zagier,
``The 1-2-3 of Modular Forms,''
(2008), Springer, doi:10.1007/978-3-540-74119-0.

\bibitem{Broedel:2018iwv}
J.~Broedel, C.~Duhr, F.~Dulat, B.~Penante and L.~Tancredi,
``Elliptic symbol calculus: from elliptic polylogarithms to iterated integrals of Eisenstein series,''
JHEP \textbf{08} (2018), 014,
doi:10.1007/JHEP08(2018)014
[arXiv:1803.10256].

\bibitem{Verlinde:1987sd}
E.~P. Verlinde and H.~L. Verlinde, ``Multiloop Calculations in Covariant
  Superstring Theory,'' Phys. Lett. B
  {\bf 192} (1987) 95--102,
  doi:10.1016/0370-2693(87)91148-8.

\bibitem{DHoker:1989cxq}
E.~D'Hoker and D.~H. Phong, ``Conformal Scalar Fields and Chiral Splitting
  on Superriemann Surfaces,''
  Commun. Math. Phys. {\bf 125} (1989) 469,
  doi:10.1007/BF01218413.

\bibitem{Geyer:2021oox}
Y.~Geyer, R.~Monteiro and R.~Stark-Much\~ao,
``Superstring Loop Amplitudes from the Field Theory Limit,''
Phys. Rev. Lett. \textbf{127} (2021) no.21, 211603,
doi:10.1103/PhysRevLett.127.211603
[arXiv:2106.03968].

\bibitem{Geyer:2024oeu}
Y.~Geyer, J.~Guo, R.~Monteiro and L.~Ren,
``Superstring amplitudes from BCJ numerators at one loop,''
JHEP \textbf{03} (2025), 017,
doi:10.1007/JHEP03(2025)017
[arXiv:2410.19663].

\bibitem{Faltings}
G.~Faltings, ``Calculus on Arithmetic Surfaces,'' Ann. Math. {\bf 119} (1984) 387,
doi:10.2307/2007043.

\bibitem{Alvarez-Gaume:1986nqf}
L.~Alvarez-Gaume, G.~W.~Moore, P.~C.~Nelson, C.~Vafa and J.~b.~Bost,
``Bosonization in Arbitrary Genus,''
Phys. Lett. B \textbf{178}, 41-47 (1986),
doi:10.1016/0370-2693(86)90466-1.

\bibitem{Baune:2024ber}
K.~Baune, J.~Broedel, E.~Im, A.~Lisitsyn and Y.~Moeckli,
``Higher-genus Fay-like identities from meromorphic generating functions,''
SciPost Phys. \textbf{18} (2025), 093,
doi:10.21468/SciPostPhys.18.3.093
[arXiv:2409.08208].

\end{thebibliography}
\end{document}